\documentclass[11pt,a4paper]{article}
\usepackage[utf8]{inputenc}

\usepackage[left=3cm,right=3cm,top=3cm,bottom=3cm]{geometry}

\author{Andreas Deuchert, Christian Hainzl, Marcel Maier (born Schaub)}

\usepackage[english]{babel}

\usepackage{fancyhdr}

\usepackage{soul}
\setuldepth{0}

\usepackage{wallpaper}

\usepackage{anyfontsize}

\usepackage[absolute]{textpos}

\usepackage{upref}

\newcommand{\superscript}[1]{\ensuremath{^{\textrm{#1}}}}

\newcommand{\tho}[0]{\superscript{th}}

\usepackage{enumerate}

\usepackage{amsmath}
\numberwithin{equation}{section}
\usepackage{mathtools}

\renewcommand{\leq}{\leqslant}
\renewcommand{\geq}{\geqslant}

\usepackage{amsfonts}
\usepackage{amssymb}

\usepackage{amsthm}
\usepackage{thmtools, thm-restate} 

\usepackage{esvect}
\usepackage{cancel}
\usepackage{nicefrac}

\usepackage{fontawesome}
\newcommand{\speaker}{{\hspace{0.7pt}\text{\raisebox{-1.7pt}{\scalebox{1.4}{\faVolumeOff}}\hspace{0.7pt}}}}

\usepackage[T1]{fontenc}

\DeclareMathAlphabet{\mathbbs}{U}{bbold}{m}{n}
\newcommand{\Idbb}{\mathbbs 1}

\usepackage[arrow, matrix, curve]{xy}

\usepackage{ifthen}

\usepackage{oldgerm}

\usepackage{eurosym}

\usepackage{float}
\usepackage[rflt]{floatflt}
\usepackage{graphicx}
\usepackage{xcolor}

\usepackage{dlfltxbcodetips}

\newtheorem{thm}{Theorem}[section]
\newtheorem{bigthm}{Theorem}

\newtheorem{lem}[thm]{Lemma}

\newtheorem{kor}[thm]{Corollary}

\newtheorem{prop}[thm]{Proposition}

\newtheoremstyle{cited}{}{}{\itshape}{}{
}{\textbf{.}}{.5em}{\textbf{#1 #2} #3}
\theoremstyle{cited}

\theoremstyle{definition}
\newtheorem{defn}[thm]{Definition}

\newtheorem{bem}[thm]{Remark}

\newtheorem{bems}[thm]{Remarks}
\newtheorem*{varbems}{Remarks}

\newtheorem{asmp}[thm]{Assumption}

\newcommand{\ra}{\rightarrow}

\newcommand{\ov}[1]{\overline{#1}}

\newcommand{\dx}{\mathrm{d}x}

\newcommand{\dy}{\mathrm{d}y}

\newcommand{\dr}{\mathrm{d}r}

\newcommand{\dd}{\mathrm{d}}

\DeclareMathOperator{\tr}{tr}
\DeclareMathOperator{\Tr}{Tr}

\DeclareMathOperator{\curl}{curl}

\DeclareMathOperator{\ran}{ran}

\DeclareMathOperator{\spec}{spec}

\newcommand{\loc}{\mathrm{loc}}

\newcommand{\Imm}{\mathrm{Im}}
\newcommand{\Rem}{\mathrm{Re}}
\let\Im\undefined
\let\Re\undefined
\DeclareMathOperator{\Im}{\Imm}
\DeclareMathOperator{\Re}{\Rem}

\renewcommand{\tilde}{\widetilde}
\renewcommand{\hat}{\widehat}

\DeclareFontFamily{U}{matha}{\hyphenchar\font45}
\DeclareFontShape{U}{matha}{m}{n}{
	<5> <6> <7> <8> <9> <10> gen * matha
	<10.95> matha10 <12> <14.4> <17.28> <20.74> <24.88> matha12
}{}
\DeclareSymbolFont{matha}{U}{matha}{m}{n}
\DeclareFontSubstitution{U}{matha}{m}{n}

\DeclareFontFamily{U}{mathx}{\hyphenchar\font45}
\DeclareFontShape{U}{mathx}{m}{n}{
	<5> <6> <7> <8> <9> <10>
	<10.95> <12> <14.4> <17.28> <20.74> <24.88>
	mathx10
}{}
\DeclareSymbolFont{mathx}{U}{mathx}{m}{n}
\DeclareFontSubstitution{U}{mathx}{m}{n}

\DeclareMathDelimiter{\vvvert}{0}{matha}{"7E}{mathx}{"17}

\DeclareFontFamily{U}{mathx}{\hyphenchar\font45}
\DeclareFontShape{U}{mathx}{m}{n}{
	<5> <6> <7> <8> <9> <10>
	<10.95> <12> <14.4> <17.28> <20.74> <24.88>
	mathx10
}{}
\DeclareSymbolFont{mathx}{U}{mathx}{m}{n}
\DeclareFontSubstitution{U}{mathx}{m}{n}
\DeclareMathAccent{\widecheck}{0}{mathx}{"71}
\DeclareMathAccent{\wideparen}{0}{mathx}{"75}

\newcommand{\Cbb}{\mathbb{C}}

\newcommand{\Hbb}{\mathbb{H}}

\newcommand{\Nbb}{\mathbb{N}}

\newcommand{\Rbb}{\mathbb{R}}

\newcommand{\Zbb}{\mathbb{Z}}

\newcommand{\Acal}{\mathcal{A}}

\newcommand{\Dcal}{\mathcal{D}}

\newcommand{\Gcal}{\mathcal{G}}
\newcommand{\Hcal}{\mathcal{H}}

\newcommand{\Jcal}{\mathcal{J}}

\newcommand{\Ocal}{\mathcal{O}}

\newcommand{\Qcal}{\mathcal{Q}}
\newcommand{\Rcal}{\mathcal{R}}
\newcommand{\Scal}{\mathcal{S}}
\newcommand{\Tcal}{\mathcal{T}}

\newcommand{\hfrak}{\mathfrak{h}}

\newcommand{\EGL}{\mathcal E^{\mathrm{GL}}_{D, B}}
\newcommand{\EGLGSE}{E^{\mathrm{GL}}(D)}
\newcommand{\FBCS}{\mathcal F^{\mathrm{BCS}}_{
			\Bbold, T
}}
\newcommand{\Abold}{\mathbf A}
\newcommand{\Bbold}{\mathbf B}

\newcommand{\Zbold}{\mathbf Z}
\newcommand{\sbold}{\mathbf s}
\newcommand{\Hmag}{H_{\mathrm{mag}}}
\newcommand{\Lmag}{L_{\mathrm{mag}}}
\newcommand{\Lsymm}{{L^2(Q_B \times \Rbb_{\mathrm s}^3)}}
\newcommand{\Hsymm}{{H^1(Q_B \times \Rbb_{\mathrm s}^3)}}
\newcommand{\Tc}{T_{\mathrm{c}}}
\newcommand{\Dc}{D_{\mathrm{c}}}
\newcommand{\betac}{\beta_{\mathrm{c}}}

\usepackage{esint} 

\usepackage{hyperref}
\newcommand{\e}{\mathrm{e}}
\renewcommand{\i}{\mathrm{i}}

\usepackage[
	sorting=nyt, 
	backend=biber, 
    giveninits=true,	
	maxbibnames=99,
	url=false,
	backref=true,
	eprint=true,
	isbn=false,
	date=year,
	]{biblatex}

\DeclareLabelalphaTemplate{
   \labelelement{
     \field[uppercase,final]{shorthand}     
     \field[uppercase,strwidth=1,strside=left,names=-5,noalphaothers]{labelname}
   }
   \labelelement{
     \field[strwidth=2,strside=right]{year}
   }
   \labelelement{
		\field[uppercase]{label}
   }
}
\renewbibmacro*{note+pages}{%
  \printfield{note}%
  \setunit{\bibpagespunct}%
  \printfield{pages}%
  \setunit{\bibpagespunct}
  \printfield{pagetotal}
  \newunit}

\DefineBibliographyStrings{english}{%
  backrefpage = {page}, 
  backrefpages = {pages}, 
}

\renewbibmacro{in:}{} 
\DeclareFieldFormat{pages}{#1}
\DeclareFieldFormat{pagetotal}{#1~pages}
\DeclareFieldFormat[article]{volume}{\mkbibbold{#1}}
\DeclareFieldFormat[article,inbook,incollection,inproceedings,patent,thesis,unpublished]{title}{#1} 

\AtEveryBibitem{\clearfield{number}}
\AtEveryBibitem{\clearfield{issue}}

\usepackage{pdfpages}
\usepackage{chngcntr}
\counterwithin{footnote}{section}
\counterwithout{footnote}{subsection}

\usepackage{ifthen}
\newcommand{\masterfile}{DHS1}

\usepackage{lmodern}

\begin{document}

\title{
Microscopic Derivation of Ginzburg--Landau Theory \\ and the BCS Critical Temperature Shift \\ in a Weak Homogeneous Magnetic Field
}

\maketitle

\begin{abstract}
%
Starting from the Bardeen--Cooper--Schrieffer (BCS) free energy functional, we derive the Ginzburg--Landau functional in the presence of a weak homogeneous magnetic field. We also provide an asymptotic formula for the BCS critical temperature as a function of the magnetic field. This extends the previous works \cite{Hainzl2012, Hainzl2014} of Frank, Hainzl, Seiringer and Solovej to the case of external magnetic fields with non-vanishing magnetic flux through the unit cell. 
\end{abstract}

\tableofcontents



\section{Introduction and Main Results}


\subsection{Introduction}

In 1950 Ginzburg and Landau (GL) introduced a phenomenological theory of superconductivity that is based on a system of nonlinear partial differential equations for a complex-valued wave function (the order parameter) and an effective magnetic field \cite{GL}. Their theory is \textit{macroscopic} in nature and contains no reference to a \textit{microscopic} mechanism behind the phenomenon of superconductivity. The GL equations show a rich mathematical structure, which has been investigated in great detail, see, e.g., \cite{Sigal1, Sigal2, Serfaty, SandierSerfaty, Correggi3, Correggi2, Giacomelli1, Giacomelli2} and references therein. They also inspired interesting new concepts beyond the realm of their original application.

The first generally accepted \textit{microscopic} theory of superconductivity was discovered seven years later by Bardeen, Cooper and Schrieffer (BCS) in \cite{BCS}. In a major breakthrough they realized that a pairing mechanism between the conduction electrons (formation of Cooper pairs) causes the resistance in certain materials to drop down to absolute zero if their temperature is sufficiently low. This pairing phenomenon at low temperatures is induced by an effective attraction between the electrons mediated by phonons, that is, by the quantized lattice vibrations of the crystal formed by the ion cores. In recognition of this contribution BCS were awarded the Nobel prize in physics in 1972. 

In the physics literature BCS theory is often formulated in terms of the gap equation, which, in the absence of external fields, is a nonlinear integral equation for a complex-valued function called the gap function (the order parameter of BCS theory). The name of the equation is related to the fact that its solution allows to determine the spectral gap of an effective \textit{quadratic} Hamiltonian that is open only in the superconducting phase. BCS theory also has a variational interpretation, where the gap equation arises as the Euler--Lagrange equation of the BCS free energy functional. This free energy functional can be obtained from a full quantum mechanical description of the system by restricting attention to quasi-free states, a point of view that was emphasized by Leggett in \cite{Leg1980}, see also \cite{de_Gennes}. In this formulation, the system is described in terms of a one-particle density matrix and a Cooper pair wave function. 

Although it was originally introduced to describe the phase transition from the normal to the superconducting state in metals and alloys, BCS theory can also be applied to describe the phase transition to the superfluid state in cold fermionic gases. In this case, the usual non-local phonon-induced interaction in the gap equation needs to be replaced by a local pair potential. From a mathematical point of view, the gap equation has been studied for interaction kernels suitable to describe the physics of conduction electrons in solids in \cite{Odeh1964, BilFan1968, Vanse1985, Yang1991, McLYang2000, Yang2005}. We refer to \cite{Hainzl2007, Hainzl2007-2, Hainzl2008-2, Hainzl2008-4, FreiHaiSei2012, BraHaSei2014, FraLe2016, DeuchertGeisinger} for works that investigate the translation-invariant BCS functional with a local pair interaction. BCS theory in the presence of external fields has been studied in \cite{HaSei2011, BrHaSei2016, FraLemSei2017, A2017, CheSi2020}.

A relation between the \textit{macroscopic} GL theory and the \textit{microscopic} BCS theory was established by Gor'kov in 1959 \cite{Gorkov}. He showed that, close to the critical temperature, where the order parameters of both models are expected to be small, GL theory arises from BCS theory when the free energy is expanded in powers of the gap function. The first mathematically rigorous proof of this relation was given by Frank, Hainzl, Seiringer and Solovej in 2012 \cite{Hainzl2012}. They showed that in the presence of weak and \textit{macroscopic} external fields, the \textit{macroscopic} variations of the Cooper pair wave function of the system are correctly described by GL theory if the temperature is close to the critical temperature of the sample in an appropriate sense. The precise parameter regime is as follows: The external electric field $W$ and the vector potential $A$ of the external magnetic field are given by $h^2 W(hx)$ and $h A(hx)$, respectively. Here $0 < h \ll 1$ denotes the ratio between the microscopic and the macroscopic length scale of the system. Such external fields change the energy by an amount of the order $h^2$ and it is therefore natural to consider temperatures $T = \Tc (1 - D h^2)$ with $D > 0$, where $\Tc$ denotes the critical temperature of the sample in the absence of external fields. Within this setup it has been shown in \cite{Hainzl2012} that the correction to the BCS free energy on the order $h^4$ is correctly described by GL theory. Moreover, the Cooper pair wave function of the system is, to leading order in $h$, given by
\begin{equation}
	\alpha(x,y) = h\, \alpha_*(x-y) \, \psi\left( \frac{h(x+y)}{2} \right).
	\label{DHS1:eq:intro}
\end{equation}
Here, $\psi$ denotes the order parameter of GL theory and $\alpha_*(x-y)$ is related to the Cooper pair wave function in the absence of external fields. 

External electric and magnetic fields may change the critical temperature of a superconductor and this shift is expected to be described by GL theory. A justification of this claim has been provided in \cite{Hainzl2014}. More precisely, it has been shown that, within the setup of \cite{Hainzl2012} described above, the critical temperature of the sample obeys the asymptotic expansion
\begin{equation}
	\Tc(h) = \Tc(1  - \Dc h^2) + o(h^2),
	\label{DHS1:eq:intro2}
\end{equation}
where the constant $\Dc$ can be computed using linearized GL theory. 

One crucial assumption in \cite{Hainzl2012} and \cite{Hainzl2014} is that the vector potential related to the external magnetic field is periodic. In this case the magnetic flux through the unit cell equals zero. An important step towards an extension of the results in \cite{Hainzl2014} to the case of a magnetic field with non-vanishing magnetic flux through the unit cell has been provided by Frank, Hainzl and Langmann in \cite{Hainzl2017}. In this article the authors consider the problem of computing the BCS critical temperature shift in the presence of a weak homogeneous magnetic field within linearized BCS theory. Heuristically, this approximation is justified by the fact that linearized GL theory is sufficient to predict the critical temperature shift, see the discussion in the previous paragraph. In the physics literature this approximation appears in \cite{Werthamer1966, WertHelHoh1966, La1990, La1991}, for instance.

The aim of the present article is to extend the results in \cite{Hainzl2012} and \cite{Hainzl2014} to a setting with an external magnetic field having non-zero flux through the unit cell. More precisely, we consider a large periodic sample of  fermionic particles subject to a weak homogeneous magnetic field $\Bbold \in \Rbb^3$. The temperatures $T$ is chosen such that $(\Tc - T)/\Tc = D |\Bbold|$ with $D \in \Rbb$. We show that the correction of the BCS free energy of the sample at the order $|\Bbold|^2$ is given by GL theory. Moreover, to leading order in $|\Bbold|$ the Cooper pair wave function of the system is given by \eqref{DHS1:eq:intro} with $h$ replaced by $|\Bbold|^{\nicefrac{1}{2}}$. We also show that the BCS critical temperature shift caused by the external magnetic field is given by \eqref{DHS1:eq:intro2} with $\Dc$ determined by linearized GL theory. Our analysis yields the same formula that was computed within the framework of linearized BCS theory in \cite{Hainzl2017}. This can be interpreted as a justification of the approximation to use linearized BCS theory to compute the BCS critical temperature shift. The main new ingredient of our proof are a priori bounds for certain low-energy states of the BCS functional that include the magnetic field.

\subsection{Gauge-periodic samples}
\label{DHS1:Magnetically_Periodic_Samples}

We consider a $3$-dimensional sample of fermionic particles described by BCS theory that is subject to an external magnetic field $\Bbold := Be_3$ with strength $B>0$, pointing in the $e_3$-direction. We choose the magnetic vector potential $\Abold(x) := \frac 12 \Bbold \wedge x$ so that $\curl \Abold = \Bbold$, where $\Bbold \wedge x\in \Rbb^3$ denotes the cross product of two vectors. The corresponding magnetic momentum operator $\pi := -\i \nabla + \Abold$ commutes with the magnetic translations $T(v)$, defined by
\begin{align}
T(v)f(x) &:= \e^{\i \frac {\Bbold} 2\cdot (v\wedge x)} f(x+v), & v &\in \Rbb^3.\label{DHS1:Magnetic_Translation}
\end{align}
The family $\{T(v)\}_{v\in \Rbb^3}$ obeys the relation $T(v + w) = \e^{\i \frac{\Bbold}{2} \cdot (v \wedge w)} \; T(v) T(w)$, that is, it is a unitary representation of the Heisenberg group. We assume that our system is periodic with respect to the Bravais lattice $\Lambda_B = \sqrt{2\pi B^{-1}} \, \Zbb^3$ with fundamental cell
\begin{align}
Q_B &:= \bigl[0, \sqrt{2\pi B^{-1}}\bigr]^3 \subseteq \Rbb^3. \label{DHS1:Fundamental_cell}
\end{align}
The magnetic flux through the unit cell $Q_B$ equals $\Bbold \cdot (b_1\wedge b_2) =2\pi$, where $b_i = \sqrt{2\pi B^{-1}} \, e_i$ are the basis vectors spanning $\Lambda_B$. This assures that the abelian subgroup $\{T(\lambda)\}_{\lambda\in \Lambda_B}$ is a unitary representation of the lattice group. 


\subsection{The BCS functional}
\label{DHS1:BCS_functional_Section}

In BCS theory a state is described by a generalized fermionic one-particle density matrix, that is, by a self-adjoint operator $\Gamma$ on $L^2(\mathbb{R}^3) \oplus L^2(\mathbb{R}^3)$ which satisfies $0\leq \Gamma\leq 1$ and is of the form
\begin{align}
\Gamma = \begin{pmatrix} \gamma & \alpha \\ \ov \alpha & 1 - \ov \gamma \end{pmatrix}. \label{DHS1:Gamma_introduction}
\end{align}
Here, $\ov \alpha = J \alpha J$ with the Riesz identification operator $J \colon L^2(\Rbb^3) \ra L^2(\Rbb^3)$, $f\mapsto \ov f$, realized by complex conjugation. The condition $\Gamma = \Gamma^*$ implies that the one-particle density matrix $\gamma$ is a self-adjoint operator. It also implies that the Cooper pair wave function $\alpha(x,y)$, the kernel of $\alpha$, is symmetric under the exchange of its coordinates. The symmetry of $\alpha$ is due to the fact that we exclude spin variables from our description and assume that Cooper pairs are in a spin singlet state. The condition $0\leq \Gamma \leq 1$ implies $0\leq \gamma \leq 1$ as well as that $\gamma$ and $\alpha$ are related through the operator inequality 
\begin{align}
\alpha \alpha^* \leq \gamma ( 1- \gamma). \label{DHS1:gamma_alpha_fermionic_relation}
\end{align}

A BCS state $\Gamma$ is called \emph{gauge-periodic} if $\mathbf T(\lambda) \, \Gamma \, \mathbf T(\lambda)^* = \Gamma$ holds for every $\lambda\in \Lambda_B$, with the magnetic translations $\mathbf T(\lambda)$ on $L^2(\Rbb^3)\oplus L^2(\Rbb^3)$ defined by
\begin{align*}
\mathbf T(v) &:= \begin{pmatrix}
T(v) & 0 \\ 0 & \ov{T(v)}\end{pmatrix}, & v &\in \Rbb^3.
\end{align*}
For $\gamma$ and $\alpha$, this implies $T(\lambda) \gamma T(\lambda)^* = \gamma$ and   $T(\lambda)\,\alpha \,\ov{T(\lambda)}^* = \alpha$ or, in terms of their kernels,
\begin{align}
\gamma(x, y) &= \e^{\i \frac \Bbold 2 \cdot (\lambda \wedge (x-y))} \; \gamma(x+\lambda,y+ \lambda), \notag\\
\alpha(x, y) &= \e^{\i \frac \Bbold 2 \cdot (\lambda \wedge (x+y))} \; \alpha(x+\lambda,y+ \lambda), & \lambda\in \Lambda_B. \label{DHS1:alpha_periodicity}
\end{align}

\begin{bem}
Since we are interested in the situation of a constant magnetic field it seems natural to consider magnetically translation-invariant BCS states, that is, states obeying $\mathbf T(v) \, \Gamma \, \mathbf T(v)^* = \Gamma$ for every $v \in \mathbb{R}^3$. However, in this case one obtains a trivial model because the Cooper pair wave function $\alpha$ of a magnetically translation-invariant state necessarily vanishes. To see this, we note that $\alpha$ satisfies $T(v)\,\alpha \,\ov{T(v)}^* = \alpha$ for all $v \in \mathbb{R}^3$. Using this and the relation $T(v+w)\,\alpha \,\ov{T(v+w)}^* =  \e^{ \i\Bbold \cdot (v \wedge w)} \, T(v) T(w) \, \alpha \, \ov{T(w)}^* \, \ov{T(v)}^* $, we conclude that $\alpha = 0$.
\end{bem}

A gauge-periodic BCS state $\Gamma$ is said to be \emph{admissible} if 
\begin{align}
\Tr \bigl[\gamma + (-\i \nabla + \Abold)^2\gamma\bigr] < \infty \label{DHS1:Gamma_admissible}
\end{align}
holds. Here $\Tr[A]$ denotes the trace per unit volume of $A$, i.e.,
\begin{align}
\Tr [A] &:= \frac{1}{|Q_B|} \Tr_{L^2(Q_B)} [\chi A \chi],  \label{DHS1:Trace_per_unit_volume_definition}
\end{align}
with the characteristic function $\chi$ of the cube $Q_B$ in \eqref{DHS1:Fundamental_cell}. By $\Tr_{L^2(Q_B)}[A]$ we denote the usual trace of an operator $A$ on $L^2(Q_B)$. The condition in \eqref{DHS1:Gamma_admissible} is meant to say that $\gamma$ and $(-\i \nabla + \Abold)^2\gamma$ are locally trace class, that is, they are trace class with respect to the trace in \eqref{DHS1:Trace_per_unit_volume_definition}. Eq.~\eqref{DHS1:Gamma_admissible}, the same inequality with $\gamma$ replaced by $\overline{\gamma}$, and the inequality in \eqref{DHS1:gamma_alpha_fermionic_relation} imply that $\alpha$, $(-\i \nabla + \Abold)\alpha$, and $(-\i \nabla + \Abold) \ov \alpha$ are locally Hilbert--Schmidt. In Section~\ref{DHS1:Preliminaries} below we will express this property in terms of $H^1$-regularity of the kernel of $\alpha$.

For any admissible BCS state $\Gamma$, we define the Bardeen--Cooper--Schrieffer free energy functional (in the following: BCS functional) at temperature $T\geq 0$ by
\begin{align}
\FBCS(\Gamma) &:= \Tr\bigl[ \bigl( (-\i \nabla + \Abold)^2 - \mu\bigr)\gamma \bigr] - T\, S(\Gamma) - \frac{1}{|Q_B|} \int_{Q_B} \dd X \int_{\Rbb^3} \dd r\; V(r) \, |\alpha(X,r)|^2,
\label{DHS1:BCS functional}
\end{align}
with the von Neumann entropy per unit volume $S(\Gamma)= - \Tr [\Gamma \ln(\Gamma)]$ and the chemical potential $\mu\in \Rbb$. The particles interact via a two-body potential $V \in L^{\nicefrac 32}(\Rbb^3) + L_{\varepsilon}^{\infty}(\mathbb{R}^3)$. The condition on $V$ guarantees that it is relatively form bounded with respect to the Laplacian and implies that $\FBCS$ is bounded from below. Furthermore, we introduced center-of-mass and relative coordinates $X = \frac{x+y}{2}$ and $r = x-y$. Here and in the following, we abuse notation slightly by writing $\alpha(X,r)\equiv \alpha(x,y)$. 
\begin{bem}
	We opt for the above set-up because the solution of the problem for the constant magnetic field already contains the main difficulties of the case of a general magnetic field. This is related to the fact that the vector potential of any magnetic field with non-zero flux through the unit cell can be written as a sum of a vector potential of a homogeneous magnetic field and a periodic vector potential, see e.g. \cite[Proposition~4.1]{Tim_Abrikosov}. The latter can be treated in some sense as a perturbation, see \cite{Hainzl2012,Hainzl2014}. However, this is not true for the constant magnetic field, see Remark~\ref{DHS1:Remarks_Main_Result}~(a) below. To solve the general case it is therefore crucial to understand the case of a homogeneous magnetic field. To keep the presentation to a reasonable length and to be able to convey the main ideas more clearly, we therefore decided to present this case first. We plan to extend our treatment to the case of a general magnetic field in a second paper. One motivation to treat general periodic magnetic fields with non-zero flux through the unit cell stems from the fact that it is an interesting and highly relevant problem to consider magnetic fields that are chosen self-consistently.
\end{bem}

The BCS functional is bounded from below and coercive on the set of admissible states. More precisely, it can be shown that the kinetic energy dominates the entropy and the interaction energy, i.e., there is a constant $C>0$ such that for all admissible $\Gamma$, we have
\begin{align}
\FBCS(\Gamma) &\geq \frac 12 \Tr \bigl[ \gamma + (-\i \nabla + \Abold)^2 \gamma\bigr] - C. \label{DHS1:BCS functional_bounded_from_below}
\end{align}

The unique minimizer of the BCS functional among admissible states with $\alpha =0$ is given by
\begin{align}
\Gamma_0 &:= \begin{pmatrix} \gamma_0 & 0  \\ 0 &  1-\ov\gamma_0 \end{pmatrix}, & \gamma_0 &:= \frac 1{1 + \e^{ ((-\i\nabla + \Abold)^2-\mu)/T}}.  \label{DHS1:Gamma0}
\end{align}
Since $\Gamma_0$ is also the unique minimizer of the BCS functional for sufficiently large temperatures $T$, it is called the normal state.
We define the BCS free energy by
\begin{align}
F^{\mathrm{BCS}}(B, T) := \inf\bigl\{ \FBCS(\Gamma) - \FBCS(\Gamma_0) : \Gamma \text{ admissible}\bigr\} \label{DHS1:BCS GS-energy}
\end{align}
and say that our system is superconducting if $F^{\mathrm{BCS}}(B, T) < 0$, that is, if the minimal energy is strictly smaller than that of the normal state. In this work we are interested in the regime of weak magnetic fields $0 < B \ll 1$. Our goal is to obtain an asymptotic expansion of $F^{\mathrm{BCS}}(B, T)$ in powers of $B$ that allows us to derive Ginzburg--Landau theory, and to show how the BCS critical temperature depends on the magnetic field $B$. For our main results to hold, we need the following assumptions concerning the regularity of the interaction potential $V$.
%
%
%

\begin{asmp}
\label{DHS1:Assumption_V}
We assume that the interaction potential $V$ is a nonnegative, radial function such that $(1+|\cdot|^2) V\in L^\infty(\Rbb^3)$.
\end{asmp}

\begin{bem}
Our main results Theorem~\ref{DHS1:Main_Result} and Theorem~\ref{DHS1:Main_Result_Tc} still hold if the assumption $V\geq 0$ is dropped. We only use it in Appendix \ref{DHS1:KTV_Asymptotics_of_EV_and_EF_Section} when we investigate the spectral properties of a certain linear operator involving $V$. These statements still hold in the case of potentials without a definite sign but their proof is longer. A proof of these statements in the general setting can be found in \cite[Chapter 6]{Diss_Marcel}. The property $V\geq 0$ does not simplify the remaining part of our analysis. We expect our results to be true also if $V$ has moderate local singularities. Furthermore, it may be possible to slightly weaken the decay assumptions of $V$. We choose to work with the assumptions above to keep the presentation at a reasonable length.
\end{bem}

\subsection{The translation-invariant BCS functional}
\label{DHS1:BCS_functional_TI_Section}

If no external fields are present, i.e. if $\Bbold =0$, we describe the system by translation-invariant states, that is, we assume that the kernels of $\gamma$ and $\alpha$ are of the form $\gamma(x-y)$ and $\alpha(x-y)$. To define the trace per unit volume we choose a cube of side length $1$. The resulting translation-invariant BCS functional and its infimum minus the free energy of the normal state are denoted by $\mathcal{F}^{\mathrm{BCS}}_{\mathrm{ti},T}$ and $F^{\mathrm{BCS}}_{\mathrm{ti}}(T)$, respectively. This functional has been studied in detail in \cite{Hainzl2007}, see also \cite{Hainzl2015} and the references therein, where it has been shown that there is a unique critical temperature $\Tc \geq 0$ such that $\mathcal{F}^{\mathrm{BCS}}_{\mathrm{ti},T}$ has a minimizer with $\alpha \neq 0$ if $T < \Tc$. For $T\geq \Tc$ the normal state in \eqref{DHS1:Gamma0} with $B=0$ is the unique minimizer. In terms of the energy, we have $F^{\mathrm{BCS}}_{\mathrm{ti}}(T) < 0$ for $T < \Tc$, while $F^{\mathrm{BCS}}_{\mathrm{ti}}(T) = 0$ if $T\geq \Tc$.

It has also been shown in \cite{Hainzl2007} that the critical temperature $\Tc$ can be characterized via a linear criterion. More precisely, the critical temperature is determined by the unique value of $T$ such that the operator 
\begin{equation*}
	K_{T} - V 
\end{equation*}
acting on $L^2_{\mathrm{sym}}(\mathbb{R}^3)$, the space of reflection-symmetric square-integrable functions, has zero as its lowest eigenvalue. Here, $K_T = K_T(-\i \nabla)$ with the symbol 
\begin{align}
	K_T(p) := \frac{p^2 - \mu}{\tanh \frac{p^2-\mu}{2T}}. \label{DHS1:KT-symbol}
\end{align}
It should be noted that the function $T \mapsto K_T(p)$ is strictly monotone increasing for fixed $p \in \mathbb{R}^3$, and that $K_T(p) \geq 2T$ if $\mu \geq 0$ and $K_T(p)\geq |\mu|/\tanh(|\mu|/(2T))$ if $\mu < 0$. Our assumptions on $V$ guarantee that the essential spectrum of the operator $K_{T} - V$ equals $[2T, \infty)$ if $\mu \geq 0$ and $[|\mu|/\tanh(|\mu|/(2T)), \infty)$ if $\mu <0$. Accordingly, an eigenvalue at zero is necessarily isolated and of finite multiplicity.

The results in \cite{Hainzl2007} have been obtained in the case where the Cooper pair wave function $\alpha(x)$ is not necessarily an even function (as opposed to our setup), which means that $K_{\Tc}(-\i \nabla) - V$ has to be understood to act on $L^2(\mathbb{R}^3)$. The results in \cite{Hainzl2007}, however, equally hold if the symmetry of $\alpha$ is enforced. That is, they hold in the same way if $V$ is reflection symmetric and if the translation-invariant BCS functional is minimized over functions $\gamma(x)$ and $\alpha(x)$ that are both assumed to be reflection symmetric. 

We are interested in the situation where (a) $\Tc > 0$ and (b) the translation-invariant BCS functional has a unique minimizer with a radial Cooper pair wave function ($s$-wave Cooper pairs) for $T$ close to $\Tc$. This is implied by the following assumption. Part~(b) should be compared to \cite[Theorem~2.8]{DeuchertGeisinger}.

\begin{asmp}
\label{DHS1:Assumption_KTc}
\begin{enumerate}[(a)]
\item We assume that $\Tc >0$. If $V \geq 0$ and it does not vanish identically this is automatically implied, see \cite[Theorem 3]{Hainzl2007}. In the case of an interaction potential without a definite sign it is a separate assumption.


\item We assume that the lowest eigenvalue of $K_{\Tc} - V$ is simple.
\end{enumerate}
\end{asmp}


In the following we denote by $\alpha_*$ the unique ground state of the operator $K_{\Tc} - V$, i.e.,
\begin{align}
	K_{\Tc} \alpha_* = V\alpha_*. \label{DHS1:alpha_star_ev-equation}
\end{align}
We choose the normalization of $\alpha_*$ such that it is real-valued and $\Vert \alpha_*\Vert_{L^2(\Rbb^3)} = 1$. Since $V$ is a radial function and $\alpha_*$ is the unique solution of \eqref{DHS1:alpha_star_ev-equation} it follows that $\alpha_*$ is radial, too.


\subsection{The Ginzburg--Landau functional}
\label{DHS1:Ginzburg-Landau-functional}

We call a function $\Psi$ on $Q_B$ \textit{gauge-periodic} 
if it is left invariant by the magnetic translations of the form
\begin{align}
	T_B(\lambda)\Psi(X) &:= \e^{\i \Bbold \cdot (\lambda \wedge X)} \; \Psi(X + \lambda ), & \lambda &\in \Lambda_B. \label{DHS1:Magnetic_Translation_Charge2}
\end{align}
The operator $T(\lambda)$ in \eqref{DHS1:Magnetic_Translation} coincides with $T_B(\lambda)$  when $\Bbold$ is replaced by $2\Bbold$.

Let 
$\Lambda_0 , \Lambda_2, \Lambda_3 >0$ and $D\in\Rbb$ be given. For $B >0$ and a gauge-periodic function $\Psi$, the Ginzburg--Landau functional is defined by 
\begin{align}
\EGL(\Psi) &:= \frac{1}{B^2} \frac{1}{|Q_B|} \int_{Q_B} \dd X \; \bigl\{ \Lambda_0 \; |(-\i\nabla + 2\Abold)\Psi(X)|^2 - DB \, \Lambda_2\, |\Psi(X)|^2 + \Lambda_3\,|\Psi(X)|^4\bigr\}. \label{DHS1:Definition_GL-functional}
\end{align}
We highlight the factor of $2$ in front of the magnetic potential in \eqref{DHS1:Definition_GL-functional} and that the definition of the magnetic translation in \eqref{DHS1:Magnetic_Translation_Charge2} differs from that in  \eqref{DHS1:Magnetic_Translation} by a factor $2$. These two factors reflect the fact that $\Psi$ describes Cooper pairs, which carry twice the charge of a single particle.
The Ginzburg--Landau energy is defined by
\begin{align*}
E^{\mathrm{GL}}(D) := \inf \bigl\{ \EGL(\Psi) : \Psi\in \Hmag^1(Q_B)\bigr\},
\end{align*}
where $\Hmag^1(Q_B)$ denotes the set of gauge-periodic functions $\Psi \in L^2(Q_B)$ that satisfy $ (-\mathrm{i} \nabla + 2 \Abold) \Psi \in L^2(Q_B)$. It is independent of $B$ by scaling. More precisely, for given $\psi$ the function
\begin{align}
\Psi(X) &:= \sqrt{B} \; \psi\bigl( \sqrt{B} \, X\bigr), & X\in \Rbb^3, \label{DHS1:GL-rescaling}
\end{align}
satisfies
\begin{align}
\EGL(\Psi) = \mathcal E_{D,1}^{\mathrm{GL}}(\psi). \label{DHS1:EGL-scaling}
\end{align}

We also define the critical parameter
\begin{align}
\Dc &:= \frac{\Lambda_0}{\Lambda_2} \inf \spec_{\Lmag^2(Q_1)} \bigl((-\i \nabla + e_3 \wedge X)^2\bigr), \label{DHS1:Dc_Definition}
\end{align}
where $\Lmag^2(Q_1)$ denotes the set of gauge-periodic square-integrable functions on $Q_1$. Its definition is motivated by the fact that $\EGLGSE < 0$ if $D > \Dc$ and $\EGLGSE =0$ if $D \leq \Dc$. The proof of this statement goes along the same lines as that of \cite[Lemma 2.5]{Hainzl2014}. In our situation with a constant magnetic field the lowest eigenvalue of the Hamiltonian in \eqref{DHS1:Dc_Definition} equals $2$, see \cite[Eq.~(6.2)]{Tim_Abrikosov}, and $\Dc$ is explicit. In the situation of \cite{Hainzl2014}, where general external fields excluding the constant magnetic field are present, the parameter $\Dc$ is not explicit.  


\subsection{Main results}
\label{DHS1:Main_Result_Section}

Our first main result concerns the asymptotics of the BCS free energy in \eqref{DHS1:BCS GS-energy} in the regime $B \ll 1$. It also contains a statement about the asymptotics of the Cooper pair wave function of states $\Gamma$, whose energy $\FBCS(\Gamma)$ has the same asymptotic behavior as the BCS free energy (approximate minimizers). The precise statement is captured in the following theorem.

\begin{bigthm}
\label{DHS1:Main_Result}
Let Assumptions \ref{DHS1:Assumption_V} and \ref{DHS1:Assumption_KTc} hold, let $D \in \mathbb{R}$, and let the coefficients $\Lambda_0, \Lambda_2, \Lambda_3 >0$ be given by \eqref{DHS1:GL-coefficient_1}-\eqref{DHS1:GL_coefficient_3} below. Then there are constants $C>0$ and $B_0 >0$ such that for all $0 < B \leq B_0$, we have
\begin{align}
F^{\mathrm{BCS}}(B,\, \Tc(1 - DB)) = B^2 \; \bigl( \EGLGSE + R \bigr), \label{DHS1:ENERGY_ASYMPTOTICS}
\end{align}
with $R$ satisfying the estimate
\begin{align}
CB \geq R \geq - \Rcal := -C B^{\nicefrac 1{12}}. \label{DHS1:Rcal_error_Definition}
\end{align}
Moreover, for any approximate minimizer $\Gamma$ of $\FBCS$ at $T = \Tc(1 - DB)$ in the sense that
\begin{align}
\FBCS(\Gamma) - \FBCS(\Gamma_0) \leq B^2 \bigl( \EGLGSE + \rho\bigr) 
\label{DHS1:BCS_low_energy}
\end{align}
holds for some $\rho \geq 0$, we have the decomposition
\begin{align}
\alpha(X, r ) = \Psi(X) \, \alpha_*(r) + \sigma(X,r) \label{DHS1:Thm1_decomposition}
\end{align}
for the Cooper pair wave function $\alpha =\Gamma_{12}$. Here, $\sigma$ satisfies
\begin{align}
\frac{1}{|Q_B|} \int_{Q_B} \dd X \int_{\Rbb^3} \dd r \; |\sigma(X, r)|^2 &\leq C B^{\nicefrac {11}6}, \label{DHS1:Thm1_error_bound}
\end{align}
$\alpha_*$ is the normalized zero energy eigenstate of $K_{\Tc}-V$, and the function $\Psi$ obeys
\begin{align}
\EGL(\Psi) \leq \EGLGSE + \rho + \Rcal. \label{DHS1:GL-estimate_Psi}
\end{align}
\end{bigthm}

Our second main result concerns the shift of the BCS critical temperature that is caused by the external magnetic field. 

\begin{bigthm}
\label{DHS1:Main_Result_Tc} \label{DHS1:MAIN_RESULT_TC}
Let Assumptions \ref{DHS1:Assumption_V} and \ref{DHS1:Assumption_KTc} hold. Then there are constants $C>0$ and $B_0 >0$ such that for all $0 < B \leq B_0$ the following holds:
\begin{enumerate}[(a)]
	\item Let $0 < T_0 < \Tc$. If the temperature $T$ satisfies
	\begin{equation}
		T_0 \leq T \leq \Tc \, ( 1 - B \, ( \Dc + C \, B^{\nicefrac 1{2}}))
		\label{DHS1:eq:lowertemp}
	\end{equation}
	with $\Dc$ in \eqref{DHS1:Dc_Definition}, then we have
	\begin{equation*}
		F^{\mathrm{BCS}}(B,T) < 0.
	\end{equation*} 
	\item If the temperature $T$ satisfies
	\begin{equation}
		T \geq \Tc \, ( 1 - B \, ( \Dc - \Rcal ) )
		\label{DHS1:eq:uppertemp}
	\end{equation}
	with $\Dc$ in \eqref{DHS1:Dc_Definition} and $\Rcal$ in \eqref{DHS1:Rcal_error_Definition}, then we have
	\begin{equation*}
		\FBCS(\Gamma) - \FBCS(\Gamma_0) > 0
	\end{equation*}
	unless $\Gamma = \Gamma_0$.
\end{enumerate}
\end{bigthm}

\begin{bems}
\label{DHS1:Remarks_Main_Result}
\begin{enumerate}[(a)]
	
\item Theorem~\ref{DHS1:Main_Result} and Theorem~\ref{DHS1:Main_Result_Tc} extend similar results in \cite[Theorem 1]{Hainzl2012} and \cite[Theorem~2.4]{Hainzl2014} to the case of a homogeneous magnetic field. Such a magnetic field has a non-periodic vector potential and a non-zero magnetic flux through the unit cell $Q_B$. The main reason why the problem with a homogeneous magnetic field is more complicated is that it cannot be treated as a perturbation of the Laplacian. More precisely, it was possible in \cite{Hainzl2012,Hainzl2014} to work with a priori bounds for low-energy states that only involve the Laplacian and not the external fields. As noticed in \cite{Hainzl2017}, see the discussion below Remark~6, this is not possible in the case of a homogeneous magnetic field. In the proof of comparable a priori estimates involving the homogeneous magnetic field, see Theorem~\ref{DHS1:Structure_of_almost_minimizers} below, we have to deal with the fact that the components of the magnetic momentum operator do not commute, which leads to significant technical difficulties.

\item If we compare Theorem~\ref{DHS1:Main_Result} to \cite[Theorem 1]{Hainzl2012} or Theorem~\ref{DHS1:Main_Result_Tc} to \cite[Theorem~2.4]{Hainzl2014} we note the following technical differences: (1) The parameter $h$ in \cite{Hainzl2012,Hainzl2014} equals $B^{\nicefrac 12}$ in our work. (2) We use microscopic coordinates while macroscopic coordinates are used in \cite{Hainzl2012,Hainzl2014}, see the discussion above Eq.~1.4 in \cite{Hainzl2012}. (3) Our free energy is normalized by a large volume factor, see \eqref{DHS1:Trace_per_unit_volume_definition} and \eqref{DHS1:BCS functional}. This is not the case in \cite{Hainzl2012,Hainzl2014}. Accordingly, the GL energy appears on the order $B^2$ in our setting and on the order $h$ in the setting in \cite{Hainzl2012}. See also point (1) for the relation between $B$ and $h$. (4) The leading order of the Cooper pair wave function in \cite[Theorem 1]{Hainzl2012} is of the form 
\begin{equation}
	\frac{1}{2} \alpha_*(x-y) (\Psi(x) + \Psi(y)).
	\label{DHS1:eq:remarksA1}
\end{equation}
This should be compared to \eqref{DHS1:Thm1_decomposition}, where relative and center-of-mass coordinates are used. Using the a priori bound for the $L^2$-norm of $\nabla \Psi$ below (5.61) in \cite{Hainzl2012}, one can see that \eqref{DHS1:eq:remarksA1} equals the first term in \eqref{DHS1:Thm1_decomposition} to leading order in $h$. The analogue in our setup does not seem to be correct.

\item The Ginzburg--Landau energy appears at the order $B^2$. This should be compared to the free energy $\FBCS(\Gamma_0)$ of the normal state, which is of order $1$.

\item To appreciate the bound in \eqref{DHS1:Thm1_error_bound}, we note that the first term in the decomposition of $\alpha$ in \eqref{DHS1:Thm1_decomposition} obeys
\begin{align*}
	\frac{1}{|Q_B|} \int_{Q_B} \dd X \int_{\Rbb^3} \dd r \; | \Psi(X)\alpha_*(r) |^2 &\sim B
\end{align*}
if $D>0$.

\item We stated Theorem~\ref{DHS1:Main_Result} with fixed $D\in \Rbb$. Our explicit error bounds show that $D$ is allowed to vary with $B$ as long as there is a $B$-independent constant $D_0>0$ such that $|D| \leq D_0$ holds. 

\item Theorem~\ref{DHS1:Main_Result_Tc} gives bounds on the range of temperatures where superconductivity is present, see \eqref{DHS1:eq:lowertemp}, or absent, see \eqref{DHS1:eq:uppertemp}. The interpretation of this theorem is that for small magnetic fields $B$ the critical temperature obeys the asymptotic expansion
\begin{equation}
	\Tc(B) = \Tc(1 - \Dc B) + o(B).
	\label{DHS1:eq:BemA1}
\end{equation}
We highlight that $\Tc$ is determined by the translation-invariant problem, and that $D_{\mathrm{c}}$ is given by the macroscopic (linearized) GL theory. The same result has been obtained in \cite[Theorem~4]{Hainzl2017} in the case of linearized BCS theory. Theorem~\ref{DHS1:Main_Result_Tc} can therefore be interpreted as a justification of this approximation. Eq.~\eqref{DHS1:eq:BemA1} allows us to compute the upper critical field $B_{c2}$. That is, the magnetic field, above which, for a given temperature $T$, superconductivity is absent. In particular, it allows us to compute the derivative of $B_{c2}$ with respect to $T$ at the critical temperature from the BCS functional. For more details we refer to \cite[Appendix~A]{Hainzl2017}. 

\item We expect that the assumption $0 < T_0 \leq T$ for some arbitrary but $B$-independent constant $T_0$ in Theorem~\ref{DHS1:Main_Result_Tc}~(a) is of technical nature. We need this assumption, which similarly appears in \cite[Theorem~4]{Hainzl2017}, because our trial state analysis in Section~\ref{DHS1:Upper_Bound} breaks down when the temperature $T$ approaches zero. This is related to the fact that the Fermi distribution function $f_T(x) = (\e^{x/T}+1)^{-1}$ cannot be represented by a Cauchy-integral uniformly in the temperature. We note that there is no such restriction in Theorem~\ref{DHS1:Main_Result_Tc}~(b). It is also not needed in \cite[Theorem~2.4]{Hainzl2014}.

\end{enumerate}
\end{bems}


\subsection{Organization of the paper and strategy of proof}

In Section~\ref{sec:HeuristicComputation} we provide a brief non-rigorous computation that shows from which terms in the BCS functional the different terms in the Ginzburg--Landau functional arise.

In Section~\ref{DHS1:Preliminaries} we complete the introduction of our mathematical setup. We recall several properties of the trace per unit volume and introduce the relevant spaces of gauge-periodic functions. 

Section~\ref{DHS1:Upper_Bound} is dedicated to a trial state analysis. We start by introducing a class of Gibbs states, whose Cooper pair wave function is given by a product of the form $\alpha_*(r) \Psi(X)$ to leading order in $B$ with $\alpha_*$ in \eqref{DHS1:alpha_star_ev-equation} and with a gauge-periodic function $\Psi$ on $Q_B$. We state and motivate several results concerning these Gibbs states and their BCS free energy, whose proofs are deferred to Section~\ref{DHS1:Proofs}. Afterwards, these statements are used to prove the upper bound on \eqref{DHS1:ENERGY_ASYMPTOTICS} as well as Theorem~\ref{DHS1:Main_Result_Tc}~(a). As will be explained below, they are also relevant for the proofs of the lower bound in \eqref{DHS1:ENERGY_ASYMPTOTICS} and of Theorem~\ref{DHS1:Main_Result_Tc}~(b) in Section~\ref{DHS1:Lower Bound Part B}. 

Section~\ref{DHS1:Proofs} contains the proof of the results concerning the Gibbs states and their BCS free energy that have been stated without proof in Section~\ref{DHS1:Upper_Bound}. Our analysis is based on an extension of the phase approximation method, which has been pioneered in the framework of linearized BCS theory in \cite{Hainzl2017}, to our nonlinear setting. The phase approximation is a well-known tool in the physics literature, see, e.g., \cite{Werthamer1966}, and has also been used in the mathematical literature to study spectral properties of Schr\"odinger operators involving a magnetic field, for instance in \cite{NenciuCorn1998,Nenciu2002}. Our approach should be compared to the trial state analysis in \cite{Hainzl2012,Hainzl2014}, where a semi-classical expansion is used. The main novelty of our trial state analysis is Lemma~\ref{DHS1:BCS functional_identity_Lemma}, where we provide an alternative way to compute a certain trace function involving the trial state. It should be compared to the related part in the proof of \cite[Theorem~2]{Hainzl2012}. While the analysis in \cite{Hainzl2012} uses a Cauchy integral representation of the function $z \mapsto \ln(1+\e^{-z})$, our approach is based on a product expansion of the hyperbolic cosine in terms of Matsubara frequencies. In this way we obtain better decay properties in the subsequent resolvent expansion, which, in our opinion, simplifies the analysis considerably.

Section~\ref{DHS1:Lower Bound Part A} contains the proof of a priori estimates for BCS states, whose BCS free energy is smaller than or equal to that of the normal state $\Gamma_0$ in \eqref{DHS1:Gamma0} plus a correction of the order $B^2$ (low-energy states). The result is captured in Theorem~\ref{DHS1:Structure_of_almost_minimizers}, which is the main novelty of the present article. It states that the Cooper pair wave function of any low-energy state in the above sense has a Cooper pair wave function, which is, to leading order in $B$, given by a product of the form $\alpha_*(r) \Psi(X)$  with $\alpha_*(r)$ in \eqref{DHS1:alpha_star_ev-equation} and with a gauge-periodic function $\Psi(X)$ on $Q_B$. Furthermore, the function $\Psi(X)$ obeys certain bounds, which show that it is slowly varying and small in an appropriate sense. As explained in Remark~\ref{DHS1:Remarks_Main_Result}~(a), the main difficulty to overcome is that our a priori bounds involve the magnetic field. Therefore, we have to deal with the non-commutativity of the components of the magnetic momentum operator. The step where this problem appears most prominently is in the proof of Proposition~\ref{DHS1:First_Decomposition_Result}. 


The proof of the lower bound on \eqref{DHS1:ENERGY_ASYMPTOTICS} and of Theorem~\ref{DHS1:Main_Result_Tc}~(b) is provided in Section~\ref{DHS1:Lower Bound Part B}, which mostly follows the strategy in \cite[Section~6]{Hainzl2012} and \cite[Section~4.2]{Hainzl2014}. Two main ingredients for the analysis in this section are the trial state analysis in Section~\ref{DHS1:Upper_Bound} and Section~\ref{DHS1:Proofs}, and the a priori bounds for low-energy states in Section~\ref{DHS1:Lower Bound Part A}. From Theorem~\ref{DHS1:Structure_of_almost_minimizers} we know that the Cooper pair wave function of any low-energy state has a product structure to leading order in $B$. The main idea of the proof of the lower bound in \eqref{DHS1:ENERGY_ASYMPTOTICS} is to construct a Gibbs state, whose Cooper pair wave function has the same asymptotics to leading order in $B$. The precise characterization of the Cooper pair wave function of the Gibbs state in Section~\ref{DHS1:Upper_Bound} and the a priori bounds in Theorem~\ref{DHS1:Structure_of_almost_minimizers} then allow us to bound the BCS free energy of the original state from below in terms of that of the Gibbs state. The latter has been computed with sufficient precision in Section~\ref{DHS1:Upper_Bound} and Section~\ref{DHS1:Proofs}.



Throughout the paper, $c$ and $C$ denote generic positive constants that change from line to line. We allow them to depend on the various fixed quantities like $B_0$, $\mu$, $\Tc$, $V$, $\alpha_*$, etc. Further dependencies are indexed.

\subsection{Heuristic computation of the terms in the Ginzburg--Landau functional}
\label{sec:HeuristicComputation}

In this section we present a brief non-rigorous computation with the trial state that we use in the proof of the upper bound for the BCS free energy in Section~\ref{DHS1:Upper_Bound}, to show from which terms in the BCS functional the different terms in the Ginzburg--Landau (GL) functional arise. The trial state (a Gibbs state) is defined by
\begin{equation}
	\begin{pmatrix} \gamma_{\Delta} & \alpha_{\Delta} \\ \overline{\alpha_{\Delta}} & 1 - \overline{\gamma_{\Delta}} \end{pmatrix} =
	 \Gamma_{\Delta} = \frac{1}{1+\e^{\beta H_{\Delta}}}  
	 \label{eq:heuristicstrialstate}
\end{equation}
with the Hamiltonian
\begin{equation*}
	H_{\Delta} = \begin{pmatrix} (-\mathrm{i} \nabla + \Abold )^2 - \mu & \Delta \\ \overline{\Delta} & -\overline{(-\mathrm{i} \nabla + \Abold )^2} + \mu \end{pmatrix},
\end{equation*}
where the operator $\Delta$ is defined via its integral kernel
\begin{equation*}
	\Delta(x,y) = -2\, V \alpha_* (x-y)\, \Psi_B \left( \frac{x+y}{2} \right)
\end{equation*}
and $\beta^{-1} = T = \Tc(1 - DB)$. We choose $\Psi_B(X)$ to be a minimizer of the Ginzburg--Landau functional. From \eqref{DHS1:GL-rescaling} it follows that $| Q_B|^{-1} \int_{Q_B} \dd X \ | \Psi(X) |^2 \sim B$ as well as that
\begin{equation}
	\Tr [ \Delta^* \Delta ] = \frac{4}{| Q_B|} \int_{Q_B} \dd X \; | \Psi(X) |^2  \int_{\mathbb{R}^3} \dd r \; | V(r) \alpha_*(r) |^2 \sim B.
	\label{eq:heurusticscalingdelta}
\end{equation}
Here $r=x-y$ and $X=(x+y)/2$ denote relative and center-of-mass coordinates. The operator $\Delta$ in \eqref{eq:heuristicstrialstate} is therefore a small perturbation of the state $\Gamma_{\Delta}$ in \eqref{eq:heuristicstrialstate} if $0 < B \ll 1$.

The BCS free energy of $\Gamma_{\Delta}$ reads
\begin{align*}
	\FBCS(\Gamma) - \FBCS(\Gamma_0) &= \frac{1}{2} \Tr\left[ H_0 (\Gamma_{\Delta} - \Gamma_0) \right] - T S(\Gamma) + T S(\Gamma_0)  \\
	&\hspace{30pt} - \frac{1}{| Q_B |} \int_{Q_B} \dd X \int_{\mathbb{R}^3} \dd r \; V(r) | \alpha_{\Delta}(X,r) |^2. 
\end{align*}
Using the identity 
\begin{equation*}
	\Tr\left[ H_0 (\Gamma_{\Delta} - \Gamma_0) \right] = \Tr\left[ H_{\Delta} \Gamma_{\Delta} - H_0 \Gamma_0 \right] - \Tr[ (H_{\Delta} - H_0 ) \Gamma_{\Delta} ],
\end{equation*}
and \cite[Eqs.~(4.3--4.5)]{Hainzl2012}, this formula can be rewritten as
\begin{align}
	&\FBCS(\Gamma) - \FBCS(\Gamma_0) = -\frac {1}{2 \beta}\Tr_0 \bigl[ \ln\bigl( 1+\exp\bigr( - \beta H_\Delta \bigr)\bigr) - \ln\bigl( 1+\exp\bigl( -\beta H_0 \bigr)\bigr)\bigr] \nonumber \\
	&+ \frac{\langle \alpha_*, V\alpha_*\rangle_{L^2(\Rbb^3)}}{|Q_B|} \int_{Q_B} \dd X \ |\Psi(X)|^2  \;   - \frac{1}{|Q_B|} \int_{Q_B} \dd X\int_{\Rbb^3} \dd r\; V(r) \, \bigl|\alpha(X,r) - \alpha_*(r) \Psi(X)\bigr|^2. \label{eq:heuristics1}
\end{align}
Here $\Tr_0[A] = \Tr[PAP] + \Tr[QAQ]$ with 
\begin{equation*}
	P = \begin{pmatrix}
		1 & 0 \\ 0 & 0 \end{pmatrix}
\end{equation*} 
and $Q = 1 - P$.

To obtain the terms in the GL functional, we need to expand the terms in \eqref{eq:heuristics1} in powers of $B$. To that end, we first expand them up to fourth order in powers of $\Delta$ (The Ginzburg--Landau functional is a fourth order polynomial in $\Psi_B$.) and afterwards expand the resulting terms in powers of $B$, keeping in mind that $T = \Tc(1 - DB)$. It turns out that the last term on the right side of \eqref{eq:heuristics1} is of the order $o(B^2)$, and therefore does not contribute to the GL energy. In our trial state analysis in Sections~\ref{DHS1:Upper_Bound}~and~\ref{DHS1:Proofs} we show that there exists a linear operator $L_{T,B}$ and a cubic map $N_{T,B}(\Delta)$ such that
\begin{align}
	-\frac {1}{2\beta}\Tr_0 \bigl[ \ln\bigl( 1+\exp\bigr( - \beta H_\Delta \bigr)\bigr) - \ln\bigl( 1+\exp\bigl( -\beta H_0 \bigr)\bigr)\bigr] & \nonumber \\
	&\hspace{-120pt} = -\frac{1}{4} \langle \Delta, L_{T,B} \Delta \rangle + \frac{1}{8} \langle \Delta, N_{T,B}(\Delta) \rangle + o(B^2). \label{eq:heuristicquadraticterms}
\end{align}
In combination with the first term in the second line of \eqref{eq:heuristics1}, the quadratic terms in \eqref{eq:heuristicquadraticterms} give the quadratic terms in the Ginzburg--Landau functional:
\begin{align*}
	&-\frac{1}{4} \langle \Delta, L_{T,B} \Delta \rangle + \frac{\langle \alpha_*, V\alpha_*\rangle_{L^2(\Rbb^3)}}{|Q_B|} \int_{Q_B} \dd X \; |\Psi(X)|^2 \\
	&\hspace{50pt} = \frac{\Lambda_0}{|Q_B|} \int_{Q_B}  \dd X \; | \Pi \Psi(X) |^2 - \frac{D B \Lambda_2}{|Q_B|} \int_{Q_B}  \dd X \; | \Pi \Psi(X) |^2 + o(B^2).
\end{align*}
From the quartic term in \eqref{eq:heuristicquadraticterms} we will extract the quartic term in the GL functional, that is,
\begin{equation*}
	\frac{1}{8} \langle \Delta, N_{T,B}(\Delta) \rangle = \frac{\Lambda_3}{|Q_B|} \int_{Q_B} \dd X \; |\Psi(X)|^4 + o(B^2),
\end{equation*}
and hence
\begin{equation*}
	\FBCS(\Gamma) - \FBCS(\Gamma_0) = B^2 \left( \EGLGSE + o(1) \right).
\end{equation*}
In the last step we used $\EGL(\Psi_B) = B^2 \EGLGSE$.


\section{Preliminaries}
\label{DHS1:Preliminaries}


\subsection{Schatten classes}
\label{DHS1:Schatten_Classes}

In our proofs we frequently use Schatten norms of periodic operators, which are defined with respect to the trace per unit volume in \eqref{DHS1:Trace_per_unit_volume_definition}. In this section we recall some basic facts about these norms.

A gauge-periodic operator $A$ belongs to the $p$\tho\ local von-Neumann--Schatten class $\Scal^p$ with $1\leq p < \infty$ if it has finite $p$-norm, that is, if $\Vert A\Vert_p^p := \Tr (|A|^p) <\infty$. By $\Scal^\infty$ we denote the set of bounded gauge-periodic operators and $\Vert \cdot \Vert_{\infty}$ is the usual operator norm. For the above norms the triangle inequality
\begin{align*}
\Vert A + B\Vert_p \leq \Vert A\Vert_p + \Vert B\Vert_p
\end{align*}
holds for $1\leq p \leq \infty$. Moreover, for $1 \leq p,q,r \leq \infty$ with $\frac{1}{r} = \frac{1}{p} + \frac{1}{q}$ we have the general Hölder inequality 
\begin{align}
\Vert AB\Vert_r \leq \Vert A\Vert_p \Vert B\Vert_q. \label{DHS1:Schatten-Hoelder}
\end{align}
It is important to note that the above norms are not monotone decreasing in the index $p$. This should be compared to the usual Schatten norms, where such a property holds. The familiar inequality
\begin{align*}
| \Tr A | \leq \Vert A \Vert_1
\end{align*}
is true also in the case of local Schatten norms.

The above inequalities can be reduced to the case of the usual Schatten norms, see, e.g., \cite{Simon05}, using the magnetic Bloch--Floquet decomposition. We refer to \cite[Section XIII.16]{Reedsimon4} for an introduction to the Bloch--Floquet transformation and to \cite{Stefan_Peierls} for a particular treatment of the magnetic case. More specifically, for a gauge-periodic operator $A$ we use the unitary equivalence 
\begin{align*}
A \cong \int^{\oplus}_{[0,\sqrt{ 2 \pi B}]^3} \mathrm{d} k \;  A_{k}
\end{align*}
to write the trace per unit volume as
\begin{equation}
\Tr A = \int_{[0,\sqrt{ 2 \pi B}]^3} \mathrm{d} k \; \Tr_{L^2(Q_B)} A_{k},
\label{DHS1:eq:ATPUV}
\end{equation}
where $\Tr_{L^2(Q_B)}$ denotes the usual trace over $L^2(Q_B)$. The inequalities for the trace per unit volume from above follow from the usual ones when we use that $(AB)_{k} = A_{k} B_{k}$ holds for two gauge-periodic operators $A$ and $B$.


\subsection{Gauge-periodic Sobolev spaces}
\label{DHS1:Periodic Spaces}

In this section we introduce several spaces of gauge-periodic functions, which will be used to describe the center-of-mass part of Cooper pair wave functions. 



For $1 \leq p < \infty$, the space $L_{\mathrm{mag}}^p(Q_B)$ consists of all $L_\loc^p(\Rbb^3)$-functions $\Psi$, which satisfy $T_B(\lambda)\Psi = \Psi$ for all $\lambda\in\Lambda_B$ with $T_B(\lambda)$ in \eqref{DHS1:Magnetic_Translation_Charge2}. The space is equipped with the usual $p$-norm per unit volume
\begin{align}
\Vert \Psi\Vert_{\Lmag^p(Q_B)}^p &:= \fint_{Q_B} \dd X \; |\Psi(X)|^p := \frac{1}{|Q_B|} \int_{Q_B} \dd X \; |\Psi(X)|^p, \label{DHS1:Periodic_p_Norm}
\end{align}
and we use the conventional abbreviation $\Vert \Psi\Vert_p$ when this does not lead to confusion.

For $m\in \Nbb_0$, the corresponding gauge-periodic Sobolev space is defined by
\begin{align}
\Hmag^m(Q_B) &:= \bigl\{ \Psi\in \Lmag^2(Q_B) :  \Pi^\nu \Psi\in \Lmag^2(Q_B) \quad \forall \nu\in \Nbb_0^3, |\nu|_1\leq m\bigr\}, \label{DHS1:Periodic_Sobolev_Space}
\end{align}
where
\begin{align*}
	\Pi := -\i \nabla + 2\Abold
\end{align*} 
denotes the magnetic momentum operator and $|\nu |_1 := \sum_{i=1}^3 \nu_i$ for $\nu\in \Nbb_0^3$. 
%
%
Equipped with the scalar product
\begin{align}
\langle \Phi, \Psi\rangle_{\Hmag^m(Q_B)} &:= \sum_{|\nu|_1\leq m} B^{-1 - |\nu|_1} \; \langle \Pi^\nu \Phi, \Pi^\nu \Psi\rangle_{\Lmag^2(Q_B)},  \label{DHS1:Periodic_Sobolev_Norm}
\end{align}
it is a Hilbert space.  We note that $\Pi^\nu \Psi$ is a gauge-periodic function if $\Psi$ is gauge-periodic because $\Pi$ commutes with the magnetic translations $T_B(\lambda)$ in \eqref{DHS1:Magnetic_Translation_Charge2}. We also note that $\Pi$ is a self-adjoint operator on $\Hmag^1(Q_B)$. 

At this point, we shall briefly explain the scaling behavior in $B$ of the norms introduced in \eqref{DHS1:Periodic_p_Norm} and \eqref{DHS1:Periodic_Sobolev_Norm} in terms of the Ginzburg--Landau scaling in \eqref{DHS1:GL-rescaling}. First, we note that if $\psi \in \Lmag^p(Q_1)$ and $\Psi$ is as in \eqref{DHS1:GL-rescaling}, then
\begin{align}
\Vert \Psi\Vert_{\Lmag^p(Q_B)} = B^{\nicefrac 12} \, \Vert \psi\Vert_{\Lmag^p(Q_1)} \label{DHS1:Periodic_p_Norm_scaling}
\end{align}
for every $1\leq p \leq \infty$. In contrast, the scaling of the norm in \eqref{DHS1:Periodic_Sobolev_Norm} is chosen such that
\begin{align*}
\Vert \Psi\Vert_{\Hmag^m(Q_B)} = \Vert \psi\Vert_{\Hmag^m(Q_1)}.
\end{align*}
This follows from \eqref{DHS1:Periodic_p_Norm_scaling} and the fact that
$\Vert \Pi^\nu\Psi\Vert_2^2$ scales as $B^{1 + |\nu|_1}$ for $\nu\in \Nbb_0^3$.

We also mention the following magnetic Sobolev inequality because it will be used frequently in the course of the paper. For any $B>0$ and any $\Psi\in \Hmag^1(Q_B)$, we have
\begin{align}
\Vert \Psi\Vert_{\Lmag^6(Q_B)}^2 &\leq  C \, B^{-1}\, \Vert \Pi\Psi\Vert_{\Lmag^2(Q_B)}^2. \label{DHS1:Magnetic_Sobolev}
\end{align}

\begin{proof}[Proof of \eqref{DHS1:Magnetic_Sobolev}]
Since $Q_1$ satisfies the cone property, \cite[Theorem 8.8]{LiebLoss} implies 
\begin{align*}
\Vert \psi \Vert_{\Lmag^6(Q_1)}^2 &\leq C\, \bigl( \Vert \psi \Vert_{\Lmag^2(Q_1)}^2 + \Vert \nabla |\psi|\, \Vert_{\Lmag^2(Q_1)}^2\bigr).
\end{align*}
From \cite[Eq.~(6.2)]{Tim_Abrikosov} we know that the bottom of the spectrum of $(-\i \nabla + e_3 \wedge X)^2$ equals~$2$. For the first term on the right side, this implies $2\Vert \psi\Vert_2^2 \leq \Vert (-\i \nabla + e_3\wedge X)\psi\Vert_2^2$. To bound the second term, we apply the  diamagnetic inequality $|\nabla |\psi(X)|| \leq |(-\i\nabla + e_3\wedge X)\psi(X)|$, see \cite[Theorem 7.21]{LiebLoss}. This proves \eqref{DHS1:Magnetic_Sobolev} for $B = 1$ and the scaling in \eqref{DHS1:GL-rescaling} yields \eqref{DHS1:Magnetic_Sobolev} for $B >0$.
\end{proof}

As indicated below \eqref{DHS1:Trace_per_unit_volume_definition}, the Cooper pair wave function $\alpha$ related to an admissible state $\Gamma$ belongs to $\Scal^2$, the Hilbert--Schmidt class introduced in Section \ref{DHS1:Schatten_Classes}. In terms of the center-of-mass and relative coordinates, the gauge-periodicity and the symmetry of the kernel of $\alpha$ in \eqref{DHS1:alpha_periodicity} read
\begin{align}
\alpha(X,r) &= \e^{\i \Bbold \cdot (\lambda \wedge X)} \; \alpha(X+ \lambda, r), \quad \lambda\in \Lambda_B; & \alpha(X,r) &= \alpha(X, -r). \label{DHS1:alpha_periodicity_COM}
\end{align}
That is, $\alpha(X,r)$ is a gauge-periodic function of the center-of-mass coordinate $X$ and a reflection-symmetric function of the relative coordinate $r \in \mathbb{R}^3$. We make use of the isometric identification of $\Scal^2$ with the space 
\begin{align*}
L^2(Q_B \times \Rbb_{\mathrm s}^3) := \Lmag^2(Q_B) \otimes L_{\mathrm{sym}}^2(\Rbb^3),
\end{align*}
the square-integrable functions obeying \eqref{DHS1:alpha_periodicity_COM}, for which the norm
\begin{align*}
\Vert \alpha\Vert_{\Lsymm}^2 := \fint_{Q_B} \dd X\int_{\Rbb^3} \dd r \; |\alpha(X, r)|^2 = \frac{1}{|Q_B|} \int_{Q_B} \dd X\int_{\Rbb^3} \dd r \; |\alpha(X, r)|^2
\end{align*} 
is finite. By \eqref{DHS1:alpha_periodicity_COM}, the identity $\Vert \alpha\Vert_2 = \Vert \alpha\Vert_{\Lsymm}$ holds. Therefore, we do not distinguish between the scalar products $\langle \cdot, \cdot\rangle$ on $\Lsymm$ and $\Scal^2$ and identify operators in $\Scal^2$ with their kernels whenever this appears convenient. 

Finally, the Sobolev space $H^1(Q_B\times \Rbb_{\mathrm s}^3)$ consists of all functions $\alpha\in L^2(Q_B\times \Rbb_{\mathrm s}^3)$ with finite $H^1$-norm given by
\begin{align}
\Vert \alpha\Vert_{H^1(Q_B\times \Rbb_{\mathrm s}^3)}^2 &:= \Vert \alpha\Vert_2^2 + \Vert \Pi_X\alpha\Vert_2^2 + \Vert \tilde \pi_r\alpha\Vert_2^2. \label{DHS1:H1-norm}
\end{align}
Here, we used the magnetic momentum operators 
\begin{align}
\Pi_X &:= -\i\nabla_X + 2 \Abold(X), & \tilde \pi_r &:= -\i\nabla_r + \frac 12 \Abold(r), \label{DHS1:Magnetic_Momenta_COM}
\end{align}
where $\Abold(x) = \frac 12 \Bbold \wedge x$. We note that the norm in \eqref{DHS1:H1-norm} is equivalent to the norm given by $\Tr [\alpha\alpha^*] + \Tr [(-\i \nabla + \Abold)\alpha \alpha^* (-\i \nabla + \Abold)]  + \Tr [(-\i \nabla + \Abold) \alpha^* \alpha (-\i \nabla + \Abold)]$, which, in turn, is given by
$\Vert \alpha\Vert_2^2 + \Vert (-\i \nabla + \Abold)\alpha\Vert_2^2 + \Vert \alpha (-\i \nabla + \Abold)\Vert_2^2$. See also the discussion below \eqref{DHS1:Trace_per_unit_volume_definition}.

\section{Trial States and their BCS Energy}
\label{DHS1:Upper_Bound} \label{DHS1:UPPER_BOUND}

The goal of this section is to provide the upper bound on \eqref{DHS1:ENERGY_ASYMPTOTICS} and the proof of Theorem~\ref{DHS1:Main_Result_Tc}~(a). Both bounds are proved with a trial state argument using Gibbs states $\Gamma_\Delta$ that are defined via a gap function $\Delta$ in the effective Hamiltonian. In Proposition~\ref{DHS1:Structure_of_alphaDelta} we show that the Cooper pair wave function $\alpha_\Delta$ of $\Gamma_\Delta$ is a product function with respect to relative and center-of-mass coordinates to leading order provided $\Delta$ is a product function that is small in a suitable sense. A representation formula for the BCS energy in terms of the energy of these states is provided in Proposition~\ref{DHS1:BCS functional_identity}. Finally, in Theorem \ref{DHS1:Calculation_of_the_GL-energy}, we show that certain parts of the BCS energy of the trial states $\Gamma_{\Delta}$ equal the terms in the Ginzburg--Landau functional in \eqref{DHS1:Definition_GL-functional} with sufficient precision provided $T = \Tc(1 - DB)$ for some fixed $D \in \mathbb{R}$. These results, whose proofs are deferred to Section \ref{DHS1:Proofs}, are combined in Section~\ref{DHS1:Upper_Bound_Proof_Section} to give the proof of the results mentioned in the beginning of this paragraph.


\subsection{The Gibbs states \texorpdfstring{$\Gamma_\Delta$}{GammaDelta}}

For any $\Psi\in \Lmag^2(Q_B)$, let us introduce the gap function $\Delta\in L^2(Q_B\times \Rbb_{\mathrm s}^3)$, given by
\begin{align}
\Delta(X,r) := \Delta_\Psi(X, r) &:= -2 \; V\alpha_*(r) \; \Psi(X).  \label{DHS1:Delta_definition}
\end{align}
In our trial state analysis, $\Psi$ is going to be a minimizer of the Ginzburg--Landau functional in \eqref{DHS1:Definition_GL-functional}. It therefore obeys the scaling in \eqref{DHS1:GL-rescaling}, which implies that the local Hilbert-Schmidt norm $\Vert \Delta\Vert_2^2$ is of the order $B$. We highlight that the $L^2(\mathbb{R}^3)$-norm of $V\alpha_*$ is of the order $1$, that is, the size of $\Vert \Delta\Vert_2^2$ is determined by $\Psi$. In the proof of the lower bound we have less information on $\Psi$. The related difficulties are discussed in Remark~\ref{DHS1:rem:Psi} below. With
\begin{align}
\hfrak_B &:=  (-\i \nabla +\Abold )^2 - \mu, \label{DHS1:hfrakB_definition}
\end{align}
we define the Hamiltonian
\begin{align}
H_{\Delta} &:= H_0 + \delta := \begin{pmatrix}
\hfrak_B & 0 \\ 0 & -\ov{\hfrak_B}
\end{pmatrix} + \begin{pmatrix}
0 & \Delta \\ \ov \Delta & 0
\end{pmatrix} = \begin{pmatrix}
\hfrak_B & \Delta \\ \ov \Delta & -\ov {\hfrak_B}
\end{pmatrix} \label{DHS1:HDelta_definition}
\end{align}
and the corresponding Gibbs state at inverse temperature $\beta = T^{-1} >0$ as
\begin{align}
\begin{pmatrix} \gamma_\Delta & \alpha_\Delta \\ \ov{\alpha_\Delta} & 1 - \ov{\gamma_\Delta}\end{pmatrix} = \Gamma_\Delta := \frac{1}{1 + \e^{\beta H_\Delta}}. \label{DHS1:GammaDelta_definition}
\end{align}
We note that the normal state $\Gamma_0$ in \eqref{DHS1:Gamma0} corresponds to setting $\Delta =0$ in \eqref{DHS1:GammaDelta_definition}.

\begin{lem}[Admissibility of $\Gamma_\Delta$]
\label{DHS1:Gamma_Delta_admissible}
Let Assumptions \ref{DHS1:Assumption_V} and \ref{DHS1:Assumption_KTc} hold. Then, for any $B>0$, any $T>0$, and any $\Psi\in \Hmag^1(Q_B)$, the state $\Gamma_\Delta$ in \eqref{DHS1:GammaDelta_definition} is admissible, where $\Delta \equiv \Delta_\Psi$ as in \eqref{DHS1:Delta_definition}.
\end{lem}

The states $\Gamma_\Delta$ are inspired by the following observation. Via variational arguments it is straightforward to see that any minimizer of $\FBCS$ in \eqref{DHS1:BCS functional} solves the nonlinear Bogolubov--de Gennes equation
\begin{align}
\Gamma &= \frac 1{1 + \e^{\beta \, \Hbb_{V\alpha}}}, & \Hbb_{V\alpha} = \begin{pmatrix} \hfrak_B & -2\, V\alpha \\ -2\, \ov{V\alpha} & -\ov{\hfrak_B}\end{pmatrix}. \label{DHS1:BdG-equation}
\end{align}
Here, $V\alpha$ is the operator given by the kernel $V(r)\alpha(X,r)$. As we look for approximate minimizers of $\FBCS$, we choose $\Gamma_\Delta$ in order to approximately solve \eqref{DHS1:BdG-equation}. As far as the leading term of $\alpha_\Delta$ is concerned this is indeed the case, as the following result shows. It should be compared to \eqref{DHS1:Thm1_decomposition}.

\begin{prop}[Structure of $\alpha_\Delta$]
\label{DHS1:Structure_of_alphaDelta} \label{DHS1:STRUCTURE_OF_ALPHADELTA}
Let Assumption \ref{DHS1:Assumption_V} and \ref{DHS1:Assumption_KTc} (a) be satisfied and let $T_0>0$ be given. Then, there is a constant $B_0>0$ such that for any $0 < B \leq B_0$, any $T\geq T_0$, and any $\Psi\in \Hmag^2(Q_B)$ the function $\alpha_\Delta$ in \eqref{DHS1:GammaDelta_definition} with $\Delta \equiv \Delta_\Psi$ as in \eqref{DHS1:Delta_definition} has the decomposition
\begin{align}
\alpha_\Delta(X,r) &= \Psi(X) \alpha_*(r) - \eta_0(\Delta)(X,r) - \eta_{\perp}(\Delta)(X,r). \label{DHS1:alphaDelta_decomposition_eq1}
\end{align}
The remainder functions $\eta_0(\Delta)$ and $\eta_\perp(\Delta)$ have the following properties:
\begin{enumerate}[(a)]
\item The function $\eta_0$ satisfies the bound
\begin{align}
\Vert \eta_0\Vert_\Hsymm^2 &\leq  C\; \bigl( B^3 + B \, |T - \Tc|^2\bigr) \; \bigl(  \Vert \Psi\Vert_{\Hmag^1(Q_B)}^6 + \Vert \Psi\Vert_{\Hmag^1(Q_B)}^2\bigr). \label{DHS1:alphaDelta_decomposition_eq2}
\end{align}

\item The function $\eta_\perp$ satisfies the bound
\begin{align}
\Vert \eta_\perp\Vert_{\Hsymm}^2 + \Vert |r|\eta_\perp\Vert_{\Lsymm}^2 &\leq C \; B^3 \; \Vert \Psi\Vert_{\Hmag^2(Q_B)}^2. \label{DHS1:alphaDelta_decomposition_eq3}
\end{align}

\item The function $\eta_\perp$ has the explicit form
\begin{align*}
\eta_\perp(X, r) &= \int_{\Rbb^3} \dd Z \int_{\Rbb^3} \dd s \; k_T(Z, r-s) \, V\alpha_*(s) \, \bigl[ \cos(Z\cdot \Pi_X) - 1\bigr] \Psi(X)
\end{align*}
with $k_T(Z,r)$ defined in Section~\ref{DHS1:Proofs} below \eqref{DHS1:MTB_definition}. For any radial $f,g\in L^2(\Rbb^3)$ the operator
\begin{align*}
\iiint_{\Rbb^9} \dd Z \dd r \dd s \; f(r) \, k_T(Z, r-s) \, g(s) \, \bigl[ \cos(Z\cdot \Pi) - 1\bigr]
\end{align*}
commutes with $\Pi^2$, and, in particular, 
if $P$ and $Q$ are two spectral projections of $\Pi^2$ with $P Q = 0$, then $\eta_\perp$ satisfies the orthogonality property
\begin{align}
	\bigl\langle f(r) \, (P \Psi)(X), \, \eta_{\perp}(\Delta_{Q\Psi}) \bigr\rangle = 0.
	\label{DHS1:alphaDelta_decomposition_eq4}
\end{align}
\end{enumerate}
\end{prop}

\begin{bem}
	\label{DHS1:rem:Psi}
The statement of Proposition \ref{DHS1:Structure_of_alphaDelta} should be read in two different ways, depending on whether we are interested in proving the upper or the lower bound for the BCS free energy. When we prove the upper bound using trial states $\Gamma_\Delta$, the bound on $\Vert |r|\eta_\perp\Vert_{\Lsymm}$ in part (b) and part (c) are irrelevant. In this case the gap function $\Delta\equiv \Delta_\Psi$ is defined with a minimizer $\Psi$ of the GL functional, whose $\Hmag^2(Q_B)$-norm is uniformly bounded, and all remainder terms can be estimated using \eqref{DHS1:alphaDelta_decomposition_eq2} and \eqref{DHS1:alphaDelta_decomposition_eq3}.

In the proof of the lower bound for the BCS free energy in Section~\ref{DHS1:Lower Bound Part B} we are forced to work with a trial state $\Gamma_{\Delta}$, whose gap function is defined via a function $\Psi$ that is related to a low-energy state of the BCS functional, see Theorem~\ref{DHS1:Structure_of_almost_minimizers} below. For such functions we only have a bound on the $\Hmag^1(Q_B)$-norm at our disposal. To obtain a function in $\Hmag^2(Q_B)$, we introduce a regularized version of $\Psi$ as in \cite[Section~6]{Hainzl2012}, \cite[Section~6]{Hainzl2014}, and \cite[Section~7]{Hainzl2017} by $\Psi_\leq := \Idbb_{[0,\varepsilon]}(\Pi^2)\Psi$ for some $B \ll \varepsilon \ll 1$, see Corollary \ref{DHS1:Structure_of_almost_minimizers_corollary}. The $\Hmag^2(Q_B)$-norm of $\Psi_\leq$ is not uniformly bounded in $B$, see \eqref{DHS1:Psileq_bounds} below. This causes a certain error term, namely the left side of \eqref{DHS1:LBpartB_3} below, to be large, a priori. 

To overcome this problem we use the bound on $\Vert |r|\eta_\perp\Vert_{\Lsymm}$ in part (b) of Proposition~\ref{DHS1:Structure_of_almost_minimizers} and part (c). Part (c) exploits the fact that the first term on the right side of \eqref{DHS1:LBpartB_3} has an explicit form that satisfies the orthogonality property in \eqref{DHS1:alphaDelta_decomposition_eq4}, 
which implies that the left side of \eqref{DHS1:LBpartB_3} is indeed small. This is the reason why we need to distinguish between $\eta_0$ and $\eta_\perp$. 
\end{bem}

\subsection{The BCS energy of the states \texorpdfstring{$\Gamma_\Delta$}{GammaDelta}}

This section pertains to the BCS energy of the states $\Gamma_\Delta$, which is given by the Ginzburg--Landau functional to leading order. We will see in Section \ref{DHS1:BCS functional_identity_proof_Section} that the BCS energy of $\Gamma_\Delta$ can be calculated in terms of
\begin{align}
\Tr_0\Bigl[\ln\bigl( \cosh\bigl( \frac \beta 2 H_\Delta\bigr)\bigr) - \ln\bigl( \cosh\bigl( \frac \beta 2 H_0\bigr)\bigr)\Bigr]. \label{DHS1:lncosh-operator}
\end{align}
Here, $\Tr_0$ is a weaker form of trace which will be introduced later in \eqref{DHS1:Weak_trace_definition}. The operator inside the trace is closely related to the relative entropy of $H_\Delta$ and $H_0$ but also incorporates the interaction energy of $\alpha_\Delta$. We refer to \eqref{DHS1:kinetic energy} for more details. In the following, we explain how the terms of the Ginzburg--Landau functional, which appear in the energy expansion in \eqref{DHS1:ENERGY_ASYMPTOTICS}, are obtained from the operator in \eqref{DHS1:lncosh-operator}. 

As pointed out in Remark \ref{DHS1:Remarks_Main_Result}, we should think of $\Delta$ as being small. In order to expand the term in \eqref{DHS1:lncosh-operator} in powers of $\Delta$, we use the fundamental theorem to formally write \eqref{DHS1:lncosh-operator} as
\begin{align}
\frac \beta 2 \Tr_0 \Bigl[ \int_0^1 \dd t \; \tanh\bigl( \frac \beta 2 H_{t \Delta} \bigr) \Bigl(\begin{matrix} 0 & \Delta \\ \ov \Delta & 0 \end{matrix}\Bigr) \Bigr]. \label{DHS1:lncosh-operator_fundamental_theorem}
\end{align}
This identity is not rigorous because it ignores the subtlety that $H_{t\Delta}$, $t\in [0,1]$, are unbounded operators which do not commute for distinct values of $t$. 
We present a rigorous version of \eqref{DHS1:lncosh-operator_fundamental_theorem} in Lemma \ref{DHS1:BCS functional_identity_Lemma} below. For the sake of the following discussion it is legitimate to assume that equality between \eqref{DHS1:lncosh-operator} and \eqref{DHS1:lncosh-operator_fundamental_theorem} holds. 

We use the Mittag-Leffler series expansion, see e.g. \cite[Eq. (7)]{Hainzl2017}, to write the hyperbolic tangent in \eqref{DHS1:lncosh-operator_fundamental_theorem} as
\begin{align}
\tanh\bigl( \frac \beta 2 z\bigr) &= -\frac{2}{\beta} \sum_{n\in \Zbb} \frac{1}{\i\omega_n - z} \label{DHS1:tanh_Matsubara}
\end{align}
with the Matsubara frequencies 
\begin{align}
\omega_n &:= \pi (2n+1) T, \qquad n \in \Zbb. \label{DHS1:Matsubara_frequencies}
\end{align}
The convergence of \eqref{DHS1:tanh_Matsubara} becomes manifest by combining the $+n$ and $-n$ terms. Thus,
\begin{align}
\tanh\bigl( \frac \beta 2H_{\Delta}\bigr) &= -\frac 2\beta \sum_{n\in \Zbb} \frac{1}{\i \omega_n - H_{\Delta}}. \label{DHS1:tanh-expansion}
\end{align}
We use this representation to expand the operator in \eqref{DHS1:lncosh-operator_fundamental_theorem} in powers of $\Delta$ using the resolvent equation. The first term obtained in this way is $\langle \Delta, L_{T,B}\Delta\rangle$ with the linear operator $L_{T,B} \colon \Lsymm \ra \Lsymm$, given by
\begin{align}
L_{T,B}\Delta &:= -\frac 2\beta \sum_{n\in \Zbb} (\i \omega_n - \hfrak_B)^{-1} \, \Delta \,  (\i \omega_n + \ov{\hfrak_B})^{-1}. \label{DHS1:LTB_definition}
\end{align}
In the temperature regime we are interested in, we will obtain the quadratic terms in the Ginzburg--Landau functional from $\langle \Delta, L_{T,B}\Delta\rangle$. 


The next term in the expansion of \eqref{DHS1:tanh-expansion} is the quartic term $\langle \Delta, N_{T,B}(\Delta)\rangle$ with the nonlinear map $N_{T,B}\colon \Hsymm \ra \Lsymm$ defined as
\begin{align}
N_{T,B}(\Delta) &:= \frac 2\beta \sum_{n\in \Zbb} (\i \omega_n - \hfrak_B)^{-1}\,  \Delta \,  (\i\omega_n + \ov{\hfrak_B})^{-1} \, \ov \Delta \,  (\i\omega_n - \hfrak_B)^{-1}\, \Delta \, (\i\omega_n + \ov{\hfrak_B})^{-1}. \label{DHS1:NTB_definition}
\end{align}
The expression $\langle \Delta, N_{T,B}(\Delta)\rangle$ will determine the quartic term in the Ginzburg--Landau functional. All higher order terms in the expansion of \eqref{DHS1:lncosh-operator_fundamental_theorem} in $\Delta$ will be summarized in a trace-class operator called $\Rcal_{T,B}(\Delta)$, whose local trace norm is small.



With the operators $L_{T,B}$ and $N_{T,B}$ at hand, we are in position to state a representation formula for the BCS functional. It serves as the fundamental equation, on which the proofs of Theorems \ref{DHS1:Main_Result} and \ref{DHS1:Main_Result_Tc} are based. In particular, it will be applied in the proofs of upper and lower bounds, and we therefore formulate the statement for a general state $\Gamma$ and not only for Gibbs states.

\begin{prop}[Representation formula for the BCS functional]
\label{DHS1:BCS functional_identity} \label{DHS1:BCS FUNCTIONAL_IDENTITY}
Let $\Gamma$ be an admissible state. For any $B>0$, let $\Psi\in \Hmag^1(Q_B)$ and let $\Delta \equiv \Delta_\Psi$ be as in \eqref{DHS1:Delta_definition}. For $T>0$ and if $V\alpha_*\in L^{\nicefrac 65}(\Rbb^3) \cap L^2(\Rbb^3)$, there is an operator $\Rcal_{T,B}(\Delta)\in \Scal^1$ such that
\begin{align}
\FBCS(\Gamma) - \FBCS(\Gamma_0)& \notag\\
&\hspace{-70pt}= - \frac 14 \langle \Delta, L_{T,B} \Delta\rangle + \frac 18 \langle \Delta, N_{T,B} (\Delta)\rangle + \Vert \Psi\Vert_{\Lmag^2(Q_B)}^2 \; \langle \alpha_*, V\alpha_*\rangle_{L^2(\Rbb^3)} \notag\\
&\hspace{-40pt}+ \Tr\bigl[\Rcal_{T,B}(\Delta)\bigr] \notag\\
&\hspace{-40pt}+ \frac{T}{2} \Hcal_0(\Gamma, \Gamma_\Delta) - \fint_{Q_B} \dd X \int_{\Rbb^3} \dd r \; V(r) \, \bigl| \alpha(X,r) - \alpha_*(r) \Psi(X)\bigr|^2, \label{DHS1:BCS functional_identity_eq}
\end{align}
where
\begin{align}
	\Hcal_0(\Gamma, \Gamma_\Delta) := \Tr_0\bigl[ \Gamma(\ln \Gamma - \ln \Gamma_\Delta) + (1 - \Gamma)(\ln(1-\Gamma) - \ln(1 - \Gamma_\Delta))\bigr] \label{DHS1:Relative_Entropy}
\end{align}
denotes the relative entropy of $\Gamma$ with respect to $\Gamma_\Delta$. Moreover, $\Rcal_{T,B}(\Delta)$ obeys the estimate
\begin{align*}
	\Vert\Rcal_{T,B}(\Delta) \Vert_1 \leq C \; T^{-5} \; B^3 \; \Vert \Psi\Vert_{\Hmag^1(Q_B)}^6.
\end{align*}
\end{prop}

The relative entropy defined in \eqref{DHS1:Relative_Entropy} is based on the weaker form of trace $\Tr_0$, whose introduction we postpone until \eqref{DHS1:Weak_trace_definition}.

The right side of \eqref{DHS1:BCS functional_identity_eq} should be read as follows. The first line yields the Ginzburg--Landau functional, see Theorem \ref{DHS1:Calculation_of_the_GL-energy} below. 
The second and third line consist of remainder terms. The second line is small in absolute value whereas the techniques used to bound the third line differ for upper and lower bounds. This is responsible for the different qualities of the upper and lower bounds in Theorems \ref{DHS1:Main_Result} and \ref{DHS1:Main_Result_Tc}, see \eqref{DHS1:Rcal_error_Definition}. For an upper bound, when choosing $\Gamma := \Gamma_\Delta$ as a trial state, the relative entropy term $\Hcal_0(\Gamma_\Delta, \Gamma_\Delta)=0$ drops out. The last term in \eqref{DHS1:BCS functional_identity_eq} is nonpositive by our assumptions on $V$ and can be dropped for an upper bound. In case of an interaction potential without a sign it can easily be estimated using Proposition~\ref{DHS1:Structure_of_alphaDelta}. In case of the proof of the lower bound for the BCS free energy, we need to estimate the third line in \eqref{DHS1:BCS functional_identity_eq} from below using a relative entropy estimate that we provide in Section \ref{DHS1:Lower Bound Part B}.

It remains to show that the first line of the right side of \eqref{DHS1:BCS functional_identity_eq} is indeed given by the Ginzburg--Landau functional.
In order to state the result, we need the function
\begin{align}
\hat{V\alpha_*}(p) := \int_{\Rbb^3} \dx\; \e^{-\i p\cdot x} \, V(x)\alpha_*(x), \label{DHS1:Gap_function}
\end{align}
which fixes our convention on the Fourier transform in this paper. 

\begin{thm}[Calculation of the GL energy]
\label{DHS1:Calculation_of_the_GL-energy} \label{DHS1:CALCULATION_OF_THE_GL-ENERGY}
Let Assumptions \ref{DHS1:Assumption_V} and \ref{DHS1:Assumption_KTc} (a) hold and let $D\in \Rbb$ be given. Then, there is a constant $B_0>0$ such that for any $0 < B \leq B_0$, any $\Psi\in \Hmag^2(Q_B)$, $\Delta \equiv \Delta_\Psi$ as in \eqref{DHS1:Delta_definition}, and $T = \Tc(1 - DB)$, we have
\begin{align}
- \frac 14 \langle \Delta, L_{T,B} \Delta\rangle + \frac 18 \langle \Delta, N_{T,B} (\Delta)\rangle + \Vert \Psi\Vert_{\Lmag^2(Q_B)}^2 \; \langle \alpha_*, V\alpha_*\rangle_{L^2(\Rbb^3)} & \notag\\
&\hspace{-40pt}= B^2\; \EGL(\Psi) + R(B). \label{DHS1:Calculation_of_the_GL-energy_eq}
\end{align}
Here,
\begin{align*}
|R(B)|\leq C \, B^3 \; \Vert\Psi\Vert_{\Hmag^2(Q_B)}^2 \;  \bigl[ 1 + \Vert \Psi\Vert_{\Hmag^1(Q_B)}^2 \bigr]
\end{align*}
and with the functions
%
\begin{align}
g_1(x) &:= \frac{\tanh(x/2)}{x^2} - \frac{1}{2x}\frac{1}{\cosh^2(x/2)}, & g_2(x) &:= \frac 1{2x} \frac{\tanh(x/2)}{\cosh^2(x/2)}, \label{DHS1:XiSigma}
\end{align}
the coefficients $\Lambda_0$, $\Lambda_2$, and $\Lambda_3$ in $\EGL$ are given by
\begin{align}
\Lambda_0 &:= \frac{\betac^2}{16} \int_{\Rbb^3} \frac{\dd p}{(2\pi)^3} \; |(-2)\hat{V\alpha_*}(p)|^2 \; \bigl( g_1 (\betac(p^2-\mu)) + \frac 23 \betac \, p^2\, g_2(\betac(p^2-\mu))\bigr), \label{DHS1:GL-coefficient_1}\\
\Lambda_2 &:= \frac{\betac}{8} \int_{\Rbb^3} \frac{\dd p}{(2\pi)^3} \; \frac{|(-2)\hat{V\alpha_*}(p)|^2}{\cosh^2(\frac{\betac}{2}(p^2 -\mu))},\label{DHS1:GL-coefficient_2} \\
\Lambda_3 &:= \frac{\betac^2}{16} \int_{\Rbb^3} \frac{\dd p}{(2\pi)^3} \; |(-2) \hat{V\alpha_*}(p)|^4 \;  \frac{g_1(\betac(p^2-\mu))}{p^2-\mu}.\label{DHS1:GL_coefficient_3}
\end{align}
%
%
\end{thm}


Let us comment on the positivity of the coefficients \eqref{DHS1:GL-coefficient_1}-\eqref{DHS1:GL_coefficient_3}. First, $\Lambda_2$ is trivially positive. 
Since $g_1(x)/x >0$ for all $x\in \Rbb$, the coefficient $\Lambda_3$ is positive as well. It cannot be immediately seen that $\Lambda_0$ is positive, however. In order to prove this, we introduce the positive function 
\begin{align*}
g_3(x) &:= \frac{2}{x^2} \frac{1}{\cosh^2(x/2)} - \frac{1}{x} \frac{1}{\tanh(x/2)} \frac{1}{\cosh^2(x/2)}
\end{align*}
and compute
\begin{align}
2\Re \langle \alpha_* , x_i (K_{\Tc} - V) x_i \alpha_*\rangle &= (2\pi)^{-3} \langle \hat{V\alpha_*} , K_{\Tc}(p)^{-1} [ -\i \partial_{p_i} , [K_{\Tc}(p) , -\i \partial_{p_i}]] K_{\Tc}(p)^{-1}  \hat {V\alpha_*}\rangle \notag\\
&= 8\, \Lambda_0 - 2 \betac^3 \int_{\Rbb^3} \frac{\dd p}{(2\pi)^3} \; |\hat{V\alpha_*}(p)|^2 \; p_i^2 \, g_3(\betac(p^2-\mu)). \label{DHS1:GL-coefficient_1_positive}
\end{align}
Since the left side is nonnegative, this proves that $\Lambda_0 >0$. The idea for this proof is borrowed from \cite[Eq. (1.22)]{Hainzl2012}.

Let us comment on the connection between \eqref{DHS1:Calculation_of_the_GL-energy_eq} and \cite{Hainzl2017}. The two-particle Birman--Schwinger operator $1 - V^{\nicefrac 12} L_{T,B} V^{\nicefrac 12}$ has been intensively studied in \cite{Hainzl2017} 
to identify temperature regimes, where the bottom of its spectrum is positive or negative. This operator also appears in \eqref{DHS1:Calculation_of_the_GL-energy_eq} because
\begin{align}
- \frac 14 \langle \Delta, L_{T,B} \Delta\rangle + \Vert \Psi\Vert_2^2 \, \langle \alpha_*, V\alpha_*\rangle = \bigl\langle V^{\nicefrac 12} \alpha_* \Psi, \bigl( 1 - V^{\nicefrac 12} L_{T,B} V^{\nicefrac 12} \bigr) V^{\nicefrac 12} \alpha_* \Psi\bigr\rangle. \label{DHS1:Birman-Schwinger_LTB}
\end{align}
That is, the question whether the bottom of the spectrum of $1 - V^{\nicefrac 12} L_{T,B} V^{\nicefrac 12}$ is positive or negative is intimately related to the sign of \eqref{DHS1:Calculation_of_the_GL-energy_eq}, and thus of \eqref{DHS1:BCS functional_identity_eq} and \eqref{DHS1:BCS GS-energy}. Accordingly, it is related to the question whether the systems displays superconductivity or not. We highlight that the operator on the right side of \eqref{DHS1:Birman-Schwinger_LTB} acts on functions in  $L^2(\mathbb{R}^6)$ in \cite{Hainzl2017}, while it acts on $\Lsymm$ in our case. Since the lowest eigenvalue of the operator $(-\i\nabla + 2 \Abold)^2$ equals $2B$ when understood to act on $L^2(\Rbb^3)$ or on $\Lmag^2(Q_B)$, we obtain the same asymptotic behavior of $\Tc(B)$ as in \cite[Theorem~4]{Hainzl2017}.

%

Theorem \ref{DHS1:Calculation_of_the_GL-energy} is valid for the precise temperature scaling $T = \Tc(1 - DB)$. In order to prove Theorem \ref{DHS1:Main_Result_Tc} (a), we also need to show that the system is superconducting for temperatures that are small compared to $\Tc (1- DB)$. This is guaranteed by the following proposition.

\begin{prop}[A priori bound on Theorem \ref{DHS1:Main_Result_Tc} (a)]
\label{DHS1:Lower_Tc_a_priori_bound}
Let Assumptions \ref{DHS1:Assumption_V} and \ref{DHS1:Assumption_KTc} (a) hold. Then, for every $T_0 > 0$ there are constants $B_0>0$ and $D_0>0$ such that for all $0 < B \leq B_0$ and all temperatures $T$ obeying
\begin{align*}
T_0 \leq T < \Tc (1 - D_0 B),
\end{align*}
there is an admissible BCS state $\Gamma$ with
\begin{align}
\FBCS(\Gamma) - \FBCS(\Gamma_0) < 0. \label{DHS1:Lower_critical_shift_2}
\end{align}
\end{prop}


\subsection{The upper bound on \texorpdfstring{(\ref{DHS1:ENERGY_ASYMPTOTICS})}{(\ref{DHS1:ENERGY_ASYMPTOTICS})} and proof of Theorem \ref{DHS1:Main_Result_Tc} (a)}
\label{DHS1:Upper_Bound_Proof_Section}

Using the results in the previous section, we provide the proofs of the upper bound on \eqref{DHS1:ENERGY_ASYMPTOTICS} and of Theorem \ref{DHS1:Main_Result_Tc} (a). The statements in the previous section, that is, Propositions \ref{DHS1:Structure_of_alphaDelta} and \ref{DHS1:BCS functional_identity}, as well as Theorem \ref{DHS1:Calculation_of_the_GL-energy} are proven in Section \ref{DHS1:Proofs}.

\begin{proof}[Proof of the upper bound on \eqref{DHS1:ENERGY_ASYMPTOTICS}]
Let $D\in \Rbb$ be given, let $D_0 := 1 +|D|$, and let $\Psi$ be a minimizer of the Ginzburg--Landau functional, i.e., $\mathcal E_{D,B}^{\mathrm{GL}}(\Psi) = \EGLGSE$. We note that $\Psi$ belongs to $\Hmag^2(Q_B)$ and has uniformly bounded $\Hmag^2(Q_B)$-norm. Let $\Delta \equiv\Delta_\Psi$ be as in \eqref{DHS1:Delta_definition} and let $T = \Tc(1 - DB)$. We apply Proposition \ref{DHS1:BCS functional_identity} with the choice $\Gamma = \Gamma_\Delta$ and find
\begin{align}
\FBCS(\Gamma_\Delta) - \FBCS(\Gamma_0) 
%
%
&\leq - \frac 14 \langle \Delta, L_{T,B} \Delta\rangle + \frac 18 \langle \Delta, N_{T,B} (\Delta)\rangle + \Vert \Psi\Vert_2^2 \, \langle \alpha_*, V\alpha_*\rangle \notag\\
&\hspace{-20pt} - \int_{Q_B} \dd X \int_{\Rbb^3} \dd r \; V(r) \, \bigl| \alpha_\Delta(X,r) - \alpha_*(r) \Psi(X)\bigr|^2 + C B^3.\label{DHS1:Upper_Bound_proof_1}
\end{align}
The first term in the last line is bounded by $\Vert V\Vert_\infty \Vert \eta\Vert_2^2$ and a bound for the $L^2$-norm of $\eta := \eta_0 + \eta_\perp$ is provided by \eqref{DHS1:alphaDelta_decomposition_eq2} and \eqref{DHS1:alphaDelta_decomposition_eq3}. In fact, by Assumption \ref{DHS1:Assumption_V}, this term is nonpositive but we do not need to use this here. By Theorem~\ref{DHS1:Calculation_of_the_GL-energy}, this implies
\begin{align*}
F^{\mathrm{BCS}}(\Tc(1 - DB), B) &\leq B^2 \, \EGLGSE + C B^3,
\end{align*}
which concludes the proof of the upper bound on \eqref{DHS1:ENERGY_ASYMPTOTICS}.
\end{proof}


\begin{proof}[Proof of Theorem~\ref{DHS1:Main_Result_Tc}~(a)]
Let $D_0 > 0$ be given and let us recall the definition of $\Dc$ in \eqref{DHS1:Dc_Definition}. We show that there is a constant $D_1>0$ and appropriate trial states such that \eqref{DHS1:Lower_critical_shift_2} holds for all temperatures $T$ obeying
\begin{align}
\Tc(1 - D_0 B) \leq T < \Tc (1 -  \Dc\, B - D_1 \, B^{\nicefrac 32}), \label{DHS1:Lower_critical_shift_3}
\end{align}
provided $B>0$ is small enough. Since Proposition \ref{DHS1:Lower_Tc_a_priori_bound} covers the remaining range of $T$, this proves Theorem~\ref{DHS1:Main_Result_Tc}~(a).

We define $D := \frac{\Tc - T}{B\Tc}$ and note that \eqref{DHS1:Lower_critical_shift_3} yields $D - \Dc > D_1B^{\nicefrac 12}$. Let $\psi \in \Hmag^2(Q_1)$ be a ground state of the linear operator in \eqref{DHS1:Dc_Definition} and let $\Psi$ be as in \eqref{DHS1:GL-rescaling}. Accordingly, we have $(\Lambda_0 / \Lambda_2) \Pi^2 \Psi = B\Dc \Psi$ and
\begin{align*}
\inf_{\theta \in \Rbb} \EGL(\theta \Psi) = - \frac{\Lambda_2^2(D - \Dc)^2 \Vert \psi\Vert_2^4}{4 \Lambda_3 \Vert \psi\Vert_4^4},
\end{align*}
where the optimal $\mathrm{\theta_c}$ satisfies $\Lambda_2 (D - \Dc) \Vert \psi\Vert_2^2 = 2\Lambda_3 \Vert \psi\Vert_4^4 \, \mathrm{\theta_c}^2$. We combine Proposition~\ref{DHS1:BCS functional_identity} and Theorem \ref{DHS1:Calculation_of_the_GL-energy} applied to $\Gamma = \Gamma_\Delta$ with $\Delta = \Delta_{\mathrm{\theta_c}\Psi}$, to see that \eqref{DHS1:Upper_Bound_proof_1} holds in this case as well. Let us note that \eqref{DHS1:Lower_critical_shift_3} implies $|T - \Tc| \leq CB$. Proposition \ref{DHS1:Structure_of_alphaDelta} and \eqref{DHS1:Upper_Bound_proof_1} therefore allow us to conclude that
\begin{align}
\FBCS(\Gamma_\Delta) - \FBCS(\Gamma_0) &\leq - \frac{\Lambda_2^2 \Vert \psi\Vert_2^4}{4 \Lambda_3\Vert \psi\Vert_4^4} \; (D - \Dc)^2\; B^2 + CB^3.
\end{align}
The right side is negative provided $D_1>0$ is chosen large enough since $D - \Dc > D_1 B^{\nicefrac 12}$. This shows \eqref{DHS1:Lower_critical_shift_2} for temperatures $T$ satisfying \eqref{DHS1:Lower_critical_shift_3} and completes the proof of Theorem~\ref{DHS1:Main_Result_Tc}~(a).
\end{proof}


\section{Proofs of the Results in Section \ref{DHS1:Upper_Bound}}
\label{DHS1:Proofs}




\subsection{Schatten norm estimates for operators given by product kernels}
\label{DHS1:Estimates_on_product_wave_functions_Section}

In this subsection we provide estimates for several norms of gauge-periodic operators with integral kernels given by product functions of the form $\tau(x-y) \Psi((x+y)/2)$, which will be used frequently in our proofs.

\begin{lem}
\label{DHS1:Schatten_estimate}
Let $B>0$, let $\Psi$ be a gauge-periodic function on $Q_B$ and let $\tau$ be an even and real-valued function on $\Rbb^3$. Moreover, let the operator $\alpha$ be defined via its integral kernel $\alpha(X,r) := \tau(r)\Psi(X)$, i.e., $\alpha$ acts as
\begin{align*}
\alpha f(x) &= \int_{\Rbb^3} \dd y \; \tau(x - y) \Psi\bigl(\frac{x+y}{2}\bigr) f(y), & f &\in L^2(\Rbb^3).
\end{align*}

\begin{enumerate}[(a)]
\item Let $p \in \{2,4,6\}$. If $\Psi\in \Lmag^p(Q_B)$ and $\tau \in L^{\frac {p}{p-1}}(\Rbb^3)$, then $\alpha \in \Scal^p$ and
\begin{align*}
\Vert \alpha\Vert_p \leq C \; \Vert \tau\Vert_{\frac{p}{p-1}} \; \Vert \Psi\Vert_p.
\end{align*}

\item For any $\nu > 3$, there is a $C_\nu >0$, independent of $B$, such that if $(1 +|\cdot|)^\nu \tau\in L^{\nicefrac 65}(\Rbb^3)$ and $\Psi\in \Lmag^6(Q_B)$, then $\alpha \in \Scal^\infty$ and
\begin{align*}
\Vert \alpha\Vert_\infty &\leq C_\nu \, B^{-\nicefrac 14} \; \max\{1 , B^{\nicefrac \nu 2}\} \; \Vert (1 + |\cdot|)^\nu \tau\Vert_{\nicefrac 65} \; \Vert \Psi\Vert_6.
\end{align*}
\end{enumerate}
\end{lem}

\begin{proof}
The case $p = 2$ of part (a) holds trivially with equality and $C = 1$. Since $\tau$ is even and real-valued, the kernel of $\alpha^*\alpha$ is given by
\begin{align*}
\alpha^* \alpha(x,y) &= \int_{\Rbb^3} \dd z \; \tau(x-z) \ov{\Psi\bigl( \frac{x+z}{2}\bigr)} \tau(z - y) \Psi\bigl( \frac{z+y}{2}\bigr).
\end{align*}
Using $\Vert \alpha\Vert_4^4 = \Vert \alpha^* \alpha\Vert_2^2$ and the change of variables $z \mapsto x-z$ and $y \mapsto x-y$, we see that
\begin{align*}
\Vert \alpha\Vert_4^4 &= \frac{1}{|Q_B|} \iint_{Q_B\times \Rbb^3} \dx \dy \; \Bigl| \int_{\Rbb^3} \dd z \; \tau(z) \, \tau(y-z) \,  \ov{\Psi\bigl( x - \frac z2\bigr)}\, \Psi\bigl( x-\frac{y+z}{2}\bigr)\Bigr|^2.
\end{align*}
By Hölder's inequality, Young's inequality and 
the fact that $|\Psi|$ is periodic, we see that
\begin{align*}
\Vert \alpha\Vert_4^4 &\leq \Vert \Psi\Vert_4^4 \; \int_{\Rbb^3} \dd y \, \Bigl| \int_{\Rbb^3} \dd z \; |\tau(z) \tau(y-z)|\Bigr|^2 \leq C  \; \Vert \tau\Vert_{\nicefrac 43}^4\; \Vert \Psi\Vert_4^4
\end{align*}
holds. This proves part (a) for $p =4$. If $p =6$ we use $\Vert \alpha\Vert_6^6 = \Vert \alpha \,\alpha^*\alpha\Vert_2^2$ and a similar change of variables to write
\begin{align*}
\Vert \alpha \alpha^* \alpha\Vert_2^2 &= \frac{1}{|Q_B|} \iint_{Q_B\times \Rbb^3} \dd x \dd y\; \Bigl| \iint_{\Rbb^3\times \Rbb^3} \dd z_1 \dd z_2 \; \tau(z_1) \, \tau(z_2 - z_1) \, \tau(y - z_2) \\
&\hspace{120pt} \times \Psi\bigl(x - \frac{z_1}{2}\bigr) \ov{ \Psi\bigl( x - \frac{x_1+ z_2}{2}\bigr)} \Psi\bigl( x - \frac{z_2 + y}{2}\bigr)\Bigr|^2.
\end{align*}
We thus obtain $\Vert \alpha\Vert_6^6 \leq \Vert \Psi\Vert_6^6 \; \Vert \tau * \tau * \tau \Vert_2^2$, which, in combination with Young's inequality, proves the claimed bound.

In case of part (b), we follow closely the strategy of the proof of \cite[Eq. (5.51)]{Hainzl2012}. Let $f,g\in L^2(\Rbb^3)$ and let $\chi_j$ denote the characteristic function of the cube with side length $\sqrt{2\pi B^{-1}}$ centered at $j\in \Lambda_B$. We estimate
\begin{align}
|\langle f, \alpha g\rangle| &\leq \sum_{j,k\in \Lambda_B}\iint_{\Rbb^3\times \Rbb^3} \dd x\dd y\; \bigl|\chi_j(x) f(x) \Psi\bigl( \frac{x+y}{2}\bigr) \tau(x - y)\chi_k(y)g(y)\bigr|. \label{DHS1:Schatten_estimate_1}
\end{align}
Let $|\cdot|_\infty$ and $| \cdot |$ denote the maximum norm and the euclidean norm on $\Rbb^3$, respectively. We observe that the estimates $|x - j|_\infty \leq \frac 12\sqrt{2\pi B^{-1}}$ and $|y - k|_\infty \leq \frac 12\sqrt{2\pi B^{-1}}$ imply $| \frac{x + y}{2} - \frac{j + k}{2}|_\infty \leq \frac 12\sqrt{2\pi B^{-1}}$. Accordingly, if $\chi_j(x)\chi_k(y)$ equals $1$, so does $\chi_{\frac{j+k}{2}}(\frac{x+y}{2})$ and we may replace $\Psi$ on the right side of \eqref{DHS1:Schatten_estimate_1} by $\chi_{\frac{j+k}{2}}\Psi$ without changing the term. The above bounds for $|x - j|_\infty$ and $|y - k|_\infty$ also imply $|j - k| \leq |x - y| + \sqrt{6\pi B^{-1}}$, which yields the lower bound
\begin{align}
|x - y| \geq \bigl[ |j -k| - \sqrt{6\pi B^{-1}}\bigr]_+. \label{DHS1:Schatten_estimate_2}
\end{align}
We choose $\nu >3$, insert the factor $(\sqrt{2\pi B^{-1}} + |x-y|)^\nu$ and its inverse in \eqref{DHS1:Schatten_estimate_1}, use \eqref{DHS1:Schatten_estimate_2} to estimate the inverse, apply Cauchy--Schwarz in the $x$-coordinate, and obtain
\begin{align*}
|\langle f, \alpha g\rangle| &\leq \sum_{j,k\in \Lambda_B} \bigl( \sqrt{2\pi B^{-1}} + \bigl[|j-k| - \sqrt{6\pi B^{-1}}\bigr]_+\bigr)^{-\nu} \; \Vert \chi_jf\Vert_2 \\
&\hspace{-20pt} \times \Bigl( \int_{\Rbb^3} \dd x \, \Bigl| \int_{\Rbb^3} \dd y\; \bigl| (\chi_{\frac{j+k}{2}} \Psi)\bigl( \frac{x+y}{2}\bigr) \bigl( \sqrt{2\pi B^{-1}} + |x-y| \bigr)^\nu \tau(x-y) \chi_k(y)g(y)\bigr| \; \Bigr|^2\Bigr)^{\nicefrac 12}.
\end{align*}
An application of Hölder's inequality in the $y$-coordinate shows that the second line is bounded by 
\begin{align*}
\Bigl\Vert \bigl|(\sqrt{2\pi B^{-1}} + |\cdot|)^\nu \tau \bigr|^{\nicefrac 65} * |\chi_kg|^{\nicefrac 65} \Bigr\Vert_{\nicefrac 53}^{\nicefrac 56} &\leq \bigl\Vert \bigl( \sqrt{2\pi B^{-1}} + |\cdot| \bigr)^\nu  \tau \bigr\Vert_{\nicefrac 65} \, \Vert \chi_kg \Vert_{2} 
%
\end{align*}
times $|Q_B|^{\nicefrac 16} \Vert \Psi\Vert_{\Lmag^6(Q_B)}$. We highlight that the $\Lmag^6(Q_B)$-norm is defined via a normalized integral, whence we needed to insert the factor of $|Q_B|^{-\nicefrac 16}$. 
Hence,
\begin{align*}
|\langle f, \alpha g\rangle| &\leq C B^{-\nicefrac 14} \, \Vert \Psi\Vert_6 \, \bigl\Vert \bigl( \sqrt{2\pi B^{-1}}  + |\cdot|\bigr)^\nu \tau\bigr\Vert_{\nicefrac 65} \\
&\hspace{50pt} \times \sum_{j,k\in \Lambda_B} \bigl( \sqrt{2\pi B^{-1}} + \bigl[ |j -k| - \sqrt{6\pi B^{-1}} \bigr]_+\bigr)^{-\nu} \; \Vert \chi_jf\Vert_2 \; \Vert \chi_kg\Vert_2 .
\end{align*}
For $\lambda >0$ we estimate $\Vert \chi_j f\Vert_2 \Vert \chi_k g\Vert_2 \leq \frac{\lambda}{2} \Vert \chi_j f\Vert_2^2 + \frac{1}{2 \lambda} \Vert \chi_kg\Vert_2^2$. In each term, we carry out one of the sums and optimize the resulting expression over $\lambda$. We find $\lambda = \Vert g\Vert_2 \; \Vert f\Vert_2^{-1}$ as well as
\begin{align*}
|\langle f, \alpha g\rangle| &\leq C B^{-\nicefrac 14}  \Vert f\Vert_2 \; \Vert g\Vert_2 \; \Vert \Psi\Vert_6 \; \frac{\Vert ( \sqrt{2\pi B^{-1}} +|\cdot|)^\nu \tau\Vert_{\nicefrac 65}}{(2\pi B^{-1})^{\nicefrac \nu 2}} \; \sum_{j\in \Zbb^3} \bigl( 1 + [ |j| - \sqrt{3}]_+\bigr)^{-\nu}.
\end{align*}
The fraction involving $\tau$ is bounded by $C_\nu\max\{1, B^{\nicefrac \nu 2}\} \Vert (1 + |\cdot|)^\nu \tau \Vert_{\nicefrac 65}$. This proves the claim.
\end{proof}


\subsection{Proof of Lemma \ref{DHS1:Gamma_Delta_admissible}}
\label{DHS1:sec:proofofadmissibility}

We recall the definition of $\Gamma_\Delta$ in \eqref{DHS1:GammaDelta_definition} and that of the normal state $\Gamma_0$ in \eqref{DHS1:Gamma0}. By definition, $\Gamma_\Delta$ is a gauge-periodic generalized fermionic one-particle density matrix. Therefore, we only have to check the trace class condition \eqref{DHS1:Gamma_admissible}.

To this end, we use the expansion \eqref{DHS1:tanh-expansion} of the hyperbolic tangent in terms of the Matsubara frequencies, the identity $(\exp(x) + 1 )^{-1}= (1 - \tanh( x/2 ))/2$, and the resolvent equation 
\begin{align}
	(z-H_\Delta)^{-1} = (z-H_0)^{-1} + (z-H_0)^{-1} \; (H_\Delta - H_0)\; (z-H_\Delta)^{-1} \label{DHS1:Resolvent_Equation}
\end{align}
to write
\begin{align}
	\Gamma_\Delta = \frac 12 - \frac 12 \tanh\bigl( \frac \beta 2 H_\Delta\bigr) = \frac 12 + \frac{1}{\beta} \sum_{n\in \Zbb} \frac{1}{\i \omega_n - H_\Delta} = \Gamma_0 + \Ocal + \Qcal_{T,B}(\Delta), \label{DHS1:alphaDelta_decomposition_1}
\end{align}
where
\begin{align}
	\Ocal &:= \frac 1\beta \sum_{n\in \Zbb} \frac{1}{ \i \omega_n - H_0} \delta \frac{1}{ \i \omega_n - H_0}, & 
	\Qcal_{T,B}(\Delta) &:= \frac 1\beta \sum_{n\in \Zbb} \frac{1}{ \i \omega_n - H_0} \delta\frac{1}{ \i \omega_n - H_0} \delta \frac{1}{ \i \omega_n - H_\Delta} \label{DHS1:alphaDelta_decomposition_2}
\end{align}
with $\delta$ in \eqref{DHS1:HDelta_definition}. Since $\Ocal$ is offdiagonal, we have $[\Ocal]_{11} =0$ and the operator $(1+ \pi^2) [\Ocal]_{11}$ is locally trace class trivially. Using \eqref{DHS1:Calculation-entry}, we see that
\begin{align*}
	\bigl[ \Qcal_{T,B}(\Delta)\bigr]_{11} = \frac 1\beta \sum_{n\in \Zbb} \frac{1}{\i\omega_n - \hfrak_B} \Delta \frac{1}{\i\omega_n + \ov{\hfrak_B}} \ov \Delta \Bigl[ \frac{1}{\i\omega_n - H_\Delta}\Bigr]_{11}.
\end{align*}
An application of Hölder's inequality shows that $(1+\pi^2)[\Qcal_{T,B}(\Delta)]_{11}$ is locally trace class. It remains to show that $(1+\pi^2) \gamma_0$ is locally trace class. But this follows from the bound $(1+x) (\exp(\beta (x - \mu))+1)^{-1} \leq C_{\beta,a} \e^{-\frac \beta 2 (x-\mu)}$ for $x \geq a$, the diamagnetic inequality for the magnetic heat kernel, see e.g. \cite[Theorem 4.4]{LiebSeiringer}, and the explicit formula for the heat kernel of the Laplacian. This concludes the proof of Lemma~\ref{DHS1:Gamma_Delta_admissible}.


\subsection{Proof of Proposition \ref{DHS1:BCS functional_identity}}
\label{DHS1:BCS functional_identity_proof_Section}

We recall the definitions of $\Delta(X, r) = -2\, V\alpha_*(r)\, \Psi(X)$ in \eqref{DHS1:Delta_definition}, the Hamiltonian $H_\Delta$ in \eqref{DHS1:HDelta_definition} and $\Gamma_\Delta = (1 + \e^{\beta H_\Delta})^{-1}$ in \eqref{DHS1:GammaDelta_definition}. Throughout this section we assume that the function $\Psi$ in the definition of $\Delta$ is in $\Hmag^1(Q_B)$. From  Lemma~\ref{DHS1:Gamma_Delta_admissible}, which is proved in Section~\ref{DHS1:sec:proofofadmissibility} below, we know that $\Gamma_{\Delta}$ is an admissible BCS state in this case. We define the anti-unitary operator
\begin{align*}
\Jcal &:= \begin{pmatrix} 0 & J \\ -J & 0\end{pmatrix} 
\end{align*}
with $J$ defined below \eqref{DHS1:Gamma_introduction}. The operator $H_\Delta$ obeys the relation $\Jcal H_\Delta \Jcal^* = - H_\Delta$, which implies $\Jcal\Gamma_\Delta \Jcal^* = 1 - \Gamma_\Delta$. Using this and the cyclicity of the trace, we write the entropy of $\Gamma_{\Delta}$ as
\begin{align}
S(\Gamma_\Delta) = \frac 12 \Tr[ \varphi(\Gamma_\Delta)] \label{DHS1:entropy matrix},
\end{align}
where $\varphi(x) := -[x\ln(x) + (1 - x) \ln(1-x)]$ for $0 \leq x \leq 1$.

In order to rewrite the BCS functional, it is useful to introduce a weaker notion of trace per unit volume. More precisely, we call a gauge-periodic operator $A$ acting on $L^2(\Rbb^3)\oplus L^2(\Rbb^3)$ weakly locally trace class if $P_0AP_0$ and $Q_0AQ_0$ are locally trace class, where
\begin{align}
P_0 = \begin{pmatrix} 1 & 0 \\ 0 & 0 \end{pmatrix} \label{DHS1:P0}
\end{align}
and $Q_0 = 1-P_0$, and we define its weak trace per unit volume by
\begin{align}
\Tr_0 (A):= \Tr\bigl( P_0AP_0 + Q_0 AQ_0\bigr). \label{DHS1:Weak_trace_definition}
\end{align}
If an operator is locally trace class then it is also weakly locally trace class but the converse need not be true. It is true, however, in case of nonnegative operators. If an operator is locally trace class then its weak trace per unit volume and its usual trace per unit volume coincide. 

Before their appearance in the context of BCS theory in \cite{Hainzl2012,Hainzl2014}, weak traces of the above kind appeared in \cite{HLS05,FLLS11}. In \cite[Lemma 1]{HLS05} it has been shown that if two weak traces $\Tr_P$ and $\Tr_{P'}$ are defined via projections $P$ and $P'$ then $\Tr_{P}(A) = \Tr_{P'}(A)$ holds for appropriate $A$ if $P - P'$ is a Hilbert--Schmidt operator. 

%

Let $\Gamma$ be an admissible BCS state and recall the normal state $\Gamma_0$ in \eqref{DHS1:Gamma0}. 
In terms of the weak trace per unit volume, the BCS functional can be written as
\begin{align}
\FBCS(\Gamma) - \FBCS(\Gamma_0) \hspace{-90pt}& \notag\\
&= \frac 12 \Tr\bigl[  (H_0\Gamma - H_0\Gamma_0) - T \varphi(\Gamma) + T \varphi(\Gamma_0)\bigr] - \fint_{Q_B} \dd X \int_{\Rbb^3} \dd r \; V(r)\, |\alpha(X,r)|^2 \notag\\
&= \frac 12 \Tr_0\bigl[ (H_\Delta\Gamma_\Delta - H_0\Gamma_0) - T\varphi(\Gamma_\Delta) + T\varphi(\Gamma_0)\bigr] \label{DHS1:kinetic energy} \\
&\hspace{55pt} + \frac 12 \Tr_0\bigl[ (H_\Delta \Gamma - H_\Delta \Gamma_\Delta) - T \varphi(\Gamma) + T\varphi(\Gamma_\Delta)\bigr] \label{DHS1:relative_entropy_term}\\
&\hspace{55pt} - \frac 12 \Tr_0 \begin{pmatrix}
0 & \Delta \\ \ov \Delta & 0 \end{pmatrix} \Gamma - \fint_{Q_B} \dd X \int_{\Rbb^3} \dd r \; V(r)\, |\alpha(X,r)|^2. \label{DHS1:interaction-term}
\end{align}
Note that we added and subtracted the first term in \eqref{DHS1:kinetic energy} and that we added and subtracted the first term in \eqref{DHS1:interaction-term} to replace the Hamiltonian $H_0$ in \eqref{DHS1:relative_entropy_term} by $H_\Delta$. The operators inside the traces in \eqref{DHS1:kinetic energy} and \eqref{DHS1:relative_entropy_term} are not necessarily locally trace class, which is the reason we introduce the weak local trace. We also note that \eqref{DHS1:relative_entropy_term} equals $\frac T2$ times the relative entropy $ \Hcal_0(\Gamma, \Gamma_\Delta)$ of $\Gamma$ with respect to $\Gamma_{\Delta}$, defined in \eqref{DHS1:Relative_Entropy}. 
%
%

The first term in \eqref{DHS1:interaction-term} can be written as
\begin{align}
-\frac  12\Tr_0 \begin{pmatrix} 0 & \Delta \\ \ov \Delta & 0\end{pmatrix} \Gamma &
=2\Re \fint_{Q_B} \dd X \int_{\Rbb^3} \dd r\; (V\alpha_*)(r) \Psi(X) \; \ov \alpha(X,r). \label{DHS1:additional-term}
\end{align}
The factors in \eqref{DHS1:interaction-term} and \eqref{DHS1:additional-term} multiplying $V$ are equal to
\begin{equation*}
-|\alpha(X,r)|^2 + 2\Re\alpha_*(r)\Psi(X) \; \ov \alpha (X,r)  = -\bigl|\alpha(X, r) - \alpha_*(r)\Psi(X)\bigr|^2 + \bigl|\alpha_*(r)\Psi(X)\bigr|^2. 
\end{equation*}
To rewrite \eqref{DHS1:kinetic energy} we need the following identities, whose proofs are straightforward computations:
\begin{align}
\Gamma_\Delta &= \frac 12 - \frac 12 \tanh\bigl( \frac \beta 2 H_\Delta\bigr), & \ln(\Gamma_\Delta) &= -\frac \beta 2 H_\Delta - \ln\bigl( 2\cosh\bigl( \frac \beta 2 H_\Delta\bigr)\bigr), \notag\\
1 - \Gamma_\Delta &= \frac 12 + \frac 12 \tanh\bigl( \frac\beta 2H_\Delta\bigr), & \ln(1 - \Gamma_\Delta) &= \frac\beta 2H_\Delta - \ln\bigl( 2\cosh\bigl( \frac \beta 2H_\Delta\bigr)\bigr).  \label{DHS1:GammaRelations}
\end{align}
Eq.~\eqref{DHS1:GammaRelations} implies
\begin{align}
\Gamma_\Delta \ln(\Gamma_\Delta) + (1-\Gamma_\Delta)\ln(1-\Gamma_\Delta) =  -\ln\bigl( 2\cosh\bigl( \frac \beta 2H_\Delta\bigr) \bigr) + \frac{\beta}{2} H_\Delta \tanh\bigl( \frac\beta 2 H_\Delta\bigr)\bigr),
\label{DHS1:GammaDeltaRelation}
\end{align}
as well as
\begin{align*}
\beta H_\Delta \Gamma_\Delta - \varphi(\Gamma_\Delta) &= \frac{\beta}{2} H_\Delta - \ln \bigl( 2 \cosh\bigl( \frac \beta 2H_\Delta\bigr)\bigr).
\end{align*}
This allows us to rewrite \eqref{DHS1:kinetic energy} as
\begin{align}
\frac 1{2\beta}\, \Tr_0\bigl[ (\beta H_\Delta\Gamma_\Delta - \beta H_0\Gamma_0) - \varphi(\Gamma_\Delta) + \varphi(\Gamma_0)\bigr] & \notag \\
&\hspace{-180pt}= \frac 14 \Tr_0 \bigl[ H_\Delta - H_0\bigr] - \frac 1{2\beta}\Tr_0 \bigl[ \ln\bigl( \cosh\bigl( \frac \beta 2H_\Delta\bigr)\bigr) - \ln\bigl( \cosh\bigl( \frac  \beta 2H_0\bigr)\bigr)\bigr]. \label{DHS1:trace-difference-ln}
\end{align}
We note that $H_\Delta - H_0$ is weakly locally trace class and that its weak trace equals $0$. This, in particular, implies that the second term on the right side of \eqref{DHS1:trace-difference-ln} is weakly locally trace class. To summarize, our intermediate result reads
\begin{align}
\FBCS(\Gamma) - \FBCS(\Gamma_0) \hspace{-60pt} & \notag\\
&= -\frac 1{2\beta}\Tr_0 \bigl[ \ln\bigl( \cosh\bigl( \frac \beta 2H_\Delta\bigr)\bigr) - \ln\bigl( \cosh\bigl( \frac  \beta 2H_0\bigr)\bigr)\bigr] \notag\\
&\hspace{30pt} + \Vert \Psi\Vert_{\Lmag^2(Q_B)}^2 \; \langle \alpha_*, V\alpha_*\rangle_{L^2(\Rbb^3)} \notag \\
&\hspace{30pt} +\frac{T}{2} \Hcal_0(\Gamma, \Gamma_\Delta) - \fint_{Q_B} \dd X\int_{\Rbb^3} \dd r\; V(r) \, \bigl|\alpha(X,r) - \alpha_*(r) \Psi(X)\bigr|^2.  \label{DHS1:BCS functional-intermediate}
\end{align}

In order to compute the first term on the right side of \eqref{DHS1:BCS functional-intermediate}, we need Lemma~\ref{DHS1:BCS functional_identity_Lemma} below. It is the main technical novelty of our trial state analysis and should be compared to the related part in the proof of \cite[Theorem~2]{Hainzl2012}. The main difference between our proof of Lemma~\ref{DHS1:BCS functional_identity_Lemma} and the relevant parts of the proof of \cite[Theorem~2]{Hainzl2012} is that we use the product representation of the hyperbolic cosine in \eqref{DHS1:cosh-Product} below instead of a Cauchy integral representation of the function $z \mapsto \ln(1+e^{-z})$. In this way we obtain better decay properties in the subsequent resolvent expansion, which simplifies the analysis considerably. 

As already noted above, the admissibility of $\Gamma_{\Delta}$ implies that the difference between the two operators in the first term on the right side of \eqref{DHS1:BCS functional-intermediate} is weakly locally trace class. We highlight that this is a nontrivial statement because each of the two operators separately does not share this property. We also highlight that our proof of Lemma~\ref{DHS1:BCS functional_identity_Lemma} does not require this as an assumption, it implies the statement independently.

In combination with \eqref{DHS1:BCS functional-intermediate}, Lemma~\ref{DHS1:BCS functional_identity_Lemma} below proves Proposition~\ref{DHS1:BCS functional_identity}.  Before we state the lemma, we recall the definitions of the operators $L_{T,B}$ and $N_{T,B}$ in \eqref{DHS1:LTB_definition} and \eqref{DHS1:NTB_definition}, respectively.


\begin{lem}
\label{DHS1:BCS functional_identity_Lemma}
Let $V\alpha_*\in L^{\nicefrac 65}(\Rbb^3)\cap L^2(\Rbb^3)$. For any $B>0$, any $\Psi\in \Hmag^1(Q_B)$, and any $T>0$, the operator
\begin{align*}
\ln\bigl( \cosh\bigl( \frac \beta 2H_\Delta\bigr)\bigr) - \ln \bigl( \cosh\bigl( \frac \beta 2H_0\bigr)\bigr) 
\end{align*}
is weakly locally trace class and its weak local trace equals
\begin{align}
-\frac 1{2\beta} \Tr_0\Bigl[ \ln\bigl( \cosh\bigl( \frac \beta 2H_\Delta\bigr)\bigr) - \ln \bigl( \cosh\bigl( \frac \beta 2H_0\bigr)\bigr) \Bigr] & \notag\\
%
%
&\hspace{-100pt}= -\frac 14\langle \Delta, L_{T,B}\Delta\rangle + \frac 18 \langle \Delta, N_{T,B}(\Delta)\rangle + \Tr\Rcal_{T,B}(\Delta). \label{DHS1:BCS functional_identity_Lemma_eq2}
\end{align}
The operator $\Rcal_{T,B}(\Delta)$ is locally trace class and its trace norm satisfies the bound
\begin{align*}
\Vert\Rcal_{T,B}(\Delta) \Vert_1 &\leq C\; T^{-5} \; B^3 \; \Vert \Psi\Vert_{\Hmag^1(Q_B)}^6.
\end{align*}
\end{lem}



\begin{proof}[Proof of Lemma \ref{DHS1:BCS functional_identity_Lemma}]
We recall the Matsubara frequencies in \eqref{DHS1:Matsubara_frequencies} and write the hyperbolic cosine in terms of the following product expansion, see \cite[Eq. (4.5.69)]{Handbook}, 
\begin{align}
\cosh\bigl(\frac \beta 2x\bigr) &= \prod_{k=0}^\infty \Bigl( 1 + \frac{x^2}{\omega_k^2}\Bigr). 
\label{DHS1:cosh-Product}
\end{align}
We have
\begin{align*}
0 &\leq \sum_{k= 0}^\infty \ln \bigl( 1 + \frac{x^2}{\omega_k^2}\bigr) = \ln \bigl( \cosh\bigl( \frac \beta 2 x\bigr)\bigr) \leq \frac\beta 2 \; |x|, & x &\in \Rbb,
\end{align*}
and accordingly
\begin{align*}
\ln\bigl( \cosh\bigl( \frac{\beta}{2} H_\Delta\bigr)\bigr) =  \sum_{k=0}^\infty \ln \bigl( 1+ \frac{H_\Delta^2}{\omega_k^2}\bigr) 
\end{align*}
holds in a strong sense on the domain of $|H_\Delta|$. Since $\Delta$ is a bounded operator by Lemma \ref{DHS1:Schatten_estimate}, the domains of $|H_\Delta|$ and $|H_0|$ coincide. The identity 
\begin{align}
\ln\bigl( \cosh\bigl( \frac{\beta}{2} H_\Delta\bigr)\bigr) - \ln\bigl( \cosh\bigl( \frac{\beta}{2} H_0\bigr)\bigr)  =  \sum_{k=0}^\infty \bigl[ \ln \bigl( \omega_k^2+ H_\Delta^2 \bigr) - \ln \bigl( \omega_k^2 + H_0^2 \bigr) \bigr] \label{DHS1:BCS functional_identity_Lemma_1}
\end{align}
therefore holds in a strong sense on the domain of $|H_0|$. Elementary arguments show that
\begin{align}
\ln\bigl( \omega^2 + H_\Delta^2\bigr) - \ln\bigl(\omega^2 + H_0^2\bigr) = -\lim_{R\to\infty} \int_\omega^R \dd u \; \Bigl[\frac{2u}{u^2 + H_\Delta^2} - \frac{2u }{u^2 + H_0^2}\Bigr] \label{DHS1:ln-integral}
\end{align}
holds for $\omega>0$ in a strong sense on the domain of $\ln(1+|H_0|)$. Therefore, by \eqref{DHS1:BCS functional_identity_Lemma_1} and \eqref{DHS1:ln-integral}, we have
\begin{align}
\ln\bigl( \cosh\bigl(\frac \beta 2 H_\Delta\bigr)\bigr) - \ln\bigl( \cosh\bigl( \frac \beta 2 H_0\bigr)\bigr) & \notag \\
&\hspace{-100pt}= -\i \sum_{k=0}^\infty \int_{\omega_k}^\infty \dd u \; \Bigl[\frac{1}{\i u - H_\Delta } - \frac{1}{\i u - H_0} + \frac{1}{\i u + H_\Delta} - \frac{1}{\i u + H_0}\Bigr] \label{DHS1:lncosh-difference}
\end{align}
in a strong sense on the domain of $|H_0|$. By a slight abuse of notation, we have incorporated the limit in \eqref{DHS1:ln-integral} into the integral.
%
%
%
%
%
%
%
%
%
%

In the next step we use the resolvent expansion in \eqref{DHS1:Resolvent_Equation} to see that the right side of \eqref{DHS1:lncosh-difference} equals
\begin{align*}
\Ocal_1 + \Dcal_2 + \Ocal_3 + \Dcal_4 + \Ocal_5 - 2\beta \,  \Rcal_{T,B}(\Delta),
\end{align*}
with two diagonal operators $\Dcal_2$ and $\Dcal_4$, three offdiagonal operators $\Ocal_1$, $\Ocal_3$ and $\Ocal_5$ and a remainder term $\Rcal_{T,B}(\Delta)$. The index of the operators reflects the number of $\delta$ matrices appearing in their definition. The diagonal operators $\Dcal_2$ and $\Dcal_4$ are given by
\begin{align*}
\Dcal_2 &:=  -\i \sum_{k=0}^\infty\int_{\omega_k}^\infty \dd u \; \Bigl[\frac{1}{\i u - H_0}\delta \frac{1}{\i u - H_0}\delta \frac{1}{\i u - H_0} + \frac{1}{\i u + H_0}\delta \frac{1}{\i u + H_0}\delta \frac{1}{\i u + H_0} \Bigr] , \\
\Dcal_4 &:=  -\i \sum_{k=0}^\infty\int_{\omega_k}^\infty \dd u \; \Bigl[\frac{1}{\i u - H_0}\delta \frac{1}{\i u - H_0}\delta \frac{1}{\i u - H_0}\delta \frac{1}{\i u - H_0}\delta \frac{1}{\i u - H_0} \\
&\hspace{120pt}+ \frac{1}{\i u + H_0}\delta \frac{1}{\i u + H_0}\delta \frac{1}{\i u + H_0}\delta \frac{1}{\i u + H_0}\delta \frac{1}{\i u + H_0}\Bigr]
\end{align*}
and the offdiagonal operators read
\begin{align*}
\Ocal_1 &:= -\i \sum_{k=0}^\infty\int_{\omega_k}^\infty \dd u \; \Bigl[\frac{1}{\i u - H_0}\delta \frac{1}{\i u - H_0} + \frac{1}{\i u + H_0}\delta \frac{1}{\i u + H_0}\Bigr], \\
\Ocal_3 &:= -\i \sum_{k=0}^\infty\int_{\omega_k}^\infty \dd u \; \Bigl[\frac{1}{\i u - H_0}\delta \frac{1}{\i u - H_0}\delta \frac{1}{\i u - H_0}\delta \frac{1}{\i u - H_0} \\
&\hspace{85pt}+ \frac{1}{\i u + H_0}\delta \frac{1}{\i u + H_0}\delta \frac{1}{\i u + H_0}\delta \frac{1}{\i u + H_0}\Bigr],\\
\Ocal_5 &:=  -\i \sum_{k=0}^\infty\int_{\omega_k}^\infty \dd u \; \Bigl[\frac{1}{\i u - H_0}\delta \frac{1}{\i u - H_0}\delta \frac{1}{\i u - H_0}\delta \frac{1}{\i u - H_0}\delta \frac{1}{\i u - H_0}\delta \frac{1}{\i u - H_0} \\
&\hspace{85pt}+ \frac{1}{\i u + H_0}\delta \frac{1}{\i u + H_0}\delta \frac{1}{\i u + H_0}\delta \frac{1}{\i u + H_0}\delta \frac{1}{\i u + H_0}\delta \frac{1}{\i u + H_0}\Bigr].
\end{align*}
Since the operators $\Ocal_1$, $\Ocal_3$, and $\Ocal_5$ are offdiagonal, they are weakly locally trace class and their weak local trace equals $0$. We also note that the operator $\Ocal_1$ is not necessarily locally trace class, which is why we need to work with the weak local trace. The operator $\Rcal_{T,B}(\Delta)$ is defined by
\begin{align*}
\Rcal_{T,B}(\Delta) & \\
&\hspace{-40pt}:= \frac{\i}{2\beta} \sum_{k=0}^\infty\int_{\omega_k}^\infty \dd u \; \Bigl[\frac{1}{\i u - H_0}\delta \frac{1}{\i u - H_0}\delta \frac{1}{\i u - H_0}\delta \frac{1}{\i u - H_\Delta}\delta \frac{1}{\i u - H_0}\delta \frac{1}{\i u - H_0}\delta \frac{1}{\i u - H_0} \\
&\hspace{50pt}+ \frac{1}{\i u + H_0}\delta \frac{1}{\i u + H_0}\delta \frac{1}{\i u + H_0}\delta \frac{1}{\i u + H_\Delta}\delta \frac{1}{\i u + H_0}\delta \frac{1}{\i u + H_0}\delta \frac{1}{\i u + H_0}\Bigr].
\end{align*}
As we show in the next paragraph, the operators $\Dcal_2$, $\Dcal_4$ and $\Rcal_{T,B}(\Delta)$ are locally trace class. This, in particular, implies that \eqref{DHS1:lncosh-difference} also holds when the expectation with respect to vectors $\psi \in L^2(\mathbb{R}^3) \oplus L^2(\mathbb{R}^3)$ satisfying either $P_0 \psi = \psi$ or $Q_0 \psi = \psi$ is taken on both sides of the equation.

It remains to compute the traces of $\Dcal_2$ and $\Dcal_4$, and to estimate the trace norm of $\Rcal_{T,B}(\Delta)$. We first consider $\Dcal_2$ and use Hölder's inequality in \eqref{DHS1:Schatten-Hoelder} to estimate
\begin{align}
  \Bigl\Vert \frac{1}{\i u \pm H_0} \delta \frac{1}{\i u \pm H_0} \delta \frac{1}{\i u \pm H_0}\Bigr\Vert_1 &\leq \Bigl\Vert \frac{1}{\i u \pm H_0} \Bigr\Vert_\infty^3 \; \Vert \delta\Vert_2^2 = \frac{2}{u^3} \; \Vert \Delta\Vert_2^2.
\end{align}
Therefore, Lemma \ref{DHS1:Schatten_estimate} shows that the combination of the series and the integral defining $\Dcal_2$ converges absolutely in local trace norm.
In particular, $\Dcal_2$ is locally trace class and we may arbitrarily interchange the trace, the sum, and the integral to compute its trace. We do this, use the cyclicity of the trace, and obtain
\begin{equation}
\Tr \Dcal_2 =  -\i \sum_{k=0}^\infty \int_{\omega_k}^\infty \dd u \; \Tr\Bigl[\Bigl(\frac{1}{\i u - H_0}\Bigr)^2 \delta \frac{1}{\i u - H_0}\delta + \Bigl(\frac{1}{\i u + H_0}\Bigr)^2\delta \frac{1}{\i u + H_0}\delta \Bigr]. \label{DHS1:eq:A7}
\end{equation}
Integration by parts shows
\begin{align*}
\int_{\omega_k}^\infty \dd u \; \Bigl(\frac{1}{\i u \pm H_0}\Bigr)^2 \delta \frac{1}{\i u \pm H_0}\delta &= -\i\; \frac{1}{\i \omega_k \pm H_0} \delta \frac{1}{\i \omega_k \pm H_0} \delta \\
&\hspace{80pt}- \int_{\omega_k}^\infty \dd u \; \frac{1}{\i u \pm H_0} \delta \Bigl( \frac{1}{\i u \pm H_0}\Bigr)^2 \delta,
\end{align*}
and another application of the cyclicity of the trace yields
\begin{align}
\Tr \int_{\omega_k}^\infty \dd u \; \Bigl(\frac{1}{\i u \pm H_0}\Bigr)^2 \delta \frac{1}{\i u \pm H_0}\delta &= -\frac \i 2 \; \Tr \frac{1}{\i \omega_k \pm H_0} \delta \frac{1}{\i \omega_k \pm H_0} \delta. \label{DHS1:BCS functional_identity_Lemma_2}
\end{align}
Note that
\begin{align}
\frac{1}{\i \omega_k \pm H_0}\,  \delta\,  \frac{1}{\i \omega_k \pm H_0} \, \delta 
&= \begin{pmatrix}
\frac{1}{\i \omega_k \pm \hfrak_B} \, \Delta \frac{1}{\i \omega_k \mp \ov{\hfrak_B}} \, \ov \Delta \\ & \frac{1}{\i \omega_k \mp \ov{\hfrak_B}}\,  \ov \Delta \frac{1}{\i \omega_k \pm \hfrak_B} \, \Delta
\end{pmatrix}. \label{DHS1:Calculation-entry}
\end{align}
We combine this with \eqref{DHS1:eq:A7} and \eqref{DHS1:BCS functional_identity_Lemma_2} and summarize the cases $\pm$ into a single sum over $n\in \Zbb$. This yields
\begin{align*}
-\frac{1}{2\beta}\Tr \Dcal_2 &= \frac{1}{2\beta}\sum_{n\in \Zbb} \Bigl\langle \Delta, \frac{1}{\i\omega_n - \hfrak_B} \Delta \frac{1}{\i\omega_n + \ov{\hfrak_B}}\Bigr\rangle =  - \frac 14\langle \Delta, L_{T,B}\Delta\rangle,
\end{align*}
where $L_{T,B}$ is the operator defined in \eqref{DHS1:LTB_definition}.

We argue as above to see that the integrand in the definition of $\Dcal_4$ is bounded by $C \Vert \Delta\Vert_4^4 \, u^{-5}$. Moreover, we have $\Vert \Delta\Vert_4^4 \leq CB^2 \Vert V\alpha_*\Vert_{\nicefrac 43}^4 \Vert \Psi\Vert_{\Hmag^1(Q_B)}^4$ by \eqref{DHS1:Magnetic_Sobolev} and Lemma \ref{DHS1:Schatten_estimate}.
Therefore, the integral and the sum in $\Dcal_4$ are absolutely convergent with respect to the local trace norm.
The trace of $\Dcal_4$ is computed similar to that of $\Dcal_2$. With $N_{T,B}$ defined in \eqref{DHS1:NTB_definition}, the result reads
\begin{align}
-\frac{1}{2\beta}\Tr \Dcal_4 &
= \frac 18\langle \Delta, N_{T,B}(\Delta)\rangle. \label{DHS1:NTB_size_bound_2}
\end{align} 
In the case of $\Rcal_{T,B}(\Delta)$, we bound the trace norm of the operator inside the integral by $u^{-7}\Vert \Delta\Vert_6^6$. Using \eqref{DHS1:Magnetic_Sobolev} and Lemma~\ref{DHS1:Schatten_estimate}, we estimate the second factor by a constant times $\Vert V\alpha_*\Vert_{\nicefrac 65}^6 B^{-3}\Vert \Pi\Psi\Vert_2^6 \leq CB^3 \Vert \Psi\Vert_{\Hmag^1(Q_B)}^6$. Finally, integration over $u$ yields the term $6\pi^{-6} T^{-6} (2k+1)^{-6}$, which is summable in $k$. This proves the claimed bound for the trace norm of $\Rcal_{T,B}(\Delta)$.
%
\end{proof}


\subsection{Proof of Theorem \ref{DHS1:Calculation_of_the_GL-energy}}
\label{DHS1:Calculation_of_the_GL-energy_proof_Section}

\subsubsection{Magnetic resolvent estimates}
\label{DHS1:Magnetic_resolvent_estimates_Section}

In this preparatory subsection, we provide estimates for the magnetic resolvent kernel
\begin{align*}
G^z_B(x,y) &:= \frac{1}{z - \hfrak_B}(x,y), & x,y &\in \Rbb^3.
\end{align*}
We also introduce the function
\begin{align}
g_B^z(x) &:= G_B^z(x,0), &x &\in \Rbb^3. \label{DHS1:gB_definition}
\end{align}
The proof of the following statement can be found in \cite[Lemma 8]{Hainzl2017}. 

\begin{lem}
\label{DHS1:gB-identities}
For all $B\geq 0$, $z\in \Cbb \setminus [B,\infty)$ and $x,y\in \Rbb^3$ we have
\begin{enumerate}[(a)]
\item $g_B^z(-x) = g_B^z(x)$,
\item $G_B^z(x,y) = \e^{\i \frac{\Bbold}{2} \cdot (x\wedge y)} \; g_B^z(x-y)$.
\end{enumerate}
\end{lem}

We start our analysis by providing a decay estimate for the $L^1$-norm of the resolvent kernel $g^z_0$ in \eqref{DHS1:gB_definition} and its gradient in the case $B=0$. For $g_0^z$ such an estimate has been provided in \cite[Lemma~9]{Hainzl2017}. Since we additionally need an estimate for $\nabla g_0^z$, we repeat some of the arguments here.

\begin{lem}
\label{DHS1:g0_decay}
Let $a > -2$. There is a constant $C_a >0$ such that for $t,\omega\in \Rbb$, we have
\begin{align}
\left \Vert \, |\cdot|^a g_0^{\i \omega + t}\right\Vert_1 &\leq C_a \; f(t, \omega)^{1+ \frac a2}, 
\label{DHS1:g0_decay_1}
\end{align}
where
\begin{align}
f(t, \omega) := \frac{|\omega| + |t + \mu|}{(|\omega| + (t + \mu)_-)^2} \label{DHS1:g0_decay_f}
\end{align}
and $x_- := -\min\{x,0\}$. Furthermore, for any $a > -1$, there is a constant $C_a >0$ with
\begin{align}
\left \Vert \, |\cdot|^a \nabla g_0^{\i\omega + t} \right\Vert_1 \leq C_a \; f(t, \omega)^{\frac 12 + \frac a2} \; \Bigl[ 1 + \frac{|\omega| + |t+ \mu|}{|\omega| + (t + \mu)_-}\Bigr]. \label{DHS1:g0_decay_2}
\end{align}
\end{lem}


\begin{proof}
The resolvent kernel $g_0^z$ is given by
\begin{align}
g_0^{z}(x) &= -\frac{1}{4\pi |x|} \; \e^{-\sqrt{-(z+ \mu)}\; |x|}, \label{DHS1:g0_definition}
\end{align}
where $\sqrt{\cdot}$ denotes the standard branch of the square root. As long as $a > - 2$ we have
\begin{align}
\left\Vert \, |\cdot|^a g_0^z\right\Vert_1 &= \int_{\Rbb^3} \dx \; |x|^a \; \Bigl| \frac{1}{4\pi |x|}\e^{-|x|\sqrt{-(z + \mu)}} \Bigr| = \frac{\Gamma(a +2)}{(\Re \sqrt{-(z +\mu)})^{a+2}}. \label{DHS1:Resolvent_kernel_integration}
\end{align}
Moreover,
\begin{align*}
( \Re \sqrt{-z})^2 = \frac 12 (|z| - \Re z) \geq \begin{cases} \frac 14 \frac{|\Im z|^2}{\Re z + |\Im z|} & \Re z\geq 0, \\ \frac 12 |z| & \Re z < 0, \end{cases} 
\end{align*}
and hence
\begin{align*}
(\Re \sqrt{-(t +\mu + \i\omega)})^2 \geq \frac 14 \frac{(|\omega| + (t+ \mu)_-)^2}{|\omega| + |t + \mu|}.
\end{align*}
This proves \eqref{DHS1:g0_decay_1}. To prove \eqref{DHS1:g0_decay_2}, we use \eqref{DHS1:g0_definition} 
and estimate
\begin{align*}
|\nabla g_0^z(x)|\leq |z+\mu|^{\nicefrac 12} \; |g_0^z(x)| +|x|^{-1}|g_0^z(x)|.
\end{align*}
This shows the second estimate for $a > - 1$.
\end{proof}

In the next step we prove estimates for the $L^1$-norms of $g_B^z$ and $g_B^z-g_0^z$ and the gradient of these functions if $B\neq 0$. Once more, some of the arguments in \cite[Lemma~10]{Hainzl2017} reappear in our proof below, ensuring self-consistency.

\begin{lem}
\label{DHS1:gB-g_decay}
For any $a\geq 0$, there are constants $\delta_a , C_a > 0$ such that for all $t, \omega\in \Rbb$ and for all $B \geq 0$ with $f(t, \omega)^2\,  B^2 \leq \delta_a$, we have 
\begin{align}
\bigl\Vert |\cdot|^a g_B^{\i\omega+ t}\bigr\Vert_1 &\leq C_a \, f(t,\omega)^{1+\frac a2}, \notag \\
\bigl\Vert |\cdot|^a \nabla g_B^{\i\omega+ t}\bigr\Vert_1 &\leq C_a \, f(t,\omega)^{\frac 12 + \frac a2} \Bigl[ 1 + \frac{|\omega| + |t+ \mu|}{|\omega| + (t + \mu)_-}\Bigr], \label{DHS1:gB-g_decay_2}
\end{align}
and
\begin{align}
\bigl\Vert \,|\cdot|^a ( g_B^{\i\omega+ t} - g_0^{\i\omega+ t} ) \bigr\Vert_1 &\leq C_a \, B^2 \, f(t,\omega)^{3+ \frac a2} , \notag \\
\bigl\Vert \,|\cdot|^a (\nabla g_B^{\i\omega+ t} - \nabla g_0^{\i\omega+ t} ) \bigr\Vert_1 &\leq C_a \, B^2 \, f(t,\omega)^{\frac 52 + \frac a2} \Bigl[ 1 + \frac{|\omega| + |t+ \mu|}{|\omega| + (t + \mu)_-}\Bigr] \label{DHS1:gB-g_decay_1}
\end{align}
with the function $f(t,\omega)$ in \eqref{DHS1:g0_decay_f}.
\end{lem}




\begin{proof}
During the proof we use the notation $z = \i \omega + t$.
We define the function
\begin{align}
h^z(x) := \frac 14\, |e_3\wedge x|^2 \, g_0^z(x) \label{DHS1:hz_definition}
\end{align}
and choose $\delta_a$ such that $2 \delta_a D_a C_2 =1$. Here $C_2$ denotes the constant in \eqref{DHS1:g0_decay_1} and $D_a := 1$ if $0 \leq a \leq 1$ and $D_a := 2^a$ if $a>1$. Lemma~\ref{DHS1:g0_decay} and the bound $\Vert h^z\Vert_1\leq \Vert |\cdot|^2g_0^z\Vert_1$ imply
\begin{align}
B^2 D_a \Vert h^z\Vert_1 \leq \frac 12 \label{DHS1:hz_estimate}
\end{align}
for all $\omega$, $t$, and $B$ that are allowed by our assumptions. We define the operator $\tilde G_B^z$ by the kernel
\begin{align*}
\tilde G_B^z(x,y) &:= \e^{\i \frac{\Bbold}{2} (x\wedge y)} g_0^z(x-y)
\end{align*}
and note that
\begin{align}
(z - \hfrak_B) \tilde G_B^z = 1 - T_B^z, \label{DHS1:gB-g_1}
\end{align}
where $T_B^z$ is the operator given by the kernel
\begin{align*}
T_B^z(x,y) &:= \e^{\i \frac \Bbold 2(x\wedge y)} \bigl[ \Bbold \wedge (x-y) (-\i\nabla_x)g_0^z(x-y) + B^2 \; h^z(x-y)\bigr].
\end{align*}
The first term in square brackets equals $0$ because $g_0^z$ is a radial function, which implies that the vector $\nabla g_0^z(x-y)$ is perpendicular to $\Bbold \wedge (x-y)$. 
Multiplication of \eqref{DHS1:gB-g_1} with $(z - \hfrak_B)^{-1}$ from the left yields
\begin{align*}
G_B^z(x,y) - \tilde G_B^z(x,y) = \int_{\Rbb^3} \dd v \; G_B^z(x,v) T_B^z(v,y).
\end{align*}
We set $y =0$, change variables $v\mapsto x - v$, and find
\begin{align}
g_B^z(x) - g_0^z(x) = B^2  \int_{\Rbb^3} \dd v \; \e^{\i \frac \Bbold 2\cdot (v\wedge x)} \; g_B^z(v) \; h^z(x-v). \label{DHS1:gB-g_2}
\end{align}
This implies
\begin{align}
\Vert g_B^z - g_0^z\Vert_1 &\leq B^2 \Vert |g_B^z| * |h^z| \Vert_1 \leq B^2 \; \Vert g_B^z - g_0^z\Vert_1 \; \Vert h^z\Vert_1 + B^2 \; \Vert g_0^z\Vert_1 \; \Vert h^z\Vert_1. \label{DHS1:gB-g_6}
\end{align}
A straightforward calculation involving \eqref{DHS1:gB-g_1} and the Neumann series shows that $g_B^z - g_0^z$ belongs to $L^1(\Rbb^3)$. Therefore, \eqref{DHS1:hz_estimate} and \eqref{DHS1:gB-g_6} imply
\begin{align}
\Vert g_B^z - g_0^z\Vert_1 &\leq \Vert g_0^z\Vert_1 \label{DHS1:gB-g_4}
\end{align}
for all $t, \omega$, and $B$ that are allowed by our assumptions.

We use this estimate as a basis to prove the bounds claimed in the lemma and start with the first bound in \eqref{DHS1:gB-g_decay_1}. By \eqref{DHS1:gB-g_2}, we have
\begin{align*}
\Vert \, |\cdot|^a (g_B^z - g_0^z)\Vert_1 &\leq D_a B^2 \bigl[ \Vert \, |\cdot|^a g_0^z\Vert_1 \, \Vert h^z\Vert_1 + \Vert \, |\cdot|^a (g_B^z - g_0^z)\Vert_1 \, \Vert h^z\Vert_1 \\
&\hspace{80pt} + \Vert g_B^z - g_0^z\Vert_1 \, \Vert |\cdot|^a h^z\Vert_1 + \Vert g_0^z\Vert_1 \, \Vert \, |\cdot|^a h^z\Vert_1 \bigr].
\end{align*}
A similar argument to the one above \eqref{DHS1:gB-g_4} shows that $|\cdot|^a (g_B^z - g_0^z)$ belongs to $L^1(\Rbb^3)$. In combination with \eqref{DHS1:gB-g_4}, we therefore obtain
\begin{align}
\Vert \, |\cdot|^a (g_B^z - g_0^z)\Vert_1 &\leq 2D_a B^2 \; \bigl[ \Vert |\cdot|^a g_0^z\Vert_1\, \Vert h^z\Vert_1 
+ 2\, \Vert g_0^z\Vert_1 \, \Vert |\cdot|^a h^z\Vert_1\bigr]. \label{DHS1:gB-g_7}
\end{align}
With the help of Lemma~\ref{DHS1:g0_decay} and \eqref{DHS1:hz_estimate}, we read off the first bound in \eqref{DHS1:gB-g_decay_1}. Moreover, the triangle inequality, Lemma~\ref{DHS1:g0_decay}, the first bound in \eqref{DHS1:gB-g_decay_1} and the bound $f(t, \omega)^2\,  B^2 \leq \delta_a$ imply the first bound in \eqref{DHS1:gB-g_decay_2}.

Next, we consider the bounds in \eqref{DHS1:gB-g_decay_2} and \eqref{DHS1:gB-g_decay_1} involving the gradient. As a preparation, the bound $|\nabla h^z (x)| \leq |x|\, |g_0^z(x)| + |x|^2\, |\nabla g_0^z(x)|$ and Lemma~\ref{DHS1:g0_decay} show
\begin{align}
\Vert \, |\cdot|^a \nabla h^z\Vert_1 &
\leq  C_a \; f(t, \omega)^{\frac 32 + \frac a2} \Bigl[ 1 + \frac{|\omega| +  |t+\mu|}{|\omega| + (t + \mu)_-}\Bigr]. \label{DHS1:gB-g_8}
\end{align}
We use $|\nabla \e^{\i \frac \Bbold 2 \cdot (v\wedge x)}| \leq B|v|$ and \eqref{DHS1:gB-g_2} to see that
\begin{align*}
|\nabla g_B^z(x) - \nabla g_0^z(x)| &\leq B^2 \int_{\Rbb^3} \dd v \; \bigl[ B\, |v|\, |g_B^z(v)| \, |h^z(x-v)| + |g_B^z(v)| \, |\nabla h^z(x-v)|\bigr]
\end{align*}
as well as
\begin{align}
\Vert\, |\cdot|^a (\nabla g_B^z - \nabla g_0^z)\Vert_1 &\leq D_a B^2  \bigl[ B\, \Vert \, |\cdot|^{a+1} g_B^z\Vert_1 \, \Vert h^z\Vert_1 + B \, \Vert \, |\cdot| g_B^z\Vert_1 \, \Vert \, |\cdot|^a h^z\Vert_1 \notag\\
&\hspace{60pt} + \Vert \, |\cdot|^a g_B^z\Vert_1 \; \Vert \nabla h^z\Vert_1 + \Vert  g_B^z\Vert_1 \; \Vert \,|\cdot|^a\nabla h^z\Vert_1\bigr].  \label{DHS1:eq:A9}
\end{align}
When we combine \eqref{DHS1:eq:A9}, the first estimates in \eqref{DHS1:gB-g_decay_2} and \eqref{DHS1:gB-g_decay_1}, the bound in \eqref{DHS1:gB-g_8} and Lemma \ref{DHS1:g0_decay}, we see that
\begin{align*}
\Vert\, |\cdot|^a (\nabla g_B^z - \nabla g_0^z)\Vert_1 &\leq C_a \, B^2 \, f(t,\omega)^{\frac 52 + \frac a2} \bigl[ 1 + B f(t,\omega)\bigr].
\end{align*}
An application of the assumption $B^2f(t,\omega)^2 \leq \delta_a$ proves the second bound in \eqref{DHS1:gB-g_decay_1}. Finally, the triangle inequality, the second bound in \eqref{DHS1:gB-g_decay_1}, and Lemma \ref{DHS1:g0_decay} show
\begin{align*}
\Vert \, |\cdot|^a \nabla g_B^z\Vert_1 
&\leq C_a \, f(t,\omega)^{\frac 12 + \frac a2} \Bigl[ 1 + \frac{|\omega| + |t + \mu|}{|\omega| + (t + \mu)_-}\Bigr] \bigl[ 1 + B^2 f(t,\omega)^2\bigr].
\end{align*}
Another application of $B^2f(t,\omega)^2 \leq \delta_a$ on the right side proves the second bound in \eqref{DHS1:gB-g_decay_2}.
\end{proof}

\subsubsection{A representation formula for \texorpdfstring{$L_{T,B}$}{LTB} and an outlook on the quadratic terms}

In this subsection we compute the terms in \eqref{DHS1:Calculation_of_the_GL-energy_eq} involving the linear operator $L_{T,B}$ defined in \eqref{DHS1:LTB_definition}. Our starting point is the representation formula for $L_{T,B}$ in \cite[Lemma 11]{Hainzl2017},
which expresses the operator explicitly in terms of the relative and the center-of-mass coordinate.

\begin{lem}
\label{DHS1:LTB_action}
The operator $L_{T,B} \colon \Lsymm \ra \Lsymm$ in \eqref{DHS1:LTB_definition} acts as
\begin{align*}
(L_{T,B}\alpha) (X,r) &= \iint_{\Rbb^3\times \Rbb^3} \dd Z \dd s \; k_{T,B}(Z, r, s) \; (\cos(Z\cdot \Pi_X)\alpha) (X,s)
\end{align*}
with
\begin{align}
k_{T,B}(Z, r, s) &:= \frac 2\beta \sum_{n\in \Zbb} k_{T,B}^n(Z, r- s) \; \e^{\frac{\i}{4} \Bbold \cdot(r\wedge s)} \label{DHS1:kTB_definition}
\end{align}
and
\begin{align}
k_{T,B}^n(Z,r) &:= g_B^{\i\omega_n} \bigl( Z - \frac{r}{2}\bigr) \; g_B^{-\i\omega_n} \bigl( Z + \frac{r}{2}\bigr). \label{DHS1:kTBn_definition}
\end{align}
\end{lem}

We analyze the operator $L_{T,B}$ in three steps. In the first two steps we introduce two operators of increasing simplicity in their dependence on $B$:
\begin{align}
L_{T,B} = (L_{T,B} - \tilde L_{T,B}) + (\tilde L_{T,B} - M_{T,B}) + M_{T,B}, \label{DHS1:LTB_decomposition}
\end{align}
where $\tilde L_{T,B}$ and $M_{T,B}$ are defined below in \eqref{DHS1:LTBtilde_definition} and \eqref{DHS1:MTB_definition}, respectively. To obtain $\tilde L_{T,B}$ we replace the functions $g_B^z$ in the definition of $L_{T,B}$ by $g_0^z$. Moreover, $M_{T,B}$ is obtained from $\tilde L_{T,B}$ when we replace $k_{T,B}$ by $k_{T,0}$, i.e., when we additionally replace the magnetic phase $\e^{\frac{\i}{4} \Bbold \cdot(r\wedge s)}$ by $1$. In Section~\ref{DHS1:Approximation_of_LTB_Section} we prove that the terms in the brackets in \eqref{DHS1:LTB_decomposition} are small in a suitable sense. The third step consists of a careful analysis of the operator $M_{T,B}$, which takes place in Section~\ref{DHS1:Analysis_of_MTB_Section}. There, we expand the operator $\cos(Z\cdot \Pi_X)$ in powers of $Z\cdot \Pi_X$ up to second order and extract the quadratic terms of the Ginzburg--Landau functional in \eqref{DHS1:Definition_GL-functional} as well as a term that cancels the last term on the left side of \eqref{DHS1:Calculation_of_the_GL-energy_eq}. In Section~\ref{DHS1:Summary_quadratic_terms_Section} we summarize our findings.

We remark that the operator $\tilde L_{T,B}$ is called $M_{T,B}$ in \cite{Hainzl2017} and that $M_{T,B}$ is called $N_{T,B}$. The reason why we did not follow the notation in \cite{Hainzl2017} is that $N_{T,B}$ is reserved for the nonlinear term in the present paper. We note that our decomposition of $L_{T,B}$ in \eqref{DHS1:LTB_decomposition} already appeared in \cite{Hainzl2017}. Parts of our analysis follow the analysis of $L_{T,B}$ in Section~4 and Section~5 in that reference. However, we additionally need $\Hsymm$-norm bounds that are not provided in \cite{Hainzl2017}. It should also be noted that $L_{T,B}$ acts on $L^2(\mathbb{R}^6)$ in \cite{Hainzl2017}, while it acts on $\Lsymm$ in our case.

\subsubsection{Approximation of \texorpdfstring{$L_{T,B}$}{LTB}}
\label{DHS1:Approximation_of_LTB_Section}

\paragraph{The operator $\tilde L_{T,B}$.} The operator $\tilde L_{T,B}$ is defined by
\begin{align}
\tilde L_{T,B}\alpha(X,r) &:= \iint_{\Rbb^3\times \Rbb^3} \dd Z \dd s \; \tilde k_{T,B}(Z, r,s) \; (\cos(Z\cdot\Pi_X)\alpha)(X,s) \label{DHS1:LTBtilde_definition}
\end{align}
with
\begin{align}
\tilde k_{T,B} (Z, r,s) := \frac 2\beta \sum_{n\in \Zbb} k_{T,0}^n(Z, r-s) \; \e^{\frac{\i}{4} \Bbold \cdot (r\wedge s)} \label{DHS1:ktildeTB_definition}
\end{align}
and $k_{T,0}^n$ in \eqref{DHS1:kTBn_definition}. In the following proposition we provide an estimate that allows us to replace $L_{T,B}$ by $\widetilde{L}_{T,B}$ in our computations.

\begin{prop}
\label{DHS1:LTB-LtildeTB}
For any $T_0>0$ there is $B_0>0$ such that for any $0 < B \leq B_0$, any $T\geq T_0$ and whenever $|\cdot|^k V\alpha_*\in L^2(\Rbb^3)$ for $k\in \{0,1\}$, $\Psi\in \Hmag^1(Q_B)$, and $\Delta \equiv \Delta_\Psi$ as in \eqref{DHS1:Delta_definition}, we have
\begin{align*}
\Vert L_{T,B} \Delta - \tilde L_{T,B}\Delta\Vert_\Hsymm^2 \leq C \; B^5 \; \bigl(\Vert V\alpha_*\Vert_2^2 + \Vert \, |\cdot|V\alpha_*\Vert_2^2\bigr) \; \Vert \Psi\Vert_{\Hmag^1(Q_B)}^2.
\end{align*}
\end{prop}

\begin{bem}
	\label{DHS1:rem:A1}
	For the proof of Theorem~\ref{DHS1:Calculation_of_the_GL-energy} we only need a bound for $\langle \Delta, (L_{T,B} - \tilde L_{T,B})\Delta\rangle$, which is easier to obtain. This bound follows directly from Proposition~\ref{DHS1:LTB-LtildeTB}, Lemma~\ref{DHS1:Schatten_estimate} and an application of the Cauchy--Schwarz inequality. The more general bound in Proposition~\ref{DHS1:LTB-LtildeTB} is needed in the proof of Proposition~\ref{DHS1:Structure_of_alphaDelta}.
\end{bem}


Before we start with the proof of Proposition~\ref{DHS1:LTB-LtildeTB} let $a \geq 0$ and introduce the functions
\begin{align}
F_{T,B}^{a} &:= \smash{\frac 2\beta \sum_{n\in \Zbb}}  \; \bigl(|\cdot|^a \, | g_B^{\i\omega_n} - g_0^{\i\omega_n}|\bigr) * |g_B^{-\i\omega_n}| + |g_B^{\i\omega_n} - g_0^{\i\omega_n}| * \bigl(|\cdot|^a \, |g_B^{-\i\omega_n}| \bigr) \notag \\
&\hspace{50pt}+ \bigl(|\cdot|^a \, |g_0^{\i\omega_n}|\bigr) * |g_B^{-\i\omega_n} - g_0^{-\i\omega_n}| + |g_0^{\i\omega_n}| * \bigl(|\cdot|^a \, |g_B^{-\i\omega_n} -  g_0^{-\i\omega_n}|\bigr) \label{DHS1:LTB-LtildeTB_FTB_definition}
\end{align}
and
\begin{align}
G_{T,B} &:= \smash{\frac 2\beta \sum_{n\in \Zbb}}\; |\nabla g_B^{\i\omega_n} - \nabla g_0^{\i\omega_n}| * |g_B^{-\i\omega_n}| + |g_B^{\i\omega_n} - g_0^{\i\omega_n}| * |\nabla g_B^{-\i\omega_n}| \notag \\
&\hspace{80pt}+ |\nabla g_0^{\i\omega_n}| * |g_B^{-\i\omega_n} - g_0^{-\i\omega_n}| + |g_0^{\i\omega_n}| * |\nabla g_B^{-\i\omega_n} - \nabla g_0^{-\i\omega_n}|  \label{DHS1:LTB-LtildeTB_GTB_definition}
\end{align}
with the Matsubara frequencies $\omega_n$ in \eqref{DHS1:Matsubara_frequencies} and the resolvent kernel $g_B^z$ in \eqref{DHS1:gB_definition}. We claim that for any $a\geq 0$ there is a constant $B_0>0$ such that for $0\leq B\leq B_0$ we have
\begin{align}
\Vert F_{T,B}^a\Vert_1 + \Vert G_{T,B}\Vert_1 \leq C_a B^2.\label{DHS1:LTB-LtildeTB_FTBGTB}
\end{align}
To prove \eqref{DHS1:LTB-LtildeTB_FTBGTB} we note that the function $f(t, \omega)$ in \eqref{DHS1:g0_decay_f} satisfies
\begin{align}
f(0, \omega_n) &\leq C \; (T^{-1} + T^{-2}) \; |2n+1|^{-1} \label{DHS1:g0_decay_f_estimate1}
\end{align}
and that
\begin{align}
\frac{|\omega_n| + |\mu|}{|\omega_n| + \mu_-} \leq C \; (1+ T^{-1}). \label{DHS1:g0_decay_f_estimate2}
\end{align}
Since $T\geq T_0>0$, Lemmas~\ref{DHS1:g0_decay} and \ref{DHS1:gB-g_decay} prove \eqref{DHS1:LTB-LtildeTB_FTBGTB}.

\begin{proof}[Proof of Proposition \ref{DHS1:LTB-LtildeTB}]
We write
\begin{align}
\Vert L_{T,B}\Delta - \tilde L_{T,B} \Delta\Vert_\Hsymm^2 & \notag\\
&\hspace{-120pt}= \Vert L_{T,B}\Delta - \tilde L_{T,B} \Delta\Vert_2^2 + \Vert \Pi_X(L_{T,B}\Delta - \tilde L_{T,B} \Delta)\Vert_2^2 + \Vert \tilde \pi_r(L_{T,B}\Delta - \tilde L_{T,B} \Delta)\Vert_2^2 \label{DHS1:LTB-LtildeTB_8}
\end{align}
and claim that
\begin{align}
\Vert L_{T,B}\Delta - \tilde L_{T,B} \Delta\Vert_2^2 &\leq 4 \; \Vert \Psi\Vert_2^2 \; \Vert F_{T,B}^0 * |V\alpha_*| \, \Vert_2^2. \label{DHS1:LTB-LtildeTB_1}
\end{align}
If this is true, Young's inequality, \eqref{DHS1:Periodic_Sobolev_Norm}, and \eqref{DHS1:LTB-LtildeTB_FTBGTB} prove
\begin{align*}
	\Vert L_{T,B}\Delta - \tilde L_{T,B} \Delta\Vert_2^2 \leq C B^5 \; \Vert V \alpha_* \Vert_2^2 \; \Vert \Psi\Vert_{\Hmag^1(Q_B)}^2.
\end{align*}
To see that \eqref{DHS1:LTB-LtildeTB_1} holds, we expand the squared modulus in the Hilbert--Schmidt norm and obtain
\begin{align}
\Vert L_{T,B}\Delta - \tilde L_{T,B}\Delta\Vert_2^2 & \leq 4 \int_{\Rbb^3} \dd r \iint_{\Rbb^3\times \Rbb^3} \dd Z\dd Z' \iint_{\Rbb^3\times \Rbb^3} \dd s\dd s' \; |V\alpha_*(s)| \; |V\alpha_*(s')|  \notag\\
&\hspace{120pt} \times | k_{T,B}(Z, r, s) - \tilde k_{T,B} (Z, r, s)|  \notag\\
&\hspace{120pt} \times |k_{T,B}(Z', r, s') - \tilde k_{T,B} (Z', r, s')|  \notag\\
&\hspace{30pt}\times \fint_{Q_B} \dd X \; |\cos(Z\cdot \Pi_X)\Psi(X)| \; |\cos(Z'\cdot \Pi_X)\Psi(X)|. \label{DHS1:Expanding_the_square}
\end{align}
The operator $\cos(Z\cdot \Pi_X)$ is bounded by $1$ and we have
\begin{align}
\fint_{Q_B} \dd X \; |\cos(Z\cdot \Pi)\Psi(X)| \; |\cos(Z'\cdot \Pi)\Psi(X)| &\leq \Vert \Psi\Vert_2^2. \label{DHS1:LTB-LtildeTB_3}
\end{align}
By \eqref{DHS1:Expanding_the_square}, this implies
\begin{align}
\Vert L_{T,B}\Delta - \tilde L_{T,B}\Delta\Vert_2^2 & \notag\\
&\hspace{-80pt}\leq 4\; \Vert \Psi\Vert_2^2 \; \int_{\Rbb^3 }\dd r\, \Bigl| \iint_{\Rbb^3\times \Rbb^3} \dd Z\dd s \; |k_{T,B}(Z, r, s) - \tilde k_{T,B} (Z, r, s)|\; |V\alpha_*(s)|\Bigr|^2, \label{DHS1:LTB-LtildeTB_2}
\end{align}
where the integrand is bounded by
\begin{align}
|k_{T,B}(Z, r, s) - \tilde k_{T,B} (Z, r, s)| &\leq \frac 2\beta\,  \smash{\sum_{n\in \Zbb}} \, \bigl[ |g_B^{\i\omega_n} - g_0^{\i\omega_n}|\bigl( Z - \frac r2\bigr) \; |g_B^{-\i\omega_n}| \bigl( Z + \frac r2\bigr) \notag\\
&\hspace{40pt}+ |g_0^{\i\omega_n}|\bigl( Z -\frac r2\bigr)\;  |g_B^{-\i\omega_n} - g_0^{-\i\omega_n}|\bigl( Z +\frac r2\bigr)\bigr]. \label{DHS1:LTB-LtildeTB_11}
\end{align}
For $a \geq 0$, we have the estimate
\begin{align}
|Z|^a &\leq 
\bigl| Z + \frac{r}{2}\bigr|^a + \bigl| Z - \frac{r}{2}\bigr|^a. \label{DHS1:Z_estimate}
\end{align}
This, \eqref{DHS1:LTB-LtildeTB_11}, and the fact that $g_B^{\pm \i\omega_n}$ is an even function imply
\begin{align}
\int_{\Rbb^3} \dd Z \; |Z|^a\; |k_{T,B} (Z, r,s) - \tilde k_{T,B}(Z, r, s)| \leq F_{T,B}^a(r-s), \label{DHS1:LTB-LtildeTB_5}
\end{align}
where $F_{T,B}^a$ is the function in \eqref{DHS1:LTB-LtildeTB_FTB_definition}.  We apply the case $a =0$ to \eqref{DHS1:LTB-LtildeTB_2} and read off \eqref{DHS1:LTB-LtildeTB_1}.

We claim that the second term on the right side of \eqref{DHS1:LTB-LtildeTB_8} is bounded by
\begin{align}
\Vert \Pi_X(L_{T,B}\Delta - \tilde L_{T,B}\Delta)\Vert_2^2 &\leq C B^2 \; \Vert \Psi\Vert_{\Hmag^1(Q_B)}^2 \; \Vert (F_{T,B}^0 + F_{T,B}^1) * |V\alpha_*| \,\Vert_2^2. \label{DHS1:LTB-LtildeTB_6}
\end{align}
If this is true, Young's inequality and \eqref{DHS1:LTB-LtildeTB_FTBGTB} show the claimed bound for this term. To prove \eqref{DHS1:LTB-LtildeTB_6}, we use \eqref{DHS1:Expanding_the_square} with $\cos(Z\cdot\Pi_X)$ replaced by $\Pi_X\cos(Z\cdot \Pi_X)$, that is, we need to replace \eqref{DHS1:LTB-LtildeTB_3} by
\begin{align*}
\fint_{Q_B} \dd X \; |\Pi \cos(Z\cdot \Pi)\Psi(X)| \; |\Pi\cos(Z'\cdot \Pi)\Psi(X)| &\leq \Vert \Pi\cos(Z\cdot \Pi)\Psi\Vert_2\; \Vert \Pi\cos(Z'\cdot \Pi)\Psi\Vert_2.
\end{align*}
In Lemma \ref{DHS1:CommutationII} in Section~\ref{DHS1:Lower Bound Part A} we prove intertwining relations for $\cos(Z\cdot \Pi)$ with various magnetic momenta. The intertwining relation \eqref{DHS1:PiXcos} therein and \eqref{DHS1:Periodic_Sobolev_Norm} show
\begin{align}
\Vert \Pi\cos(Z\cdot \Pi)\Psi\Vert_2 &\leq \Vert \Pi\Psi\Vert_2 + 2B |Z| \; \Vert \Psi\Vert_2 \leq C\, B \; \Vert \Psi\Vert_{\Hmag^1(Q_B)} \; (1 + |Z|), \label{DHS1:PiXcos_estimate}
\end{align}
which yields
\begin{align}
\Vert \Pi_X(L_{T,B}\Delta - \tilde L_{T,B}\Delta)\Vert_2^2 &\leq C \,  B^2 \; \Vert \Psi\Vert_{\Hmag^1(Q_B)}^2  \notag\\
&\hspace{-80pt} \times  \int_{\Rbb^3} \dd r \, \Bigl| \iint_{\Rbb^3\times \Rbb^3} \dd Z\dd s \; (1+ |Z|) \; |k_{T,B} (Z, r,s) - \tilde k_{T,B}(Z, r,s)| \; |V\alpha_*(s)|\Bigr|^2. \label{DHS1:LTB-LtildeTB_9}
\end{align}
We apply the cases $a=0$ and $a=1$ of \eqref{DHS1:LTB-LtildeTB_5} to this and obtain \eqref{DHS1:LTB-LtildeTB_6}.


Concerning the third term on the right side of \eqref{DHS1:LTB-LtildeTB_8} we claim that
\begin{align}
\Vert \tilde \pi_r (L_{T,B}\Delta - \tilde L_{T,B} \Delta)\Vert_2^2 &\leq C \; \Vert \Psi\Vert_2^2 \; \bigl\Vert \bigl(G_{T,B} + F_{T,B}^1\bigr) * |V\alpha_*| + F_{T,B}^0 * |\cdot|\,|V\alpha_*| \, \bigr\Vert_2^2. \label{DHS1:LTB-LtildeTB_4}
\end{align}
If this is true, Young's inequality, \eqref{DHS1:LTB-LtildeTB_FTBGTB}, and \eqref{DHS1:Periodic_Sobolev_Norm} show the  relevant bound for this term.
To prove \eqref{DHS1:LTB-LtildeTB_4}, we estimate
\begin{align}
\Vert \tilde\pi_r (L_{T,B}\Delta - \tilde L_{T,B}\Delta)\Vert_2^2 & \notag \\
&\hspace{-90pt}\leq 4 \;\Vert \Psi\Vert_2^2  \int_{\Rbb^3} \dd r \, \Bigl| \iint_{\Rbb^3\times \Rbb^3} \dd Z\dd s \; |\tilde \pi_r k_{T,B}(Z, r, s) - \tilde \pi_r\tilde k_{T,B} (Z, r, s)| \;|V\alpha_*(s)| \Bigr|^2.  \label{DHS1:LTB-LtildeTB_10}
\end{align}
Using $\frac 14 |\Bbold \wedge r| \leq B(|r-s| + |s|)$ we see that the integrand on the right side is bounded by
\begin{align}
|\tilde \pi_r k_{T,B}(Z, r, s) - \tilde \pi_r\tilde k_{T,B} (Z, r, s)| &\leq \frac 2\beta \sum_{n\in \Zbb} \Big[ |\nabla_r k_{T,B}^n(Z, r-s) - \nabla_r k_{T,0}^n(Z, r-s)| \notag \\
&\hspace{30pt} + B |r-s|\;|k_{T,B}^n(Z, r-s) -  k_{T,0}^n(Z, r-s)| \notag \\
&\hspace{30pt} + B |s|\;|k_{T,B}^n(Z, r-s) -  k_{T,0}^n(Z, r-s)| \Big]. \label{DHS1:LTB-LtildeTB_13}
\end{align}
We also have
\begin{align*}
|\nabla_r k_{T,B}^n(Z, r) - \nabla_r k_{T,0}^n(Z, r)| &\leq |\nabla g_B^{\i\omega_n} - \nabla g_0^{\i\omega_n}|\bigl( Z + \frac r2\bigr) \;|g_B^{-\i\omega_n}| \bigl( Z - \frac r2\bigr) \\
&\hspace{30pt}+ |g_B^{\i\omega_n} - g_0^{\i\omega_n}|\bigl( Z + \frac r2\bigr)\; |\nabla g_B^{-\i\omega_n}|\bigl( Z -\frac r2\bigr) \\
&\hspace{30pt}+ |\nabla g_0^{\i\omega_n}|\bigl( Z +\frac r2\bigr)\; |g_B^{-\i\omega_n} - g_0^{-\i\omega_n}|\bigl( Z -\frac r2\bigr)\\
&\hspace{30pt}+ |g_0^{\i\omega_n}|\bigl( Z + \frac r2\bigr) |\nabla g_B^{-\i\omega_n} - \nabla g_0^{-\i\omega_n}|\bigl( Z - \frac r2\bigr).
\end{align*}
Since $g_B^{\pm \i \omega_n}$ is an even function this implies
\begin{align}
\frac 2\beta\sum_{n\in\Zbb}\int_{\Rbb^3} \dd Z \; |\nabla k_{T,B}^n(Z, r) - \nabla k_{T,0}^n(Z, r)| \leq G_{T,B}(r). \label{DHS1:LTB-LtildeTB_12}
\end{align}
Moreover, the estimate
\begin{align}
|r-s| &= \bigl| \frac{r-s}{2} + Z + \frac{r-s}{2} - Z\bigr| \leq \bigl| Z - \frac{r-s}{2}\bigr| + \bigl| Z + \frac{r-s}{2}\bigr| \label{DHS1:r-s_estimate}
\end{align}
shows that $ |r-s|\, F_{T,B}^0(r-s) \leq F_{T,B}^1(r-s)$. We conclude the estimate
\begin{align}
\smash{\int_{\Rbb^3}} \dd Z \; |\tilde \pi_r k_{T,B}(Z, r, s) - \tilde \pi_r\tilde k_{T,B} (Z, r, s)| \notag \\
&\hspace{-80pt}\leq G_{T,B}(r-s) + B\;F_{T,B}^1(r-s) + B\;F_{T,B}^0(r-s) \;|s|. \label{DHS1:LTB-LtildeTB_7}
\end{align}
From \eqref{DHS1:LTB-LtildeTB_7} we deduce \eqref{DHS1:LTB-LtildeTB_4}, which proves the claim.
\end{proof}

%

\paragraph{The operator $M_{T,B}$.} The operator $M_{T,B}$ is defined by
\begin{align}
M_{T,B} \alpha(X,r) := \iint_{\Rbb^3\times \Rbb^3} \dd Z \dd s \; k_T(Z, r-s) \;(\cos(Z\cdot \Pi_X)\alpha)(X,s), \label{DHS1:MTB_definition}
\end{align}
where $k_T(Z, r) := k_{T,0}(Z, r, 0)$ with $k_{T,0}$ in \eqref{DHS1:kTB_definition}. The following proposition allows us to replace $\tilde L_{T,B}$ by $M_{T,B}$ in our computations.

\begin{prop}
\label{DHS1:LtildeTB-MTB}
For any $T_0>0$ there is $B_0>0$ such that for any $0< B \leq B_0$, any $T\geq T_0$, and whenever  $|\cdot|^k V\alpha_*\in L^2(\Rbb^3)$ for $k\in \{0,1\}$,  $\Psi\in \Hmag^1(Q_B)$, and $\Delta \equiv \Delta_\Psi$ as in \eqref{DHS1:Delta_definition}, we have
\begin{align}
	\Vert \tilde L_{T,B}\Delta - M_{T,B}\Delta \Vert_\Hsymm^2 &\leq C\;B^3 \;\bigl( \Vert V\alpha_*\Vert_2^2 + \Vert \, |\cdot|V\alpha_*\Vert_2^2\bigr) \;\Vert \Psi\Vert_{\Hmag^1(Q_B)}^2. \label{DHS1:LtildeTB-MTB_2}
\end{align}
If instead $|\cdot|^2 V\alpha_*\in L^2(\Rbb^3)$ then
\begin{align}
	|\langle \Delta, \tilde L_{T,B}\Delta - M_{T,B}\Delta\rangle| &\leq C \;B^3 \;\Vert\,  |\cdot|^2 V\alpha_*\Vert_2^2 \;\Vert \Psi\Vert_{\Hmag^1(Q_B)}^2. \label{DHS1:LtildeTB-MTB_3}
\end{align}
\end{prop}

\begin{bem}
	The $\Hsymm$-norm bound in \eqref{DHS1:LtildeTB-MTB_2} is needed for the proof of Proposition \ref{DHS1:Structure_of_alphaDelta} and the quadratic form bound in \eqref{DHS1:LtildeTB-MTB_3} is needed for the proof of Theorem~\ref{DHS1:Calculation_of_the_GL-energy}. We highlight that the bound in \eqref{DHS1:LtildeTB-MTB_2} is insufficient as far as the proof of Theorem~\ref{DHS1:Calculation_of_the_GL-energy} is concerned. More precisely, if we apply Cauchy--Schwarz to the left side of \eqref{DHS1:LtildeTB-MTB_3}, and use \eqref{DHS1:LtildeTB-MTB_2} as well as the Lemma~\ref{DHS1:Schatten_estimate} to estimate $\Vert \Delta\Vert_2$ we obtain a bound of the order $B^2$ only. This is not good enough because $B^2$ is the order of the Ginzburg--Landau energy.

	To obtain the desired quality for the quadratic form bound \eqref{DHS1:LtildeTB-MTB_3}, we exploit the fact that $V\alpha_*$ is real-valued, which allows us to replace the magnetic phase factor $\exp(\frac{\i}{4} \Bbold (r\wedge s))$ in $\tilde k_{T,B}$ in \eqref{DHS1:ktildeTB_definition} by $\cos(\frac 14 \Bbold (r\wedge s))$. This improves the error estimate by an additional factor of $B$.
\end{bem}

Before we start with the proof of Proposition \ref{DHS1:LtildeTB-MTB}, let $a\in \Nbb_0$ and define the functions
\begin{align}
F_T^{a} := \frac 2\beta \sum_{n\in \Zbb} \sum_{b = 0}^a \binom ab \; \bigl(|\cdot|^{b}\,  |g_0^{\i\omega_n}|\bigr) * \bigl(|\cdot|^{a-b} \, |g_0^{-\i\omega_n}| \bigr) \label{DHS1:LtildeTB-MTB_FT_definition}
\end{align}
and
\begin{align}
G_T &:= \smash{\frac 2\beta \sum_{n\in \Zbb}} \bigl( |\cdot| \, |\nabla g_0^{\i \omega_n}| \bigr) * |g_0^{-\i\omega_n}| + |\nabla g_0^{\i\omega_n}| * \bigl( |\cdot| \, |g_0^{-\i\omega_n}|\bigr) \notag\\
&\hspace{110pt} + \bigl( |\cdot| \, |g_0^{\i\omega_n}| \bigr) *  |\nabla g_0^{-\i\omega_n}| + |g_0^{\i\omega_n}| * \bigl( |\cdot| \, |\nabla g_0^{-\i\omega_n}|\bigr). \label{DHS1:LtildeTB-MTB_GT_definition}
\end{align}
For $T \geq T_0 > 0$ and $a\in \Nbb_0$, by Lemma~\ref{DHS1:g0_decay},  \eqref{DHS1:g0_decay_f_estimate1}, and \eqref{DHS1:g0_decay_f_estimate2}, we have
\begin{align}
\Vert F_{T}^a\Vert_1 + \Vert G_{T}\Vert_1 &\leq C_{a}. \label{DHS1:LtildeTB-MTB_FTBGTB}
\end{align}

\begin{proof}[Proof of Proposition \ref{DHS1:LtildeTB-MTB}]
We start with the proof of \eqref{DHS1:LtildeTB-MTB_2}, which is similar to the proof of Proposition \ref{DHS1:LTB-LtildeTB}. We claim that
\begin{align}
\Vert \tilde L_{T,B}\Delta - M_{T,B}\Delta\Vert_2^2 &\leq 4 \;B^2 \;\Vert \Psi\Vert_2^2 \; \Vert F_T^1 * |\cdot|\, |V\alpha_*| \, \Vert_2^2. \label{DHS1:LtildeTB-MTB_1}
\end{align}
If this is true Young's inequality, \eqref{DHS1:Periodic_Sobolev_Norm}, and \eqref{DHS1:LtildeTB-MTB_FTBGTB} prove the claimed bound for this term.
To see that \eqref{DHS1:LtildeTB-MTB_1} holds, we argue as in \eqref{DHS1:Expanding_the_square}-\eqref{DHS1:LTB-LtildeTB_2} and find
\begin{align*}
\Vert \tilde L_{T,B}\Delta - M_{T,B}\Delta\Vert_2^2 & \\
&\hspace{-60pt}\leq 4\Vert \Psi\Vert_2^2 \int_{\Rbb^3} \dd r\,  \Bigl|\frac{2}{\beta } \sum_{n\in \Zbb} \iint_{\Rbb^3\times \Rbb^3}\dd Z \dd s \;  \bigl| k_{T,0}^n (Z, r-s) \bigl[ \e^{\frac{\i}{4} \Bbold \cdot (r\wedge s)} - 1\bigr] \bigr| \;|V\alpha_*(s)|\Bigr|^2.
\end{align*}
Since $|r\wedge s| \leq |r - s|\, |s|$, we have  $|\e^{\frac{\i}{4}\Bbold \cdot (r\wedge s)} - 1| \leq B\, |r-s|\, |s|$ as well as
\begin{align*}
|k_{T,0}^n(Z, r-s)| \bigl|\e^{\frac{\i}{4} \Bbold \cdot (r\wedge s)} - 1\bigr| &\leq B \;|g_0^{\i\omega_n}|\bigl( Z - \frac{r-s}{2}\bigr) \;|g_0^{-\i\omega_n}|\bigl( Z + \frac{r-s}{2}\bigr) \; |r-s|\; |s|.
\end{align*}
In combination with the estimate for $|r-s|$ in \eqref{DHS1:r-s_estimate} and the bound for $|Z|^a$ in \eqref{DHS1:Z_estimate}, this proves 
\begin{align}
\frac{2}{\beta} \sum_{n\in \Zbb} \int_{\Rbb^3} \dd Z \; |Z|^a \, |k_{T,0}^n(Z, r-s)| \, \bigl| \e^{\frac{\i}{4} \Bbold \cdot (r\wedge s)} - 1\bigr| &\leq B\;F_T^{a+1} (r - s)\;|s| \label{DHS1:LtildeTB-MTB_4}
\end{align}
for $a \in \Nbb_0$. The case $a = 0$ implies \eqref{DHS1:LtildeTB-MTB_1}. A computation similar to the one leading to \eqref{DHS1:LTB-LtildeTB_9} shows 
\begin{align*}
\Vert \Pi_X (\tilde L_{T,B}\Delta - M_{T,B}\Delta)\Vert_2^2 &\leq C\, B^4 \; \Vert \Psi\Vert_{\Hmag^1(Q_B)}^2 \; \Vert (F_T^1 + F_T^2) * |\cdot|\, |V\alpha_*|\, \Vert_2^2.
\end{align*}
To obtain the result we also used \eqref{DHS1:PiXcos_estimate} and \eqref{DHS1:LtildeTB-MTB_4}. We apply Young's inequality and use \eqref{DHS1:LtildeTB-MTB_FTBGTB} to prove the claimed bound for this term.
%
Finally, a computation similar to the one that leads to \eqref{DHS1:LTB-LtildeTB_10} shows
\begin{align*}
\Vert \tilde \pi_r (\tilde L_{T,B}\Delta - M_{T,B}\Delta)\Vert_2^2 & \\
&\hspace{-80pt}\leq 4\, \Vert \Psi\Vert_2^2 \int_{\Rbb^3} \dd r \, \Bigl| \iint_{\Rbb^3\times \Rbb^3}\dd Z \dd s \; \frac{2}{\beta } \sum_{n\in \Zbb} \bigl| \tilde \pi_r k_{T,0}^n (Z, r-s) \bigl[ \e^{\frac{\i}{4} \Bbold (r\wedge s)} - 1\bigr] \bigr| \; |V\alpha_*(s)|\Bigr|^2.
\end{align*}
We argue as in the proof of \eqref{DHS1:LTB-LtildeTB_7} to see that
\begin{align*}
\smash{\frac 2\beta \sum_{n\in \Zbb} \int_{\Rbb^3} \dd Z}\; \bigl| \tilde \pi_r k_{T,0}^n (Z, r-s) \bigl[ \e^{\frac{\i}{4} \Bbold (r\wedge s)} - 1\bigr] \bigr| &\\
&\hspace{-100pt}\leq C\, B \; \bigl( G_T(r-s)\; |s| + F_T^1(r-s) + F_T^0(r-s) \; |s|\bigr) .
\end{align*}
With the help of Young's inequality and \eqref{DHS1:LtildeTB-MTB_FTBGTB}, these considerations prove \eqref{DHS1:LtildeTB-MTB_2}.

It remains to prove \eqref{DHS1:LtildeTB-MTB_3}. The term we need to estimate reads
\begin{align}
\langle \Delta, \tilde L_{T,B} \Delta - M_{T,B} \Delta \rangle &= 4 \iint_{\Rbb^3\times \Rbb^3} \dd r \dd s \; \bigl( \e^{\frac \i 4 \Bbold \cdot (r\wedge s)} - 1 \bigr) V\alpha_*(r) V\alpha_*(s)  \notag\\
&\hspace{-20pt} \times \int_{\Rbb^3} \dd Z \; \frac{2}{\beta} \sum_{n\in\Zbb} k_{T,0}^n(Z, r-s)  \fint_{Q_B} \dd X \; \ov{\Psi(X)} \cos(Z\cdot \Pi_X)\Psi(X). \label{DHS1:LtildeTB-MTB_5}
\end{align}
Except for the factor $\e^{\frac \i 4 \Bbold (r\wedge s)}$, the right side is symmetric under the exchange of the coordinates $r$ and $s$. The exponential factor acquires a minus sign in the exponent when this transformation is applied. When we add the right side of \eqref{DHS1:LtildeTB-MTB_5} and the same term with the roles of $r$ and $s$ interchanged, we get
%
\begin{align}
\langle \Delta, \tilde L_{T,B} \Delta - M_{T,B} \Delta \rangle & = -8 \iint_{\Rbb^3\times \Rbb^3} \dd r \dd s \; \sin^2\bigl( \frac 1 8 \, \Bbold\cdot (r\wedge s)\bigr) \; V\alpha_*(r) V\alpha_*(s) \notag \\
&\hspace{-20pt}\times \int_{\Rbb^3} \dd Z \; \frac{2}{\beta} \sum_{n\in\Zbb} k_{T,0}^n(Z, r-s)  \fint_{Q_B} \dd X \; \ov{\Psi(X)} \cos(Z\cdot \Pi_X)\Psi(X). \label{DHS1:eq:A12}
\end{align}
To obtain \eqref{DHS1:eq:A12} we also used $\cos(x) -1 =- 2\sin^2(\frac x2)$. The operator norm of $\cos(Z\cdot\Pi_X)$ is bounded by $1$ and we have $\sin^2(\frac 18 \Bbold \cdot (r\wedge s)) \leq \frac 18 B^2 |r|^2 |s|^2$. Therefore, \eqref{DHS1:eq:A12} proves
\begin{align}
|\langle \Delta, \tilde L_{T,B}\Delta - M_{T,B}\Delta\rangle | &\leq B^2 \; \Vert \Psi\Vert_2^2 \; \bigl\Vert |\cdot|^2|V\alpha_*| \; \bigl( |\cdot |^2|V\alpha_*| * F_T^0\bigr) \bigr\Vert_1.
\label{DHS1:eq:A13}
\end{align}
Finally, we use Young's inequality, \eqref{DHS1:Periodic_Sobolev_Norm}, and \eqref{DHS1:LtildeTB-MTB_FTBGTB} and obtain \eqref{DHS1:LtildeTB-MTB_3}. This proves Proposition~\ref{DHS1:LtildeTB-MTB}.
\end{proof}

%

\subsubsection{Analysis of \texorpdfstring{$M_{T,B}$}{MTB} and calculation of the quadratic terms}
\label{DHS1:Analysis_of_MTB_Section}

We decompose $M_{T,B} = M_T^{(1)} + M_{T,B}^{(2)} + M_{T,B}^{(3)}$, where
\begin{align}
M_T^{(1)} \alpha(X,r) &:= \iint_{\Rbb^3\times \Rbb^3} \dd Z \dd s \; k_T(Z, r-s) \; \alpha(X,s), \label{DHS1:MT1_definition}\\
M_{T,B}^{(2)} \alpha(X, r) &:=  \iint_{\Rbb^3\times \Rbb^3} \dd Z \dd s\; k_T(Z, r-s) \; \bigl( -\frac 12\bigr) (Z\cdot \Pi_X)^2 \; \alpha(X, s), \label{DHS1:MTB2_definition}\\
M_{T,B}^{(3)} \alpha(X,r) &:= \iint_{\Rbb^3\times \Rbb^3} \dd Z \dd s\; k_T(Z, r-s) \; \Rcal(Z\cdot \Pi_X) \; \alpha(X, s), \label{DHS1:MTB3_definition}
\end{align}
and $\Rcal(x) = \cos(x) - 1 + \frac 12 x^2$.

\paragraph{The operator $M_T^{(1)}$.} The expression $\langle \Delta, M_T^{(1)} \Delta \rangle$ contains a term that cancels the last term on the left side of \eqref{DHS1:Calculation_of_the_GL-energy_eq} as well as the quadratic term without magnetic gradient in the Ginzburg--Landau functional in \eqref{DHS1:Definition_GL-functional}. The following result allows us to extract these terms. We recall that $\Delta \equiv \Delta_\Psi = -2 V\alpha_* \Psi$.

\begin{prop}
\label{DHS1:MT1}
Assume that $V\alpha_*\in L^2(\Rbb^3)$ and let $\Psi \in \Lmag^2(Q_B)$ and $\Delta \equiv \Delta_\Psi$ as in \eqref{DHS1:Delta_definition}.
\begin{enumerate}[(a)]
\item We have $M_{\Tc}^{(1)} \Delta (X, r) = -2\, \alpha_* (r) \Psi(X)$.

\item For any $T_0>0$ there is a constant $c>0$ such that for $T_0 \leq T \leq \Tc$ we have
\begin{align*}
\langle \Delta, M_T^{(1)} \Delta - M_{\Tc}^{(1)} \Delta \rangle \geq  c \, \frac{\Tc - T}{\Tc} \; \Vert \Psi\Vert_2^2.
\end{align*}

\item Given $D\in \Rbb$ there is $B_0>0$ such that for $0< B\leq B_0$, and $T = \Tc (1 - DB)$ we have
\begin{align*}
	\langle \Delta, M_T^{(1)} \Delta -  M_{\Tc}^{(1)} \Delta\rangle  = 4\; DB \; \Lambda_2 \; \Vert \Psi\Vert_2^2 + R(\Delta)
\end{align*}
with the coefficient $\Lambda_2$ in \eqref{DHS1:GL-coefficient_2}, and
\begin{align*}
	|R(\Delta)| &\leq C \; B^2 \; \Vert V\alpha_*\Vert_2^2\; \Vert \Psi\Vert_{2}^2.
\end{align*}

\item Assume additionally that $| \cdot | V\alpha_*\in L^2(\Rbb^3)$. There is $B_0>0$ such that for any $0< B\leq B_0$, any $\Psi\in \Hmag^1(Q_B)$, and any $T \geq T_0 > 0$ we have 
\begin{align*}
	\Vert M_T^{(1)}\Delta - M_{\Tc}^{(1)}\Delta\Vert_{\Hsymm}^2 &\leq C \, B \, | T - \Tc |^2 \,  \bigl( \Vert V\alpha_*\Vert_2^2 + \Vert \, |\cdot| V\alpha_*\Vert_2^2\bigr) \Vert\Psi\Vert_{\Hmag^1(Q_B)}^2.
\end{align*}

\end{enumerate}
\end{prop}

\begin{bem}
	The above bound for the remainder term implies
	\begin{equation*}
	| R(\Delta) | \leq C \; B^3 \; \Vert V\alpha_*\Vert_2^2\; \Vert \Psi\Vert_{\Hmag^1(Q_B)}^2.
	\end{equation*}
	Part~(b) in the Proposition is needed for the proof of Proposition~\ref{DHS1:Lower_Tc_a_priori_bound}. Part~(d) is needed in the proof of Proposition~\ref{DHS1:Structure_of_alphaDelta}.
\end{bem}

Before we give the proof of the above proposition, we introduce the function
\begin{align}
	F_{T,\Tc} &:= \frac{2}{\beta} \sum_{n\in \Zbb} |2n+1| \bigl[  |g_0^{\i\omega_n^T}| * |g_0^{\i\omega_n^{\Tc}}| * |g_0^{-\i\omega_n^T}|  + |g_0^{\i\omega_n^{\Tc}}| *  |g_0^{-\i\omega_n^T}| * |g_0^{-\i\omega_n^{\Tc}}| \bigr], \label{DHS1:MT1_FTTc_definition}
\end{align}
where we indicated the $T$-dependence of the Matusubara frequencies in \eqref{DHS1:Matsubara_frequencies} because different temperatures appear in the formula. As long as $T \geq T_0 > 0$, Lemma~\ref{DHS1:g0_decay} and \eqref{DHS1:g0_decay_f_estimate1} imply the bound  
\begin{equation}
	\Vert F_{T,\Tc} \Vert_1 \leq C.
	\label{DHS1:MT1_FTTc_bound}
\end{equation}

\begin{proof}[Proof of Proposition \ref{DHS1:MT1}]
We start with the proof of part (a). First of all, we recall that $k_T(Z, r) = k_{T,0}(Z, r, 0)$ with $k_{T,B}(Z,r,s)$ in \eqref{DHS1:kTB_definition}. In Fourier space the convolution operator $g_0^{\pm \i \omega_n}(x-y)$ equals multiplication with $(\pm\i\omega_n + \mu - k^2)^{-1}$. This allows us to write
\begin{align*}
	k_T(Z, r) &= \frac{2}{\beta} \sum_{n\in\Zbb} \iint_{\Rbb^3\times \Rbb^3} \frac{\dd k}{(2\pi)^3} \frac{\dd \ell}{(2\pi)^3} \; \frac{\e^{\i k\cdot (Z - \frac r2)}}{\i\omega_n  + \mu - k^2} \frac{\e^{\i \ell \cdot (Z + \frac r2)}}{-\i\omega_n + \mu - \ell^2} \\
	&= -\frac 2\beta \sum_{n\in \Zbb} \iint_{\Rbb^3\times \Rbb^3} \frac{\dd p}{(2\pi)^3} \frac{\dd q}{(2\pi)^3} \; \e^{\i Z\cdot q} \e^{-\i r\cdot p} \; \frac{1}{\i\omega_n + \mu - (p+\frac  q2)^2} \frac{1}{\i\omega_n - \mu + (p - \frac q 2)^2},
\end{align*}
where we applied the change of variables $q= k+\ell$ and $p = \frac{k-\ell}{2}$. We use the partial fraction expansion
\begin{align*}
	\frac{1}{E + E'} \Bigl( \frac{1}{\i\omega_n - E} - \frac{1}{\i\omega_n + E'} \Bigr) = \frac{1}{\i\omega_n - E} \frac{1}{\i\omega_n + E'}
\end{align*}
and the representation formula of the hyperbolic tangent in \eqref{DHS1:tanh_Matsubara} to see that
\begin{equation}
	k_T(Z, r) = \iint_{\Rbb^3 \times \Rbb^3} \frac{\dd p}{(2\pi)^3} \frac{\dd q}{(2\pi)^3} \; \e^{\i Z\cdot q} \e^{-\i r\cdot p} \; L_T\bigl( p+ \frac{q}{2}, p - \frac{q}{2}\bigr), \label{DHS1:kT_Fourier_representation2}
\end{equation}
where
\begin{align}
	L_T(p,q) := \frac{\tanh(\frac \beta 2(p^2-\mu)) + \tanh(\frac \beta 2 (q^2-\mu))}{p^2-\mu + q^2-\mu}. \label{DHS1:LT_definition}
\end{align}
In particular,
\begin{align}
	\int_{\Rbb^3} \dd Z \; k_T(Z, r) = \int_{\mathbb{R}^3} \frac{\dd p}{(2\pi)^3} \; \e^{-\i r\cdot p}  L_T(p, p) = \int_{\mathbb{R}^3} \frac{\dd p}{(2\pi)^3} \; \e^{-\i r\cdot p} K_T(p)^{-1} \label{DHS1:LT_integration_KT}
\end{align}
with $K_T(p)$ in \eqref{DHS1:KT-symbol}. Therefore, we have
\begin{align*}
M_T^{(1)}\Delta(X, r) = K_T^{-1} V\alpha_*(r) \, \Psi(X),
\end{align*}
which together with $K_{\Tc} \alpha_* = V \alpha_*$ proves part (a).
To prove part (b), we use \eqref{DHS1:LT_integration_KT} to write
\begin{align}
\langle \Delta, M_T^{(1)} \Delta - M_{\Tc}^{(1)} \Delta\rangle =  \int_{\Rbb^3} \frac{\dd p}{(2\pi)^3} \; \bigl[ K_T(p)^{-1} - K_{\Tc}(p)^{-1}\bigr] \, |(-2)V\alpha_*(p)|^2 \, \Vert \Psi\Vert_2^2. \label{DHS1:MT1_3}
\end{align}
With the help of the first order Taylor expansion
\begin{align}
K_T(p)^{-1} - K_{T_c}(p)^{-1} &
= \frac 12\int_{T}^{T_c} \dd T' \; \frac{1}{(T')^2} \frac{1}{\cosh^2( \frac{p^2-\mu}{2T'})} \label{DHS1:MT1_1}
\end{align}
we see that
\begin{align*}
	K_T(p)^{-1} - K_{\Tc}(p)^{-1} \geq \frac 12 \frac{\Tc - T}{\Tc^2} \, \frac{1}{\cosh^2(\frac{p^2-\mu}{2T_0})}
\end{align*}
holds for $T_0 \leq T \leq \Tc $. This and \eqref{DHS1:MT1_3} prove part (b).

To prove part (c), we expand \eqref{DHS1:MT1_1} to second order in $T-T_c$ and find
\begin{align*}
	\Bigl|\int_{\Rbb^3} \frac{\dd p}{(2\pi)^3} \; [K_T(p)^{-1} - K_{\Tc}(p)^{-1} ] \; |(-2)\hat{V\alpha_*}(p)|^2 - 4\, \Lambda_2\; \frac{\Tc - T}{\Tc}  \Bigr| \leq C\, |T- \Tc|^2 \; \Vert V\alpha_*\Vert_2^2
\end{align*}
with $\Lambda_2$ in \eqref{DHS1:GL-coefficient_2}. By \eqref{DHS1:MT1_3}, this proves part (c).

It remains to prove part (d). We use the resolvent identity to see that
\begin{align}
g_0^{\pm \i\omega_n^T} - g_0^{\pm \i \omega_n^{\Tc}}  = \mp \i (\omega_n^T - \omega_n^{\Tc}) \;  g_0^{\pm \i \omega_n^T} * g_0^{\pm \i \omega_n^{\Tc}} . \label{DHS1:gTgTc}
\end{align}
Using \eqref{DHS1:gTgTc}, it is straightforward to see that
\begin{align*}
\Vert M_T^{(1)}\Delta - M_{\Tc}^{(1)}\Delta\Vert_2^2 &\leq C \, |T - \Tc|^2 \, \Vert F_{T,\Tc} \Vert_1 \,  \Vert V\alpha_*\Vert_2^2 \, \Vert\Psi\Vert_2^2
\end{align*}
holds with $F_{T,\Tc}$ in \eqref{DHS1:MT1_FTTc_definition}. In combination with \eqref{DHS1:MT1_FTTc_bound} this proves the claimed bound for this term. The estimates for the terms $\Vert \tilde \pi_r (M_T^{(1)}\Delta - M_{\Tc}^{(1)}\Delta)\Vert_2^2$ and $\Vert \Pi_X(M_T^{(1)}\Delta - M_{\Tc}^{(1)}\Delta)\Vert_2^2$ are proved similarly. We omit the details.
\end{proof}

\paragraph{The operator $M_{T,B}^{(2)}$.} The term $\langle \Delta, M_{T,B}^{(2)} \Delta \rangle$ with $M_{T,B}^{(2)}$ in \eqref{DHS1:MTB2_definition} contains the kinetic term in the Ginzburg--Landau functional in \eqref{DHS1:Definition_GL-functional}. The following proposition allows us to compare the two.

\begin{prop}
\label{DHS1:MTB2}
Assume that the function $V\alpha_*$ is radial and belongs to $L^2(\Rbb^3)$. For any $B>0$, $\Psi\in \Hmag^1(Q_B)$, and $\Delta \equiv \Delta_\Psi$ as in \eqref{DHS1:Delta_definition}, we have
\begin{align}
\langle \Delta, M_{\Tc,B}^{(2)} \Delta\rangle = - 4\; \Lambda_0 \; \Vert \Pi\Psi\Vert_2^2 \label{DHS1:MTB2_1}
\end{align}
with $\Lambda_0$ in \eqref{DHS1:GL-coefficient_1}. Moreover, for any $T \geq T_0 > 0$ we have
\begin{align}
	|\langle \Delta,  M_{T,B}^{(2)} \Delta - M_{\Tc,B}^{(2)}\Delta\rangle| \leq C\; B^2 \; |T - \Tc| \; \Vert V\alpha_*\Vert_2^2 \; \Vert \Psi\Vert_{\Hmag^1(Q_B)}^2. \label{DHS1:MTB2_2}
\end{align}
\end{prop}

Before we give the proof of Proposition~\ref{DHS1:MTB2}, let us introduce the function
\begin{align}
F_{T,\Tc}^a &:= \smash{\frac{2}{\beta} \sum_{n\in \Zbb} \sum_{\substack{a_1,a_2,a_3\in \Nbb_0 \\ a_1+ a_2 + a_3 = a}}}  |2n+1| \bigl[ \bigl(|\cdot|^{a_1} \, |g_0^{\i\omega_n^T}|\bigr) * \bigl(|\cdot|^{a_2} \, |g_0^{\i\omega_n^{\Tc}}|\bigr) * \bigl(|\cdot|^{a_3} \, |g_0^{-\i\omega_n^T}|\bigr) \notag\\
&\hspace{110pt} + \bigl(|\cdot|^{a_1} \, |g_0^{\i\omega_n^{\Tc}}|\bigr) * \bigl(|\cdot|^{a_2} \, |g_0^{-\i\omega_n^T}|\bigr)* \bigl(|\cdot|^{a_3} \, |g_0^{-\i\omega_n^{\Tc}}|\bigr)\bigr], \label{DHS1:MTB2_FTTc_definition}
\end{align}
where $a\in \Nbb_0$ and where we indicated the $T$-dependence of the Matusubara frequencies in \eqref{DHS1:Matsubara_frequencies} because different temperatures appear in the formula. 
As long as $T \geq T_0 > 0$, Lemma~\ref{DHS1:g0_decay} and \eqref{DHS1:g0_decay_f_estimate1} imply the bound $\Vert F_{T,\Tc}^a \Vert_1 \leq C_a$.

\begin{proof}[Proof of Proposition \ref{DHS1:MTB2}]
We have
\begin{align}
\langle \Delta, M_{\Tc, B}^{(2)} \Delta\rangle &= -2 \iint_{\Rbb^3\times\Rbb^3} \dd r\dd s \; V\alpha_*(r)V\alpha_*(s) \int_{\Rbb^3} \dd Z\; k_{\Tc}(Z, r-s) \; \langle \Psi, (Z\cdot \Pi_X)^2 \Psi\rangle
\label{DHS1:eq:A10}
\end{align}
and 
\begin{align*}
\langle \Psi, (Z \cdot \Pi)^2 \Psi\rangle &= \sum_{i,j=1}^3 Z_iZ_j \; \langle \Pi^{(i)} \Psi, \Pi^{(j)} \Psi\rangle. 
\end{align*}
The integration over $Z$ in \eqref{DHS1:eq:A10} defines a $3\times 3$ matrix with matrix elements
\begin{align*}
	\int_{\Rbb^3} \dd Z \, k_{\Tc}(Z, r) \; Z_iZ_j &=   \iint_{\Rbb^3\times \Rbb^3} \frac{\dd p}{(2\pi)^3} \frac{\dd q}{(2\pi)^3} \int_{\Rbb^3} \dd Z \; \e^{-\i Z\cdot q} \e^{-\i p\cdot r} \; L_{\Tc}\bigl( p + \frac q2, p -  \frac q2\bigr)Z_iZ_j,
\end{align*}
which we have written in terms of the Fourier representation of $k_{\Tc}(Z,r)$ in \eqref{DHS1:kT_Fourier_representation2}. We use $Z_i Z_j \e^{-\i Z\cdot q} = -\partial_{q_i} \partial_{q_j} \e^{-\i Z\cdot q}$, integrate by parts twice, and find
\begin{align*}
	\int_{\Rbb^3} \frac{\dd q}{(2\pi)^3} \int_{\Rbb^3} \dd Z \; \e^{-\i Z\cdot q} \; L_{\Tc}\bigl( p + \frac q2, p - \frac q2\bigr) \; Z_iZ_j &= -\Bigl[ \frac{\partial}{\partial q_i} \frac{\partial}{\partial q_j} L_{\Tc} \big(p + \frac q2,p - \frac q2\bigr) \Bigr]_{q=0}.
\end{align*}
A tedious but straightforward computation shows that the right side of the above equation can be written in terms of the functions $g_1$ and $g_2$ in \eqref{DHS1:XiSigma} as
\begin{align*}
	-\Bigl[ \frac{\partial}{\partial q_i} \frac{\partial}{\partial q_j}  L_{\Tc} \bigl( p + \frac q2, p - \frac q2\bigr) \Bigr]_{q=0}  =  \frac{\beta_c^2}{2} \bigl[ g_1 (\beta_c(p^2-\mu)) \delta_{ij} + 2\beta_c \; g_2(\beta_c(p^2-\mu)) \; p_ip_j\bigr],
\end{align*}
and hence
\begin{equation*}
	\int_{\Rbb^3} \dd Z \; k_{\Tc}(Z, r) \, Z_iZ_j = \frac{\beta_c^2}{2}  \int_{\Rbb^3} \frac{\dd p}{(2\pi)^3} \;  \e^{-\i p \cdot r} \bigl[ g_1 (\beta_c(p^2-\mu)) \delta_{ij} + 2\beta_c \; g_2(\beta_c(p^2-\mu)) \; p_ip_j\bigr].
\end{equation*}
Let us denote the term in the bracket on the right side by $A_{ij}(p)$. When we insert the above identity into \eqref{DHS1:eq:A10} we find
\begin{equation}
	\langle \Delta, M_{\Tc, B}^{(2)} \Delta\rangle = -\frac{\beta_{\mathrm{c}}^2}{4} \sum_{i,j = 1}^3 \langle \Pi^{(i)} \Psi, \Pi^{(j)} \Psi \rangle \int_{\mathbb{R}^3} \frac{\dd p}{(2\pi)^3} \; |(-2)\hat{V\alpha_*}(p)|^2 A_{ij}(p).
\end{equation}
We use that $V\alpha_*$ is a radial function to see that the integral of the term proportional to $p_i p_j$ equals zero unless $i = j$. Since the angular average of $p_i^2$ equals $\frac 13 p^2$ this proves \eqref{DHS1:MTB2_1}.

It remains to prove \eqref{DHS1:MTB2_2}. To this end, we estimate
\begin{align}
|\langle \Delta, M_{T,B}^{(2)}\Delta - M_{\Tc,B}^{(2)}\Delta \rangle | & \notag\\
&\hspace{-100pt}\leq 2 \iiint_{\Rbb^3\times \Rbb^3\times \Rbb^3} \dd r\dd s\dd Z \; |V\alpha_*(r)|\; |V\alpha_*(s)| \; |k_T(Z, r-s)-k_{\Tc}(Z, r-s)|  \notag\\
&\hspace{160pt} \times |\langle \Psi, (Z\cdot \Pi)^2\Psi\rangle|. \label{DHS1:MTB2_3}
\end{align}
%
%
%
%
%
%
For general operators $A, B, C$, we have $|A + B + C|^2 \leq 3(|A|^2 + |B|^2 + |C|^2)$.
This implies
\begin{align}
	(Z\cdot \Pi)^2 &\leq 3 \; \bigl( Z_1^2 \; (\Pi^{(1)})^2 + Z_2^2 \; (\Pi^{(2)})^2 + Z_3^2 \; (\Pi^{(3)})^2\bigr) \leq 3 \; Z^2 \; \Pi^2, \label{DHS1:ZPiX_inequality}
\end{align}
and, in particular,
\begin{align}
|\langle \Psi, (Z\cdot \Pi)^2 \Psi\rangle | \leq 3 \, |Z|^2 \, \Vert \Pi\Psi\Vert_2^2 \leq 3\, B^2\, |Z|^2 \, \Vert \Psi\Vert_{\Hmag^1(Q_B)}^2. \label{DHS1:MTB2_4}
\end{align}
Moreover, \eqref{DHS1:gTgTc}
%
%
and the estimate for $|Z|^2$ in \eqref{DHS1:Z_estimate} show
\begin{align}
\int_{\Rbb^3} \dd Z \; |Z|^2 \, |k_T(Z, r) - k_{\Tc}(Z, r)| &\leq C\, |T-\Tc|\; F_{T,\Tc}^2(r) \label{DHS1:MTB2_5}
\end{align}
with $F_{T,\Tc}^2$ in \eqref{DHS1:MTB2_FTTc_definition}.
%
%
With the help of \eqref{DHS1:MTB2_3}, \eqref{DHS1:MTB2_4}, and \eqref{DHS1:MTB2_5}, we deduce 
\begin{align*}
|\langle \Delta, M_{T,B}^{(2)} \Delta - M_{\Tc,B}^{(2)}\Delta\rangle | &\leq C\, B^2 \; |T - \Tc| \; \Vert \Psi\Vert_{\Hmag^1(Q_B)}^2 \; \bigl\Vert |V\alpha_*| \; \bigl( |V\alpha_*| * F_{T,\Tc}^2\bigr) \bigr\Vert_1 .
%
\end{align*}
An application of Young's inequality completes the proof.
\end{proof}

\paragraph{The operator $M_{T,B}^{(3)}$.} The term $\langle \Delta, M_{T,B}^{(3)} \Delta \rangle$ with $M_{T,B}^{(3)}$ in \eqref{DHS1:MTB3_definition} is the remainder of our expansion of $\langle \Delta, M_{T,B} \Delta \rangle$ in powers of $B$. In contrast to the previous estimates, we need the $\Hmag^2(Q_B)$-norm of $\Psi$ to control its size. 

\begin{prop}
\label{DHS1:MTB3}
For any $T_0>0$ there is $B_0>0$ such that for any $0 < B \leq B_0$, any $T\geq T_0$, and whenever $V\alpha_*\in L^2(\Rbb^3)$, $\Psi \in \Hmag^2(Q_B)$, and $\Delta \equiv \Delta_\Psi$ as in \eqref{DHS1:Delta_definition}, we have
\begin{align*}
|\langle \Delta,  M_{T,B}^{(3)} \Delta\rangle| &\leq C \; B^3 \; \Vert V\alpha_*\Vert_2^2 \; \Vert \Psi\Vert_{\Hmag^2(Q_B)}^2.
\end{align*}
\end{prop}

Before we give the proof of Proposition~\ref{DHS1:MTB3}, let us introduce the function
\begin{align}
F_T(r) := \frac 1\beta \sum_{n\in \Zbb}\; \bigl(|\cdot|^4 |g_0^{\i\omega_n}| \bigr) * |g_0^{-\i\omega_n}| + |g_0^{\i\omega_n}| * \bigl(|\cdot|^4 |g_0^{-\i\omega_n}|\bigr). \label{DHS1:MTB3_FT_definition}
\end{align}
As long as $T \geq T_0 > 0$, Lemma~\ref{DHS1:g0_decay} and \eqref{DHS1:g0_decay_f_estimate1} imply $\Vert F_T\Vert_1\leq C$.

\begin{proof}[Proof of Proposition \ref{DHS1:MTB3}]
We have
\begin{align}
\langle \Delta, M_{T_B}^{(3)} \Delta \rangle &= 4 \iiint_{\Rbb^3\times \Rbb^3\times \Rbb^3} \dd r\dd s\dd Z \; V\alpha_*(r) V\alpha_*(s)\, k_T(Z, r-s) \; \langle \Psi , \Rcal(Z\cdot \Pi_X)\Psi\rangle, \label{DHS1:MTB3_1}
\end{align}
where the function $\Rcal(x) = \cos(x) - 1 + \frac{x^2}{2}$ obeys the bound $0\leq\Rcal(x) \leq \frac{1}{24} x^4$. We claim that
\begin{align}
|Z\cdot \Pi|^4 &\leq 9\; |Z|^4 \; ( \Pi^4 + 8 B^2),\label{DHS1:ZPiX_inequality_quartic}
\end{align}
which implies
\begin{align}
\langle \Psi, \Rcal(Z\cdot \Pi) \Psi\rangle &\leq \frac{9}{24} \; |Z|^4 \; \bigl( \Vert \Pi^2\Psi\Vert_2^2 + 8 B^2 \Vert \Psi\Vert_2^2\bigr) \leq C\; B^3 \; |Z|^4 \; \Vert \Psi\Vert_{\Hmag^2(Q_B)}^2. \label{DHS1:MTB3_2}
\end{align}
To see that \eqref{DHS1:ZPiX_inequality_quartic} is true, we note that $[\Pi^{(1)}, \Pi^{(2)}] = -2\i B$ implies
\begin{align}
\Pi \; \Pi^2 = \Pi^2 \; \Pi + 4\i B \;  (-\Pi^{(2)} , \Pi^{(1)} , 0)^t, \label{DHS1:PiX-operator-equality_1}
\end{align}
and hence
\begin{align}
\Pi \; \Pi^2\; \Pi = \Pi^4+ 8B^2. \label{DHS1:PiPi2Pi_equality}
\end{align}
We also have $[Z\cdot \Pi, \Pi] = -2\i \, \Bbold \wedge Z$, which implies $(Z\cdot \Pi)\Pi^2(Z\cdot \Pi)= \Pi (Z\cdot \Pi)^2\Pi$. We combine this with the operator inequality \eqref{DHS1:ZPiX_inequality} for $(Z\cdot \Pi)^2$ and \eqref{DHS1:PiPi2Pi_equality} and get $(Z\cdot \Pi)\Pi^2(Z\cdot \Pi) \leq 3|Z|^2(\Pi^4+8B^2)$. Finally, we write $|Z\cdot \Pi|^4 = (Z\cdot \Pi) (Z\cdot \Pi)^2 (Z\cdot \Pi)$, apply \eqref{DHS1:ZPiX_inequality} again, and obtain \eqref{DHS1:ZPiX_inequality_quartic}. 

Using the estimate \eqref{DHS1:Z_estimate} on $|Z|^4$ and \eqref{DHS1:MTB3_2}, we argue as in the proof of \eqref{DHS1:LTB-LtildeTB_5} to see that
\begin{align}
\int_{\Rbb^3} \dd Z\; |Z|^4 \; |k_T(Z,r)| &\leq F_T(r) \label{DHS1:MTB3_3}
\end{align}
with $F_T$ in \eqref{DHS1:MTB3_FT_definition}. The bound on the $L^1(\mathbb{R}^3)$-norm of $F_T$ below \eqref{DHS1:MTB3_FT_definition}, \eqref{DHS1:MTB3_1}, \eqref{DHS1:MTB3_2}, and \eqref{DHS1:MTB3_3} prove the claim.
\end{proof}

\subsubsection{Summary: The quadratic terms}
\label{DHS1:Summary_quadratic_terms_Section}

Let us summarize the results concerning the quadratic terms in $\Delta \equiv \Delta_\Psi$ that are relevant for the proof of Theorem~\ref{DHS1:Calculation_of_the_GL-energy} and provide an intermediate statement that is needed for the proof of Proposition \ref{DHS1:Lower_Tc_a_priori_bound}.

\begin{prop}
\label{DHS1:Rough_bound_on_BCS energy}
Given $T_0> 0$ there is a constant $B_0>0$ such that for any $T_0 \leq T\leq \Tc$, any $0 < B \leq B_0$, and whenever $|\cdot|^k V\alpha_*\in L^2(\Rbb^3)$ for $k\in \{0,1,2\}$, $\Psi \in \Hmag^1(Q_B)$, and $\Delta \equiv \Delta_\Psi$ as in \eqref{DHS1:Delta_definition}, we have
\begin{align}
- \frac 14 \langle \Delta, L_{T,B} \Delta\rangle + \Vert \Psi\Vert_2^2 \; \langle \alpha_*, V\alpha_*\rangle & \leq c  \, \frac{T - \Tc}{\Tc}\, \Vert \Psi\Vert_2^2 + C B^2  \, \Vert \Psi\Vert_{\Hmag^1(Q_B)}^2. \label{DHS1:Rough_bound_on_BCS energy_eq1}
\end{align}
\end{prop}

\begin{proof}
By Lemma \ref{DHS1:LTB_action}, the decomposition \eqref{DHS1:LTB_decomposition} of $L_{T,B}$, as well as Propositions~\ref{DHS1:LTB-LtildeTB} and \ref{DHS1:LtildeTB-MTB}, we have
\begin{align}
- \frac 14 \langle \Delta, L_{T,B} \Delta\rangle + \Vert \Psi\Vert_2^2 \; \langle \alpha_*, V\alpha_*\rangle &\notag\\
&\hspace{-100pt} = - \frac 14 \langle \Delta, M_T^{(1)}\Delta- M_{\Tc}^{(1)} \Delta\rangle - \frac 14 \langle \Delta, M_{T, B} \Delta - M_T^{(1)}\Delta\rangle + R_1(\Delta), \label{DHS1:Rough_bound_on_BCS energy_proof_1}
\end{align}
where
\begin{align*}
	| R_1(\Delta) | \leq C \; B^3 \; \Vert \Psi\Vert_{\Hmag^1(Q_B)}^2
\end{align*}
and, by Proposition \ref{DHS1:MT1}, 
\begin{align*}
-\frac 14 \langle \Delta, M_T^{(1)}\Delta- M_{\Tc}^{(1)} \Delta\rangle \leq  c \; \frac{T - \Tc}{\Tc} \; \Vert \Psi \Vert_2^2.
\end{align*}
We claim that
\begin{align}
|\langle \Delta, M_{T,B}\Delta - M_T^{(1)}\Delta\rangle| &\leq C\; B^2\; \Vert V\alpha_*\Vert_2^2 \; \Vert \Psi\Vert_{\Hmag^1(Q_B)}^2. \label{DHS1:MTB-MT1_eq1}
\end{align}
The proof of \eqref{DHS1:MTB-MT1_eq1} goes along the same lines as that of Proposition~\ref{DHS1:LtildeTB-MTB} and uses the operator inequality \eqref{DHS1:ZPiX_inequality} on $(Z\cdot \Pi)^2$ to estimate
\begin{align}
|\langle \Psi, [\cos(Z\cdot\Pi) - 1] \Psi \rangle| 
&\leq C\; B^2 \; |Z|^2 \; \Vert \Psi\Vert_{\Hmag^1(Q_B)}^2. \label{DHS1:MTB-MT1_1}
\end{align}
We omit the details. This proves \eqref{DHS1:Rough_bound_on_BCS energy_eq1}.
\end{proof}

Let the assumptions of Theorem~\ref{DHS1:Calculation_of_the_GL-energy} hold. We combine \eqref{DHS1:Rough_bound_on_BCS energy_proof_1} with the results of 
Propositions \ref{DHS1:MT1}, \ref{DHS1:MTB2}, and \ref{DHS1:MTB3} to see that for $T = \Tc(1-DB)$ with $D \in \mathbb{R}$ we have
\begin{equation}
	-\frac{1}{4} \langle \Delta, L_{T,B} \Delta \rangle + \Vert \Psi \Vert_2^2 \, \langle \alpha_*, V \alpha_* \rangle = \Lambda_0 \; \Vert \Pi\Psi\Vert_2^2 - DB \; \Lambda_2 \; \Vert \Psi\Vert_2^2 +  R_2(\Delta),
	\label{DHS1:eq:A15}
\end{equation}
where
\begin{equation*}
	| R_2(\Delta) | \leq C\,  B^3 \, \Vert \Psi\Vert_{\Hmag^2(Q_B)}^2.
\end{equation*}
This concludes our analysis of the operator $L_{T,B}$.

\subsubsection{A representation formula for the operator \texorpdfstring{$N_{T,B}$}{NTB}}


Let us introduce the notation $\Zbold$ for the vector $(Z_1,Z_2,Z_3)$ with $Z_1, Z_2, Z_3 \in \Rbb^3$. We also denote $\dd \Zbold = \dd Z_1 \dd Z_2 \dd Z_3$.
Remarkably, the strategy of the analysis we used for $L_{T,B}$ carries over to the nonlinear operator $N_{T,B}$ in \eqref{DHS1:NTB_definition}. As in the case of $L_{T,B}$, we start with a representation formula for the operator $N_{T,B}$ and note the analogy to Lemma \ref{DHS1:LTB_action}.

\begin{lem}
\label{DHS1:NTB_action}
The operator $N_{T,B} \colon \Hsymm \ra \Lsymm$ in \eqref{DHS1:NTB_definition} acts as
\begin{align*}
N_{T,B}(\alpha) (X, r) &= \iiint_{\Rbb^9} \dd \Zbold \iiint_{\Rbb^9} \dd \sbold \; \ell_{T,B}(\Zbold, r, \sbold)\; \Acal(X, \Zbold, \sbold)
\end{align*}
with
\begin{align}
\Acal(X, \Zbold, \sbold) &:= \e^{\i Z_1\cdot \Pi_X} \alpha(X, s_1) \; \ov{\e^{\i Z_2\cdot \Pi_X}\alpha(X, s_2)} \; \e^{\i Z_3\cdot \Pi_X} \alpha(X,s_3) \label{DHS1:NTB_alpha_definition}
\end{align}
and
\begin{align}
\ell_{T,B}(\Zbold, r, \sbold) &:= \frac 2\beta \sum_{n\in \Zbb} \ell_{T,B}^n(\Zbold, r, \sbold) \; \e^{\i \frac \Bbold 2 \cdot \Phi(\Zbold, r, \sbold)}, \label{DHS1:lTB_definition}
\end{align}
where
\begin{align}
\ell_{T,B}^n(\Zbold, r, \sbold) &:= g_B^{\i\omega_n} \bigl(Z_1 - \frac{r-s_1}{2}\bigr) \; g_B^{-\i\omega_n} \bigl( Z_1 - Z_2 - \frac{s_1 + s_2}{2}\bigr) \notag \\
&\hspace{50pt} \times g_B^{\i\omega_n} \bigl( Z_2 - Z_3 - \frac{s_2 + s_3}{2} \bigr) \; g_B^{-\i\omega_n} \bigl( Z_3 + \frac{r-s_3}{2}\bigr) \label{DHS1:lTBn_definition}
\end{align}
with $g_B^{\pm \i\omega_n}$ in \eqref{DHS1:gB_definition} and
\begin{align}
\Phi(\Zbold, r, \sbold) &:= \frac r2 \wedge \bigl( Z_1 - \frac{r - s_1}{2}\bigr) + \frac r2 \wedge \bigl( Z_3 + \frac{r-s_3}{2}\bigr) \notag \\
&\hspace{-35pt} + \bigl( Z_2 - Z_3 - \frac{s_2 + s_3}{2}\bigr) \wedge \bigl( Z_1 - Z_2 - \frac{s_1 + s_2}{2}\bigr) \notag \\
&\hspace{-35pt}+ \bigl( Z_3 + \frac{r - s_3}{2}\bigr) \wedge \bigl( Z_1 - Z_2 - \frac{s_1 + s_2}{2}\bigr)+ \bigl( s_2 + s_3 - \frac r2\bigr) \wedge \bigl( Z_1 - Z_2 - \frac{s_1 + s_2}{2}\bigr) \notag \\
&\hspace{-35pt}+ \bigl( Z_3 + \frac{r - s_3}{2} \bigr) \wedge \bigl( Z_3 - Z_2 + \frac{s_2 + s_3}{2}\bigr)+ \bigl( s_3 - \frac r2\bigr) \wedge \bigl( Z_3 - Z_2 + \frac{s_2 + s_3}{2}\bigr). 
%
%
%
%
\label{DHS1:PhiB_definition}
\end{align}
\end{lem}

\begin{bem}
	We highlight that the formula \eqref{DHS1:PhiB_definition} for the phase function $\Phi$ only involves the coordinates that appear in \eqref{DHS1:lTBn_definition} and the relative coordinates $r$ and $\sbold$. This structure allows us to remove the magnetic phase factor in \eqref{DHS1:lTB_definition} with techniques that are similar to the ones used in the analysis of $L_{T,B}$.
\end{bem}

\begin{proof}[Proof of Lemma \ref{DHS1:NTB_action}]
The integral kernel of $N_{T,B}$ reads
\begin{align*}
N_{T,B} (\alpha)(X,r) &= \smash{\frac 2\beta \sum_{n\in \Zbb} \iiint_{\Rbb^{9}} \dd \mathbf u \iiint_{\Rbb^9}\dd \mathbf v} \; G_B^{\i\omega_n} (\zeta_X^r, u_1)\, \alpha(u_1,v_1)\, G_{B}^{-\i\omega_n} (u_2,v_1) \, \ov{\alpha(u_2,v_2)}  \\
&\hspace{150pt} \times G_B^{\i\omega_n}(v_2,u_3)\, \alpha(u_3,v_3)\,  G_{B}^{-\i\omega_n}(\zeta_X^{-r}, v_3),
\end{align*}
where we used the short-hand notation $\zeta_X^r := X+\frac r2$. We also used that
\begin{align}
\frac{1}{\i\omega_n + \ov {\hfrak_B}} (x,y) = -G_B^{-\i\omega_n}(y,x), \label{DHS1:Kernel_of_complex_conjugate}
\end{align}
which follows from $\ov{ A^*(x,y) } = A(y,x)$ and
\begin{align*}
\frac{1}{z - \ov{\hfrak_B}} = \ov{\Bigl(\frac{1}{z - \hfrak_B}\Bigr)^*}.
\end{align*}
We hereby correct a typo in the analogue of \eqref{DHS1:Kernel_of_complex_conjugate} in the proof of \cite[Lemma~11]{Hainzl2017}.

Let us define the coordinates $\Zbold$ and $\sbold$ by
\begin{align*}
\mathbf u &= X + \Zbold + \frac \sbold 2, & \mathbf v &= X + \Zbold - \frac \sbold 2,
%
%
\end{align*}
and note that we interpret them as relative and center-of-mass coordinates. For $N_{T,B}$ this implies
\begin{align*}
N_{T,B}(\alpha) (X,r) &= \iiint_{\Rbb^9} \dd \Zbold \iiint_{\Rbb^9} \dd \sbold \; \e^{-\i \Bbold \cdot (X\wedge Z_1)} \, \e^{\i \Bbold \cdot (X \wedge Z_2)} \, \e^{-\i \Bbold \cdot (X\wedge Z_3)} \; \Acal(X,\Zbold,\sbold)  \\
&\hspace{-60pt} \times \frac{2}{\beta} \sum_{n\in \Zbb}  G_B^{\i\omega_n}(\zeta_X^r,\zeta_{Z_1 + X}^{s_1}) \, G_B^{-\i\omega_n}(\zeta_{Z_2+X}^{s_2}, \zeta_{Z_1 + X}^{-s_1}) \,  G_B^{\i\omega_n}(\zeta_{Z_2+X}^{-s_2}, \zeta_{Z_3+X}^{s_3}) \,  G_B^{-\i\omega_n}(\zeta_X^{-r}, \zeta_{Z_3+X}^{-s_3})
\end{align*}
with $\Acal(X, \Zbold, \sbold)$ in \eqref{DHS1:NTB_alpha_definition}. Here, we used $\Bbold \cdot (X\wedge Z) = Z \cdot (\Bbold \wedge X)$ and that $Z\cdot (\Bbold \wedge X)$ commutes with $Z\cdot (-\i\nabla_X)$, which implies 
\begin{align*}
\alpha(X + Z, s) = \e^{\i Z \cdot (-\i\nabla_X)} \, \alpha(X, s) = \e^{-\i \Bbold \cdot (X\wedge Z)} \; \e^{\i Z \cdot \Pi_X}\alpha(X, s).
\end{align*}
A tedious but straightforward computation that uses Lemma~\ref{DHS1:gB-identities}~(b) shows
\begin{align}
 G_B^{\i\omega_n}(\zeta_X^r,\zeta_{Z_1 + X}^{s_1}) \; G_B^{-\i\omega_n}(\zeta_{Z_2+X}^{s_2}, \zeta_{Z_1 + X}^{-s_1}) \; G_B^{\i\omega_n}(\zeta_{Z_2+X}^{-s_2}, \zeta_{Z_3+X}^{s_3}) \; G_B^{-\i\omega_n}(\zeta_X^{-r}, \zeta_{Z_3+X}^{-s_3}) &\notag \\
&\hspace{-300pt} =\e^{\i \Bbold \cdot (X\wedge Z_1)} \e^{-\i \Bbold \cdot (X \wedge Z_2)}\e^{\i \Bbold \cdot (X\wedge Z_3)} \; \e^{\i  \frac \Bbold 2 \cdot \Phi(\Zbold, r, \sbold)} \; \ell_{T,B}^n(\Zbold, r, \sbold). \label{DHS1:NTB_representation_1}
\end{align}
This proves the claim.
\end{proof}

As in the case of $L_{T,B}$, we analyze the operator $N_{T,B}$ by introducing several steps of simplification. Namely, we write
\begin{align}
N_{T,B} = (N_{T,B} - \tilde N_{T,B}) + (\tilde N_{T,B} - N_{T,B}^{(1)}) + (N_{T,B}^{(1)} - N_T^{(2)}) + N_{T}^{(2)}. \label{DHS1:NTB_decomposition}
\end{align}
with $\tilde N_{T,B}$ in \eqref{DHS1:NtildeTB_definition}, $N_{T,B}^{(1)}$ in \eqref{DHS1:NTB1_definition}, and $N_T^{(2)}$ in \eqref{DHS1:NT2_definition}. To obtain $\tilde N_{T,B}$ we replace the functions $g_B^z$ by  $g_0^z$ in $N_{T,B}$. When we replace $\ell_{T,B}$ by $\ell_{T,0}$, we obtain $N_{T,B}^{(1)}$, and $N_T^{(2)}$ is obtained from $N_{T,B}^{(1)}$ by replacing the magnetic translations $\e^{\i Z_i \cdot \Pi_X}$ by $1$. Using arguments that are comparable to the ones applied in the analysis of the operator $L_{T,B}$, we show in Section~\ref{DHS1:sec:approxNTB} below that the contributions from the terms in the parentheses in \eqref{DHS1:NTB_decomposition} can be treated as remainders. In Section~\ref{DHS1:sec:compquarticterm} we prove a proposition that allows us to extract the quartic term in the Ginzburg--Landau functional from the term $\langle \Delta, N_{T}^{(2)}(\Delta) \rangle$. Finally, we summarize our findings in Section~\ref{DHS1:sec:quarticterms}.

\subsubsection{Approximation of \texorpdfstring{$N_{T,B}$}{NTB}}
\label{DHS1:sec:approxNTB}

\paragraph{The operator $\tilde N_{T,B}$.} The operator $\tilde N_{T,B}$ is defined by
\begin{align}
\tilde N_{T,B}(\alpha) (X,r) &:= \iiint_{\Rbb^{9}} \dd \Zbold \iiint_{\Rbb^{9}} \dd \sbold \; \tilde \ell_{T,B} (\Zbold, r, \sbold) \; \Acal(X, \Zbold , \sbold) \label{DHS1:NtildeTB_definition}
\end{align}
with
\begin{align*}
\tilde \ell_{T,B}(\Zbold, r, \sbold) &:= \frac 2\beta \sum_{n\in\Zbb} \ell_{T,0}^n (\Zbold, r, \sbold) \; \e^{\i  \frac \Bbold 2 \cdot \Phi(\Zbold, r, \sbold)},
\end{align*}
$\mathcal{A}$ in \eqref{DHS1:NTB_alpha_definition}, $\ell_{T,0}^n$ in \eqref{DHS1:lTBn_definition} and $\Phi$ in \eqref{DHS1:PhiB_definition}. The following proposition quantifies the error that we make when we replace $N_{T,B}(\Delta)$ by $\tilde N_{T,B}(\Delta)$ in our computations.

\begin{prop}
\label{DHS1:NTB-NtildeTB}
Assume that $V\alpha_*\in L^{\nicefrac 43}(\Rbb^3)$. For every $T_0>0$ there is $B_0>0$ such that for any $0 < B \leq B_0$, any $T\geq T_0$, any $\Psi \in \Hmag^1(Q_B)$, and $\Delta \equiv \Delta_\Psi$ as in \eqref{DHS1:Delta_definition}, we have
\begin{align*}
|\langle \Delta,  N_{T,B}(\Delta) - \tilde N_{T,B}(\Delta)\rangle| &\leq C \; B^4 \; \Vert V\alpha_*\Vert_{\nicefrac 43}^4 \; \Vert \Psi\Vert_{\Hmag^1(Q_B)}^4.
\end{align*}
\end{prop}

Before we give the proof of Proposition~\ref{DHS1:NTB-NtildeTB}, let us introduce the function
\begin{align}
F_{T,B} &:= \smash{\frac 2\beta \sum_{n\in \Zbb}} \; |g_B^{\i\omega_n} - g_0^{\i\omega_n}| * |g_B^{-\i\omega_n} | * |g_B^{\i\omega_n}| * |g_B^{-\i\omega_n}| \notag \\
&\hspace{100pt}+|g_0^{\i\omega_n}| * |g_B^{-\i\omega_n}  - g_0^{-\i\omega_n}| * |g_B^{\i\omega_n}| * |g_B^{-\i\omega_n}| \notag \\
&\hspace{100pt} +|g_0^{\i\omega_n}| * |g_0^{-\i\omega_n} | * |g_B^{\i\omega_n} - g_0^{\i\omega_n}| * |g_B^{-\i\omega_n}| \notag \\
&\hspace{100pt} +|g_0^{\i\omega_n}| * |g_0^{-\i\omega_n} | * |g_0^{\i\omega_n}| * |g_B^{-\i\omega_n} - g_0^{-\i\omega_n}| \label{DHS1:NTB-NtildeTB_FTB_definition}.
\end{align}
By Lemmas~\ref{DHS1:g0_decay} and \ref{DHS1:gB-g_decay} as well as \eqref{DHS1:g0_decay_f_estimate1} we have
\begin{align}
	\Vert F_{T,B}\Vert_1 \leq C \, B^2
	\label{DHS1:eq:A16}
\end{align}
for $T \geq T_0 > 0$. 

\begin{proof}[Proof of Proposition \ref{DHS1:NTB-NtildeTB}]
The function $|\Psi|$ is periodic and \eqref{DHS1:Magnetic_Sobolev} therefore implies
\begin{align}
\Vert \e^{\i Z \cdot\Pi}\Psi\Vert_6^2 = \Vert \Psi\Vert_6^2 \leq C\,  B \, \Vert \Psi\Vert_{\Hmag^1(Q_B)}^2. \label{DHS1:NTB-NtildeTB_3}
\end{align}
Consequently, we have
\begin{align}
\fint_{Q_B} \dd X \; |\Psi(X)|\; \prod_{i=1}^3  |\e^{\i Z_i\cdot \Pi}\Psi(X)| &\leq \Vert \Psi\Vert_2 \; \prod_{i=1}^3  \Vert \e^{\i Z_i\cdot \Pi}\Psi\Vert_6 \leq C \, B^2 \, \Vert \Psi\Vert_{\Hmag^1(Q_B)}^4 \label{DHS1:NTB-NtildeTB_2}
\end{align}
as well as
\begin{align}
|\langle \Delta,  N_{T,B}(\Delta)- \tilde N_{T,B}(\Delta)\rangle| & \notag\\
&\hspace{-100pt}\leq C\, B^2 \, \Vert \Psi\Vert_{\Hmag^1(Q_B))}^4  \int_{\Rbb^3} \dd r \,  \iiint_{\Rbb^9} \dd \sbold\; |V\alpha_*(r)|\;  |V\alpha_*(s_1)| \; |V\alpha_*(s_2)| \; |V\alpha_*(s_3)|  \notag\\
&\hspace{60pt}\times \iiint_{\Rbb^9} \dd \Zbold \; |\ell_{T,B}(\Zbold, r,\sbold) - \tilde \ell_{T,B}(\Zbold, r, \sbold)|.  \label{DHS1:NTB-NtildeTB_1}
\end{align}
We use the change of variables 
\begin{align}
Z_1' - Z_2' &:= Z_1 - Z_2 - \frac{s_1 +s_2}{2}, & Z_2' - Z_3' &:= Z_2 - Z_3 - \frac{s_2 + s_3}{2}, & Z_3' &:= Z_3 + \frac{r-s_3}{2}, \label{DHS1:NTB_change_of_variables_1}
\end{align}
whence
\begin{align}
Z_1 - \frac{r-s_1}{2} = Z_1' - (r - s_1 - s_2 - s_3). \label{DHS1:NTB_change_of_variables_2}
\end{align}
As in the proof of \eqref{DHS1:LTB-LtildeTB_5}, we conclude
\begin{align*}
\iiint_{\Rbb^9} \dd \Zbold\; |\ell_{T,B}(\Zbold , r, \sbold) - \tilde \ell_{T,B}(\Zbold, r, \sbold)| \leq F_{T,B}(r - s_1 - s_2 - s_3)
\end{align*}
with $F_{T,B}$ in \eqref{DHS1:NTB-NtildeTB_FTB_definition}. We insert the above bound in \eqref{DHS1:NTB-NtildeTB_1} and use
\begin{align*}
\bigl\Vert V\alpha_* \; \bigl( V\alpha_* * V\alpha_* * V\alpha_* * F_{T,B}\bigr) \bigr\Vert_1 &\leq C \; \Vert V\alpha_*\Vert_{\nicefrac 43}^4 \; \Vert F_{T,B}\Vert_1
\end{align*}
as well as \eqref{DHS1:eq:A16}, which proves the claim.
\end{proof}

\paragraph{The operator $N_{T,B}^{(1)}$.} The operator $N_{T,B}^{(1)}$ is defined by
\begin{align}
N_{T,B}^{(1)}(\alpha)(X,r) &:= \iiint_{\Rbb^9} \dd \Zbold \iiint_{\Rbb^9} \dd \sbold \; \ell_{T,0} (\Zbold, r, \sbold) \; \Acal (X, \Zbold , \sbold) \label{DHS1:NTB1_definition}
\end{align}
with $\mathcal{A}$ in \eqref{DHS1:NTB_alpha_definition} and $\ell_{T,0}$ in \eqref{DHS1:lTB_definition}. The following proposition allows us to replace $\langle \Delta, \widetilde{N}_{T,B}(\Delta) \rangle$ by $\langle \Delta, N_{T,B}^{(1)}(\Delta) \rangle$ in our computations.

\begin{prop}
\label{DHS1:NtildeTB-NTB1}
Assume that $|\cdot|^kV\alpha_* \in L^{\nicefrac 43}(\Rbb^3)$ for $k\in \{0,1\}$. For every $T \geq T_0 > 0 $, every $B>0$, every $\Psi \in \Hmag^1(Q_B)$ and $\Delta \equiv \Delta_\Psi$ as in \eqref{DHS1:Delta_definition}, we have 
\begin{align*}
|\langle \Delta, \tilde N_{T,B}(\Delta) - N_{T,B}^{(1)}(\Delta)\rangle| &\leq C \; B^3 \; \bigl( \Vert V\alpha_*\Vert_{\nicefrac 43}^4 + \Vert \, |\cdot|V\alpha_*\Vert_{\nicefrac 43}^4\bigr)\; \Vert \Psi\Vert_{\Hmag^1(Q_B)}^4.
\end{align*}
\end{prop}

Before we start with the proof of Proposition~\ref{DHS1:NtildeTB-NTB1}, we introduce the functions
\begin{align}
	F_T^{(1)} &:= \smash{\frac 2\beta \sum_{n\in \Zbb}} \, |g_0^{\i\omega_n}| * \bigl(|\cdot|\, |g_0^{-\i\omega_n}|\bigr) * \bigl(|\cdot|\, |g_0^{\i\omega_n}|\bigr) * |g_0^{-\i\omega_n}| \notag \\
	&\hspace{100pt}+ |g_0^{\i\omega_n}| * \bigl(|\cdot|\, |g_0^{-\i\omega_n}|\bigr) * |g_0^{\i\omega_n}| * \bigl(|\cdot|\, |g_0^{-\i\omega_n}|\bigr) \notag \\
	&\hspace{100pt}+ |g_0^{\i\omega_n}| * |g_0^{-\i\omega_n}| * \bigl(|\cdot|\, |g_0^{\i\omega_n}|\bigr) * \bigl(|\cdot|\, |g_0^{-\i\omega_n}|\bigr) \label{DHS1:NtildeTB-NTB1_FT1_definition}
\end{align}
and
\begin{align}
	F_T^{(2)} &:= \smash{\frac 2\beta \sum_{n\in \Zbb}} \,
	\bigl( |\cdot| \, |g_0^{\i\omega_n}|\bigr) *  |g_0^{-\i\omega_n}| * |g_0^{\i\omega_n}| * |g_0^{-\i\omega_n}| 
	+ |g_0^{\i\omega_n}| * \bigl(|\cdot|\, |g_0^{-\i\omega_n}|\bigr) * |g_0^{\i\omega_n}| * |g_0^{-\i\omega_n}| \notag\\
	&\hspace{30pt}+ |g_0^{\i\omega_n}| * |g_0^{-\i\omega_n}| * \bigl(|\cdot|\, |g_0^{\i\omega_n}|\bigr) * |g_0^{-\i\omega_n}| 
	+ |g_0^{\i\omega_n}| * |g_0^{-\i\omega_n}| * |g_0^{\i\omega_n}| * \bigl(|\cdot|\, |g_0^{-\i\omega_n}|\bigr). \label{DHS1:NtildeTB-NTB1_FT2_definition}
\end{align}
As long as $T \geq T_0 > 0$, Lemma~\ref{DHS1:g0_decay} and \eqref{DHS1:g0_decay_f_estimate1} imply the bound
\begin{align}
	\Vert F_T^{(1)} \Vert_1 + \Vert F_T^{(2)} \Vert_1 \leq C. \label{DHS1:NtildeTB-NTB1_FT1-2_estimate}
\end{align}

\begin{proof}[Proof of Proposition \ref{DHS1:NtildeTB-NTB1}]
We use \eqref{DHS1:NTB-NtildeTB_2} and estimate
\begin{align}
|\langle \Delta, \tilde N_{T,B}(\Delta) - N_{T,B}^{(1)}(\Delta)\rangle| &  \notag\\
&\hspace{-100pt}\leq C\, B^2 \, \Vert \Psi\Vert_{\Hmag^1(Q_B)}^4 \int_{\Rbb^3} \dd r  \iiint_{\Rbb^9} \dd \sbold \; |V\alpha_*(r)| \; |V\alpha_*(s_1)| \; |V\alpha_*(s_2)| \; |V\alpha_*(s_3)|  \notag \\
&\hspace{30pt}\times \frac{2}{\beta} \sum_{n\in \Zbb} \iiint_{\Rbb^9} \dd \Zbold \; |\ell_{T,0}^n(\Zbold, r, \sbold)| \; \bigl| \e^{\i \Bbold \, \Phi(\Zbold, r, \sbold)} - 1\bigr|. \label{DHS1:eq:A17}
\end{align}
In terms of the coordinates in \eqref{DHS1:NTB_change_of_variables_1} and with \eqref{DHS1:NTB_change_of_variables_2}, the phase function $\Phi$ in \eqref{DHS1:PhiB_definition} can be written as
\begin{align}
\Phi(\Zbold, r,\sbold) &= (Z_2' - Z_3') \wedge (Z_1' - Z_2') + Z_3' \wedge (Z_1' - Z_2') + Z_3' \wedge (Z_3'- Z_2') \notag\\
& \hspace{50pt} + \frac r2 \wedge \bigl( Z_1' - (r - s_1 - s_2 - s_3)\bigr) + \bigl( s_2 + s_3 - \frac r2\bigr) \wedge (Z_1 ' - Z_2')\notag \\
&\hspace{50pt} + \bigl( s_3 - \frac r2\bigr) \wedge (Z_3' - Z_2') + \frac r2 \wedge Z_3'.
\end{align}
We use the estimate $|\e^{\i  \frac \Bbold 2 \cdot \Phi(\Zbold, r, \sbold)} - 1| \leq B \, |\Phi(\Zbold, r, \sbold)|$, \eqref{DHS1:eq:A17}, and argue as in the proof of \eqref{DHS1:LtildeTB-MTB_4} to see that
\begin{align*}
&\frac{2}{\beta} \sum_{n\in \Zbb} \iiint_{\Rbb^9} \dd \Zbold \; |\ell_{T,0}^n(\Zbold, r, \sbold)| \; \bigl| \e^{\i  \frac \Bbold 2 \cdot \Phi(\Zbold, r, \sbold)} - 1\bigr| \\
&\hspace{20pt} \leq CB \; \bigl[ F_T^{(1)} (r - s_1 - s_2 - s_3) + F_T^{(2)}(r - s_1 - s_2 - s_3) \; \bigl(1 + |r| + |s_1| + |s_2| + |s_3|\bigr)\bigr] 
\end{align*}
with $F_T^{(1)}$ in \eqref{DHS1:NtildeTB-NTB1_FT1_definition} and $F_T^{(2)}$ in \eqref{DHS1:NtildeTB-NTB1_FT2_definition}. Young's inequality then implies
\begin{align*}
|\langle \Delta, \tilde N_{T,B}(\Delta) - N_{T,B}^{(1)}(\Delta)\rangle| &\\
&\hspace{-100pt}\leq C\; B^3 \, \Vert \Psi\Vert_{\Hmag^1(Q_B)}^4  \; \bigl( \Vert V\alpha_*\Vert_{\nicefrac 43}^4 + \Vert \, |\cdot| V\alpha_*\Vert_{\nicefrac 43}^4 \bigr) \bigl( \Vert F_T^{(1)}\Vert_1 + \Vert F_T^{(2)}\Vert_1 \bigr).
\end{align*}
Finally, an application of \eqref{DHS1:NtildeTB-NTB1_FT1-2_estimate} 
proves the claim.
\end{proof}

\paragraph{The operator $N_T^{(2)}$.} The operator $N_{T}^{(2)}$ is defined by
\begin{align}
N_T^{(2)}(\alpha) (X, r) &:= \iiint_{\Rbb^9} \dd \Zbold \iiint_{\Rbb^9} \dd \sbold \; \ell_{T,0} (\Zbold, r, \sbold) \, \Acal(X, 0 , \sbold) \label{DHS1:NT2_definition}
\end{align}
with $\mathcal{A}$ in \eqref{DHS1:NTB_alpha_definition} and $\ell_{T,0}$ in \eqref{DHS1:lTB_definition}.

The following proposition allows us to replace $\langle \Delta, N_{T,B}^{(1)}(\Delta) \rangle$ by $\langle \Delta, N_{T}^{(2)}(\Delta) \rangle$ in our computations. We highlight that the $\Hmag^2(Q_B)$-norm of $\Psi$ is needed to bound the difference between the two terms.

\begin{prop}
\label{DHS1:NTB1-NT2}
Assume that $|\cdot|^kV\alpha_* \in L^{\nicefrac 43}(\Rbb^3)$ for $k\in \{0,2\}$. For any $T \geq T_0 >0$, any $B>0$, any $\Psi \in \Hmag^2(Q_B)$, and $\Delta\equiv \Delta_\Psi$ as in \eqref{DHS1:Delta_definition}, we have
\begin{align*}
|\langle \Delta, N_{T, B}^{(1)}(\Delta) - N_{T}^{(2)}(\Delta) \rangle|  &\leq C \; B^3 \; \bigl( \Vert V\alpha_*\Vert_{\nicefrac 43}^4 + \Vert \, |\cdot|^2 V\alpha_*\Vert_{\nicefrac 43}^4\bigr) \\
&\hspace{130pt}  \times  \Vert \Psi\Vert_{\Hmag^1(Q_B)}^3 \; \Vert \Psi\Vert_{\Hmag^2(Q_B)}.
\end{align*}
\end{prop}

Before we prove the above proposition, let us introduce the functions
\begin{align*}
	F_T^{(1)} &:= \frac 2\beta \sum_{n\in \Zbb} |g_0^{\i\omega_n}| *  |g_0^{-\i\omega_n}| * |g_0^{\i\omega_n}| * |g_0^{-\i\omega_n}|
\end{align*}
and
\begin{align*}
	F_T^{(2)} &:= \smash{\frac 2\beta \sum_{n\in \Zbb}} \; |g_0^{\i\omega_n}| *  \bigl(|\cdot|^2 \,|g_0^{-\i\omega_n}| \bigr) * |g_0^{\i\omega_n}| * |g_0^{-\i\omega_n}| \\
	&\hspace{100pt}+ |g_0^{\i\omega_n}| *  \,|g_0^{-\i\omega_n}| *   \bigl(|\cdot|^2 \, |g_0^{\i\omega_n}|\bigr) * |g_0^{-\i\omega_n}| \\
	&\hspace{100pt}+ |g_0^{\i\omega_n}| * |g_0^{-\i\omega_n}| * |g_0^{\i\omega_n}| * \bigl(|\cdot|^2 \, |g_0^{-\i\omega_n}|\bigr).
\end{align*}
For $T \geq T_0 > 0$, an application of Lemma~\ref{DHS1:g0_decay} and the estimate \eqref{DHS1:g0_decay_f_estimate1} on $f(t,\omega)$ show
\begin{align}
	\Vert F_T^{(1)} \Vert_1 + \Vert F_T^{(2)} \Vert_1 \leq C. \label{DHS1:eq:A18}
\end{align}

\begin{proof}
We have
\begin{align}
\langle \Delta, N_{T,B}^{(1)}(\Delta) - N_T^{(2)}(\Delta)\rangle &\notag\\
&\hspace{-115pt}=  16 \int_{\Rbb^3} \dd r \iiint_{\Rbb^9} \dd \sbold \; V\alpha_*(r)V\alpha_*(s_1)V\alpha_*(s_2)V\alpha_*(s_3) \iiint_{\Rbb^9} \dd \Zbold \; \ell_{T,0} (\Zbold, r,\sbold)  \notag \\
&\hspace{-115pt} \times\fint_{Q_B} \dd X \; \ov{\Psi(X)} \; \bigl( \e^{\i Z_1\cdot \Pi_X}\Psi(X) \, \ov{\e^{\i Z_2 \cdot \Pi_X} \Psi(X)} \,  \e^{\i Z_3 \cdot \Pi_X}\Psi(X) - \Psi(X)\ov{\Psi(X)} \Psi(X)\bigr). \label{DHS1:NTB1-NTB2_2}
\end{align}
Apart from the exponential factors, this expression is symmetric under the simultaneous replacement of $(Z_1,Z_2,Z_3)$ by $(-Z_1,-Z_2,-Z_3)$. 
When we expand the magnetic translations in cosine and sine functions of $Z_i \cdot \Pi_X$, $i=1,2,3$, the above symmetry implies that all terms with an odd number of sine functions vanish. Accordingly, we may replace the bracket in the last line of \eqref{DHS1:NTB1-NTB2_2} by 
\begin{align}
& \bigl(\cos(Z_1\cdot \Pi_X) \Psi(X) \; \ov{\cos(Z_2\cdot \Pi_X) \Psi(X)} \; \cos(Z_3\cdot \Pi_X) \Psi(X) - \Psi(X) \overline{ \Psi(X) } \Psi(X) \bigr) \notag \\
&\hspace{10pt}+ \cos(Z_1\cdot \Pi_X) \Psi(X) \; \ov{\i\sin(Z_2\cdot \Pi_X) \Psi(X)}  \;  \i \sin(Z_3\cdot \Pi_X) \Psi(X) \notag \\ 
&\hspace{10pt}+ \i \sin(Z_1\cdot \Pi_X) \Psi(X) \; \ov{ \cos(Z_2\cdot \Pi_X) \Psi(X)} \; \i \sin(Z_3\cdot \Pi_X) \Psi(X) \notag \\
&\hspace{10pt}+ \i \sin(Z_1\cdot \Pi_X) \Psi(X) \; \ov{\i\sin(Z_2\cdot \Pi_X) \Psi(X)}\;  \cos(Z_3\cdot \Pi_X) \Psi(X).  \label{DHS1:NTB1-NTB2_1}
\end{align}
Let us consider the first term in \eqref{DHS1:NTB1-NTB2_1}. We use $|\cos(x)  - 1|^2= 4|\sin^4(\frac x2)| \leq \frac{1}{4} |x|^4$ and the operator inequality in \eqref{DHS1:ZPiX_inequality_quartic} to see that $|\cos(Z\cdot \Pi)-1|^2 \leq C\cdot|Z|^4\; (\Pi^4 + B^2)$ holds. In particular,
\begin{align}
\Vert [ \cos(Z\cdot \Pi) - 1] \Psi \Vert_2^2 &
\leq C \; B^3 \; |Z|^4  \;  \Vert \Psi\Vert_{\Hmag^2(Q_B)}^2. \label{DHS1:NTB1-NTB2_5}
\end{align}
In combination with the estimate \eqref{DHS1:NTB-NtildeTB_3} on $\Vert \e^{\i Z\cdot \Pi}\Psi\Vert_6$, this implies
\begin{align}
&\fint_{Q_B} \dd X \; |\Psi(X)| \bigl| \cos( Z_1 \cdot \Pi)\Psi(X) \; \ov{\cos( Z_2  \cdot \Pi)\Psi(X)} \; \cos( Z_3 \cdot \Pi)\Psi(X) - \Psi(X)\ov{\Psi(X)}\Psi(X)\bigr| \notag\\
&\hspace{15pt}\leq  \Vert \Psi\Vert_6 \;  \Vert (\cos( Z_1\cdot \Pi) - 1)\Psi\Vert_2 \; \Vert \cos(Z_2\cdot \Pi)\Psi\Vert_6 \; \Vert \cos( Z_3\cdot \Pi)\Psi\Vert_6 \notag \\
&\hspace{30pt}+ \Vert \Psi\Vert_6^2 \; \Vert (\cos(Z_2\cdot \Pi) - 1)\Psi\Vert_2 \; \Vert \cos( Z_3\cdot \Pi)\Psi\Vert_6 + \Vert \Psi\Vert_6^3 \; \Vert (\cos(Z_3\cdot \Pi) - 1)\Psi\Vert_2 \notag \\
&\hspace{15pt}\leq C \, B^3 \, \Vert \Psi\Vert_{\Hmag^1(Q_B)}^3\; \Vert \Psi\Vert_{\Hmag^2(Q_B)} \; \bigl(|Z_1|^2 + |Z_2|^2 + |Z_3|^2\bigr). \label{DHS1:NTB1-NTB2_6}
%
\end{align}

To treat the other terms in \eqref{DHS1:NTB1-NTB2_1} we use the operator inequality in \eqref{DHS1:ZPiX_inequality} to see that
\begin{align*}
\Vert \sin(Z \cdot \Pi)\Psi\Vert_2^2  = \langle \Psi, \sin^2(Z\cdot \Pi) \, \Psi\rangle \leq C \, B^2\, |Z|^2 \, \Vert \Psi\Vert_{\Hmag^1(Q_B)}^2, 
%
\end{align*}
which yields 
\begin{align}
&\fint_{Q_B} \dd X \; |\Psi(X)| \; |\cos(Z_i\cdot \Pi)\Psi(X)| \; |\sin(Z_j\cdot \Pi)\Psi(X)|\; |\sin(Z_k\cdot \Pi)\Psi(X)| \notag \\
&\hspace{180pt}\leq C \, B^3 \, \bigl( |Z_j|^2 + |Z_k|^2\bigr) \; \Vert \Psi\Vert_{\Hmag^1(Q_B)}^4. \label{DHS1:NTB1-NTB2_7}
\end{align}
We gather \eqref{DHS1:NTB1-NTB2_2}, \eqref{DHS1:NTB1-NTB2_1}, \eqref{DHS1:NTB1-NTB2_6}, and \eqref{DHS1:NTB1-NTB2_7} and find
\begin{align*}
|\langle \Delta, N_{T,B}^{(1)}(\Delta)- N_T^{(2)}(\Delta)\rangle| &\leq C\, B^3 \, \Vert \Psi\Vert_{\Hmag^1(Q_B)}^3 \; \Vert \Psi\Vert_{\Hmag^2(Q_B)}  \\
&\hspace{-30pt} \times \int_{\Rbb^3} \dd r\iiint_{\Rbb^9}\dd \sbold \; |V\alpha_*(r)|\; |V\alpha_*(s_1)|\; |V\alpha_*(s_2)| \; |V\alpha_*(s_3)| \\
&\hspace{60pt} \times \iiint_{\Rbb^9} \dd \Zbold \; |\ell_{T,0}(\Zbold, r, \sbold)| \; \bigl(|Z_1|^2 + |Z_2|^2 + |Z_3|^2\bigr).
\end{align*}
When we write the coordinates $Z_i$, $i=1,2,3$, in terms of the coordinates in \eqref{DHS1:NTB_change_of_variables_1} and \eqref{DHS1:NTB_change_of_variables_2} plus linear combinations of $r$ and $s_i$, $i=1,2,3$, we see that
\begin{align*}
|Z_1| &\leq |Z_1' -Z_2'| + |Z_2' - Z_3'| + |Z_3'| + |r| + |s_1| + |s_2| + |s_3|,\\
|Z_2| &\leq |Z_2 ' - Z_3'| + |Z_3'| + |r| + |s_2| + |s_3|,\\
|Z_3| &\leq |Z_3'| + |r| + |s_3|.
\end{align*}
We use this and argue as in the proof of \eqref{DHS1:LtildeTB-MTB_4}, which yields
%
%
\begin{align*}
&\iiint_{\Rbb^9} \dd \Zbold \; |\ell_{T,0}(\Zbold, r, \sbold)| \; \bigl(|Z_1|^2 + |Z_2|^2 + |Z_3|^2\bigr)  \\
&\hspace{30pt} \leq C\bigl( F_T^{(1)}(r-s_1-s_2-s_3) \; (|r|^2 + |s_1|^2 + |s_2|^2 + |s_3|^2) + F_T^{(2)}(r - s_1 - s_2 - s_3)\bigr).
\end{align*}
In particular,
\begin{align*}
|\langle \Delta, N_{T,B}^{(1)}(\Delta)- N_T^{(2)}(\Delta)\rangle| &\leq C \;  B^3 \; \Vert \Psi\Vert_{\Hmag^1(Q_B)}^3 \; \Vert \Psi\Vert_{\Hmag^2(Q_B)} \\
&\hspace{40pt} \times \bigl( \Vert V\alpha_*\Vert_{\nicefrac 43}^4 + \Vert \, |\cdot|^2V\alpha_*\Vert_{\nicefrac 43}^4  \bigr) \bigl( \Vert F_T^{(1)}\Vert_1 + \Vert F_T^{(2)}\Vert_1\bigr).
\end{align*}
In combination with \eqref{DHS1:eq:A18}, this proves the claim.
\end{proof}

\subsubsection{Calculation of the quartic term in the Ginzburg--Landau functional}
\label{DHS1:sec:compquarticterm}

The following proposition allows us to extract the quartic term in the Ginzburg--Landau functional in \eqref{DHS1:Definition_GL-functional} from $\langle \Delta, N_T^{(2)}(\Delta) \rangle$.

\begin{prop}
\label{DHS1:NTc2}
Assume $V\alpha_* \in L^{\nicefrac 43}(\Rbb^3)$. For any $B>0$, any $\Psi\in \Hmag^1(Q_B)$, and $\Delta \equiv \Delta_\Psi$ as in \eqref{DHS1:Delta_definition}, we have
\begin{align*}
\langle \Delta, N_{\Tc}^{(2)}(\Delta)\rangle = 8\; \Lambda_3 \; \Vert \Psi\Vert_4^4
\end{align*}
with $\Lambda_3$ in \eqref{DHS1:GL_coefficient_3}. Moreover, for any $T \geq T_0 > 0$, we have
\begin{align*}
	|\langle \Delta,  N_T^{(2)}(\Delta) - N_{\Tc}^{(2)}(\Delta)\rangle| &\leq C\; B^2  \; |T - \Tc| \; \Vert V\alpha_*\Vert_{\nicefrac 43}^4 \; \Vert \Psi\Vert_{\Hmag^1(Q_B)}^4.
\end{align*}
\end{prop}

Before we prove the above proposition, let us introduce the function
\begin{align}
	F_{T,\Tc} &:= \smash{\frac 2\beta \sum_{n\in \Zbb} |2n+1| \bigl[} |g_0^{\i\omega_n^T}| * |g_0^{\i\omega_n^{\Tc}}| * |g_0^{-\i\omega_n^T}| * |g_0^{\i\omega_n^T}| * |g_0^{-\i\omega_n^T}|  \phantom {\sum^i} \notag \\
	&\hspace{120pt}+|g_0^{\i\omega_n^{\Tc}}| *  |g_0^{-\i\omega_n^T}| * |g_0^{-\i\omega_n^{\Tc}}| * |g_0^{\i\omega_n^T}| * |g_0^{-\i\omega_n^T}| \notag \\
	&\hspace{120pt}+|g_0^{\i\omega_n^{\Tc}}| *  |g_0^{-\i\omega_n^{\Tc}}| * |g_0^{\i\omega_n^T}| * |g_0^{\i\omega_n^{\Tc}}| * |g_0^{-\i\omega_n^T}| \notag \\
	&\hspace{120pt}+|g_0^{\i\omega_n^{\Tc}}| * |g_0^{-\i\omega_n^{\Tc}}| * |g_0^{\i\omega_n^{\Tc}}| * |g_0^{-\i\omega_n^T}| * |g_0^{-\i\omega_n^{\Tc}}| \bigr], \label{DHS1:eq:A19}
\end{align}
where we have included the $T$-dependence of the Matsubara frequencies in our notation once more because different temperatures appear in the formula. As long as $T \geq T_0 > 0$, Lemma~\ref{DHS1:g0_decay} and \eqref{DHS1:g0_decay_f_estimate1} imply
\begin{equation}
	\Vert F_{T,\Tc} \Vert_1 \leq C.
	\label{DHS1:eq:A21}
\end{equation}

\begin{proof}[Proof of Proposition \ref{DHS1:NTc2}]
Set
\begin{align*}
\ell_T(\Zbold, r) &:= \frac 2\beta \sum_{n\in \Zbb} g_0^{\i\omega_n}(r - Z_1) \, g_0^{-\i\omega_n}(Z_1 - Z_2) \, g_0^{\i\omega_n}(Z_2 - Z_3) \, g_0^{-\i\omega_n}(Z_3).
\end{align*}
Then, by the change of variables \eqref{DHS1:NTB_change_of_variables_1} and \eqref{DHS1:NTB_change_of_variables_2}, we have
\begin{align*}
\iiint_{\Rbb^9} \dd \Zbold \; \ell_{T,0}(\Zbold, r,\sbold) = \iiint_{\Rbb^9} \dd \Zbold \; \ell_T(\Zbold, r-s_1 - s_2 - s_3).
\end{align*}
We use that $(\pm \i \omega_n + \mu -p^2)^{-1}$ is the Fourier transform of $g_0^{\pm \i\omega_n}(x)$, which yields
\begin{align*}
	\ell_T(\Zbold, r) &= \frac 2\beta \sum_{n\in \Zbb} \iiiint_{\Rbb^{12}} \frac{\dd \mathbf p}{(2\pi)^{12}}  \; \frac{\e^{\i p_1 \cdot (r - Z_1)}}{\i\omega_n + \mu - p_1^2} \frac{\e^{\i p_2\cdot (Z_1 - Z_2)}}{-\i\omega_n + \mu - p_2^2 } \frac{\e^{\i p_3\cdot (Z_2 - Z_3)}}{\i\omega_n + \mu - p_3^2} \frac{\e^{\i p_4\cdot Z_3}}{-\i\omega_n + \mu - p_4^2}.
\end{align*}
Integration over $\Zbold$ gives
\begin{align*}
	\iiint_{\Rbb^9} \dd \Zbold \; \ell_T(\Zbold, r) = \frac 2\beta \sum_{n\in \Zbb} \int_{\Rbb^3} \frac{\dd p}{(2\pi)^3} \; \e^{\i p\cdot r} \frac{1}{(\i\omega_n + \mu - p^2)^2 (\i\omega_n - \mu + p^2)^2}. 
\end{align*}
In view of the partial fraction expansion
\begin{align*}
\frac{1}{(\i\omega_n - E)^2(\i\omega_n + E)^2} = \frac{1}{4E^2} \Bigl[ \frac{1}{(\i\omega_n - E)^2} + \frac{1}{(\i\omega_n + E)^2}\Bigr] - \frac{1}{4E^3} \Bigl[ \frac{1}{\i\omega_n - E} - \frac{1}{\i\omega_n + E}\Bigr]
\end{align*}
and the identity
\begin{align}
\frac{\beta}{2} \frac{1}{\cosh^2(\frac \beta 2z)} = \frac{\dd}{\dd z} \tanh\bigl( \frac \beta 2 z\bigr) = - \frac{2}{\beta} \sum_{n\in \Zbb} \frac{1}{(\i\omega_n - z)^2}, \label{DHS1:cosh2_Matsubara}
\end{align}
which follows from \eqref{DHS1:tanh_Matsubara}, we have
\begin{align*}
\frac{2}{\beta} \sum_{n\in \Zbb} \frac{1}{(\i\omega_n - E)^2(\i\omega_n + E)^2} = \frac{\beta^2}{2} \; \frac{g_1(\beta E)}{E}
\end{align*}
with the function $g_1$ in \eqref{DHS1:XiSigma}. We conclude that
\begin{align*}
\iiint_{\Rbb^9} \dd \Zbold \; \ell_T(\Zbold,r) &= \frac{\beta^2}{2} \int_{\Rbb^3} \frac{\dd p}{(2\pi)^3}  \; \e^{\i p\cdot r} \; \frac{g_1(\beta (p^2-\mu))}{p^2 - \mu}.
\end{align*}
For the term we are interested in, this implies
\begin{align}
\langle \Delta, N_{\Tc}^{(2)}(\Delta)\rangle &= 16\, \Vert \Psi\Vert_4^4 \; \frac{\beta_c^2}{2} \int_{\Rbb^3} \dd r \iiint_{\Rbb^9} \dd \sbold \;  V\alpha_*(r)V\alpha_*(s_1)V\alpha_*(s_2)V\alpha_*(s_3)\notag\\
&\hspace{100pt} \times \int_{\Rbb^3} \frac{\dd p}{(2\pi)^3} \; \e^{\i p\cdot (r-s_1-s_2-s_3)} \; \frac{g_1(\beta_c(p^2-\mu))}{p^2-\mu}\notag\\
&= 8 \,  \Vert \Psi\Vert_4^4 \, \frac{\beta_c^2}{16} \int_{\Rbb^3} \frac{\dd p}{(2\pi)^3} \; |(-2)\hat{V\alpha_*}(p)|^4 \; \frac{g_1(\beta_c(p^2-\mu))}{p^2-\mu} =  8\; \Lambda_3 \; \Vert \Psi\Vert_4^4 \label{DHS1:NTc2_quartic_term_result}
\end{align}
with $\Lambda_3$ in \eqref{DHS1:GL_coefficient_3}. This proves the first claim.

To prove the second claim, we note that
\begin{align}
	 \langle \Delta, N_T^{(2)}(\Delta) - N_{\Tc}^{(2)}(\Delta)& \rangle = 16 \int_{\Rbb^3} \dd r \iiint_{\Rbb^9} \dd \sbold \; V\alpha_*(r)V\alpha_*(s_1)V\alpha_*(s_2)V\alpha_*(s_3) \notag \\
	 & \times \iiint_{\Rbb^9} \dd \Zbold \; \left( \ell_{T,0} - \ell_{\Tc,0} \right) (\Zbold, r,\sbold) \fint_{Q_B} \dd X \; |\Psi(X)|^4. \label{DHS1:eq:A22}
\end{align}
Afterwards, we argue as in the proof of \eqref{DHS1:MTB2_2}, that is, we use the resolvent equation \eqref{DHS1:gTgTc} as well as the change of variables in \eqref{DHS1:NTB_change_of_variables_1} and \eqref{DHS1:NTB_change_of_variables_2} and obtain
\begin{align}
\int_{\Rbb^9} \dd \Zbold \; | \ell_{T,0}(\Zbold,r , \sbold ) - \ell_{\Tc,0}(\Zbold, r,\sbold) | &\leq C\, |T - \Tc| \; F_{T,\Tc}(r-s_1-s_2-s_3) 
\label{DHS1:eq:A20}
\end{align}
%
with the function $F_{T,\Tc}$ in \eqref{DHS1:eq:A19}. 
%
Together with \eqref{DHS1:eq:A22}, this implies
\begin{align*}
|\langle \Delta, N_T^{(2)}(\Delta) - N_{\Tc}^{(2)}(\Delta)\rangle| &\leq C \, |T-\Tc| \,  \Vert \Psi\Vert_6^3 \Vert \Psi\Vert_2 \, \bigl\Vert V\alpha_* \; \bigl( V\alpha_* * V\alpha_* * V\alpha_* * F_{T,\Tc}\bigr)\bigr\Vert_1. 
%
\end{align*}
Finally, an application of \eqref{DHS1:Magnetic_Sobolev}, Young's inequality, and \eqref{DHS1:eq:A21} concludes the proof.
\end{proof}

\subsubsection{Summary: The quartic terms and proof of Theorem~\ref{DHS1:Calculation_of_the_GL-energy}}
\label{DHS1:sec:quarticterms}

Let the assumptions of Theorem~\ref{DHS1:Calculation_of_the_GL-energy} hold. We collect the results of Lemma~\ref{DHS1:NTB_action}, as well as Propositions~\ref{DHS1:NTB-NtildeTB}, \ref{DHS1:NtildeTB-NTB1}, \ref{DHS1:NTB1-NT2}, and \ref{DHS1:NTc2}, which yield
\begin{equation}
	\frac{1}{8} \langle \Delta, N_{T,B}(\Delta) \rangle = \; \Lambda_3 \; \Vert \Psi\Vert_4^4 + R(B)
	\label{DHS1:eq:A28}
\end{equation}
with
\begin{equation*}
	| R(B) | \leq C \; B^3 \; \Vert \Psi \Vert_{\Hmag^1(Q_B)}^3 \; \Vert \Psi \Vert_{\Hmag^2(Q_B)}.
\end{equation*}
Together with \eqref{DHS1:eq:A15}, this completes the proof of  Theorem~\ref{DHS1:Calculation_of_the_GL-energy}.

\subsection{Proof of Proposition \ref{DHS1:Structure_of_alphaDelta}}
\label{DHS1:sec:strcturealphadelta}

We use $\alpha_\Delta = [\Gamma_\Delta]_{12}$, the resolvent equation \eqref{DHS1:Resolvent_Equation} and \eqref{DHS1:alphaDelta_decomposition_1} to see that
\begin{align*}
\alpha_\Delta &= [\Ocal]_{12} + [\Qcal_{T,B}(\Delta)]_{12} = [\Ocal]_{12} + \Rcal_{T,B}(\Delta), 
\end{align*}
with $\Ocal$ in \eqref{DHS1:alphaDelta_decomposition_2},
and
\begin{align*}
\Rcal_{T,B}(\Delta) &:=  \frac 1\beta \sum_{n\in \Zbb} \Bigl[ \frac{1}{ \i \omega_n - H_0} \delta\frac{1}{ \i \omega_n - H_0} \delta\frac{1}{ \i \omega_n - H_\Delta} \delta \frac{1}{ \i \omega_n - H_0}\Bigr]_{12}.
\end{align*}
The definition of $L_{T,B}$ in \eqref{DHS1:LTB_definition} implies $[\Ocal]_{12} = -\frac 12 L_{T,B}\Delta$, and we define
\begin{align}
\eta_0(\Delta) &:= \frac 12 \bigl(L_{T,B}\Delta - M_{T,B}\Delta\bigr) + \frac 12 \bigl( M_T^{(1)}\Delta - M_{\Tc}^{(1)}\Delta\bigr) + \Rcal_{T,B}(\Delta), \notag \\
\eta_\perp(\Delta) &:= \frac 12 \bigl( M_{T,B}\Delta - M_{T}^{(1)}\Delta\bigr), \label{DHS1:eta_perp_definition}
\end{align}
with $M_{T,B}$ in \eqref{DHS1:MTB_definition} and $M_T^{(1)}$ in \eqref{DHS1:MT1_definition}. Proposition~\ref{DHS1:MT1} implies that $-\frac 12 M_{\Tc}^{(1)} \Delta = \Psi\alpha_*$, so these definitions allow us to write $\alpha_{\Delta}$ as in \eqref{DHS1:alphaDelta_decomposition_eq1}. It remains to prove the properties of $\eta_0$ and $\eta_\perp$ that are listed in Proposition~\ref{DHS1:Structure_of_alphaDelta}.

We start with the proof of \eqref{DHS1:alphaDelta_decomposition_eq2}, and note that 
\begin{align*}
\Rcal_{T,B}(\Delta) &= \frac 1\beta \sum_{n\in\Zbb} \frac 1{\i\omega_n - \hfrak_B} \, \Delta \,  \frac 1{\i\omega_n + \ov{\hfrak_B}}\,  \ov \Delta\,  \Bigl[ \frac{1}{\i \omega_n - H_\Delta}\Bigr]_{11}\,  \Delta \, \frac 1{\i\omega_n + \ov{\hfrak_B}}.
\end{align*}
Using Hölder's inequality, we immediately see that $\Vert \Rcal_{T,B}(\Delta)\Vert_2 \leq C \beta^3 \Vert \Delta\Vert_6^3$. Furthermore, we estimate
\begin{equation*}
	\Vert \pi \Rcal_{T,B}(\Delta) \Vert_2 \leq \frac 1\beta \sum_{n\in\Zbb} \Bigl\Vert \pi \frac 1{\i\omega_n - \hfrak_B} \Bigr\Vert_{\infty} \Bigl\Vert \frac 1{\i\omega_n + \ov{\hfrak_B}} \Bigr\Vert_{\infty}^2 \Bigl\Vert \Bigl[ \frac{1}{\i \omega_n - H_\Delta}\Bigr]_{11} \Bigr\Vert_{\infty} \Vert \Delta \Vert_6^3.
\end{equation*}
With the help of $\Vert A\Vert_\infty^2 = \Vert A^*A\Vert_\infty$ for a general operator $A$, the first norm on the right side is bounded by
\begin{align*}
\Bigl\Vert \pi \frac 1{\i\omega_n - \hfrak_B} \Bigr\Vert_{\infty} &\leq \Bigl\Vert \frac 1{-\i\omega_n - \hfrak_B}\Bigr\Vert_\infty^{\nicefrac 12} \Bigl\Vert\pi^2 \frac 1{\i\omega_n - \hfrak_B} \Bigr\Vert_{\infty}^{\nicefrac 12} \leq C \,  |\omega_n|^{-\nicefrac 12}.
\end{align*}
Hence,
\begin{equation}
	\Vert \pi \Rcal_{T,B}(\Delta) \Vert_2 \leq C\, \beta^{\nicefrac 52}\,\Vert \Delta \Vert_6^3.
	\label{DHS1:eq:A25}
\end{equation}
With a similar argument, we see that $\Vert \Rcal_{T,B}(\Delta) \pi \Vert_2$ is bounded by the right side of \eqref{DHS1:eq:A25}, too. An application of Lemma~\ref{DHS1:Schatten_estimate} and of \eqref{DHS1:Magnetic_Sobolev} on the right side of \eqref{DHS1:eq:A25} finally shows
\begin{equation*}
	\Vert \Rcal_{T,B}(\Delta) \Vert_{H^1(Q_B \times \mathbb{R}^3_{\mathrm{s}})}^2 \leq C \; B^3 \; \Vert \Psi \Vert_{\Hmag^1(Q_B)}^6. 
\end{equation*}
The remaining terms in $\eta_0(\Delta)$ can be estimated with the help of Propositions~\ref{DHS1:LTB-LtildeTB}, \ref{DHS1:LtildeTB-MTB}, and \ref{DHS1:MT1}, which establishes \eqref{DHS1:alphaDelta_decomposition_eq2}. 

It remains to prove \eqref{DHS1:alphaDelta_decomposition_eq3} and \eqref{DHS1:alphaDelta_decomposition_eq4}. We start with the proof of \eqref{DHS1:alphaDelta_decomposition_eq3} and write
\begin{align}
	\eta_\perp(\Delta)(X,r) = \iint_{\Rbb^3\times \Rbb^3} \dd Z \dd s\; k_T(Z, r-s) \; [ \cos(Z\cdot \Pi_X) - 1 ]  \; \Delta(X, s).
	\label{DHS1:eq:A27}
\end{align}
Using \eqref{DHS1:NTB1-NTB2_5} we see that
\begin{align}
	\Vert \eta_\perp\Vert_2^2 &\leq C\; B^3  \; \Vert F_T^{(2)}\Vert_1^2 \; \Vert V\alpha_*\Vert_2^2\;  \Vert \Psi\Vert_{\Hmag^2(Q_B)}^2,
	\label{DHS1:eq:A26}
\end{align}
with the function $F_T^{(2)}$ in \eqref{DHS1:LtildeTB-MTB_FT_definition}. The $L^1(\Rbb^3$)-norm of this function was estimated in \eqref{DHS1:NtildeTB-NTB1_FT1-2_estimate}. We use this bound and conclude the claimed bound for the $L^2(Q_B \times \mathbb{R}^3_{\mathrm{s}})$-norm of $\eta_{\perp }$. Bounds for $\Vert \tilde \pi_r \eta_\perp \Vert_2$ and $\Vert |r|\eta_\perp \Vert_2$ can be proved similarly and we leave the details to the reader. 

To prove the claimed bound for $\Vert \Pi_X\eta_\perp \Vert_2$, we need to replace $[\cos(Z\cdot\Pi_X)- 1]\Psi(X)$ by $\Pi_X[\cos(Z\cdot \Pi_X) - 1]\Psi(X)$ in the proof of \eqref{DHS1:eq:A26}. Using the intertwining relation \eqref{DHS1:PiXcos} in Lemma~\ref{DHS1:CommutationII} below, the operator inquality \eqref{DHS1:ZPiX_inequality} for $(Z\cdot \Pi)^2$, and the equality \eqref{DHS1:PiPi2Pi_equality} for $\Pi \, \Pi^2\, \Pi$, we see that
\begin{align}
	\Vert \Pi [\cos(Z\cdot\Pi) - 1]\Psi \Vert^2 &\leq C \; B^3 \; |Z|^2 \; \Vert \Psi\Vert_{\Hmag^2(Q_B)}^2 \label{DHS1:alphaDelta_decomposition_4}
\end{align}
holds. The claimed bound for $\Vert \Pi_X\eta_\perp \Vert_2$ follows from \eqref{DHS1:eq:A27} and \eqref{DHS1:alphaDelta_decomposition_4}, which, in combination with the previous considerations, proves \eqref{DHS1:alphaDelta_decomposition_eq3}. 

To prove \eqref{DHS1:alphaDelta_decomposition_eq4}, we note that for any two radial functions $f,g\in L^2(\Rbb^3)$ the function
\begin{equation}
	\iint_{\Rbb^6} \dd r \dd s \; f(r) \, k_T(Z, r-s) \, g(s) 
\end{equation}
is radial in $Z$. We claim that this implies that the operator
\begin{equation}
	\iiint_{\Rbb^9} \dd Z \dd s \dd r \; f(r)  k_T(Z, r-s) g(s) [ \cos(Z\cdot \Pi) - 1 ] 
	\label{DHS1:eq:A29}
\end{equation}
equals $h( \Pi^2 )$ for some function $h \colon [0,\infty) \to \mathbb{R}$. To prove this, let us denote by $\tilde \Pi$ the same operator $\Pi$ but understood to act on $L^2(\mathbb{R}^3)$ instead of $\Lmag^2(Q_B)$. From \cite[Lemma~28]{Hainzl2017} we know that the above statement is true when $\Pi$ is replaced by $\tilde \Pi$. To reduce our claim to this case, we use the unitary Bloch--Floquet transformation 
\begin{equation}
	( \mathcal{U}_{\mathrm{BF}} \Psi )(k,X) := \sum_{\lambda \in \Lambda_B} \e^{-\i k \cdot (X-\lambda)} (T_B(\lambda) \Psi)(X) 
	\label{DHS1:eq:ABF1}
\end{equation}
with $T_B(\lambda)$ in \eqref{DHS1:Magnetic_Translation_Charge2} and inverse
\begin{equation}
	( \mathcal{U}^*_{\mathrm{BF}} \Phi )(X) = \int_{[0,\sqrt{2 \pi B}]^3} \mathrm{d} k \; \e^{\i k\cdot X} \Phi(k,X).
	\label{DHS1:eq:ABF2}
\end{equation}
The magnetic momentum operator $\tilde \Pi$ obeys the identity
\begin{equation}
	\mathcal{U}_{\mathrm{BF}}\,  \tilde \Pi\,  \mathcal{U}^*_{\mathrm{BF}} = \int^{\oplus}_{[0,\sqrt{ 2 \pi B}]^3} \mathrm{d}k \; \tilde \Pi(k)
	\label{DHS1:eq:ABF3}
\end{equation}
with $\tilde \Pi(k) = \Pi + k$ acting on $\Lmag^2(Q_B)$. The claim follows when we conjugate both sides of the equation
\begin{equation*}
	\iiint_{\Rbb^9} \dd Z \dd s \dd r \; f(r) k_T(Z, r-s) g(s) [ \cos(Z\cdot \tilde{\Pi}) - 1 ] = h(\tilde{\Pi}^2)
\end{equation*}
with the Bloch--Floquet transformation and use that $\tilde \Pi(0) = \Pi$. Eq.~\eqref{DHS1:alphaDelta_decomposition_eq4} is a direct consequence of the fact that the operator in \eqref{DHS1:eq:A29} equals $h( \Pi^2 )$. This proves Proposition~\ref{DHS1:Structure_of_alphaDelta}.

\subsection{Proof of Proposition \ref{DHS1:Lower_Tc_a_priori_bound}}
\label{DHS1:Lower_Tc_a_priori_bound_proof_Section}

Let the assumptions of Proposition~\ref{DHS1:Lower_Tc_a_priori_bound} hold. We show that there are constants $D_0>0$ and $B_0>0$ such that for $0 < B \leq B_0$ and temperatures $T$ obeying
\begin{align*}
	0 < T_0 \leq T < \Tc (1 - D_0 B)
\end{align*}
there is a function $\Psi \in \Hmag^2(Q_B)$, such that the Gibbs state $\Gamma_{\Delta}$ in \eqref{DHS1:GammaDelta_definition} built upon the gap function $\Delta(X,r) = -2 V\alpha_*(r) \Psi(X)$ obeys \eqref{DHS1:Lower_critical_shift_2}.

To prove this, we choose $\psi \in \Hmag^2(Q_1)$ with $\Vert \psi\Vert_{\Hmag^2(Q_B)}=1$ and $\Psi \in \Hmag^2(Q_B)$ as in \eqref{DHS1:GL-rescaling}. This, in particular, implies $\Vert \Psi\Vert_{\Hmag^2(Q_B)}=1$. We collect the results of Propositions~\ref{DHS1:Structure_of_alphaDelta}, \ref{DHS1:BCS functional_identity}, \ref{DHS1:Rough_bound_on_BCS energy}, as well as \eqref{DHS1:Magnetic_Sobolev} and \eqref{DHS1:eq:A28}, and conclude that
\begin{align*}
	\FBCS(\Gamma_\Delta) - \FBCS(\Gamma_0) &< B \, \bigl( - cD_0 \, \Vert \psi\Vert_2^2 + C \bigr)
\end{align*}
holds as long as $B$ is small enough. We remark that this argument can be carried out without the assumption of $\Hmag^2(Q_1)$-regularity of $\psi$ by instead using the sign of $V$. Compare this to the discussion below \eqref{DHS1:Upper_Bound_proof_1}. Choosing $D_0 = \frac{C}{c \Vert \psi\Vert_2^2}$ ends the proof of Proposition~\ref{DHS1:Lower_Tc_a_priori_bound}.

\section{The Structure of Low-Energy States}
\label{DHS1:Lower Bound Part A}

In Section~\ref{DHS1:Upper_Bound} we use a Gibbs state to show that the BCS free energy is bounded from above by the Ginzburg--Landau energy plus corrections of lower order. The Gibbs state has a Cooper pair wavefunction which is given by a product of the form $\alpha_*(r) \Psi(X)$ to leading order, where $\Psi$ is a minimizer of the Ginzburg--Landau functional in \eqref{DHS1:Definition_GL-functional} and $\alpha_*$ is the unique solution of the gap equation \eqref{DHS1:alpha_star_ev-equation}.
Moreover, close to the critical temperature the Cooper pair wave function is small in an appropriate sense, which allows us to expand the BCS functional in powers of $\Psi$ and to obtain the terms in the Ginzburg--Landau functional.

Our proof of a matching lower bound for the BCS free energy in Section~\ref{DHS1:Lower Bound Part B} is based on the fact that certain low-energy states of the BCS functional have a Cooper pair wave function with a similar structure. The precise statement is provided in Theorem~\ref{DHS1:Structure_of_almost_minimizers} below, which is the main technical novelty of this paper. This section is devoted to its proof.

%

We recall the definition of the generalized one-particle density matrix $\Gamma$ in \eqref{DHS1:Gamma_introduction}, its offdiagonal entry $\alpha$, as well as the normal state $\Gamma_0$ in \eqref{DHS1:Gamma0}.

\begin{thm}[Structure of low-energy states]
\label{DHS1:Structure_of_almost_minimizers}
Let Assumptions \ref{DHS1:Assumption_V} and \ref{DHS1:Assumption_KTc} hold. For given $D_0, D_1 \geq 0$, there is a constant $B_0>0$ such that for all $0 <B \leq B_0$ the following holds: If $T>0$ obeys $T - \Tc \geq -D_0B$ and if $\Gamma$ is a gauge-periodic state with low energy, that is,
\begin{align}
\FBCS(\Gamma) - \FBCS(\Gamma_0) \leq D_1B^2, \label{DHS1:Second_Decomposition_Gamma_Assumption}
\end{align}
then there are $\Psi\in \Hmag^1(Q_B)$ and $\xi\in \Hsymm$ such that
\begin{align}
\alpha(X,r) = \Psi(X)\alpha_*(r) + \xi(X,r), \label{DHS1:Second_Decomposition_alpha_equation}
\end{align}
where
\begin{align}
\sup_{0< B\leq B_0} \Vert \Psi\Vert_{\Hmag^1(Q_B)}^2 &\leq C, &  \Vert \xi\Vert_{\Hsymm}^2 &\leq CB^2 \bigl( \Vert \Psi\Vert_{\Hmag^1(Q_B)}^2 + D_1\bigr). \label{DHS1:Second_Decomposition_Psi_xi_estimate}
\end{align}
\end{thm}

\begin{varbems}
\begin{enumerate}[(a)]
\item Equation \eqref{DHS1:Second_Decomposition_Psi_xi_estimate}
proves that, despite $\Psi$ being dependent on $B$, it is a macroscopic quantity in the sense that its $\Hmag^1(Q_B)$-norm scales as that of the function in \eqref{DHS1:GL-rescaling}.

\item We highlight that, in contrast to the $\Hmag^1(Q_B)$-norm of $\Psi$, the $\Hsymm$-norm of $\xi$ is not scaled with additional factors of $B$, see \eqref{DHS1:H1-norm}. The unscaled $\Lmag^2(Q_B)$-norm of $\Psi$ is of the order $B^{\nicefrac 12}$, whence it is much larger than that of $\xi$.
\item Theorem~\ref{DHS1:Structure_of_almost_minimizers} should be compared to \cite[Eq.~(5.1)]{Hainzl2012} and \cite[Theorem~22]{Hainzl2017}.
\end{enumerate}
\end{varbems}

Theorem~\ref{DHS1:Structure_of_almost_minimizers} contains the natural a priori bounds for the Cooper pair wave function $\alpha$ of a low-energy state $\Gamma$ in the sense of \eqref{DHS1:Second_Decomposition_Gamma_Assumption}. However, in Section~\ref{DHS1:Lower Bound Part B} we are going to need more regularity of $\Psi$ than is provided by Theorem \ref{DHS1:Structure_of_almost_minimizers}. More precisely, we are going to use the function $\Psi$ from this decomposition to construct a Gibbs state $\Gamma_{\Delta_{\Psi}}$ and apply our trial state analysis provided by Propositions \ref{DHS1:Structure_of_alphaDelta} and \ref{DHS1:BCS functional_identity} as well as Theorem~\ref{DHS1:Calculation_of_the_GL-energy} to extract the Ginzburg--Landau energy. In order to control the errors during this analysis, we need the $\Hmag^2(Q_B)$-norm of $\Psi$. The following corollary provides us with a decomposition of $\alpha$ in terms of a center-of-mass Cooper pair wave function $\Psi_\leq$ with $\Hmag^2(Q_B)$-regularity.



\begin{kor}
\label{DHS1:Structure_of_almost_minimizers_corollary}
Let the assumptions of Theorem~\ref{DHS1:Structure_of_almost_minimizers} hold and let $\varepsilon \in [B, B_0]$. Let $\Psi$ be as in 
\eqref{DHS1:Second_Decomposition_alpha_equation} and define
\begin{align}
	\Psi_\leq &:= \Idbb_{[0,\varepsilon]}(\Pi^2) \Psi, &  \Psi_> &:= \Idbb_{(\varepsilon,\infty)}(\Pi^2) \Psi. \label{DHS1:PsileqPsi>_definition}
\end{align}
Then, we have
\begin{align}
	\Vert \Psi_\leq\Vert_{\Hmag^1(Q_B)}^2 &\leq \Vert \Psi\Vert_{\Hmag^1(Q_B)}^2, \notag \\ 
	\Vert \Psi_\leq \Vert_{\Hmag^k(Q_B)}^2 &\leq C\, (\varepsilon B^{-1})^{k-1} \, \Vert \Psi\Vert_{\Hmag^1(Q_B)}^2, \qquad k\geq 2,  \label{DHS1:Psileq_bounds}
\end{align}
as well as 
\begin{align}
\Vert \Psi_>\Vert_2^2 &\leq C \varepsilon^{-1}B^2 \, \Vert \Psi\Vert_{\Hmag^1(Q_B)}^2, & \Vert \Pi\Psi_>\Vert_2^2 &\leq CB^2 \, \Vert \Psi\Vert_{\Hmag^1(Q_B)}^2. \label{DHS1:Psi>_bound}
\end{align}
Furthermore,
\begin{align}
	\sigma_0(X,r) := \Psi_>(X)\alpha_*(r) \label{DHS1:sigma0}
\end{align}
satisfies
\begin{align}
	\Vert \sigma_0\Vert_{\Hsymm}^2 &\leq C\varepsilon^{-1}B^2 \, \Vert \Psi\Vert_{\Hmag^1(Q_B)}^2 \label{DHS1:sigma0_estimate}
\end{align}
and, with $\xi$ in \eqref{DHS1:Second_Decomposition_alpha_equation}, the function
\begin{align}
	\sigma :=  \xi + \sigma_0 \label{DHS1:sigma}
\end{align}
obeys
\begin{align}
	\Vert \sigma\Vert_{\Hsymm}^2 \leq CB^2 \bigl( \varepsilon^{-1}\Vert \Psi\Vert_{\Hmag^1(Q_B)}^2 + D_1\bigr). \label{DHS1:Second_Decomposition_sigma_estimate}
\end{align}
In terms of these functions, the Cooper pair wave function $\alpha$ of the low-energy state $\Gamma$ in \eqref{DHS1:Second_Decomposition_Gamma_Assumption} admits the decomposition
\begin{align}
	\alpha(X,r) = \Psi_\leq (X)\alpha_*(r) + \sigma(X,r). \label{DHS1:Second_Decomposition_alpha_equation_final}
\end{align}
\end{kor}


%
%

\begin{proof}
The bounds for $\Psi_\leq$ and $\Psi_>$ in \eqref{DHS1:Psileq_bounds} and \eqref{DHS1:Psi>_bound} are a direct consequence of their definition in \eqref{DHS1:PsileqPsi>_definition}.
%
The bound \eqref{DHS1:Psi>_bound} immediately implies \eqref{DHS1:sigma0_estimate}. Moreover, $\sigma$ obeys \eqref{DHS1:Second_Decomposition_sigma_estimate} by \eqref{DHS1:Second_Decomposition_Psi_xi_estimate} and \eqref{DHS1:sigma0_estimate}. Finally, \eqref{DHS1:Second_Decomposition_alpha_equation_final} follows from \eqref{DHS1:Second_Decomposition_alpha_equation}.
\end{proof}


\subsection{A lower bound for the BCS functional}

We start the proof of Theorem \ref{DHS1:Structure_of_almost_minimizers} with the following lower bound on the BCS functional.


\begin{lem}
Let $\Gamma_0$ be the normal state in \eqref{DHS1:Gamma0}. We have the lower bound
\begin{align}
\FBCS(\Gamma) - \FBCS(\Gamma_0) \geq  \Tr\bigl[ (K_{T,B} - V) \alpha  \alpha^*\bigr] + \frac{4T}{5} \Tr\bigl[ (\alpha^* \alpha)^2\bigr], \label{DHS1:Lower_Bound_A_3}
\end{align}
where $K_{T,B} = K_T(\pi)$ and $V\alpha(x,y) = V(x-y) \alpha(x,y)$.
\end{lem}

\begin{proof}
The statement follows from Eqs.~(5.3)--(5.12) in \cite{Hainzl2012} with the evident replacements. The argument uses the relative entropy inequality \cite[Lemma~1]{Hainzl2012}, which is a refinement of the bound \cite[Theorem~1]{Hainzl2008_Lewin}.
\end{proof}

In Proposition \ref{DHS1:KTV_Asymptotics_of_EV_and_EF} in Appendix~\ref{DHS1:KTV_Asymptotics_of_EV_and_EF_Section} we show that the magnetic field can lower the lowest eigenvalue zero of $K_{\Tc} - V$
at most by a constant times $B$. This information is used in the following lemma to bound $K_{T,B} - V$ from below by a nonnegative operator, up to a correction of the size $CB$. The inequality \eqref{DHS1:KTB_Lower_bound_eq} below is stated for $K_{T,B} - V$ as a one-particle operator but it holds equally for the operator $K_{T,B} - V(x-y)$ in \eqref{DHS1:Lower_Bound_A_3} because $V$ intertwines as $T(y)^* V(x) T(y) = V(x-y)$ with the magnetic translations $T(y)$ in \eqref{DHS1:Magnetic_Translation}.

\begin{lem}
\label{DHS1:KTB_Lower_bound}
Let Assumptions \ref{DHS1:Assumption_V} and \ref{DHS1:Assumption_KTc} be true.
For any $D_0 \geq 0$, there are constants $B_0>0$ and $T_0>0$ such that for $0< B\leq B_0$ and $T>0$ with $T - \Tc \geq -D_0B$, the estimate
\begin{align}
K_{T, B} - V &\geq c \; (1 - P) (1 + \pi^2) (1- P) + c \, \min \{ T_0, (T - \Tc)_+\} - CB \label{DHS1:KTB_Lower_bound_eq}
\end{align}
holds. Here, $P = |\alpha_*\rangle\langle \alpha_*|$ is the orthogonal projection onto the ground state $\alpha_*$ of $K_{\Tc} - V$.
\end{lem}

\begin{proof}
We prove two lower bounds on $K_{T,B} - V$, which we add up to etablish \eqref{DHS1:KTB_Lower_bound_eq}.

\emph{Step 1.} We claim that there are $T_0, c, C >0$ such that
\begin{align}
K_{T,B} - V \geq c \; \min \{ T_0, (T - \Tc)_+\} - CB. \label{DHS1:KTB_Lower_bound_5}
\end{align}
To prove \eqref{DHS1:KTB_Lower_bound_5}, we note that the derivative of the symbol $K_T$ in \eqref{DHS1:KT-symbol} with respect to $T$ equals
\begin{align}
\frac{\dd}{\dd T} K_T(p) = 2\;  \frac{( \frac{p^2 - \mu}{2T})^2}{\sinh^2(\frac{p^2-\mu}{2T})} \label{DHS1:KTc_bounded_derivative}
\end{align}
and is bounded from above by 2. If $T \leq \Tc$, we infer $K_{T,B} - K_{\Tc,B} \geq -2D_0B$ as an operator inequality, which, in combination with Proposition~\ref{DHS1:KTV_Asymptotics_of_EV_and_EF} in the appendix,  proves \eqref{DHS1:KTB_Lower_bound_5} in this case. To treat the case $T \geq \Tc$, we denote by $e_0^{T,B}$ and $e_1^{T,B}$ the lowest and the second lowest eigenvalue of the operator $K_{T,B} - V$, respectively. Also let $P_{T,B}$ be the spectral projection corresponding to $e_0^{T,B}$ and define $Q_{T,B} = 1 - P_{T,B}$. We have 
\begin{align*}
	K_{T,B} - V \geq e_0^{T,B} P_{T,B} + e_{1}^{T,B} Q_{T,B}.
\end{align*}
Since $K_T(p) - K_{\Tc}(p) \geq 0$ for all $p \in \mathbb{R}^3$, which follows from \eqref{DHS1:KTc_bounded_derivative}, we know the lower bound $e_{1}^{T,B} \geq e_{1}^{\Tc,B} \geq \kappa$ for some $\kappa > 0$. Here, the second inequality follows from Proposition~\ref{DHS1:KTV_Asymptotics_of_EV_and_EF}. From Proposition~\ref{DHS1:KTV_Asymptotics_of_EV_and_EF} we also know that the lowest eigenvalue of $K_{T,B} - V $ is simple. According to \eqref{DHS1:KTc_bounded_derivative}, the function $T \mapsto K_T(p)$ is increasing and has a non-vanishing derivative for each $p \in \mathbb{R}^3$. Analytic perturbation theory therefore implies the lower bound $e_0^{T,B} \geq e_0^{\Tc,B} + c (T - \Tc)$ for some $c > 0$ as long as $|T - \Tc |$ is small enough. Since Proposition~\ref{DHS1:KTV_Asymptotics_of_EV_and_EF} shows $e_0^{\Tc,B} \geq -CB$ these consideration prove \eqref{DHS1:KTB_Lower_bound_5} in the case $T \geq \Tc$.

\emph{Step 2.} We claim there are $c,C>0$ such that
\begin{align}
K_{T,B} - V \geq c \; (1 - P) (1 + \pi^2) (1- P) - CB. \label{DHS1:KTB_Lower_bound_9}
\end{align}
From the arguments in Step~1 we know that we can replace $T$ by $\Tc$ for a lower bound if we allow for a remainder of the size $-CB$. To prove \eqref{DHS1:KTB_Lower_bound_9}, we choose $0 < \eta < 1$ and write
\begin{align}
K_{\Tc,B}-V = e_0^BP_B + (1-P_B) [(1-\eta) K_{\Tc,B} -V](1-P_B) + \eta (1-P_B) K_{\Tc,B} (1-P_B), \label{DHS1:KTB_Lower_bound_1}
\end{align}
where $e_0^B$ denotes the ground state energy of $K_{\Tc, B} - V$ and $P_B = |\alpha_*^B\rangle \langle \alpha_*^B|$ is the spectral projection onto the corresponding unique ground state vector $\alpha_*^B$. From Proposition~\ref{DHS1:KTV_Asymptotics_of_EV_and_EF} we know that the first term on the right side of \eqref{DHS1:KTB_Lower_bound_1} is bounded from below by $-CB$. The lowest eigenvalue of $K_{\Tc} - V$ is simple and isolated from the rest of the spectrum. Proposition~\ref{DHS1:KTV_Asymptotics_of_EV_and_EF} therefore implies that the second term in \eqref{DHS1:KTB_Lower_bound_1} is nonnegative as long as $\eta$ is, independently of $B$, chosen small enough, and can be dropped for a lower bound.
To treat the third term, we note that the symbol $K_T(p)$ in \eqref{DHS1:KT-symbol} satisfies the inequality $K_{\Tc}(p) \geq c' (1 + p^2)$ for some constant $c'$, and hence $K_{\Tc,B} \geq c' (1 + \pi^2)$. In combination, the above considerations prove
\begin{align*}
K_{\Tc,B}-V \geq  c' \; (1-P_B)(1+\pi^2)(1-P_B) - CB.
\end{align*}
It remains to replace $P_B$ by $P = |\alpha_*\rangle\langle \alpha_*|$. To this end, we write
\begin{align}
(1-P_B)(1+\pi^2)(1-P_B) - (1-P)(1+\pi^2)(1-P) &\notag \\
&\hspace{-160pt}= (P-P_B) +  (P - P_B)\pi^2(1-P_B) + (1-P)\pi^2(P-P_B) \label{DHS1:KTB_Lower_bound_4}.
\end{align}
From Proposition \ref{DHS1:KTV_Asymptotics_of_EV_and_EF} we know that $\Vert P_B - P\Vert_\infty \leq CB$ and $\Vert \pi^2(P_B- P)\Vert_\infty \leq CB$. Hence, the norm of the operator on the right side of \eqref{DHS1:KTB_Lower_bound_4} is bounded by a constant times $B$. This shows \eqref{DHS1:KTB_Lower_bound_9} and concludes our proof.
\end{proof}

We deduce two corollaries from \eqref{DHS1:Lower_Bound_A_3} and Lemma \ref{DHS1:KTB_Lower_bound}. The first statement is an a priori bound that we use in the proof of Theorem \ref{DHS1:Main_Result_Tc} (b).

\begin{kor}
\label{DHS1:TcB_First_Upper_Bound}
Let Assumptions \ref{DHS1:Assumption_V} and \ref{DHS1:Assumption_KTc} be true. Then, there are constants $B_0>0$ and $C>0$ such that for all $0 < B \leq B_0$ and all temperatures $T\geq \Tc(1 + CB)$, we have $\FBCS(\Gamma) - \FBCS(\Gamma_0) >0$ unless $\Gamma = \Gamma_0$.
\end{kor}

\begin{proof}
Let $D_0>0$ and assume that $T \geq \Tc (1 + D_0B)$. From \eqref{DHS1:Lower_Bound_A_3} and Lemma~\ref{DHS1:KTB_Lower_bound} we know that
\begin{align}
\FBCS(\Gamma) - \FBCS(\Gamma_0) \geq (c \, \min\{T_0, \Tc D_0B\}  - CB) \, \Vert \alpha\Vert_2^2. \label{DHS1:TcB_First_Upper_Bound_1}
\end{align}
For the choice $D_0 = \frac{2 C}{ c \Tc}$ and $B_0 = \frac{T_0}{D_0 \Tc} $ the right side of \eqref{DHS1:TcB_First_Upper_Bound_1} is strictly positive unless $\alpha = 0$. We conclude that $\Gamma_0$ is the unique minimizer of $\FBCS$, which proves the claim. 
\end{proof}

The second corollary provides a bound for the Cooper pair wave functions of low-energy BCS states in the sense of \eqref{DHS1:Second_Decomposition_Gamma_Assumption}. It is based upon \eqref{DHS1:Lower_Bound_A_3} and to state it we need to introduce the operator
\begin{align}
U  &:= \e^{-\i \frac r2 \Pi_X}. \label{DHS1:U_definition}
\end{align}
We highlight that it acts on both, the relative coordinate $r = x-y$ and the center-of-mass coordinate $X = \frac{x+y}{2}$ of a function $\alpha(x,y)$. 

\begin{kor}
\label{DHS1:cor:lowerbound}
Let Assumptions \ref{DHS1:Assumption_V} and \ref{DHS1:Assumption_KTc} be true. For any $D_0, D_1 \geq 0$, there is a constant $B_0>0$ such that if $\Gamma$ satisfies \eqref{DHS1:Second_Decomposition_Gamma_Assumption}, if $0 < B\leq B_0$, and if $T$ is such that $T - \Tc \geq -D_0B$, then $\alpha = \Gamma_{12}$ obeys
\begin{align}
&\langle \alpha, [U(1 - P)(1 + \pi_r^2)(1 - P)U^* + U^*(1 - P)(1 + \pi_r^2)(1 - P)U] \alpha \rangle \notag\\
&\hspace{200pt} + \Tr\bigl[(\alpha^* \alpha)^2\bigr] \leq C B \Vert \alpha\Vert_2^2 + D_1B^2, \label{DHS1:Lower_Bound_A_2}
\end{align}
where $P = | \alpha_* \rangle \langle \alpha_* |$ and $\pi_r = -\i \nabla_r + \frac 12\Bbold \wedge r$ both act on the relative coordinate.
\end{kor}

In the statement of the corollary and in the following, we refrain from equipping the projection $P = |\alpha_*\rangle \langle \alpha_*|$ with an index $r$ although it acts on the relative coordinate. This should not lead to confusion and keeps the formulas readable.

\begin{proof}
We recall that the operator $V$ acts by multiplication with $V(x-y)$ and that $K_T(p)$ is defined in \eqref{DHS1:KT-symbol}. Using $\alpha(x,y) = \alpha(y,x)$ we write
\begin{align}
\Tr \bigl[ (K_{T,B} - V) \alpha \alpha^*\bigr] &= \frac 12 \fint_{Q_B} \dd x \int_{\Rbb^3} \dd y \; \overline{\alpha(x,y)} \bigl[ (K_T(\pi_x) - V) + (K_T(\pi_y) - V) \bigr] \alpha(x,y). \label{DHS1:Lower_Bound_A_4}
\end{align}
We note that $\pi_x = \frac 12 \Pi_X + \tilde \pi_r = U \pi_rU^*$ and $\pi_y = \frac 12 \Pi_X - \tilde \pi_r = -U^*\pi_r U$, with $\tilde \pi_r$ and $\Pi_X$ in \eqref{DHS1:Magnetic_Momenta_COM}. Using the above identities we see that
\begin{align}
K_T(\pi_x) - V(r) &= U ( K_{T}(\pi_r) - V(r) ) U^*, \notag \\ 
K_T(\pi_y) - V(r) &= U^* ( K_{T}(\pi_r) - V(r) ) U. \label{DHS1:eq:idK_T^r} 
\end{align}
The result follows from a short computation or from Lemma~\ref{DHS1:CommutationI} below.
We combine \eqref{DHS1:Second_Decomposition_Gamma_Assumption}, \eqref{DHS1:Lower_Bound_A_3}, \eqref{DHS1:Lower_Bound_A_4} and \eqref{DHS1:eq:idK_T^r} to show the inequality
\begin{align*}
\frac 12\langle \alpha, [U (K_{T}(\pi_r) - V(r))U^* + U^* (K_{T}(\pi_r) - V(r))U]\alpha\rangle + c \Tr \bigl[ (\alpha^* \alpha)^2\bigr] \leq D_1 B^2. 
\end{align*}
Finally, we apply Lemma \ref{DHS1:KTB_Lower_bound} to the first term on the left side and obtain \eqref{DHS1:Lower_Bound_A_2}.
\end{proof}


\subsection{The first decomposition result}

The proof of Theorem~\ref{DHS1:Structure_of_almost_minimizers} is based on Corollary~\ref{DHS1:cor:lowerbound} and is given in two steps. In the first step we drop the second term on the left side of \eqref{DHS1:Lower_Bound_A_2} for a lower bound, and investigate the implications of the resulting inequality for $\alpha$. The result of the corresponding analysis is summarized in Proposition~\ref{DHS1:First_Decomposition_Result} below. The second term on the left side of \eqref{DHS1:Lower_Bound_A_2} is used later in Lemma~\ref{DHS1:Bound_on_psi}. 

\begin{prop}
\label{DHS1:First_Decomposition_Result}
Given $D_0, D_1 \geq 0 $, there is $B_0>0$ with the following properties. If, for some $0< B\leq B_0$, the wave function $\alpha\in \Lsymm$ satisfies
\begin{align}
\langle \alpha , [U^* (1-P)(1 + \pi_r^2)(1-P)U + U (1-P) (1 + \pi_r^2)(1-P)U^*] \alpha \rangle & \leq D_0 B \Vert \alpha\Vert_2^2 + D_1 B^2 , \label{DHS1:First_Decomposition_Result_Assumption}
\end{align}
then there are $\Psi\in \Hmag^1(Q_B)$ and $\xi_0\in \Hsymm$ such that
\begin{align}
\alpha(X,r) =  \alpha_*(r)\cos\bigl( \frac r2 \cdot \Pi_X\bigr) \Psi(X) + \xi_0(X,r) \label{DHS1:First_Decomposition_Result_Decomp}
\end{align}
with
\begin{align}
\langle \Psi, \Pi^2 \Psi\rangle + \Vert \xi_0\Vert_{\Hsymm}^2 \leq C\bigl( B \Vert \Psi \Vert_2^2 + D_1 B^2 \bigr).  \label{DHS1:First_Decomposition_Result_Estimate}
\end{align}
\end{prop}
Before we give the proof of the a priori estimates in Proposition \ref{DHS1:First_Decomposition_Result}, we define the decomposition of $\alpha$, explain the idea behind it, and discuss relations to the existing literature. For this purpose, let the operator $A \colon \Lsymm \ra \Lmag^2(Q_B)$ be given by
\begin{align}
(A\alpha)(X) := \int_{\Rbb^3} \dd r\; \alpha_*(r) \; \cos\bigl( \frac r2\cdot \Pi_X\bigr) \alpha(X,r). \label{DHS1:Def_A}
\end{align}
A short computation shows that its adjoint $A^*\colon \Lmag^2(Q_B) \ra \Lsymm$ is given by
\begin{align}
(A^*\Psi) (X,r) = \alpha_*(r) \cos\bigl( \frac r2\cdot \Pi_X\bigr) \Psi(X). \label{DHS1:Def_Astar}
\end{align}
We highlight that this is the form of the first term in \eqref{DHS1:First_Decomposition_Result_Decomp}. For a given Cooper pair wave function $\alpha$, we use these operators to define the two functions $\Psi$ and $\xi_0$ by
\begin{align}
	\Psi &:= (AA^*)^{-1} A\alpha, & \xi_0 &:= \alpha - A^*\Psi. \label{DHS1:Def_Psixi}
\end{align}
Lemma~\ref{DHS1:AAstar_Positive} below guarantees that $AA^*$ is invertible, and we readily check that  \eqref{DHS1:First_Decomposition_Result_Decomp} holds with these definitions. Moreover, this decomposition of $\alpha$ is orthogonal in the sense that $\langle A^* \Psi, \xi_0 \rangle = 0$ holds. The claimed orthogonality follows from
\begin{align}
	A\xi_0 =0, \label{DHS1:fundamental_property}
\end{align}
which is a direct consequence of \eqref{DHS1:Def_Psixi}. In the following we motivate our choice for $\Psi$ and $\xi_0$ and comment on its appearance in the literature.

The decomposition of $\alpha$ is motivated by the minimization problem for the low-energy operator $2 -UPU^* - U^*PU$, that is, the operator in \eqref{DHS1:First_Decomposition_Result_Assumption} with $\pi_r^2$ replaced by zero. The operators $UPU^*$ and $U^*PU$ act as $A^*A$ on the space $\Lsymm$ of reflection symmetric functions in the relative coordinate. If $\Pi$ is replaced by $P_X$ in the definition of $U$ then $A^*A$ can be written as 
\begin{equation}
	A^* A \cong \int^{\oplus}_{[0,\sqrt{ 2 \pi B}]^3} \mathrm{d}P_X \; | a_{P_X} \rangle \langle a_{P_X} |, \label{DHS1:eq:directintegralA}
\end{equation}
with $| a_{P_X} \rangle \langle a_{P_X} |$ the orthogonal projection onto the function $a_{P_X}(r) = \cos(r/2 \cdot P_X) \alpha_*(r)$. Here the variable $P_X$ is the dual of the center-of-mass coordinate $X$ in the sense of Fourier transformation and $r$ denotes the relative coordinate. That is, the function $a_{P_X}(r)$ minimizes $1-A^* A$ in each fiber, whence it is the eigenfunction with respect to the lowest eigenvalue of $1 - A^* A = 1 - (UPU^* + U^*PU)/2$. This discussion should be compared to \cite[Eq. (5.47)]{Hainzl2012} and the discussion before Lemma~20 in \cite{ProceedingsSpohn}.

If we replace $P_X$ by the magnetic momentum operator $\Pi$ again the above picture changes because the components of $\Pi$ cannot be diagonalized simultaneously (they do not commute), and hence \eqref{DHS1:eq:directintegralA} has no obvious equivalent in this case. The decomposition of $\alpha$ in terms of the operators $A$ and $A^*$ above has been introduced in \cite{Hainzl2017} in order to study the operator $1 - V^{\nicefrac 12} L_{T,B} V^{\nicefrac 12}$ with $L_{T,B}$ in \eqref{DHS1:LTB_definition}, see also the discussion below Theorem~\ref{DHS1:Calculation_of_the_GL-energy}. The situation in this work is comparable to our case with $\pi_r^2$ replaced by zero in \eqref{DHS1:First_Decomposition_Result_Assumption}. Our analysis below shows that the ansatz \eqref{DHS1:Def_Psixi} is useful even if the full range of energies is considered, that is, if $\pi_r^2$ is present in \eqref{DHS1:First_Decomposition_Result_Assumption}. 

In the following lemma we collect useful properties of the operator $A A^*$. It should be compared to \cite[Lemma~27]{Hainzl2017}.
%
%
%
%
%
%

\begin{lem}
\label{DHS1:AAstar_Positive}
The operators
\begin{align*}
AA^* &= \int_{\Rbb^3} \dd r\; \alpha_*(r)^2 \cos^2\bigl( \frac{r}{2} \cdot \Pi\bigr), & 1-AA^* &= \int_{\Rbb^3} \dd r \;\alpha_*(r)^2 \sin^2\bigl( \frac r2 \cdot \Pi\bigr) 
\end{align*}
on $\Lmag^2(Q_B)$ are both bounded nonnegative functions of $\Pi^2$ and satisfy the following properties:
\begin{enumerate}[(a)]
\item $0\leq AA^*\leq 1$ and $0 \leq 1 - AA^*\leq 1$.
\item There is a constant $c>0$ such that $AA^* \geq c$ and $1 - AA^* \geq c \; \Pi^2\; (1 + \Pi^2)^{-1}$.
\end{enumerate}
In particular, $AA^*$ and $1 - AA^*$ are boundedly invertible on $\Lmag^2(Q_B)$.
\end{lem}


\begin{proof}
Part (a) is a direct consequence of the fact that $\Vert \alpha_* \Vert_2 = 1$. In the following we reduce the proof of part (b) to known results in \cite{Hainzl2017}. To this end, we introduce the operator
\begin{align}
	R := \int_{\Rbb^3} \dd r \; \alpha_*(r)^2 \; \cos( r \cdot \Pi)
	\label{DHS1:eq:R}
\end{align}
and note that
\begin{align*}
	AA^* &= \frac 12 (1 + R), & 1 - AA^* &= \frac 12 (1 - R).
\end{align*}
Let us also denote by $\widetilde{R}$ the operator in \eqref{DHS1:eq:R} but with $\Pi$ replaced by $\tilde \Pi$, which is the same operator but understood to act on $L^2(\mathbb{R}^3)$ instead of $\Lmag^2(Q_B)$. In Lemma~28 in \cite{Hainzl2017} it has been shown that $\widetilde{R}$ is a function of $\tilde \Pi^2$. Moreover, the statement of Lemma~27 in \cite{Hainzl2017} is equivalent to
\begin{align}
	1 - \widetilde{R}^2 \geq c\,  \frac{\tilde \Pi^2}{1 + \tilde \Pi^2}
	\label{DHS1:eq:R1}
\end{align}
for some $0<c<1$, and Eq.~(55) in the same reference implies
\begin{align}
	| \widetilde{R} | \leq 1 - c.
	\label{DHS1:eq:R2}
\end{align}
In combination, \eqref{DHS1:eq:R1}, \eqref{DHS1:eq:R2}, and $ \widetilde{R} \leq 1$ show
\begin{align}
	 1+\widetilde{R} &\geq c, & 1 - \widetilde{R} \geq \frac{1 - \widetilde{R}^2}{2} &\geq \frac{c}{2} \frac{\tilde \Pi^2}{1 + \tilde \Pi^2} .
	 \label{DHS1:eq:R3} 
\end{align}
It remains to argue that $R$ is a function of $\Pi^2$ and that a version of \eqref{DHS1:eq:R3} with $\tilde R$ and $\tilde \Pi$ replaced by $R$ and $\Pi$ holds. 

The fact that $R$ is a function of $\Pi^2$ follows from the argument that we used to show that the same statement is true for the operator in \eqref{DHS1:eq:A29}. To show that \eqref{DHS1:eq:R3} with $\tilde R$ and $\tilde \Pi$ replaced by $R$ and $\Pi$ holds, we conjugate both sides of the inequalities with the Bloch-Floquet transformation in \eqref{DHS1:eq:ABF1} and \eqref{DHS1:eq:ABF2}, and use \eqref{DHS1:eq:ABF3}. The inequalities in \eqref{DHS1:eq:R3} therefore hold equally in any fiber, that is, with $\tilde \Pi$ on the left and on the right sides replaced by $\Pi(k) = \Pi + k$ acting on $\Lmag^2(Q_B)$. Since $\tilde \Pi(0) = \Pi$, this proves the claim. 
\end{proof}

The remainder of this subsection is devoted to the proof of Proposition~\ref{DHS1:First_Decomposition_Result}. We start with a lower bound on the operator in \eqref{DHS1:First_Decomposition_Result_Assumption} when it acts on wave functions of the form $A^*\Psi$, see Lemma~\ref{DHS1:MainTerm} below. 

\subsubsection{Step one -- lower bound on the range of \texorpdfstring{$A^*$}{A*}}

The main result of this subsection is the following lemma.

\begin{lem}
\label{DHS1:MainTerm}
For any $\Psi\in \Lmag^2(Q_B)$, with $A$ and $A^*$ given by \eqref{DHS1:Def_A} and \eqref{DHS1:Def_Astar}, with $U$ given by \eqref{DHS1:U_definition}, and $P = |\alpha_*\rangle \langle \alpha_*|$ with $\alpha_*$ from \eqref{DHS1:alpha_star_ev-equation} acting on the relative coordinate, we have
\begin{align}
\frac 12 \langle A^* \Psi, [ U^* (1 - P) (1 + \pi_r^2) (1- P) U + U(1 - P) (1 + \pi_r^2)(1-P) U^* ] A^*\Psi\rangle \hspace{-340pt}& \notag\\
&=  \langle \Psi, AA^* (1 - AA^*)(1 + \Pi^2) \Psi\rangle \notag\\
&\hspace{10pt} +  \fint_{Q_B} \dd X \int_{\Rbb^3} \dd r \; \ov{(1 - AA^*)\Psi(X)} \; |\nabla \alpha_*(r)|^2 \; \cos^2 \bigl(\frac r2 \Pi_X\bigr) \; (1 - AA^*)\Psi(X) \notag \\
&\hspace{10pt}+ \fint_{Q_B} \dd X \int_{\Rbb^3} \dr \; \ov{AA^*\Psi(X)} \; |\nabla \alpha_*(r)|^2\; \sin^2 \bigl(\frac r2  \Pi_X\bigr) \; AA^*\Psi(X)\notag \\
&\hspace{10pt} + \frac 14 \fint_{Q_B}\dd X \int_{\Rbb^3} \dd r \; \ov{(1 - AA^*)\Psi(X)} \; |\Bbold\wedge r|^2 \alpha_*(r)^2 \; \sin^2 \bigl(\frac r2 \Pi_X\bigr) \; (1 - AA^*)\Psi(X) \notag\\
&\hspace{10pt} + \frac 14\fint_{Q_B}\dd X\int_{\Rbb^3} \dr \; \ov{AA^*\Psi(X)} \; |\Bbold\wedge r|^2 \alpha_*(r)^2 \;  \cos^2 \bigl(\frac r2 \Pi_X\bigr) \; AA^*\Psi(X). \label{DHS1:MainTerm_5}
\end{align}
In particular, we have the lower bound
\begin{align}
\frac 12 \langle A^* \Psi, [ U^* (1 - P) (1 + \pi_r^2) (1- P) U + U(1 - P) (1 + \pi_r^2)(1-P) U^* ] A^*\Psi\rangle 
%
\geq c\, \langle \Psi, \Pi^2\Psi\rangle . \label{DHS1:MainTerm_LowerBound}
\end{align}
\end{lem}


\begin{bem}
Let us replace $\pi_r^2$ on the left side of \eqref{DHS1:MainTerm_5} by zero for the moment. In this case, the substitute of \eqref{DHS1:MainTerm_5} reads
\begin{align}
\frac 12 \langle A^*\Psi, [U^*(1 - P)U + U(1 - P)U^* ] A^*\Psi\rangle = \langle \Psi , AA^*(1 - AA^*) \Psi\rangle. \label{DHS1:Low_Energy_Operator}
\end{align}
It follows from Lemma~\ref{DHS1:AAstar_Positive} that the operator $AA^*(1 - AA^*)$ is bounded from below by $\Pi^2$ only for small values of $\Pi^2$, which is not enough for the proof of Proposition \ref{DHS1:First_Decomposition_Result}. This justifies the term ``low-energy operator'' for $1 - A^*A$, which we used earlier in the discussion below \eqref{DHS1:Def_Astar}. The additional factor $1 + \Pi^2$ in the first term on the right side of \eqref{DHS1:MainTerm_5} compensates for the problematic behavior of \eqref{DHS1:Low_Energy_Operator} for high energies. The expression on the right side of \eqref{DHS1:Low_Energy_Operator} also appears in \cite{Hainzl2017}. 
\end{bem}

Before we give the proof of Lemma~\ref{DHS1:MainTerm}, we prove two technical lemmas, which provide intertwining relations for various magnetic momentum operators with $U$ and linear combinations of $U$. A part of the relations in the first lemma can be found in \cite[Lemma~24]{Hainzl2017}.

\begin{lem}
\label{DHS1:CommutationI}
Let $p_r := -\i \nabla_r$, $\pi_r = p_r + \frac 12\Bbold \wedge r$ and $\tilde \pi_r$ and $\Pi_X$ be given by \eqref{DHS1:Magnetic_Momenta_COM}. With $U$ in \eqref{DHS1:U_definition}, we have the following intertwining relations:
\begin{align*}
\begin{split}
U \Pi_X U^* &= \Pi_X - \Bbold\wedge r,\phantom{\frac 12}  \\  U^* \Pi_X U &= \Pi_X + \Bbold\wedge r, \phantom{\frac 12}
\end{split} &
\begin{split}
U \pi_r U^* &= \tilde \pi_r + \frac{1}{2}\Pi_X, \\ U^* \pi_r U &= \tilde \pi_r - \frac{1}{2}\Pi_X, \phantom{\frac 12}
\end{split} &
\begin{split}
U \tilde \pi_r U^* &= p_r + \frac{1}{2}\Pi_X, \\ U^* \tilde \pi_r U &= p_r - \frac{1}{2}\Pi_X.
\end{split}
\end{align*}
\end{lem}

\begin{proof}
Let us denote $P_X := -\i \nabla_X$. We use the fact that $r\cdot P_X$ commutes with $r\cdot (\Bbold \wedge X)$ to see that
\begin{align}
U^* = \e^{\i \frac \Bbold 2 \cdot (X\wedge r)} \; \e^{\i \frac r2P_X} \label{DHS1:representation_Ustar}
\end{align}
holds. To prove the first intertwining relation with $\Pi_X$, we compute
\begin{align*}
\Pi_XU^* &= (P_X + \Bbold\wedge X)U^* =  \e^{\i \frac \Bbold 2 \cdot (X\wedge r)} \Bigl[ P_X - \frac 12 \Bbold \wedge r + \Bbold\wedge X\Bigr] \e^{\i \frac r2P_X} \\
&= U^* \Bigl[ P_X - \frac 12 \Bbold \wedge r + \Bbold \wedge \bigl(X -\frac r2\bigr)\Bigr] = U^* [ \Pi_X - \Bbold \wedge r].
\end{align*}
Here we used that $f(X)\, \e^{\i \frac r2P_X} = \e^{\i \frac r2 P_X}\, f(X-\frac r2)$. The second intertwining relation with $\Pi_X$ is obtained by replacing $r$ by $-r$. 

Next we consider the first intertwining relation with $\pi_r$ and compute
\begin{align*}
\pi_r U^* &= \bigl( p_r + \frac 12\Bbold \wedge r\bigr) U^* = \e^{\i \frac \Bbold 2 \cdot (X\wedge r)}\Bigl[ p_r + (-\i)\i \frac 12 \Bbold \wedge X + \frac 12 \Bbold \wedge r\Bigr] \e^{\i \frac r2P_X} \\
&= U^* \Bigl[ p_r + \frac{P_X}{2} + \frac 12 \Bbold \wedge \bigl( X - \frac r2\bigr) + \frac 12 \Bbold \wedge r\Bigr] = U^* \Bigl[ \tilde \pi_r + \frac{\Pi_X}{2}\Bigr].
\end{align*}
The remaining relations can be proved similarly and we skip the details.
\end{proof}

\begin{lem}
\label{DHS1:CommutationII}
\begin{enumerate}[(a)]
\item We have the following intertwining relations for $\Pi_X$:
\begin{align}
\Pi_X \cos\bigl( \frac r2 \Pi_X\bigr) &=  \cos\bigl( \frac r2 \Pi_X\bigr)\Pi_X - \i \sin\bigl( \frac r2 \Pi_X\bigr) \Bbold\wedge r, \label{DHS1:PiXcos} \\
\Pi_X \sin\bigl( \frac r2 \Pi_X\bigr) &= \sin\bigl( \frac r2 \Pi_X\bigr) \Pi_X + \i \cos\bigl( \frac r2 \Pi_X\bigr) \Bbold\wedge r. \label{DHS1:PiXsin}
\end{align}

\item The operators $p_r$, $\tilde \pi_r$ and $\pi_r$ intertwine according to
\begin{align}
\tilde \pi_r \cos\bigl( \frac r2 \Pi_X\bigr) &= \cos\bigl( \frac r2 \Pi_X\bigr) p_r + \i \sin\bigl( \frac r2 \Pi_X\bigr) \frac{\Pi_X}{2}, \label{DHS1:pirtilde_cos_pr}\\
\tilde \pi_r \cos\bigl( \frac r2 \Pi_X\bigr) &= \cos\bigl( \frac r2 \Pi_X\bigr)\pi_r + \i \frac{\Pi_X}{2} \sin\bigl( \frac r2 \Pi_X\bigr), \label{DHS1:pirtilde_cos_pir}
\end{align}
and
\begin{align}
\tilde \pi_r \sin\bigl( \frac r2 \Pi_X\bigr)  &= \sin\bigl( \frac r2 \Pi_X\bigr) p_r - \i \cos\bigl( \frac r2 \Pi_X\bigr) \frac{\Pi_X}{2}, \label{DHS1:pirtilde_sin_pr}\\
\tilde \pi_r \sin\bigl( \frac r2 \Pi_X\bigr) &= \sin\bigl( \frac r2 \Pi_X\bigr) \pi_r - \i \frac{\Pi_X}{2} \cos\bigl( \frac r2 \Pi_X\bigr) \label{DHS1:pirtilde_sin_pir}
\end{align}
\end{enumerate}
\end{lem}

It will be useful in the proof of Lemma \ref{DHS1:MainTerm} to have displayed both, \eqref{DHS1:pirtilde_cos_pr} and \eqref{DHS1:pirtilde_cos_pir} as well as \eqref{DHS1:pirtilde_sin_pr} and \eqref{DHS1:pirtilde_sin_pir}, even though they follow trivially from each other and \eqref{DHS1:PiXcos} or \eqref{DHS1:PiXsin}.

\begin{proof}
The proof is a direct consequence of the representations
\begin{align}
	\cos\bigl( \frac r2 \Pi_X\bigr) &= \frac 12 (U^* + U), & \sin\bigl( \frac r2 \Pi_X\bigr) &= \frac 1{2\i} (U^*-U),
\end{align}
and the intertwining relations in Lemma~\ref{DHS1:CommutationII}. We omit the details.
\end{proof}

\begin{proof}[Proof of Lemma \ref{DHS1:MainTerm}]
The proof is a tedious computation that is based on the intertwining relations in Lemma \ref{DHS1:CommutationII}. 
We start by defining
\begin{align}
\Tcal_1 &:= U^* \pi_r^2 U + U\pi_r^2 U^* = 2\, \tilde \pi_r^2 + \frac 12 \, \Pi_X^2, & \Tcal_2 &:= U^*P\pi_r^2PU + UP\pi_r^2PU^*, \notag \\
\Tcal_3 &:= U^* P\pi_r^2U + UP\pi_r^2U^*, & \Tcal_4 &:= U^* \pi_r^2 PU + U\pi_r^2PU^*. \label{DHS1:MainTerm_6}
\end{align}
Then, \eqref{DHS1:MainTerm_5} can be written as
\begin{align}
\langle A^* \Psi, [ U^* (1 - P) (1 + \pi_r^2) (1- P) U + U(1 - P) (1 + \pi_r^2)(1-P) U^* ] A^*\Psi\rangle  & = \notag\\
&\hspace{-330pt}= 2\langle A^*\Psi, (1 - A^*A)A^*\Psi\rangle \notag \\
&\hspace{-300pt} + \langle A^*\Psi, \Tcal_1 A^*\Psi\rangle + \langle A^*\Psi, \Tcal_2 A^*\Psi\rangle - \langle A^*\Psi, \Tcal_3 A^*\Psi\rangle - \langle A^*\Psi, \Tcal_4 A^*\Psi\rangle. \label{DHS1:Tcal_first_line}
\end{align}
The first term on the right side equals twice the term in \eqref{DHS1:Low_Energy_Operator}, which is in its final form. 

We start by computing the $\Pi_X^2$ term in $\Tcal_1$, which reads
\begin{align*}
\langle A^* \Psi , \Pi_X^2 A^* \Psi \rangle = \fint_{Q_B} \dd X \int_{\Rbb^3} \dd r \; \ov{\Psi(X)} \alpha^*(r) \; \cos\bigl( \frac r2 \Pi_X\bigr) \Pi_X^2 \cos\bigl( \frac r2 \Pi_X\bigr) \; \alpha_*(r) \Psi(X).
\end{align*}
Our goal is to move $\Pi_X^2$ to the right. To that end, we apply \eqref{DHS1:PiXcos} twice and obtain
\begin{align*}
\Pi_X^2 \cos\bigl( \frac r2 \Pi_X\bigr) = \cos\bigl( \frac r2 \Pi_X\bigr)\Pi_X^2 - \i \sin\bigl( \frac r2 \Pi_X\bigr) \Pi_X\Bbold \wedge r - \i \Pi_X \sin\bigl( \frac r2 \Pi_X\bigr)\Bbold \wedge r.
\end{align*}
We multiply this from the left with $\cos( \frac r2 \Pi_X)$, use \eqref{DHS1:PiXcos} to commute $\Pi_X$ to the left in the last term, and find
\begin{align}
\cos\bigl( \frac r2 \Pi_X\bigr) \Pi_X^2\cos\bigl( \frac r2 \Pi_X\bigr) &= \cos^2\bigl( \frac r2 \Pi_X\bigr) \Pi_X^2 + \sin^2\bigl( \frac r2 \Pi_X\bigr) |\Bbold\wedge r|^2 \notag\\
&\hspace{-60pt}- \i \bigl[ \Pi_X\cos\bigl( \frac r2 \Pi_X\bigr)\sin\bigl( \frac r2 \Pi_X\bigr) + \cos\bigl( \frac r2 \Pi_X\bigr)\sin\bigl( \frac r2 \Pi_X\bigr) \Pi_X\bigr]\Bbold\wedge r. \label{DHS1:PiX-term_1}
\end{align}
The operator $| \Bbold \wedge r |^2$ in the second term on the right side commutes with $\sin^2( \frac r2 \Pi_X)$. The operator in square brackets is self-adjoint and commutes with $\Bbold \wedge r$. When we add \eqref{DHS1:PiX-term_1} and its own adjoint, we obtain
\begin{align}
\cos\bigl( \frac r2 \Pi_X\bigr) \Pi_X^2\cos\bigl( \frac r2 \Pi_X\bigr) & \notag\\
&\hspace{-50pt}= \frac 12 \cos^2\bigl( \frac r2 \Pi_X\bigr) \Pi_X^2 + \frac 12 \, \Pi_X^2 \cos^2\bigl( \frac r2 \Pi_X\bigr) + \sin^2\bigl( \frac r2 \Pi_X\bigr) |\Bbold\wedge r|^2. \label{DHS1:PiX-term_2}
\end{align}
We evaluate \eqref{DHS1:PiX-term_2} in the inner product with $\alpha_*\Phi$ and $\alpha_*\Psi$ on the left and right side, respectively, use the fact that $AA^*$ commutes with $\Pi^2$, see Lemma \ref{DHS1:AAstar_Positive}, and obtain
\begin{align}
\langle A^*\Phi, \Pi_X^2A^*\Psi\rangle &= \langle \Phi, AA^* \Pi^2\Psi\rangle \notag\\
&\hspace{10pt} + \fint_{Q_B}\dd X\int_{\Rbb^3} \dd r \; \ov{\Phi(X)} \; |\Bbold\wedge r|^2  \alpha_*(r)^2 \; \sin^2\bigl( \frac r2 \Pi_X\bigr)  \; \Psi(X). \label{DHS1:PiX-term}
\end{align}
When we choose $\Phi = \Psi$ we obtain the result for the term proportional to $\Pi_X^2$ in $\Tcal_1$.

Next, we investigate the term proportional to $\tilde\pi_r^2$ in $\Tcal_1$. We use \eqref{DHS1:pirtilde_cos_pir} to move the operators $\tilde \pi_r$ from the middle to the outer positions and find
\begin{align}
\langle A^*\Psi, \tilde \pi_r^2A^*\Psi\rangle &= \fint_{Q_B} \dd X\int_{\Rbb^3} \dd r\; \ov{\Psi(X)} \alpha_*(r)\left[ \pi_r\cos\bigl( \frac r2 \Pi_X\bigr) -\i \sin\bigl( \frac r2 \Pi_X\bigr) \frac{\Pi_X}{2}\right] \notag\\
&\hspace{80pt} \times \left[ \cos\bigl( \frac r2 \Pi_X\bigr)\pi_r + \i \frac{\Pi_X}{2}\sin\bigl( \frac r2 \Pi_X\bigr)\right]\alpha_*(r) \Psi(X). \label{DHS1:tildepir_eq1}
\end{align}
We multiply out the brackets and obtain four terms. The terms proportional to $\cos^2$ and $\sin^2$ read
\begin{align}
&\fint_{Q_B} \dd X \int_{\Rbb^3}\dd r\; \ov{\Psi(X)}\;  |\pi_r\alpha_*(r)|^2\;  \cos^2\bigl( \frac r2 \Pi_X\bigr) \; \Psi(X) \notag\\
&\hspace{20pt} + \fint_{Q_B}\dd X\int_{\Rbb^3} \dd r \; \ov{\Psi(X)} \; \alpha_*(r)\;  \sin\bigl( \frac r2 \Pi_X\bigr) \frac{\Pi_X^2}{4} \sin\bigl( \frac r2 \Pi_X\bigr) \; \alpha_*(r) \; \Psi(X).\label{DHS1:tildepir_eq4}
\end{align}
For the moment the second line remains untouched. It is going to be canceled by a term in \eqref{DHS1:tildepir_eq5} below. The term in the first line equals
\begin{align}
%
&\fint_{Q_B} \dd X \int_{\Rbb^3}\dd r\; \ov{\Psi(X)} \; |\nabla\alpha_*(r)|^2 \; \cos^2\bigl( \frac r2 \Pi_X\bigr) \; \Psi(X) \notag \\ 
&\hspace{50pt} + \frac 14\fint_{Q_B} \dd X \int_{\Rbb^3}\dd r\; \ov{\Psi(X)} \; |\Bbold \wedge r|^2 \alpha_*(r)^2 \; \cos^2\bigl( \frac r2 \Pi_X\bigr) \; \Psi(X). \label{DHS1:tildepir_eq2}
\end{align}
To obtain this result, we used $(\nabla \alpha_*)(r) \cdot \Bbold \wedge r =0$, which holds because $\alpha_*$ is radial. This term is in its final form.

Now we have a closer look at the terms proportional to $\sin $ times $\cos$ in \eqref{DHS1:tildepir_eq1}. The operator inside the relevant quadratic form is given by
\begin{align}
\i \pi_r\cos\bigl( \frac r2 \Pi_X\bigr) \frac{\Pi_X}{2} \sin\bigl( \frac r2 \Pi_X\bigr) - \i \sin\bigl( \frac r2 \Pi_X\bigr)\frac{\Pi_X}{2} \cos\bigl( \frac r2 \Pi_X\bigr)\pi_r. \label{DHS1:MainTerm_1}
\end{align}
We intend to interchange $\sin( \frac r2 \Pi_X)$ and $\cos( \frac r2 \Pi_X)$ in the first term. To do this, we use \eqref{DHS1:PiXsin} to move $\Pi_X$ out of the center so that the first term equals
\begin{align*}
\i \pi_r\cos\bigl( \frac r2 \Pi_X\bigr) \sin\bigl( \frac r2 \Pi_X\bigr)\frac{\Pi_X}{2} -\frac 12\pi_r\cos^2\bigl( \frac r2 \Pi_X\bigr) \Bbold \wedge r.
\end{align*}
%
In the first term we may now commute the sine and the cosine and use \eqref{DHS1:PiXcos} and \eqref{DHS1:pirtilde_sin_pir} to bring $\pi_r$ and $\Pi_X$ in the center again. We also move $\pi_r$ into the center in the second term in \eqref{DHS1:MainTerm_1}. As a result, \eqref{DHS1:MainTerm_1} equals
\begin{align}
%
&\cos\bigl( \frac r2 \Pi_X\bigr) \frac{\Pi_X^2}{4}\cos\bigl( \frac r2 \Pi_X\bigr) - \sin\bigl( \frac r2 \Pi_X\bigr) \frac{\Pi_X^2}{4}\sin\bigl( \frac r2 \Pi_X\bigr) \notag \\
&\hspace{40pt}+ \i \cos\bigl( \frac r2 \Pi_X\bigr) \frac{\Pi_X}{4} \sin\bigl( \frac r2 \Pi_X\bigr) \Bbold \wedge r \notag \\ 
&\hspace{80pt}- \frac 12 \pi_r \cos^2\bigl( \frac r2 \Pi_X\bigr)\Bbold \wedge r - \frac 12 \sin\bigl( \frac r2 \Pi_X\bigr)\tilde \pi_r \sin\bigl( \frac r2 \Pi_X\bigr)\Bbold \wedge r. \label{DHS1:MainTerm_3}
\end{align}
We use \eqref{DHS1:pirtilde_sin_pir} to move $\tilde \pi_r$ to the left in the last term in \eqref{DHS1:MainTerm_3}. One of the terms we obtain in this way cancels the third term in \eqref{DHS1:MainTerm_3}. We also use $\cos( \frac r2 \Pi_X)^2 + \sin( \frac r2 \Pi_X)^2=1$ to rewrite the fourth term in \eqref{DHS1:MainTerm_3}. In combination, these considerations imply that the terms in \eqref{DHS1:MainTerm_3} equal
\begin{align}
\begin{split}
%
\cos\bigl( \frac r2 \Pi_X\bigr) \frac{\Pi_X^2}{4}\cos\bigl( \frac r2 \Pi_X\bigr) - \sin\bigl( \frac r2 \Pi_X\bigr)\frac{\Pi_X^2}{4}\sin\bigl( \frac r2 \Pi_X\bigr) - \frac 12 \pi_r\Bbold \wedge r. \end{split} \label{DHS1:tildepir_eq5}
\end{align}
The expectation of the second term with respect to $\alpha_*(r) \Psi(X)$ cancels the second term in \eqref{DHS1:tildepir_eq4}. We multiply the last term from the left and from the right with $\alpha_*(r)$, integrate over $r$ and find 
\begin{align}
\frac 12\int_{\Rbb^3} \dd r\; \alpha_*(r) \pi_r \Bbold \wedge r \alpha_*(r) = \frac 12 \int_{\Rbb^3} \ov{p_r\alpha_*(r)} \Bbold \wedge r\alpha_*(r) + \frac 14\int_{\Rbb^3} \dd r\; |\Bbold \wedge r|^2 \alpha_*(r)^2. \label{DHS1:Tcal3_eq2}
\end{align}
The first term on the right side vanishes because $\alpha_*$ is radial, see the remark below \eqref{DHS1:tildepir_eq4}. Let us summarize where we are. We combine \eqref{DHS1:tildepir_eq1}-\eqref{DHS1:Tcal3_eq2} to see that
\begin{align}
\langle A^*\Psi, \tilde \pi_r^2A^*\Psi\rangle &= \fint_{Q_B} \dd X \int_{\Rbb^3}\dd r\; \ov{\Psi(X)} \; |\nabla\alpha_*(r)|^2 \; \cos^2\bigl( \frac r2 \Pi_X\bigr) \; \Psi(X) \notag \\
&\hspace{-30pt} + \frac 14 \fint_{Q_B} \dd X \int_{\Rbb^3}\dd r\; \ov{\Psi(X)} \; |\Bbold \wedge r|^2 \alpha_*(r)^2 \; \bigl( \cos^2 \bigl( \frac r2 \Pi_X\bigr) -1 \bigr) \; \Psi(X) \notag \\
&\hspace{-30pt} + \frac 14 \fint_{Q_B} \dd X \int_{\Rbb^3}\dd r \; \ov{\alpha_*(r) \Psi(X)} \; \cos\bigl( \frac r2 \Pi_X\bigr) \Pi_X^2 \cos\bigl( \frac r2 \Pi_X\bigr) \; \alpha_*(r) \Psi(X). \label{DHS1:eq:1}
\end{align}
The term in the last line equals $\langle A^*\Psi, \Pi_X^2A^*\Psi\rangle$ and we use \eqref{DHS1:PiX-term} to rewrite it. This yields
\begin{align*}
\langle A^*\Psi, \tilde \pi_r^2A^*\Psi\rangle &= \frac 14 \langle \Psi, AA^* \Pi^2\Psi\rangle + \fint_{Q_B} \dd X \int_{\Rbb^3} \dd r\;  \ov{\Psi(X)}\;  |\nabla \alpha_*(r)|^2\;  \cos^2\bigl( \frac r2 \Pi_X\bigr)\; \Psi(X).
\end{align*}
In combination with \eqref{DHS1:MainTerm_6} and \eqref{DHS1:PiX-term}, this yields
\begin{align}
\langle A^* \Psi, \Tcal_1 A^*\Psi\rangle &= \langle \Psi , AA^*\Pi^2\Psi\rangle \notag \\
&\hspace{10pt}+ 2\fint_{Q_B} \dd X\int_{\Rbb^3} \dd r\; \ov{\Psi(X)}\;  |\nabla \alpha_*(r)|^2 \; \cos^2\bigl( \frac r2 \Pi_X\bigr)\;  \Psi(X) \notag \\
&\hspace{10pt} + \frac 12 \fint_{Q_B}\dd X\int_{\Rbb^3} \dd r \; \ov{\Psi(X)} \; |\Bbold\wedge r|^2 \alpha_*(r)^2\; \sin^2\bigl( \frac r2 \Pi_X\bigr)  \; \Psi(X)  \label{DHS1:Tcal1-Result}
\end{align}
and completes our computation of the term involving $\Tcal_1$.

A short computation shows that
\begin{equation}
\langle A^*\Psi , \Tcal_2A^*\Psi\rangle = 2\, \langle AA^*\Psi , AA^*\Psi\rangle \Bigl[ \Vert \nabla \alpha_*\Vert_2^2 + \frac 14 \int_{\Rbb^3} \dd r \; |\Bbold \wedge r|^2 \alpha_*(r)^2\Bigr]. \label{DHS1:Tcal2_result}
\end{equation}
It remains to compute the terms in \eqref{DHS1:Tcal_first_line} involving the operators $\Tcal_3$ and $\Tcal_4$, where $\Tcal_4^* = \Tcal_3$.

In the following we compute the term with $\Tcal_3$. A short computation, which uses the fact that $\alpha_*$ is radial, shows
\begin{align}
\langle A^* \Psi, \Tcal_3A^*\Psi\rangle &= \langle \alpha_* A A^*\Psi, \pi_r^2 (U^* + U) A^*\Psi \rangle \notag\\
&\hspace{-30pt}= 2\fint_{Q_B}\dd X\int_{\Rbb^3} \dd r \; \ov{AA^*\Psi(X)} \; \ov{p_r\alpha_*(r)} \; p_r\cos^2\bigl( \frac r2 \Pi_X\bigr) \; \alpha_*(r) \; \Psi(X) \notag\\
&\hspace{-10pt} + \frac 12 \fint_{Q_B} \dd X \int_{\Rbb^3} \dd r\; \ov{AA^*\Psi(X)} \; |\Bbold \wedge r|^2 \alpha_*(r)^2\;  \cos^2\bigl( \frac r2 \Pi_X\bigr) \; \Psi(X). \label{DHS1:Tcal3_eq3}
\end{align}
%
The second term on the right side is in its final form and will be canceled by a term below. We continue with the first term, use \eqref{DHS1:pirtilde_cos_pr} twice to commute $p_r$ with the squared cosine, as well as \eqref{DHS1:PiXcos} and \eqref{DHS1:PiXsin} to commute $\Pi_X$ to the center in the emerging terms, and find
\begin{align}
p_r\cos^2\bigl( \frac r2 \Pi_X\bigr) 
&= \cos^2\bigl( \frac r2 \Pi_X\bigr)p_r + \i \sin\bigl( \frac r2 \Pi_X\bigr) \Pi_X \cos\bigl( \frac r2 \Pi_X\bigr) - \frac 12 \Bbold \wedge r. \label{DHS1:Tcal3_eq1}
\end{align}
We note that the last term, when inserted back into \eqref{DHS1:Tcal3_eq3}, vanishes because $\alpha_*$ is radial. The first term is final and its quadratic form with $p_r \alpha_* AA^*\Psi$ and $\alpha_*\Psi$ reads
\begin{align}
2 \fint_{Q_B} \dd X\int_{\Rbb^3} \dd r \; \ov{AA^*\Psi(X)} \; |\nabla \alpha_*(r)|^2 \; \cos^2\bigl( \frac r2\Pi_X\bigr) \; \Psi(X). \label{DHS1:MainTerm_4}
\end{align}
Let us continue with the second term on the right side of \eqref{DHS1:Tcal3_eq1}. We multiply it with $p_r$ from the left and use \eqref{DHS1:pirtilde_cos_pr} and \eqref{DHS1:pirtilde_sin_pr} to commute $p_r$ to the right. In the two emerging terms we bring $\Pi_X$ to the center and obtain
\begin{align*}
p_r \sin\bigl( \frac r2 \Pi_X\bigr) \Pi_X\cos\bigl( \frac r2 \Pi_X\bigr) &= \sin\bigl( \frac r2 \Pi_X\bigr)\Pi_X\cos\bigl( \frac r2 \Pi_X\bigr) p_r + \i \sin\bigl( \frac r2 \Pi_X\bigr)\frac{\Pi_X^2}{2} \sin\bigl( \frac r2 \Pi_X\bigr) \\
&\hspace{145pt}- \i \cos\bigl( \frac r2 \Pi_X\bigr) \frac{\Pi_X^2}{2}\cos\bigl( \frac r2 \Pi_X\bigr).
\end{align*}
We plug the second term of \eqref{DHS1:Tcal3_eq1}, written in this form, back into \eqref{DHS1:Tcal3_eq3} and obtain
\begin{align}
2 \i \fint_{Q_B}\dd X\int_{\Rbb^3} \dd r\; \ov{AA^*\Psi(X)} \; \ov{p_r\alpha_*(r)} \; \sin\bigl( \frac r2 \Pi_X\bigr) \Pi_X  \cos\bigl( \frac r2 \Pi_X\bigr)\; \alpha_*(r)\; \Psi(X) \hspace{-350pt}& \notag\\
&= 2 \i \fint_{Q_B}\dd X\int_{\Rbb^3} \dd r\; \ov{AA^* \Psi(X)} \; \alpha_*(r) \; \sin\bigl( \frac r2 \Pi_X\bigr) \Pi_X \cos\bigl( \frac r2 \Pi_X\bigr) \; p_r\alpha_*(r)\; \Psi(X) \notag \\
&\hspace{20pt} + \fint_{Q_B}\dd X\int_{\Rbb^3} \dd r \; \ov{AA^*\Psi(X)} \;  \alpha_*(r) \; \cos\bigl( \frac r2 \Pi_X\bigr)\Pi_X^2 \cos\bigl( \frac r2 \Pi_X\bigr) \; \alpha_*(r)\; \Psi(X) \notag \\
&\hspace{20pt} - \fint_{Q_B}\dd X\int_{\Rbb^3} \dd r \; \ov{AA^*\Psi(X)} \; \alpha_*(r) \; \sin\bigl( \frac r2 \Pi_X\bigr)\Pi_X^2 \sin\bigl( \frac r2 \Pi_X\bigr) \; \alpha_*(r) \; \Psi(X).\label{DHS1:MainTerm_7}
\end{align}
Notice that the first term on the right side equals $(-1)$ times the term on the left side. Thus, the left side equals $\frac 12$ times the third line plus the fourth line. To compute the third line of \eqref{DHS1:MainTerm_7} we use \eqref{DHS1:PiX-term} with the choice $\Phi = AA^*\Psi$. A short computation shows that \eqref{DHS1:PiX-term_2} holds equally with $\cos$ and $\sin$ interchanged. Accordingly, $(-1)$ times the fourth line of \eqref{DHS1:MainTerm_7} equals
\begin{equation*}
	\langle AA^*\Psi, (1 -AA^*) \Pi^2\Psi\rangle + \fint_{Q_B}\dd X\int_{\Rbb^3} \dd r \; \ov{AA^*\Psi(X)} \; |\Bbold\wedge r|^2  \alpha_*(r)^2 \; \cos^2\bigl( \frac r2 \Pi_X\bigr)  \; \Psi(X). 
\end{equation*}
In combination, these considerations imply that the left side of \eqref{DHS1:MainTerm_7} is given by
\begin{align}
&\frac 12 \langle AA^*\Psi, AA^* \Pi^2\Psi\rangle - \frac 12 \langle AA^*\Psi, (1 -AA^*) \Pi^2\Psi\rangle \notag \\
&\hspace{20pt}+ \frac 12 \fint_{Q_B}\dd X\int_{\Rbb^3} \dd r \; \ov{AA^*\Psi(X)} \; |\Bbold\wedge r|^2  \alpha_*(r)^2 \; \sin^2\bigl( \frac r2 \Pi_X\bigr)  \; \Psi(X) \notag\\
&\hspace{20pt} - \frac 12 \fint_{Q_B}\dd X\int_{\Rbb^3} \dd r \; \ov{AA^*\Psi(X)} \; |\Bbold\wedge r|^2  \alpha_*(r)^2 \; \cos^2\bigl( \frac r2 \Pi_X\bigr)  \; \Psi(X). \label{DHS1:Tcal3_eq4}
\end{align}
We note that the third term in \eqref{DHS1:Tcal3_eq4} cancels the second term in \eqref{DHS1:Tcal3_eq3}. Adding all this to \eqref{DHS1:MainTerm_4}, we find
\begin{align}
\langle A^*\Psi, \Tcal_3 A^*\Psi\rangle &= \frac 12 \langle \Psi, AA^* AA^* \Pi^2 \Psi\rangle - \frac 12 \langle \Psi, AA^* (1 - AA^*) \Pi^2 \Psi\rangle  \notag\\
&\hspace{-10pt} + 2 \fint_{Q_B} \dd X\int_{\Rbb^3} \dd r\; \ov{AA^*\Psi(X)} \; |\nabla \alpha_*(r)|^2 \; \cos^2\bigl( \frac r2 \Pi_X\bigr) \; \Psi(X) \notag \\ 
&\hspace{-10pt}+ \frac 12 \fint_{Q_B} \dd X \int_{\Rbb^3} \dd r \; \ov{AA^*\Psi(X)} \; |\Bbold \wedge r|^2 \alpha_*(r)^2 \;  \sin^2\bigl( \frac r2 \Pi_X\bigr)\; \Psi(X). \label{DHS1:Tcal3_result}
\end{align}
The corresponding result for $\langle A^*\Psi, \Tcal_4 A^*\Psi\rangle$ is obtained by taking the complex conjugate of the right side of \eqref{DHS1:Tcal3_result}, which amounts to interchanging the roles of $AA^*\Psi$ and $\Psi$ in the last two lines.

We are now prepared to collect our results and to provide the final formula for \eqref{DHS1:Tcal_first_line}. We need to collect the terms in \eqref{DHS1:Tcal1-Result}, \eqref{DHS1:Tcal2_result}, \eqref{DHS1:Tcal3_result} and the complex conjugate of \eqref{DHS1:Tcal3_result}. The terms involving $|\nabla \alpha_*|^2$ read
\begin{align}
&2\fint_{Q_B} \dd X\int_{\Rbb^3} \dd r\; \ov{\Psi(X)}\;  |\nabla \alpha_*(r)|^2 \; \cos^2\bigl( \frac r2 \Pi_X\bigr) \; \Psi(X) + 2\langle AA^*\Psi, AA^*\Psi\rangle \Vert \nabla \alpha_*\Vert_2^2 \notag \\
&\hspace{50pt}-2 \fint_{Q_B} \dd X\int_{\Rbb^3} \dd r\; \ov{AA^*\Psi(X)} \; |\nabla \alpha_*(r)|^2 \; \cos^2\bigl( \frac r2 \Pi_X\bigr) \; \Psi(X) \notag \\
&\hspace{50pt}- 2 \fint_{Q_B} \dd X\int_{\Rbb^3} \dd r\; \ov{\Psi(X)} \; |\nabla \alpha_*(r)|^2 \; \cos^2\bigl( \frac r2 \Pi_X\bigr) \;  AA^*\Psi(X). \label{DHS1:nablaterms}
\end{align}
When we insert the factor $1 = \cos^2(\frac r2 \Pi_X) + \sin^2(\frac r2\Pi_X)$ in the second term, we obtain the final result for the terms proportional to $|\nabla \alpha_*|^2$.


The terms proportional to $\alpha_*^2$ with magnetic fields read
\begin{align*}
&\frac 12 \fint_{Q_B}\dd X\int_{\Rbb^3} \dd r \; \ov{\Psi(X)} \; |\Bbold\wedge r|^2 \alpha_*(r)^2\; \sin^2\bigl( \frac r2 \Pi_X\bigr)  \; \Psi(X) \\
&\hspace{30pt} + \frac 12\langle AA^*\Psi, AA^*\Psi\rangle \int_{\Rbb^3} \dd r \; |\Bbold \wedge r|^2 \alpha_*(r)^2 \\
&\hspace{30pt}-  \frac 12 \fint_{Q_B} \dd X \int_{\Rbb^3} \dd r \; \ov{AA^*\Psi(X)} \; |\Bbold \wedge r|^2 \alpha_*(r)^2 \; \sin^2\bigl( \frac r2 \Pi_X\bigr) \; \Psi(X)  \\
&\hspace{30pt}-  \frac 12 \fint_{Q_B} \dd X \int_{\Rbb^3} \dd r \; \ov{\Psi(X)} \; |\Bbold \wedge r|^2\alpha_*(r)^2\; \sin^2\bigl( \frac r2 \Pi_X\bigr) \; AA^*\Psi(X).
\end{align*}
When we insert $1 = \cos^2(\frac r2 \Pi_X) + \sin^2(\frac r2\Pi_X)$ in the second term we can bring these terms in the claimed form.


Finally, we collect the terms proportional to $\alpha_*^2$ but without magnetic field. Taking into account the first term in \eqref{DHS1:Tcal_first_line}, we find
\begin{align*}
&2\langle \Psi, AA^*(1 - AA^*)\Psi\rangle + \langle \Psi, AA^*\Pi^2\Psi\rangle + \langle \Psi, AA^*(1 - AA^*)\Pi^2\Psi\rangle - \langle \Psi, AA^* AA^* \Pi^2 \Psi\rangle \\
&\hspace{50pt}= 2\langle \Psi, AA^*(1 - AA^*)(1 + \Pi^2)\Psi\rangle.
\end{align*}
To obtain the result, we used that the terms coming from $\Tcal_3$ and $\Tcal_4$ are actually the same because $AA^*$ and $1 - AA^*$ commute with $\Pi^2$, see Lemma \ref{DHS1:AAstar_Positive}. This proves \eqref{DHS1:MainTerm_5} and the lower bound \eqref{DHS1:MainTerm_LowerBound} is implied by the operator bounds in Lemma \ref{DHS1:AAstar_Positive}.
%
\end{proof}

\subsubsection{Step two -- estimating the cross terms}

In the second step of the proof of Proposition \ref{DHS1:First_Decomposition_Result} we estimate the cross terms that we obtain when the decomposition in \eqref{DHS1:First_Decomposition_Result_Decomp} with $\Psi$ and $\xi_0$ in \eqref{DHS1:Def_Psixi} is inserted into the left side of \eqref{DHS1:First_Decomposition_Result_Assumption}. 

%
\begin{lem}
\label{DHS1:Decomp_Low_Momenta_Crossterms}
Given $D_0, D_1 \geq 0$, there is $B_0>0$ with the following properties. If, for some $0< B\leq B_0$, the wave function $\alpha\in \Lsymm$ satisfies
\begin{align*}
\frac 12 \langle \alpha , [U^* (1 - P)U + U(1 - P)U^* ]\alpha \rangle \leq D_0 B\, \Vert \alpha \Vert_2^2 + D_1B^2,
\end{align*}
then $\Psi$ and $\xi_0$ in \eqref{DHS1:Def_Psixi} satisfy the estimates
\begin{align}
	\Vert \alpha \Vert_2^2 \leq C \bigl( \Vert \Psi \Vert_2^2 + D_1 B^2 \bigr)
	\label{DHS1:eq:new1}
\end{align}
and
\begin{align}
\langle \Psi, AA^*(1 - AA^*)\Psi\rangle + \Vert \xi_0\Vert_2^2 \leq C \bigl( B  \Vert \Psi \Vert_2^2 +D_1 B^2\bigr). \label{DHS1:Estimate_Low_Momenta}
\end{align}
Furthermore, for any $\eta >0$ we have
\begin{align}
|\langle \xi_0, [U(1-P)(1+\pi_r^2)(1-P)U^* + U^* (1-P)(1+\pi_r^2)(1-P)U] A^*\Psi\rangle| & \notag \\
&\hspace{-180pt}\leq \eta \, \Vert \Pi\Psi\Vert_2^2 + C \left(1+ \eta^{-1} \right)\, \bigl( B \Vert \Psi \Vert_2^2 + D_1B^2\bigr). \label{DHS1:Tcal_CrossTerms} 
\end{align}
\end{lem}

\begin{proof}
We start by noting that $A\xi_0 =0$ implies $\langle \xi_0, A^*\Psi\rangle =0$, and hence
\begin{align}
\Vert \alpha\Vert_2^2 = \Vert A^*\Psi\Vert_2^2 + \Vert \xi_0\Vert_2^2 \leq \Vert \Psi\Vert_2^2 + \Vert \xi_0\Vert_2^2. \label{DHS1:Decomp_Low_Momenta_1}
\end{align}
We use $\alpha \in \Lsymm$ and $A(1 - A^*A)A^* = AA^*(1 - AA^*)$ to see that
\begin{align}
D_0 B \Vert \alpha\Vert_2^2 + D_1 B^2 &\geq \frac 12 \langle \alpha, [U^* (1 - P)U + U(1 - P)U^* ]\alpha \rangle = \langle \alpha, (1 - A^*A) \alpha\rangle \notag\\
&= \langle A^*\Psi, (1 - A^*A)A^*\Psi\rangle + \langle \xi_0, (1- A^*A) A^*\Psi\rangle + \langle A^*\Psi, (1- A^*A)\xi_0\rangle \notag \\
&\hspace{230pt} + \langle \xi_0 , (1- A^*A)\xi_0\rangle \notag \\
&= \langle \Psi, AA^*(1 - AA^*) \Psi\rangle + \Vert \xi_0\Vert_2^2. \label{DHS1:Decomp_Low_Momenta_9}
\end{align}
From Lemma~\ref{DHS1:AAstar_Positive} we know that the first term on the right side is nonnegative and hence
\begin{equation*}
	\Vert \xi_0\Vert_2^2 \leq D_0 B \Vert \alpha\Vert_2^2 + D_1 B^2.
\end{equation*}
Together with \eqref{DHS1:Decomp_Low_Momenta_1}, this also proves $(1 - D_0B)\Vert \alpha\Vert_2^2 \leq \Vert \Psi\Vert_2^2 + D_1 B^2$, that is, \eqref{DHS1:eq:new1}. Finally, \eqref{DHS1:eq:new1} and \eqref{DHS1:Decomp_Low_Momenta_9} prove \eqref{DHS1:Estimate_Low_Momenta}.

Next we prove \eqref{DHS1:Tcal_CrossTerms}. Let us define 
\begin{align}
\Tcal := U^* (1 - P)(1 + \pi_r^2)(1 - P)U + U(1 - P)(1 + \pi_r^2)(1 - P)U^* \label{DHS1:Tcal_Op_Def}
\end{align}
and consider $\langle \xi_0, \Tcal A^*\Psi\rangle$. We note that $A\xi_0 =0$ implies $PU \xi_0 = 0 = PU^*\xi_0$, where the projection $P$ is understood to act on the relative coordinate. In combination with \eqref{DHS1:MainTerm_6} this allows us to see that
\begin{align}
\langle \xi_0, \Tcal A^*\Psi\rangle &= \Bigl\langle \xi_0, \Bigl[2\tilde \pi_r^2 + \frac{\Pi_X^2}{2} \Bigr] A^*\Psi\Bigr\rangle - \langle \xi_0, (U^* + U) \pi_r^2 \, \alpha_* AA^*\Psi\rangle \label{DHS1:Decomp_Low_Momenta_2}
\end{align}
holds. We use \eqref{DHS1:PiXcos} and \eqref{DHS1:PiXsin} to commute $\Pi_X^2$ in the first term on the right side of \eqref{DHS1:Decomp_Low_Momenta_2} to the right and find
\begin{align}
\frac 12\langle \xi_0, \Pi_X^2 A^*\Psi\rangle &= \frac 12\langle \xi_0, A^*\Pi_X^2\Psi\rangle \notag \\
&\hspace{10pt}- \i \fint_{Q_B} \dd X\int_{\Rbb^3} \dd r\; \ov{\xi_0(X,r)} \; \sin\bigl( \frac r2 \Pi_X\bigr) \; \Bbold \wedge r\; \alpha_*(r) \; \Pi_X \Psi(X) \notag \\
&\hspace{10pt}+ \frac 12\fint_{Q_B} \dd X \int_{\Rbb^3} \dd r \; \ov{\xi_0(X,r)}  \; \cos\bigl( \frac r2 \Pi_X\bigr) \;  |\Bbold\wedge r|^2 \alpha_*(r) \; \Psi(X). \label{DHS1:Decomp_Low_Momenta_4}
\end{align}
The first term on the right side vanishes because $A\xi_0 =0$. Similarly, we apply \eqref{DHS1:pirtilde_cos_pr} and \eqref{DHS1:pirtilde_sin_pr} to commute $\tilde \pi_r^2$ in the first term in \eqref{DHS1:Decomp_Low_Momenta_2} to the right and find
\begin{align}
2\langle \xi_0, \tilde \pi_r^2A^*\Psi\rangle &= 2 \fint_{Q_B} \dd X \int_{\Rbb^3} \dd r\; \ov{\xi_0(X,r)} \; \cos\bigl( \frac r2 \Pi_X\bigr) \; p^2\alpha_*(r) \; \Psi(X) \notag \\ 
&\hspace{20pt} + 2\i \fint_{Q_B} \dd X \int_{\Rbb^3} \dd r \, \ov{\xi_0(X,r)} \;  \sin\bigl( \frac r2 \Pi_X\bigr) \; p\alpha_*(r) \; \Pi_X\Psi(X). \label{DHS1:Decomp_Low_Momenta_6}
\end{align}
When we combine $\pi_r^2 \alpha_*(r) = p_r^2\alpha_*(r) + \frac 14|\Bbold \wedge r|^2\alpha_*(r)$, which holds because $\alpha_*$ is radial, \eqref{DHS1:Decomp_Low_Momenta_2}, \eqref{DHS1:Decomp_Low_Momenta_4} and \eqref{DHS1:Decomp_Low_Momenta_6}, we obtain
\begin{align*}
\langle \xi_0, \Tcal A^*\Psi\rangle &= 2 \fint_{Q_B} \dd X \int_{\Rbb^3} \dd r\; \ov{\xi_0(X,r)} \; \cos\bigl( \frac r2 \Pi_X\bigr) \; \pi_r^2\alpha_*(r) \; (1 - AA^*)\Psi(X) \\
&\hspace{30pt}+ 2\i \fint_{Q_B}\dd X\int_{\Rbb^3} \dd r\; \ov{\xi_0(X,r)} \; \sin\bigl( \frac r2 \Pi_X\bigr) \; \bigl[ p - \frac 12\Bbold \wedge r\bigr] \alpha_*(r) \; \Pi_X\Psi(X).
\end{align*}
Using Cauchy-Schwarz, we bound the absolute value of this by
\begin{align}
|\langle \xi_0, \Tcal A^*\Psi\rangle| \leq 2\Vert \xi_0\Vert_2 \; \bigl[ \Vert \pi_r^2\alpha_*\Vert_2 \; \Vert (1 - AA^*)\Psi\Vert_2 + \bigl(\Vert \nabla \alpha_*\Vert_2 + B \Vert \, | \cdot  |\alpha_*\Vert_2\bigr) \Vert \Pi\Psi\Vert_2\bigr], \label{DHS1:Decomp_Low_Momenta_7}
\end{align}
and with the decay properties of $\alpha_*$ in \eqref{DHS1:Decay_of_alphastar} we see that the norms of $\alpha_*$ on the right side are bounded uniformly in $0 \leq B \leq B_0$. Moreover, Lemma~\ref{DHS1:AAstar_Positive} and \eqref{DHS1:Estimate_Low_Momenta} imply that there is a constant $c>0$ such that
\begin{equation}
	\Vert (1 - AA^*)\Psi\Vert_2^2 \leq \langle \Psi, (1-AA^*) \Psi \rangle \leq \frac{1}{c} \langle \Psi, A A^* (1-AA^*) \Psi \rangle \leq C \bigl( B  \Vert \Psi \Vert_2^2 +D_1 B^2\bigr).
\end{equation}
For $\eta >0$ we thus obtain
\begin{align}
|\langle \xi_0, \Tcal A^*\Psi\rangle| \leq C \bigl[ \eta\,  \Vert \Pi\Psi\Vert_2^2 + \eta^{-1} \, \Vert \xi_0 \Vert_2^2 + \bigl( B \Vert \Psi \Vert_2^2 +D_1 B^2\bigr) \bigr] \label{DHS1:Decomp_Low_Momenta_8}
\end{align}
and an application of \eqref{DHS1:Estimate_Low_Momenta} proves the claim.
\end{proof}

\subsubsection{Proof of Proposition \ref{DHS1:First_Decomposition_Result}}

We recall the decomposition $\alpha = A^*\Psi + \xi_0$ with $\Psi$ and $\xi_0$ in \eqref{DHS1:Def_Psixi} as well as $\Tcal$ in \eqref{DHS1:Tcal_Op_Def}. From \eqref{DHS1:First_Decomposition_Result_Assumption} and \eqref{DHS1:eq:new1} we know that
\begin{align}
C \bigl( B \Vert\Psi\Vert_2^2 + D_1 B^2 \bigr)  \geq \langle A^*\Psi, \Tcal A^*\Psi\rangle + 2\Re\langle \xi_0, \Tcal A^*\Psi\rangle + \langle \xi_0, \Tcal\xi_0\rangle . \label{DHS1:First_Decomposition_Result_1}
\end{align}
With the help of Lemma \ref{DHS1:CommutationI}, the identities $PU\xi_0 = 0 = PU^*\xi_0$ imply
\begin{align}
\langle \xi_0, \Tcal \xi_0 \rangle = \Bigl\langle \xi_0, \bigl(2 + \frac{\Pi_X^2}{2} + 2\tilde \pi_r^2\bigr) \xi_0\Bigr\rangle \geq \frac 12 \, \Vert \xi_0\Vert_\Hsymm^2. \label{DHS1:First_Decomposition_Result_2}
\end{align}
Lemma \ref{DHS1:AAstar_Positive} guarantees the existence of a constant $\rho>0$ such that
\begin{align*}
AA^*(1 - AA^*)(1 + \Pi^2) \geq \rho\; \Pi^2.
\end{align*}
Therefore, \eqref{DHS1:MainTerm_LowerBound} implies 
\begin{align}
\langle A^* \Psi, \Tcal A^*\Psi\rangle \geq 2\, \langle \Psi, AA^*(1 - AA^*) (1 + \Pi^2)\Psi\rangle \geq 2\rho\, \langle \Psi, \Pi^2\Psi\rangle. \label{DHS1:First_Decomposition_Result_3}
\end{align}

To estimate the second term on the right side of \eqref{DHS1:First_Decomposition_Result_1}, we note that $\Tcal$ is bounded from below by $U(1 - P)U^* + U^*(1-P)U$. Therefore, we may apply Lemma \ref{DHS1:Decomp_Low_Momenta_Crossterms} with $\eta = \frac \rho2$ and find
\begin{align*}
2\Re \langle \xi_0, \Tcal A^*\Psi\rangle \geq -2\, |\langle \xi_0, \Tcal A^*\Psi\rangle| \geq - \rho \, \Vert \Pi\Psi\Vert_2^2 - C\bigl( B\Vert \Psi\Vert_2^2 + D_1 B^2\bigr).
\end{align*}
In combination with \eqref{DHS1:First_Decomposition_Result_1}, \eqref{DHS1:First_Decomposition_Result_2} and \eqref{DHS1:First_Decomposition_Result_3}, we thus obtain
\begin{align*}
C\bigl( B\Vert \Psi\Vert_2^2 + D_1B^2 \bigr)\geq \rho \, \Vert \Pi\Psi\Vert_2^2 + \frac 12 \, \Vert \xi_0\Vert_\Hsymm^2.
\end{align*}
This proves \eqref{DHS1:First_Decomposition_Result_Estimate}. 


\subsection{Uniform estimate on \texorpdfstring{$\Vert\Psi\Vert_2$}{Psi}}

Up to now we neglected the nonlinear term on the left side of \eqref{DHS1:Lower_Bound_A_2}. This term provides the inequality
\begin{align}
		\Tr\bigl[(\alpha^* \alpha)^2\bigr] \leq C \left( B \Vert \alpha\Vert_2^2 + B^2 \right). \label{DHS1:Lower_Bound_A_2b}
\end{align}
In this section we will take this term and \eqref{DHS1:Lower_Bound_A_2b} into account and show that it can be combined with Proposition~\ref{DHS1:First_Decomposition_Result} to obtain a bound for $\Vert \Psi\Vert_2$. This will afterwards allow us to prove Theorem~\ref{DHS1:Structure_of_almost_minimizers}.

\begin{lem}
\label{DHS1:Bound_on_psi}
Given $D_0\geq0$, there is $B_0>0$ such that for all $0 < B \leq B_0$ the following holds. If the wave function $\alpha\in \Lsymm$ obeys \eqref{DHS1:Lower_Bound_A_2} then $\Psi$ in \eqref{DHS1:Def_Psixi} satisfies
\begin{align}
\Vert \Psi\Vert_2^2 &\leq CB. \label{DHS1:Bound_on_psi_result}
\end{align}
\end{lem}

\begin{proof}
We recall the decomposition $\alpha = A^*\Psi + \xi_0$ with $\Psi$ and $\xi_0$ in \eqref{DHS1:Def_Psixi}. Eq.~\eqref{DHS1:Lower_Bound_A_2b} and an application of the triangle inequality imply
\begin{align}
C\bigl( B\Vert \Psi\Vert_2^2 + B^2\bigr)^{\nicefrac 14} \geq \Vert \alpha\Vert_4 \geq \Vert A^*\Psi\Vert_4 - \Vert \xi_0\Vert_4. \label{DHS1:Nonlinearity_8}
\end{align}
Thus, it suffices to prove an upper bound for $\Vert \xi_0\Vert_4$ and a lower bound for $\Vert A^*\Psi\Vert_4$. Our proof follows closely the proof of \cite[Eq. (5.48)]{Hainzl2012}. 

\emph{Step 1.} Let us start with the upper bound on $\Vert \xi_0\Vert_4$. We claim the estimate
\begin{align}
\Vert \xi_0\Vert_4 \leq C \bigl( B^{\nicefrac 14} \Vert \Psi\Vert_2^{\nicefrac 12} + B^{\nicefrac 18} \Vert \Psi\Vert_2 + B^{\nicefrac 12}\bigr).
\label{DHS1:Nonlinearity_1}
\end{align}
To see this, we first use Hölder's inequality to estimate $\Vert \xi_0\Vert_4^4 \leq \Vert \xi_0\Vert_2^2 \; \Vert \xi_0\Vert_\infty^2$. From Proposition~\ref{DHS1:First_Decomposition_Result} we know that $\Vert \xi_0\Vert_2^2 \leq C(B\Vert \Psi\Vert_2^2+ B^2)$, and it thus remains to prove a bound for $\Vert \xi_0\Vert_\infty$. We claim that for any $\nu > 3$
\begin{align}
\Vert \xi_0\Vert_\infty \leq  1 + C_\nu\, B^{-\nicefrac 14} \, \Vert (1 + |\cdot |)^\nu \alpha_*\Vert_{\nicefrac 65} \; \Vert \Psi \Vert_6, \label{DHS1:Claim_xi_infty}
\end{align}
where the right side is finite by the decay properties of $\alpha_*$ in \eqref{DHS1:Decay_of_alphastar}. To prove \eqref{DHS1:Claim_xi_infty}, we first note that \eqref{DHS1:gamma_alpha_fermionic_relation} implies $\Vert \alpha\Vert_\infty\leq 1$, and hence $\Vert \xi_0\Vert_\infty \leq 1 + \Vert A^*\Psi\Vert_\infty$. We apply Lemma \ref{DHS1:Schatten_estimate} (b) to $A^*\Psi$ and obtain \eqref{DHS1:Claim_xi_infty}. We also combine \eqref{DHS1:Magnetic_Sobolev} with Proposition \ref{DHS1:First_Decomposition_Result} and obtain $
\Vert \Psi\Vert_6 \leq C( \Vert \Psi\Vert_2 + B^{\nicefrac 12})$. In combination, these considerations imply \eqref{DHS1:Nonlinearity_1}. 

\emph{Step 2.} We claim that
\begin{align}
\Vert A^*\Psi\Vert_4^4 \geq \frac 1{16} \; \Vert \hat \alpha_* \Vert_4^4 \; \Vert \Psi\Vert_4^4 - C \bigl( B^{\nicefrac 18} \Vert \Psi\Vert_2 + B^{\nicefrac 58} \bigr)^4 \label{DHS1:Nonlinearity_5}
\end{align}
holds. To prove \eqref{DHS1:Nonlinearity_5}, we write $\Vert A^*\Psi\Vert_4^4 = \tr ((A^*\Psi)^* A^*\Psi)^2$. The fact that $\alpha_*$ is real-valued implies
\begin{align*}
\Vert A^*\Psi\Vert_4^4 &= \fint_{Q_B} \dd x \int_{\Rbb^3} \dd y \; \Bigl| \int_{\Rbb^3} \dd z \; \alpha_*(x - z)\ov{ \left[ \cos\bigl( \frac{x - z}{2}\Pi_{\frac{x+z}{2}}\bigr)  \Psi\bigl( \frac{x+z}{2}\bigr) \right] } \\
&\hspace{150pt} \times \alpha_*(z-y) \left[ \cos\bigl( \frac{z-y}{2} \Pi_{\frac{z+y}{2}}\bigr)  \Psi\bigl( \frac{z+y}{2}\bigr) \right] \Bigr|^2.
\end{align*}
We use $\cos(x) = 1 - 2\sin^2(\frac x2)$ twice and find
\begin{align}
\Vert A^*\Psi\Vert_4^4 &\geq \frac 14 \Tcal_* - C \, (\Tcal_1 + \Tcal_2), \label{DHS1:Nonlinearity_2}
\end{align}
where
\begin{align*}
\Tcal_* &:= \fint_{Q_B} \dx \int_{\Rbb^3} \dy \;\Bigl| \int_{\Rbb^3} \dd z \;  \alpha_*(x - z) \ov{\Psi\bigl(\frac{x+z}{2}\bigr)} \alpha_*(z - y) \Psi\bigl( \frac{z+y}{2}\bigr)\Bigr|^2
\end{align*}
and 
\begin{align}
\Tcal_1 &:= \fint_{Q_B} \dx \int_{\Rbb^3} \dy \;\Bigl| \int_{\Rbb^3} \dd z \;  \alpha_*(x - z) \ov{\Psi\bigl( \frac{x+z}{2}\bigr)} \notag\\
&\hspace{140pt} \times \alpha_*(z - y) \left[ \sin^2\bigl( \frac{z-y}{4} \Pi_{\frac{z+y}{2}}\bigr)  \Psi\bigl( \frac{z +y}{2}\bigr) \right] \Bigr|^2, \notag\\
\Tcal_2 &:= \fint_{Q_B} \dx \int_{\Rbb^3} \dy \; \Bigl| \int_{\Rbb^3} \dd z\; \alpha_*(x - z)\ov{ \left[ \sin^2\bigl( \frac{x -z}{4} \Pi_{\frac{x+z}{2}}\bigr) \Psi\bigl( \frac{x + z}{2}\bigr) \right]} \notag \\
&\hspace{140pt} \times\alpha_*(z - y) \left[ \cos\bigl( \frac{z - y}{2}\Pi_{\frac{z+y}{2}}\bigr)  \Psi\bigl( \frac{z+y}{2}\bigr) \right] \Bigr|^2. \label{DHS1:Nonlinearity_3}
\end{align}
In the following we derive a lower bound on $\Tcal_*$ and an upper bound on $\Tcal_1$ and $\Tcal_2$.

\emph{Lower bound on $\Tcal_*$.} We change variables $z\mapsto z + x$ and $y\mapsto y + x$ and afterwards replace $x$ by~$X$, which allows us to write
\begin{align}
\Tcal_* = \fint_{Q_B} \dd X \int_{\Rbb^3} \dy \; \Bigl| \int_{\Rbb^3} \dd z \;  \alpha_*(z) \ov{\Psi\bigl(X + \frac z2\bigr)} \alpha_*(z - y) \Psi\bigl(X + \frac{z+y}{2}\bigr)\Bigr|^2. \label{DHS1:Nonlinearity_9}
\end{align}
Next, we combine $\Psi(X + \frac z2) = \e^{\i \frac z2P_X} \Psi(X)$ and the identity $\e^{\i \frac r2P_X} = \e^{\i \frac \Bbold 2 \cdot (r\wedge X)}\e^{\i \frac r2\Pi_X}$ in \eqref{DHS1:representation_Ustar} to write $\Psi(X + \frac z2) = \e^{\i \frac \Bbold 2 \cdot (z\wedge X)} \e^{\i \frac z2\Pi_X} \Psi(X)$. 
We conclude that
\begin{align*}
\ov{\Psi\bigl( X + \frac z2\bigr)} \Psi\bigl(X + \frac{z+y}{2}\bigr) = \e^{\i \frac{\Bbold}{2} \cdot (y\wedge X)} \; \ov{\left[ \e^{\i \frac{z}{2}\Pi_X}\Psi(X) \right]} \; \left[ \e^{\i \frac{z+y}{2}\Pi_X} \Psi(X) \right],
\end{align*}
as well as
\begin{align*}
\Tcal_* = \fint_{Q_B} \dd X \int_{\Rbb^3} \dd y\; \Bigl| \int_{\Rbb^3} \dd z \; \alpha_*(-z) \ov{ \left[ \e^{\i \frac z2\Pi_X} \Psi(X) \right]} \alpha_*(z - y) \left[ \e^{\i \frac{z+y}{2}\Pi_X} \Psi(X) \right] \Bigr|^2.
\end{align*}
This also implies
\begin{align}
\Tcal_* \geq \frac 14\Tcal_*^* - C(\Tcal_*^{(1)} + \Tcal_*^{(2)})
\label{DHS1:eq:A1}
\end{align}
with
\begin{align*}
\Tcal_*^* := \fint_{Q_B}\dd X\int_{\Rbb^3}\dd y\; \Bigl| \int_{\Rbb^3} \dd z \; \alpha_*(z) \ov{\Psi(X)} \alpha_*(z - y) \Psi(X)\Bigr|^2 = \Vert \Psi\Vert_4^4 \; \Vert \alpha_* * \alpha_*\Vert_2^2
\end{align*}
and
\begin{align}
\Tcal_*^{(1)} &:= \fint_{Q_B} \dd X\int_{\Rbb^3} \dd y\; \Bigl| \int_{\Rbb^3} \dd z \; \alpha_*(z) \ov{\bigl[\e^{\i \frac{z}{2}\Pi_X}\Psi(X)\bigr]} \alpha_*(z - y) \bigl[\bigl( \e^{\i \frac{z+y}{2}\Pi_X} - 1\bigr) \Psi(X)\bigr]\Bigr|^2, \notag \\
\Tcal_*^{(2)} &:= \fint_{Q_B} \dd X\int_{\Rbb^3} \dd y\; \Bigl| \int_{\Rbb^3} \dd z \;\alpha_*(z)  \ov{\bigl[\bigl( \e^{\i \frac z2\Pi_X} - 1\bigr) \Psi(X)\bigr]} \alpha_*(z - y) \Psi(X)\Bigr|^2. \label{DHS1:Nonlinearity_10}
\end{align}

\emph{Upper bound on $\Tcal_*^{(1)}$ and $\Tcal_*^{(2)}$.} We start with $\Tcal_*^{(1)}$, expand the square and estimate
\begin{align}
&\Tcal_*^{(1)} \leq  \int_{\Rbb^3} \dd y  \int_{\Rbb^3} \dd z  \int_{\Rbb^3} \dd z' \;  | \alpha_*(z) \, \alpha_*(z') \, \alpha_*(z - y) \, \alpha_*(z' - y)| \label{DHS1:Nonlinearity_13} \\
& \times \fint_{Q_B} \dd X \; \Bigl| \bigl[ \e^{\i \frac z2\Pi_X} \Psi(X) \bigr]  \bigl[ \e^{\i \frac {z'}2\Pi_X} \Psi(X) \bigr]  \bigl[ \bigl( \e^{\i \frac{z+y}{2}\Pi_X}-1\bigr) \Psi(X) \bigr]  \bigl[ \bigl( \e^{\i \frac{z'+y}{2}\Pi_X}-1\bigr) \Psi(X) \bigr] \Bigr|. \nonumber
\end{align}
When we use Hölder's inequality, \eqref{DHS1:NTB-NtildeTB_3}, \eqref{DHS1:ZPiX_inequality}, and \eqref{DHS1:Magnetic_Sobolev}, we see that the integral in the second line can be bounded by
\begin{align}
\Vert \e^{\i \frac{z}{2}\Pi} \Psi\Vert_6^2 \; \Vert (\e^{\i \frac{z+y}{2}\Pi} - 1) \Psi\Vert_6 \; \Vert (\e^{\i \frac{z+y}{2}\Pi} - 1) \Psi\Vert_2 \leq C\; \bigl|\frac{z+y}{2}\bigr| \; B^{-\nicefrac 32} \; \Vert \Pi\Psi\Vert_2^4. \label{DHS1:Nonlinearity_11}
\end{align}
%
%
%
%
%
%
%
%
Proposition~\ref{DHS1:First_Decomposition_Result} provides us with a bound for $\Vert \Pi \Psi \Vert_2$. In combination with \eqref{DHS1:Nonlinearity_13}, \eqref{DHS1:Nonlinearity_11}, Young's inequality, and the bound $|z + y| \leq 2|z| + |z- y|$, this implies
\begin{align}
\Tcal_*^{(1)} &\leq C \, B^{-\nicefrac 32}\,  \left( B^2 \Vert \Psi \Vert_2^4 + D_1 B^4 \right) \int_{\Rbb^3} \dd y \; \Bigl| \int_{\Rbb^3} \dd z\;  |z+y|\; |  \alpha_*(z) \alpha_*(y-z)|\Bigr|^2 \nonumber \\
&\leq C \left( B^{-\nicefrac 12} \Vert \Psi \Vert_2^4 + D_1 B^{ \nicefrac 52 } \right) \, \Vert \alpha_*\Vert_{\nicefrac 43} \, \Vert \, |\cdot|\alpha_*\Vert_{\nicefrac 43}, \label{DHS1:Nonlinearity_12} 
\end{align}
where the right side is finite by \eqref{DHS1:Decay_of_alphastar}. Similarly, we see that $\Tcal_*^{(2)}$ is bounded by the right side of \eqref{DHS1:Nonlinearity_12}.

\emph{Upper bound on $\Tcal_1$ and $\Tcal_2$ in \eqref{DHS1:Nonlinearity_3}.} Bounds for $\Tcal_1$ and $\Tcal_2$ can be obtain along the same lines as the bound for $\Tcal_*^{(1)}$. We apply the same change of variables as above and use estimates similar to the ones in \eqref{DHS1:Nonlinearity_13}. In case of $\Tcal_1$, the bound in \eqref{DHS1:Nonlinearity_11} needs to be replaced by
\begin{align}
\Vert \Psi\Vert_6^2 \, \bigl\Vert \sin^2\bigl( \frac{z - y}{2}  \Pi\bigr) \Psi \bigr\Vert_6 \, \bigl\Vert \sin^2\bigl( \frac{z - y}{2}  \Pi\bigr) \Psi \bigr\Vert_2 \leq C \; \frac{|z-y|}{2} \; B^{-\nicefrac 32}\,  \Vert \Pi\Psi\Vert_2^4. \label{DHS1:Nonlinearity_14}
\end{align}
Here, we used $\sin^2(x) \leq |x|$ and the operator inequality in  \eqref{DHS1:ZPiX_inequality} to estimate the third factor. For the first and the second factor, we used
\begin{align*}
\sin^2\bigl( \frac{z-y}{4} \Pi\bigr) = - \frac{1}{4} \bigl( 2+ \e^{\i\frac{z-y}{2} \Pi} + \e^{-\i \frac{z-y}{2}\Pi}\bigr)
\end{align*}
and \eqref{DHS1:Magnetic_Sobolev} or \eqref{DHS1:NTB-NtildeTB_3}, respectively. A bound for $\Tcal_2$ can be proved analogously. The final estimate we obtain in this way reads
\begin{equation}
	\Tcal_1 + \Tcal_2 \leq C \left( B^{-\nicefrac 12} \Vert \Psi \Vert_2^4 + D_1 B^{ \nicefrac 52 } \right). 
	\label{DHS1:eq:A2}
\end{equation}
In combination with \eqref{DHS1:Nonlinearity_2}, \eqref{DHS1:eq:A1}, and \eqref{DHS1:Nonlinearity_12}, this proves \eqref{DHS1:Nonlinearity_5}.

\emph{Step 3.} We denote $c:= \frac 1{2} \Vert \hat\alpha_*\Vert_4$, insert \eqref{DHS1:Nonlinearity_1} and \eqref{DHS1:Nonlinearity_5} into \eqref{DHS1:Nonlinearity_8} and obtain
\begin{align}
C B^{\frac 14}\Vert \Psi\Vert_2^{\nicefrac 12}  \geq c \Vert \Psi\Vert_4 - CB^{\frac 18} \Vert \Psi\Vert_2 - CB^{\frac 12}, \label{DHS1:Nonlinearity_6}
\end{align}
which holds for $B$ small enough. For $\eta >0$ the left side is bounded from above by a constant times $\eta \Vert \Psi\Vert_2 + \eta^{-1} B^{\frac 12}$ and Hölder's inequality implies $\Vert \Psi\Vert_4 \geq \Vert \Psi\Vert_2$. 
Accordingly,
\begin{equation}
C \left( \eta \Vert \Psi\Vert_2 + \eta^{-1} B^{\frac 12} \right) \geq (c - CB^{\frac 18}) \Vert \Psi\Vert_2 - CB^{\frac 12}.
\label{DHS1:eq:A3}
\end{equation}
When we choose $\eta$ and $B$ in \eqref{DHS1:eq:A3} small enough, this proves the claim.
%
%
%
\end{proof}


\subsection{Proof of Theorem \ref{DHS1:Structure_of_almost_minimizers}}

The assumption \eqref{DHS1:Second_Decomposition_Gamma_Assumption} in Theorem~\ref{DHS1:Structure_of_almost_minimizers}, Corollary~\ref{DHS1:cor:lowerbound}, Proposition~\ref{DHS1:First_Decomposition_Result}, \ifthenelse{\equal\masterfile{Diss}}{as well as}{and} Lemma~\ref{DHS1:Bound_on_psi} imply the decomposition $\alpha = A^*\Psi + \xi_0$, where $\Psi$ and $\xi_0$ in \eqref{DHS1:Def_Psixi} obey
\begin{align}
\Vert \Psi\Vert_{\Hmag^1(Q_B)}^2 = B^{-1} \Vert \Psi\Vert_2^2 + B^{-2} \Vert \Pi\Psi\Vert_2^2 \leq C  \label{DHS1:Psi_estimate}
\end{align}
and
\begin{align}
\Vert \xi_0\Vert_{\Hsymm}^2 \leq C B^2 \bigl( \Vert \Psi\Vert_{\Hmag^1(Q_B)}^2 + D_1\bigr). \label{DHS1:xi0_estimate}
\end{align}
Define
\begin{align}
\xi &:= \xi_0 + \bigl( \cos\bigl( \frac r2 \Pi_X\bigr) - 1\bigr) \alpha_*(r) \Psi(X). \label{DHS1:xi_definition}
\end{align}
Then, \eqref{DHS1:Second_Decomposition_alpha_equation} holds and we claim that $\xi$ satisfies \eqref{DHS1:Second_Decomposition_Psi_xi_estimate}. To prove this, we estimate the second term in \eqref{DHS1:xi_definition}. We use $1 - \cos(x) \leq |x|$ and \eqref{DHS1:ZPiX_inequality} to bound
\begin{align*}
\bigl\Vert \bigl( \cos\bigl( \frac r2\Pi_X\bigr) - 1\bigr) \Psi \alpha_*\bigr\Vert_2^2 &\leq  C\; \Vert |\cdot|\alpha_*\Vert_2^2 \; \Vert \Pi\Psi\Vert_2^2 \leq CB^2 \, \Vert \Psi\Vert_{\Hmag^1(Q_B)}^2,
\end{align*}
where the right side is finite by the decay properties of $\alpha_*$ in \eqref{DHS1:Decay_of_alphastar}. Using additionally \eqref{DHS1:PiXcos}, we also see that
\begin{align*}
\bigl\Vert \Pi_X \bigl( \cos\bigl( \frac r2\Pi_X\bigr) -1\bigr) \alpha_* \Psi \bigr\Vert_2^2 &\leq \bigl\Vert \bigl[ \cos\bigl(\frac r2\Pi_X\bigr) - 1 \bigr] \Pi_X \alpha_* \Psi \bigr\Vert_2^2 + \bigl\Vert \sin\bigl( \frac r2\Pi_X\bigr) \Bbold \wedge r \alpha_* \Psi \bigr\Vert_2^2\\
&\leq C \bigl(\Vert \Pi\Psi\Vert_2^2 + B^2 \Vert \Psi \Vert_2^2\bigr) \leq CB^3 \, \Vert \Psi\Vert_{\Hmag^1(Q_B)}^2
\end{align*}
holds. Finally, an application of \eqref{DHS1:pirtilde_cos_pir} and \eqref{DHS1:ZPiX_inequality} allows us to estimate
\begin{align*}
\bigl\Vert \pi_r \bigl( \cos\bigl( \frac r2\Pi_X\bigr) - 1\bigr) \Psi \alpha_*\bigr\Vert_2^2 &= \bigl\Vert \bigl[ \bigl( \cos\bigl(\frac r2\Pi_X\bigr) - 1\bigr) \tilde \pi_r + \frac \i 2 \sin\bigl( \frac r2\Pi_X\bigr)\Pi_X\bigr] \Psi \alpha_*\bigr\Vert_2^2 \\
&\leq C \bigl( \Vert \Pi\Psi\Vert_2^2 + B^2 \Vert \Psi \Vert_2^2\bigr) \leq CB^2 \, \Vert \Psi\Vert_{\Hmag^1(Q_B)}^2.
\end{align*}
This proves that $\xi$ obeys \eqref{DHS1:Second_Decomposition_Psi_xi_estimate} and ends the proof of Theorem \ref{DHS1:Structure_of_almost_minimizers}.


\section{The Lower Bound on \texorpdfstring{(\ref{DHS1:ENERGY_ASYMPTOTICS})}{(\ref{DHS1:ENERGY_ASYMPTOTICS})} and Proof of Theorem \ref{DHS1:Main_Result_Tc} (b)}
\label{DHS1:Lower Bound Part B}


\subsection{The BCS energy of low-energy states}

In this section, we provide the lower bound on \eqref{DHS1:ENERGY_ASYMPTOTICS} and the proof of Theorem~\ref{DHS1:Main_Result_Tc}~(b), and thereby complete the proof of Theorems \ref{DHS1:Main_Result} and \ref{DHS1:Main_Result_Tc}. Let $D_1\geq 0$ and $D\in \Rbb$ be given and assume that $\Gamma$ is a gauge-periodic state at temperature $T = \Tc(1 - DB)$ that satisfies \eqref{DHS1:Second_Decomposition_Gamma_Assumption}.
Corollary~\ref{DHS1:Structure_of_almost_minimizers_corollary} provides us with a decomposition of the Cooper pair wave function $\alpha = [\Gamma]_{12}$ in terms of $\Psi_\leq$ in \eqref{DHS1:PsileqPsi>_definition} and $\sigma$ in \eqref{DHS1:sigma}, where $\Vert \Psi_\leq \Vert_{\Hmag^1(Q_B)} \leq C$ and where the bound
\begin{align}
	\Vert \Psi_\leq\Vert_{\Hmag^2(Q_B)}^2 &\leq C \, \varepsilon B^{-1} \, \Vert \Psi\Vert_{\Hmag^1(Q_B)}^2 \label{DHS1:Lower_Bound_B_Psileq}
\end{align}
holds in terms of the function $\Psi$ in Theorem \ref{DHS1:Structure_of_almost_minimizers}. With the function $\Psi_\leq$ we construct a Gibbs state $\Gamma_{\Delta}$ with the gap function $\Delta \equiv \Delta_{\Psi_\leq}$ as in \eqref{DHS1:Delta_definition}. Using Proposition~\ref{DHS1:BCS functional_identity}, we write the BCS free energy of $\Gamma$ as
\begin{align*}
	\FBCS(\Gamma) - \FBCS(\Gamma_0) &= - \frac 14 \langle \Delta, L_{T,B} \Delta\rangle + \frac 18 \langle \Delta, N_{T,B} (\Delta)\rangle + \Vert \Psi_\leq \Vert_2^2 \;  \langle \alpha_*, V\alpha_*\rangle \\
	&\hspace{10pt}+ \Tr\bigl[\Rcal_{T,B}(\Delta)\bigr] + \frac T2 \Hcal_0(\Gamma, \Gamma_\Delta) - \fint_{Q_B} \dd X\int_{\Rbb^3} \dd r\; V(r) \, |\sigma(X,r)|^2,
\end{align*}
where
\begin{equation*}
	\Vert \Rcal_{T,B}(\Delta) \Vert_1 \leq C \; B^3 \; \Vert \Psi \Vert_{\Hmag^1(Q_B)}^6.
\end{equation*}
We also apply Theorem~\ref{DHS1:Calculation_of_the_GL-energy} to compute the terms in the first line on the right side, and find the lower bound
\begin{align}
		\FBCS(\Gamma) - \FBCS(\Gamma_0) &\geq B^2 \; \EGL(\Psi_\leq) -C \bigl( B^3 + \varepsilon B^2 \bigr) \Vert \Psi\Vert_{\Hmag^1(Q_B)}^2 \notag \\
		&\hspace{30pt}+ \frac T2 \Hcal_0(\Gamma, \Gamma_\Delta) - \fint_{Q_B} \dd X\int_{\Rbb^3} \dd r\; V(r) \, |\sigma(X,r)|^2. \label{DHS1:Lower_Bound_B_2}
\end{align}
The relative entropy is nonnegative and the last term on the right side is nonpositive. In the next section we show that their sum is negligible.

\subsection{Estimate on the relative entropy}

In this section we prove a lower bound for the second line in \eqref{DHS1:Lower_Bound_B_2}, showing that it is negligible. We start with the following lower bound for the relative entropy.

\begin{lem}
For all admissible BCS states $\Gamma$, we have
\begin{align}
T\Hcal_0(\Gamma, \Gamma_\Delta) &\geq \Tr \Bigl[ (\Gamma - \Gamma_\Delta) \frac{H_\Delta}{\tanh(\frac \beta 2H_\Delta)} (\Gamma - \Gamma_\Delta)\Bigr]. \label{DHS1:LBpartB_1}
\end{align}
\end{lem}

\begin{proof}
The proof is given in \cite[Lemma 5]{Hainzl2012} and uses the fact that $\Gamma_\Delta$ is admissible, which follows from Lemma \ref{DHS1:Gamma_Delta_admissible}.
\end{proof}

To be able to combine the term on the right side of \eqref{DHS1:LBpartB_1} and the last term on the right side of \eqref{DHS1:Lower_Bound_B_2}, we first need to replace the operator $H_{\Delta}$ in the second factor on the right side of \eqref{DHS1:LBpartB_1} by $H_0$. To that end, we note that the estimate $H_\Delta^2\geq (1 - \delta)H_0^2 - \delta^{-1} \Vert \Delta\Vert_\infty^2$ holds for $0 < \delta < 1$ and we rewrite it as
\begin{align}
H_0 \leq (1 - \delta)^{-1} \bigl( H_\Delta^2 + \delta^{-1} \Vert \Delta\Vert_\infty^2\bigr). \label{DHS1:LBpartB_4}
\end{align}
Furthermore, we note that the series expansion 
\begin{align*}
	\frac{x}{\tanh(\frac x2)} = 2+ \sum_{k=1}^\infty \bigl( 2 - \frac{8k^2\pi^2}{x^2 + 4k^2\pi^2}\bigr),
\end{align*}
see \cite[Eq. (5.14)]{Hainzl2012}, shows that the function $x \mapsto \frac{\sqrt{x}}{\tanh(\frac \beta 2\sqrt{x})}$ is operator monotone. We use this together with \eqref{DHS1:LBpartB_4} and the monotonicity of the map $x \mapsto \tanh(x)$, which yields
 \begin{align*}
	\begin{pmatrix}
		K_{T,B} & 0 \\ 0 & \ov{K_{T,B}}
	\end{pmatrix} = \frac{H_0}{\tanh(\frac \beta 2 H_0)}
&\leq (1 - \delta)^{-\nicefrac 12} \frac{\sqrt{H_\Delta^2 + \delta^{-1}\Vert \Delta\Vert_\infty^2}}{\tanh(\frac \beta 2 \sqrt{H_\Delta^2 + \delta^{-1} \Vert\Delta\Vert_\infty^2})}.
\end{align*}
When we apply a first order Taylor expansion on the right side, the above inequality can be written as
\begin{align*}
\begin{pmatrix}
K_{T,B} & 0 \\ 0 & \ov{K_{T,B}}
\end{pmatrix} 
&\leq (1 - \delta)^{-\nicefrac 12} \Bigl[\frac{H_\Delta}{\tanh(\frac \beta 2H_\Delta)} + \frac{\beta}{4} \int_0^{\delta^{-1} \Vert \Delta\Vert_\infty^2} \dd t \; g\bigl( \frac \beta 2 \sqrt{H_\Delta^2 + t}\bigr)\Bigr]
\end{align*}
with the nonnegative function 
\begin{align*}
g(x) &:= \frac{1}{x} \frac{1}{\tanh(x)} - \frac{1}{\tanh^2(x)} \frac{1}{\cosh^2(x)}.
\end{align*}
Using $\sup_{x \geq 0} g(x) \leq 1$ and $1 \leq \frac{x}{\tanh(x)}$, we conclude that
\begin{align*}
	\begin{pmatrix}
		K_{T,B} & 0 \\ 0 & \ov{K_{T,B}}
	\end{pmatrix} \leq (1 - \delta)^{-\nicefrac 12} \bigl( 1 +  \frac{\delta^{-1} \beta^2}{8} \;  \Vert \Delta\Vert_\infty^2\bigr) \frac{H_\Delta}{\tanh(\frac \beta 2 H_\Delta)}
\end{align*}
holds. We choose $\delta := \Vert \Delta\Vert_\infty$ and note that Lemma~\ref{DHS1:Schatten_estimate} and \eqref{DHS1:Magnetic_Sobolev} imply 
\begin{equation}
	\Vert \Delta\Vert_\infty \leq C \; B^{\nicefrac 14} \; \Vert \Psi \Vert_{\Hmag^1(Q_B)}.
	\label{DHS1:eq:ABDelta}
\end{equation} 
In particular, $\delta < 1 $ as long as $B>0$ is sufficiently small, and we have
\begin{align}
	\frac{H_\Delta}{\tanh(\frac \beta 2 H_\Delta)} \geq \frac{\sqrt{1 - \Vert \Delta\Vert_\infty}}{1 + \frac{\beta^2}{8} \Vert \Delta\Vert_\infty} \; \begin{pmatrix}
		K_{T,B} & 0 \\ 0 & \ov{K_{T,B}}
	\end{pmatrix} 
	\geq (1 - C\Vert \Delta\Vert_\infty) \; \begin{pmatrix}
		K_{T,B} & 0 \\ 0 & \ov{K_{T,B}}
	\end{pmatrix}.
	\label{DHS1:LBpartB_2}
\end{align}
In combination, \eqref{DHS1:LBpartB_1} and \eqref{DHS1:LBpartB_2} prove
\begin{align*}
\frac 12 \Tr \Bigl[(\Gamma - \Gamma_\Delta)\frac{H_\Delta}{\tanh(\frac \beta 2H_\Delta)} (\Gamma - \Gamma_\Delta)\Bigr] & \\
&\hspace{-160pt} \geq 
%
(1 - C\Vert \Delta\Vert_\infty) \langle \alpha - \alpha_\Delta , K_{T,B} (\alpha - \alpha_\Delta)\rangle + (1 - C\Vert \Delta\Vert_\infty) \Tr[ (\gamma - \gamma_\Delta) K_{T,B}(\gamma - \gamma_\Delta)],
\end{align*}
where we can drop the last term for a lower bound because it is nonnegative if $B$ is sufficiently small. This is the lower bound for the relative entropy of $\Gamma$ with respect to $\Gamma_{\Delta}$ we were looking for. 

It remains to combine the first term on the right side and the interaction term on the right side of \eqref{DHS1:Lower_Bound_B_2}.
%
Let us define the function $\eta := \alpha_* \Psi_\leq - \alpha_\Delta$. By Corollary~\ref{DHS1:Structure_of_almost_minimizers_corollary} we have $\alpha - \alpha_\Delta = \sigma + \eta$ as well as
\begin{align}
	&\frac T2 \Hcal_0(\Gamma, \Gamma_\Delta) - \fint_{Q_B} \dd X \int_{\Rbb^3} \dd r \; V(r)|\sigma(X, r)|^2 \notag\\
	&\hspace{20pt} \geq (1 - C\Vert \Delta\Vert_\infty) \langle \sigma + \eta, K_{T,B} (\sigma + \eta)\rangle - \langle \sigma , V\sigma\rangle \notag\\
	&\hspace{20pt} \geq (1 - C\Vert \Delta\Vert_\infty) \langle \sigma, (K_{T,B} - V) \sigma \rangle - C\Vert \Delta\Vert_\infty \Vert V\Vert_\infty \Vert \sigma \Vert_2^2 - 2 \, | \langle\eta, K_{T,B} \sigma\rangle|. \label{DHS1:Lower_Bound_2_intermediate}
\end{align}
From \eqref{DHS1:KTB_Lower_bound_5} we know that the lowest eigenvalue of $K_{T,B} - V$ is bounded from below by $-CB$. In combination with \eqref{DHS1:Second_Decomposition_sigma_estimate} and \eqref{DHS1:eq:ABDelta}, this implies that the first two terms on the right side of \eqref{DHS1:Lower_Bound_2_intermediate} are bounded from below by 
\begin{align}
	-C \varepsilon^{-1}B^{\nicefrac 94} \Vert \Psi\Vert_{\Hmag^1(Q_B)} \; \bigl(  \Vert \Psi\Vert_{\Hmag^1(Q_B)}^2 + D_1\bigr)^{\nicefrac 12}. \label{DHS1:Lower_Bound_B_3}
\end{align}
To estimate the last term on the right side of \eqref{DHS1:Lower_Bound_2_intermediate}, we use \eqref{DHS1:KTc_bounded_derivative} to replace $K_{T, B}$ by $ K_{\Tc, B}$, which yields the estimate
\begin{align*}
	|\langle \eta ,(K_{T,B} - K_{\Tc,B} )\sigma\rangle| &\leq 2D_0 B \, \Vert \sigma\Vert_2 \, \Vert \eta\Vert_2 \leq C \, B^3 \, \Vert \Psi\Vert_{\Hmag^1(Q_B)} \; \bigl(  \Vert \Psi\Vert_{\Hmag^1(Q_B)}^2 + D_1\bigr)^{\nicefrac 12}.
\end{align*}
To obtain this result we also used \eqref{DHS1:Second_Decomposition_sigma_estimate}, 
Proposition~\ref{DHS1:Structure_of_alphaDelta} and \eqref{DHS1:Lower_Bound_B_Psileq}. Next, we decompose $\eta = \eta_0 + \eta_\perp$ with $\eta_0(\Delta)$ and $\eta_{\perp}(\Delta)$ as in Proposition~\ref{DHS1:Structure_of_alphaDelta} and write
\begin{align}
	\langle \eta, K_{\Tc,B} \sigma \rangle &= \langle \eta_0 , K_{\Tc,B} \sigma\rangle + \langle \eta_\perp, K_{\Tc,B} (\sigma - \sigma_0)\rangle + \langle \eta_\perp , K_{\Tc,B} \sigma_0\rangle. \label{DHS1:Lower_Bound_B_1}
\end{align}
Using \eqref{DHS1:alphaDelta_decomposition_eq2} and \eqref{DHS1:Second_Decomposition_sigma_estimate}, we see that the first term on the right side of \eqref{DHS1:Lower_Bound_B_1} is bounded by
\begin{align}
	|\langle \eta_0, K_{\Tc,B} \sigma \rangle| &\leq \bigl\Vert \sqrt{K_{\Tc,B}}\, \eta_0\bigr\Vert_{2} \bigl\Vert \sqrt{K_{\Tc,B}}\,\sigma\bigr\Vert_{2} \notag \\
	&\leq C \varepsilon^{-\nicefrac 12} B^{\nicefrac 52} \, \Vert \Psi\Vert_{\Hmag^1(Q_B)} \; \bigl(  \Vert \Psi\Vert_{\Hmag^1(Q_B)}^2 + D_1\bigr)^{\nicefrac 12}.
	\label{DHS1:eq:A30}
\end{align}
We note that $\sigma - \sigma_0 = \xi$ and use \eqref{DHS1:alphaDelta_decomposition_eq3}, \eqref{DHS1:Second_Decomposition_Psi_xi_estimate}, and \eqref{DHS1:Lower_Bound_B_Psileq} to estimate
\begin{align}
	|\langle \eta_\perp, K_{\Tc,B} \xi \rangle| &\leq \bigl\Vert \sqrt{K_{\Tc,B}} \, \eta_\perp\bigr\Vert_{2} \bigl\Vert \sqrt{K_{\Tc,B}} \, \xi \bigr\Vert_{2} \notag \\
	&\leq C\varepsilon^{\nicefrac 12} B^2 \, \Vert \Psi\Vert_{\Hmag^1(Q_B)} \; \bigl(  \Vert \Psi\Vert_{\Hmag^1(Q_B)}^2 + D_1\bigr)^{\nicefrac 12}. \label{DHS1:eq:A31} 
\end{align}
It remains to estimate the last term on the right side of \eqref{DHS1:Lower_Bound_B_1}, which we write as 
\begin{align}
		\langle \eta_\perp, K_{\Tc,B} \sigma_0\rangle &= \langle \eta_\perp, K_{\Tc}^r  \sigma_0\rangle + \langle \eta_\perp, [K_{\Tc,B}^r - K_{\Tc}^r ] \sigma_0\rangle + \langle \eta_\perp, (U-1)K_{\Tc,B}^r \sigma_0\rangle \notag \\
		&\hspace{190pt}  + \langle \eta_\perp, UK_{\Tc,B}^r (U^* - 1) \sigma_0\rangle. \label{DHS1:LBpartB_3}
\end{align}
with the unitary operator $U$ in \eqref{DHS1:U_definition}. We recall that the operators $K_{\Tc,B}^r$ and $K_{\Tc}^r$ act on the relative coordinate $r = x-y$.

Since $\Delta(X,r) = - 2 V(r) \alpha_*(r) \Psi_{\leq}(X)$ and $\sigma_{0}(X,r) = \alpha_*(r) \Psi_>(X)$ we know from Proposition~\ref{DHS1:Structure_of_alphaDelta}~(c) that the first term on the right side of \eqref{DHS1:LBpartB_3} vanishes. A bound for the remaining terms is provided by the following lemma. Its proof will be given in Section~\ref{DHS1:sec:A1} below.
\begin{lem}
	\label{DHS1:Lower_Bound_B_remainings}
	We have the following estimates on the remainder terms of \eqref{DHS1:LBpartB_3}:
	\begin{enumerate}[(a)]
		\item $|\langle \eta_\perp, [K_{\Tc,B}^r - K_{\Tc}^r]\sigma_0\rangle| \leq C B^3 \, \Vert \Psi\Vert_{\Hmag^1(Q_B)}^2$,
		
		\item $|\langle \eta_\perp, (U - 1) K_{\Tc,B}^r  \sigma_0\rangle| \leq C \varepsilon^{\nicefrac 12} B^2 \, \Vert \Psi\Vert_{\Hmag^1(Q_B)}^2$,
		
		\item $|\langle \eta_\perp, UK_{\Tc,B}^r (U^* - 1)\sigma_0\rangle| \leq C \varepsilon^{\nicefrac 12} B^2 \, \Vert \Psi\Vert_{\Hmag^1(Q_B)}^2$.
	\end{enumerate}
\end{lem} 

Accordingly, we have
\begin{align*}
|\langle \sigma, K_{T,B} \eta \rangle | &\leq C\bigl( \varepsilon^{-\nicefrac 12} B^{\nicefrac 52} + \varepsilon^{\nicefrac 12} B^2 \bigr) \, \Vert \Psi\Vert_{\Hmag^1(Q_B)} \; \bigl(  \Vert \Psi\Vert_{\Hmag^1(Q_B)}^2 + D_1\bigr)^{\nicefrac 12}.
\end{align*}
We combine this with \eqref{DHS1:Lower_Bound_B_2}, \eqref{DHS1:Lower_Bound_2_intermediate}, and \eqref{DHS1:Lower_Bound_B_3} to see that
\begin{align}
\FBCS(\Gamma) - \FBCS(\Gamma_0) &  \geq B^2\; \EGL(\Psi_\leq) \notag\\
&\hspace{-70pt} - C\bigl( \varepsilon^{-\nicefrac 12} B^{\nicefrac 52} + \varepsilon^{\nicefrac 12} B^2 + \varepsilon^{-1} B^{\nicefrac 94} \bigr) \Vert \Psi\Vert_{\Hmag^1(Q_B)} \bigl( \Vert \Psi\Vert_{\Hmag^1(Q_B)} + D_1 \bigr)^{\nicefrac 12}.  \label{DHS1:Lower_Bound_B_8}
\end{align}
The optimal choice $\varepsilon = B^{\nicefrac 16}$ in \eqref{DHS1:Lower_Bound_B_8} yields
\begin{align}
	&\FBCS(\Gamma) - \FBCS(\Gamma_0) \notag \\
	&\hspace{2cm}\geq B^2 \, \bigl(\EGL(\Psi_\leq) - C \,  B^{\nicefrac {1}{12}} \Vert \Psi\Vert_{\Hmag^1(Q_B)} \; \bigl(  \Vert \Psi\Vert_{\Hmag^1(Q_B)}^2 + D_1\bigr)^{\nicefrac 12} \bigr). \label{DHS1:Lower_Bound_B_5}
\end{align}
This proves the lower bound for the BCS free energy in Theorem~\ref{DHS1:Main_Result}.


\subsection{Conclusion}

Using \eqref{DHS1:Lower_Bound_B_5}, we now finish the proofs of Theorem~\ref{DHS1:Main_Result} and Theorem~\ref{DHS1:Main_Result_Tc}, and we start with the former. Let $\Gamma$ be an approximate minimizer of the BCS functional, i.e., let \eqref{DHS1:Second_Decomposition_Gamma_Assumption} hold with
\begin{align}
D_1 := \EGLGSE + \rho \label{DHS1:Lower_Bound_B_4}
\end{align}
and $\rho\geq 0$. Since $\Vert \Psi\Vert_{\Hmag^1(Q_B)} \leq C$ by \eqref{DHS1:Second_Decomposition_Psi_xi_estimate}, \eqref{DHS1:Lower_Bound_B_5} implies
\begin{align*}
B^2 \bigl(\EGLGSE + \rho\bigr) \geq \FBCS(\Gamma) - \FBCS(\Gamma_0) \geq B^2 \bigl( \EGL(\Psi_\leq) - C\, B^{\nicefrac 1{12}}\bigr).
\end{align*}
This proves the claimed bound for the Cooper pair wave function of an approximate minimizer of the BCS functional in Theorem~\ref{DHS1:Main_Result}. 


We turn to the proof of Theorem~\ref{DHS1:Main_Result_Tc}. Let the temperature $T$ obey
\begin{align}
\Tc ( 1 - B (\Dc - D_0 B^{\nicefrac 1{12}})) < T \leq \Tc(1 + CB) \label{DHS1:TcB_Upper_Fine_1}
\end{align}
with $\Dc$ in \eqref{DHS1:Dc_Definition} and $D_0 > 0$. We claim that the normal state $\Gamma_0$ minimizes the BCS functional for such temperatures $T$ if $D_0$ is chosen sufficiently large. Since Corollary \ref{DHS1:TcB_First_Upper_Bound} takes care of the remaining temperature range, this implies part (b) of Theorem \ref{DHS1:Main_Result_Tc} and completes its proof.

To see that the above claim is true, we start with the lower bound in \eqref{DHS1:Lower_Bound_B_5} and assume that \eqref{DHS1:Second_Decomposition_Gamma_Assumption} holds with $D_1 = 0$. We drop the nonnegative quartic term in the Ginzburg--Landau functional for a lower bound and obtain
\begin{align*}
	\EGL(\Psi_\leq) &\geq B^{-2}  \langle \Psi_\leq , (\Lambda_0 \, \Pi^2 - DB \Lambda_2) \Psi_\leq\rangle \geq  \Lambda_2 \, ( \Dc - D) \; B^{-1}  \Vert \Psi_\leq \Vert_2^2, 
\end{align*}
with $\Lambda_0$ in \eqref{DHS1:GL-coefficient_1}, $\Lambda_2$ in \eqref{DHS1:GL-coefficient_2}, and with $D \in \mathbb{R}$ defined by $T = \Tc(1-D B)$. We combine \eqref{DHS1:Psi>_bound} and \eqref{DHS1:First_Decomposition_Result_Estimate} and estimate
\begin{align*}
\Vert \Psi_\leq \Vert_2 \geq \Vert \Psi\Vert_2 - \Vert \Psi_>\Vert_2 \geq c \; B^{\nicefrac 12} \; \Vert \Psi \Vert_{\Hmag^1(Q_B)} ( 1 - C \, B^{\nicefrac{5}{12}} ).
\label{DHS1:eq:A32}
\end{align*}
When we insert our findings in the lower bound for the BCS energy in \eqref{DHS1:Lower_Bound_B_5}, this gives
\begin{equation}
0\geq \FBCS(\Gamma) - \FBCS(\Gamma_0) \geq c \; B^2 \; \Vert \Psi \Vert_{\Hmag^1(Q_B)}^2 \bigl( (\Dc - D)  - CB^{\nicefrac {1}{12}}\bigr), \label{DHS1:eq:A33}
\end{equation}
We note that the lower bound in \eqref{DHS1:TcB_Upper_Fine_1} is equivalent to
\begin{align}
\Dc - D > D_0 B^{\nicefrac 1{12}}. \label{DHS1:TcB_Upper_Fine_2}
\end{align}
When we choose $D_0 > C$ with $C>0$ in \eqref{DHS1:eq:A33} and use \eqref{DHS1:TcB_Upper_Fine_2} to obtain a lower bound for the right side of \eqref{DHS1:eq:A33}, we conclude that $\Psi =0$. By \eqref{DHS1:Second_Decomposition_alpha_equation} and \eqref{DHS1:Second_Decomposition_Psi_xi_estimate}, this implies that $\alpha = 0$ whence $\Gamma$ is a diagonal state. Therefore, $\Gamma_0$ is the unique minimizer of $\FBCS$ if $T$ satisfies \eqref{DHS1:TcB_Upper_Fine_1} with our choice of $D_0$. As explained below \eqref{DHS1:TcB_Upper_Fine_1}, this proves Theorem \ref{DHS1:Main_Result_Tc}.


\subsection{Proof of Lemma~\ref{DHS1:Lower_Bound_B_remainings}}
\label{DHS1:sec:A1}

In this section we prove Lemma~\ref{DHS1:Lower_Bound_B_remainings}.
%
%
Our proof of part (a) uses a Cauchy integral representation for the operator $K_{\Tc,B}- (\pi^2 -\mu)$, which is provided in Lemma~\ref{DHS1:KT_integral_rep} below. Let us start by defining the contour for the Cauchy integral.

\begin{defn}[Speaker path]
	\label{DHS1:speaker path}
	Let $R>0$, assume that $\mu \geq -1$ and define the following complex paths 
	\begin{align*}
		\begin{split}
			u_1(t) &:= \frac{\pi\i}{2\betac} + (1 + \i)t\\
			u_2(t) &:= \frac{\pi\i}{2\betac} - (\mu + 1)t\\
			u_3(t) &:= -\frac{\pi\i}{2\betac}t - (\mu + 1)\\
			u_4(t) &:= -\frac{\pi\i}{2\betac} - (\mu + 1)(1-t) \\
			u_5(t) &:= -\frac{\pi\i }{2\betac} + (1 - \i)t
		\end{split}
		&
		\begin{split}
			\phantom{ \frac \i\betac }t&\in [0,R], \\
			\phantom{ \frac \i\betac }t &\in [0,1], \\
			\phantom{ \frac \i\betac }t&\in [-1,1],\\
			\phantom{ \frac \i\betac }t &\in [0,1],\\
			\phantom{ \frac \i \betac }t&\in [0,R].
		\end{split}
		& \begin{split} 
			\text{\includegraphics[width=6cm]{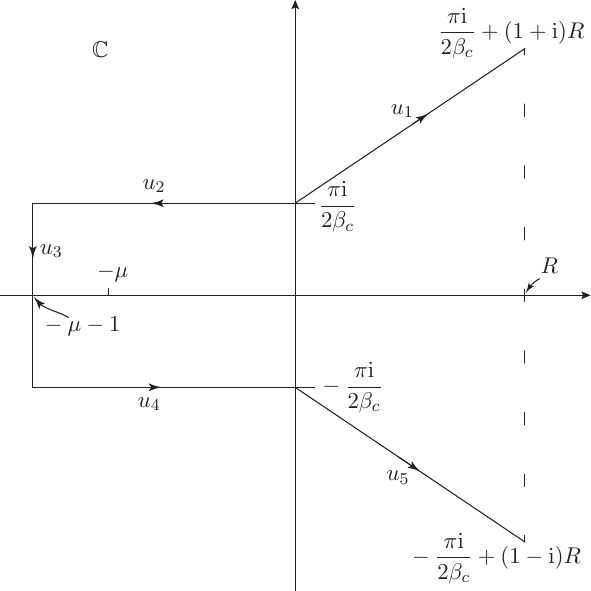}}
		\end{split} 
	\end{align*}
	The speaker path is defined as the union of paths $u_i$, $i=1, \ldots, 5$, with $u_1$ taken in reverse direction, i.e.,
	\begin{align*}
		\speaker_R := \mathop{\dot -}u_1 \mathop{\dot +} u_2 \mathop{\dot +} u_3 \mathop{\dot +} u_4 \mathop{\dot +} u_5.
	\end{align*}
	If $\mu < -1$ we choose the same path as in the case $\mu = -1$.
\end{defn} 
This path has the property that certain norms of the resolvent kernel of $\pi^2$ are uniformly bounded for $z \in \speaker_R$ and $R > 0$. More precisely, Lemma~\ref{DHS1:gB-g_decay} implies
\begin{align}
	\sup_{0\leq B\leq B_0} \sup_{R>0} \sup_{w\in \speaker_R} \bigl[ \left\Vert \, |\cdot |^a g_B^w\right\Vert_1 + \left \Vert \, |\cdot|^a \nabla g_B^w\right\Vert_1 \bigr] < \infty. \label{DHS1:g0_decay_along_speaker}
\end{align}
We could also choose a path parallel to the real axis in Lemma~\ref{DHS1:KT_integral_rep} below. In this case the above norms would depend on $R$. Although our analysis also works in this case, we decided to use the path $\speaker_R$ because of the more elegant bound in \eqref{DHS1:g0_decay_along_speaker}. 
With the above definition at hand, we are prepared to state the following lemma.

\begin{lem}
	\label{DHS1:KT_integral_rep}
	Let $H\colon \Dcal(H)\ra \Hcal$ be a self-adjoint operator on a separable Hilbert space~$\Hcal$ with $H\geq -\mu$ and let $\beta > 0$. Then, we have
	\begin{align}
		\frac{H}{\tanh(\frac{\beta}{2} H)} = H + \lim_{R\to\infty} \int_{\speaker_R} \frac{\dd z}{2\pi\i} \Bigl( \frac{z}{\tanh(\frac{\beta}{2} z)} - z \Bigr) \frac{1}{z - H}, \label{DHS1:KT_integral_rep_eq}
	\end{align}
	with the speaker path $\speaker_R$ in Definition~\ref{DHS1:speaker path}. The above integral including the limit is understood as an improper Riemann integral with respect to the uniform operator topology.  
\end{lem}

\begin{proof}
	We have that the function $f(z) = \frac{z}{\tanh(\frac{\beta }{2}z)} - z = \frac{2z}{\e^{\beta z} - 1}$ is analytic in the open domain $\Cbb \setminus 2\pi T \i \Zbb_{\neq 0}$. The construction of the Riemann integral over the path $\speaker_R$ with respect to the uniform operator topology is standard. The fact that the limit $R \to \infty$ exists in the same topology follows from the exponential decay of the function $f(z)$ along the speaker path. To check the equality in \eqref{DHS1:KT_integral_rep_eq}, we evaluate both sides in the inner product with two vectors in $\ran \Idbb_{(-\infty, K]}(H)$ for $K > 0$, use the functional calculus, the Cauchy integral formula, and the fact that $\bigcup_{K>0} \ran \Idbb_{(-\infty, K]}(H)$ is a dense subset of $\Hcal$. This proves the claim.
\end{proof}

Henceforth, we use the symbol $\int_{\speaker}$ to denote the integral on the right side of \eqref{DHS1:KT_integral_rep_eq} including the limit and we denote $\speaker = \bigcup_{R > 0} \speaker_R$.

\begin{proof}[Proof of Lemma \ref{DHS1:Lower_Bound_B_remainings}]
	We apply Cauchy-Schwarz to estimate
	\begin{align}
		|\langle \eta_\perp, (K_{\Tc,B}^r - K_{\Tc}^r)\sigma_0\rangle| \leq \Vert \eta_\perp\Vert_2 \, \Vert (K_{\Tc,B}^r- K_{\Tc}^r)\sigma_0\Vert_2
		\label{DHS1:eq:A34}
	\end{align}
	and claim that 
	\begin{align}
		\Vert [K_{{\Tc},B}^r - K_{\Tc}^r]\sigma_0\Vert_2 \leq C \varepsilon^{-\nicefrac 12} B^2 \, \Vert \Psi\Vert_{\Hmag^1(Q_B)} \label{DHS1:Lower_Bound_B_remainings_3}
	\end{align}
	holds. To see this, we apply Lemma~\ref{DHS1:KT_integral_rep} and write
	\begin{align}
		K_{{\Tc},B}^r - K_{\Tc}^r = \pi_r^2 - p_r^2 + \int_\speaker \frac{\dd w}{2\pi\i} \; f(w) \;  \frac{1}{w + \mu - \pi_r^2} [\pi_r^2 - p_r^2] \frac{1}{w + \mu - p_r^2}, \label{DHS1:Lower-bound-final_2}
	\end{align}
	where $\pi_r^2 - p_r^2 = \i \, \Bbold\wedge r\;  p_r + \frac 14 |\Bbold\wedge r|^2$. Using \eqref{DHS1:Psi>_bound} and \eqref{DHS1:Decay_of_alphastar}, we estimate the first term on the right side of \eqref{DHS1:Lower-bound-final_2} by
	\begin{align}
		\Vert [\pi_r^2 - p_r^2]\sigma_0\Vert_2 &\leq B \, \Vert \, |\cdot|\nabla \alpha_*\Vert_2 \Vert \Psi_>\Vert_2 + B^2 \Vert \,|\cdot|^2\alpha_*\Vert_2 \Vert \Psi_>\Vert_2 \notag\\
		&\leq C\varepsilon^{-\nicefrac 12} B^2 \, \Vert \Psi\Vert_{\Hmag^1(Q_B)}. \label{DHS1:Lower_Bound_B_remainings_1}
	\end{align}
	To estimate the second term in \eqref{DHS1:Lower-bound-final_2}, we use Hölder's inequality in \eqref{DHS1:Schatten-Hoelder} and find
	\begin{align*}
		\Bigl\Vert \int_\speaker \frac{\dd w}{2\pi\i} \, f(w)\;  \frac{1}{w + \mu - \pi_r^2} [\pi_r^2 - p_r^2]\frac{1}{w+\mu - p_r^2}\sigma_0\Bigr\Vert_2 &\\
		&\hspace{-180pt}\leq \int_\speaker \frac{\dd |w|}{2\pi}\,  |f(w)| \, \Bigl\Vert \frac{1}{w + \mu- \pi_r^2} \Bigr\Vert_\infty \Bigl\Vert [\pi_r^2 - p_r^2]\frac{1}{w + \mu - p_r^2}\sigma_0\Bigr\Vert_2,
	\end{align*}
	where $\dd |w| = \dd t \; |w'(t)|$. Eq.~\eqref{DHS1:g0_decay_along_speaker} implies that the operator norm of the magnetic resolvent is uniformly bounded for $w \in \speaker$. Since the function $f$ is exponentially decaying along the speaker path it suffices to prove a bound on the last factor that is uniform for $w \in \speaker$. We have
	\begin{align*}
		[\pi_r^2 - p_r^2] \frac{1}{w + \mu - p_r^2} \sigma_0(X,r) &= \int_{\Rbb^3} \dd s \; [\pi_r^2 - p_r^2] g_0^w(r - s) \alpha_*(s) \Psi_>(X),
	\end{align*}
	which implies
	\begin{align}
		\Bigl\Vert [\pi_r^2 - p_r^2] \frac{1}{w + \mu - p_r^2}\sigma_0\Bigr\Vert_2^2 &
		%
		\leq \Vert \Psi_>\Vert_2^2 \int_{\Rbb^3} \dd r \, \Bigl| \int_{\Rbb^3} \dd s \; |[\pi_r^2 - p_r^2] g_0^w(r - s) \alpha_*(s)|\Bigr|^2. \label{DHS1:Lower_Bound_B_remainings_2}
	\end{align}
	Moreover,
	\begin{align*}
		\int_{\Rbb^3} \dd r \; \Bigl| \int_{\Rbb^3} \dd s \; |[\pi_r^2 - p_r^2] g_0^w(r - s) \alpha_*(s)|\Bigr|^2 & \\
		&\hspace{-195pt} \leq CB^2 \bigl( \Vert \, |\cdot| \nabla g_0^w\Vert_1^2 \; \Vert \alpha_*\Vert_2^2 + \Vert \nabla g_0^w\Vert_1^2 \; \Vert \, |\cdot|\alpha_*\Vert_2^2 + \Vert \,  |\cdot|^2 g_0^w\Vert_1^2 \;  \Vert\alpha_*\Vert_2^2 + \Vert g_0^w\Vert_1^2 \;  \Vert \, |\cdot|^2\alpha_*\Vert_2^2\bigr).
	\end{align*}
	The right side is uniformly bounded for $w \in \speaker$ by \eqref{DHS1:g0_decay_along_speaker} and \eqref{DHS1:Decay_of_alphastar}. In combination with \eqref{DHS1:Psi>_bound} and \eqref{DHS1:Lower_Bound_B_remainings_2}, this implies
	\begin{align*}
		\Bigl\Vert [\pi_r^2 -p_r^2]\frac{1}{w + \mu - p_r^2} \sigma_0\Bigr\Vert_2^2 & \leq CB^2 \;  \Vert\Psi_>\Vert_2^2\leq C \varepsilon^{-1} B^4 \, \Vert \Psi\Vert_{\Hmag^1(Q_B)}^2.
	\end{align*}
	Using this and \eqref{DHS1:Lower_Bound_B_remainings_1}, we read off \eqref{DHS1:Lower_Bound_B_remainings_3}. Finally, we apply Proposition~\ref{DHS1:Structure_of_alphaDelta} to estimate $\Vert \eta_\perp\Vert_2$ in \eqref{DHS1:eq:A34}, which proves part (a).
	
	
	To prove part (b), we start by noting that
	\begin{align*}
		|\langle \eta_\perp, (U-1) K_{\Tc,B}^r\sigma_0 \rangle| &\leq \Vert \,  |r| \eta_\perp\Vert_2 \;  \Vert\, |r|^{-1} (U - 1) K_{\Tc,B}^r \, \sigma_0 \Vert_2. 
	\end{align*}
	A bound for the left factor on the right side is provided by Proposition~\ref{DHS1:Structure_of_alphaDelta}. To estimate the right factor, we use \eqref{DHS1:Psi>_bound}, \eqref{DHS1:Decay_of_alphastar} and the operator inequality in \eqref{DHS1:ZPiX_inequality}, which implies $|U- 1|^2 \leq 3 r^2 \Pi_X^2$, and find
	\begin{align*}
		\Vert\, |r|^{-1} (U - 1) K_{\Tc,B}^r \, \sigma_0 \Vert_2 &\leq C \Vert K_{\Tc,B}^r \alpha_*\Vert_2 \; \Vert \Pi\Psi_>\Vert_2 \leq CB \, \Vert \Psi\Vert_{\Hmag^1(Q_B)}.
	\end{align*}
	This proves part (b).
	
	For part (c), we estimate
	\begin{align}
		|\langle \eta_\perp, UK_{\Tc,B}^r (U^* - 1) \sigma_0\rangle|  &\leq \bigl\Vert \sqrt{K_{\Tc,B}^r} \, U^* \eta_\perp \bigr\Vert_2 \;  \bigl\Vert \sqrt{K_{\Tc,B}^r}\, (U^* - 1)\sigma_0\bigr\Vert_2
		\label{DHS1:eq:A35}
	\end{align}
	and note that $K_{\Tc,B}^r \leq C(1+ \pi_r^2)$ implies
	\begin{align}
		\bigl\Vert \sqrt{K_{\Tc,B}^r} (U^* - 1)\sigma_0\bigr\Vert_2^2 &= \langle \sigma_0, (U - 1) K_{\Tc,B}^r (U^* - 1)\sigma_0\rangle \notag\\
		&\leq C \Vert (U^* - 1)\sigma_0\Vert_2^2 + C\Vert \pi_r (U^* - 1)\sigma_0\Vert_2^2.\label{DHS1:Lower-bound-final_3}
	\end{align}
	Using the bound for $|U-1|^2$ in part (b), \eqref{DHS1:Psi>_bound} and \eqref{DHS1:Decay_of_alphastar}, we see that the first term is bounded by $C\Vert |r|\alpha_* \Pi_X\Psi_>\Vert^2 \leq CB^2$. Lemma~\ref{DHS1:CommutationI} allows us to write
	\begin{align*}
		\pi_r (U^* - 1) &= (U^* - 1) \tilde \pi_r + \frac 12 U^*\Pi_X - \frac 14 \Bbold \wedge r.
	\end{align*}
	Accordingly, we have
	\begin{align*}
		\Vert \pi_r (U^* - 1)\sigma_0\Vert_2^2 &\\
		&\hspace{-50pt}\leq C\bigl( \Vert \, |r| p_r\alpha_*\Pi_X\Psi_>\Vert_2^2 + B \, \Vert \, |r|^2\alpha_* \Pi_X \Psi_> \Vert_2^2 + \Vert \alpha_*\Pi_X\Psi_>\Vert_2^2 + B \Vert \, |r|\alpha_* \Psi_>\Vert_2^2\bigr)\\
		&\hspace{-50pt}\leq C \bigl( B^2 + \varepsilon^{-1} B^3 \bigr) \leq CB^2 \, \Vert \Psi\Vert_{\Hmag^1(Q_B)}^2.
	\end{align*}
	We conclude that the right side of \eqref{DHS1:Lower-bound-final_3} is bounded by $CB^2 \Vert \Psi\Vert_{\Hmag^1(Q_B)}^2$. 
	
	With $K_T(p) \leq C (1+p^2)$ we see that the first factor on the right side of \eqref{DHS1:eq:A35} is bounded by
	\begin{align*}
		\bigl\Vert \sqrt{K_{\Tc,B}^r} \, U^* \eta_\perp\bigr\Vert_2^2 &= \langle \eta_\perp , U K_{\Tc,B}^r U^* \eta_\perp\rangle \leq C \Vert \eta_\perp\Vert_2^2 + C\Vert \pi_rU^*\eta_\perp\Vert_2^2.
	\end{align*}
	From Lemma~\ref{DHS1:CommutationI} we know that $\pi_r U^* = U^* [\tilde\pi_r + \frac 12 \Pi_X]$, and hence
	\begin{align*}
		\bigl\Vert \sqrt{K_{\Tc,B}^r} \, U^* \eta_\perp\bigr\Vert_2^2 &\leq C \bigl( \Vert \eta_\perp\Vert_2^2 + \Vert \tilde \pi_r\eta_\perp\Vert_2^2 + \Vert \Pi_X\eta_\perp\Vert_2^2\bigr) \leq C \varepsilon B^2 \, \Vert \Psi\Vert_{\Hmag^1(Q_B)}^2. 
	\end{align*}
	This proves part (c) and ends the proof of the Lemma~\ref{DHS1:Lower_Bound_B_remainings}.
\end{proof}

\newpage
\appendix

\begin{center}
\huge \textsc{--- Appendix ---}
\end{center}


\section{Estimates on Eigenvalues and Eigenfunctions of \texorpdfstring{$K_{\Tc,B}-V$}{KTcB-V}}
\label{DHS1:KTV_Asymptotics_of_EV_and_EF_Section}

In this section, we investigate the low lying eigenvalues of $K_{\Tc,B} - V$ and its ground state wave function. Our analysis is carried out at $T = \Tc$ and we omit $\Tc$ from the notation throughout the appendix. The goal is to prove the following result.

\begin{prop}
\label{DHS1:KTV_Asymptotics_of_EV_and_EF}
Let Assumptions \ref{DHS1:Assumption_V} and \ref{DHS1:Assumption_KTc} hold. There is a constant $B_0>0$ such that for any $0 \leq B\leq B_0$ the following holds. Let $e_0^B$ and $e_1^B$ denote the lowest and next to lowest eigenvalue of $K_{\Tc,B} - V$. Then:
\begin{enumerate}[(a)]
\item $|e_0^B| \leq CB$,


\item $K_{\Tc,B}- V$ has a uniform spectral gap above $e_0^B$, i.e., $e_1^B - e_0^B \geq \kappa > 0$.

\item Let $\alpha_*$ be the eigenfunction in \eqref{DHS1:alpha_star_ev-equation} and let $\alpha_*^B$ be an eigenfunction corresponding to $e_0^B$ such that $\langle \alpha_*^B , V\alpha_*\rangle$ is real and nonnegative for all $0\leq B \leq B_0$. Then,
\begin{align*}
\Vert \alpha_*^B - \alpha_*\Vert_2 + \Vert \pi^2(\alpha_*^B - \alpha_*)\Vert_2 \leq CB.
\end{align*}

\item With $P_B := |\alpha_*^B\rangle \langle \alpha_*^B|$ and $P := |\alpha_*\rangle \langle \alpha_*|$ and with $\alpha_*^B$ and $\alpha_*$ as in part (c), we have 
\begin{align*}
\Vert P_B- P\Vert_\infty  + \Vert \pi^2(P_B - P)\Vert_\infty \leq CB.
\end{align*}
\end{enumerate}
\end{prop}

\begin{bem}
	We emphasize that this appendix is the only place in the paper where the assumption $V\geq 0$ is used. It simplifies our analysis because it implies that the Birman--Schwinger operator $V^{\nicefrac 12} [K_{\Tc,B} - e]^{-1} V^{\nicefrac 12}$ is self-adjoint. However, for the statement of Proposition \ref{DHS1:KTV_Asymptotics_of_EV_and_EF} to be true, it is not necessary that $V$ has a sign. In fact, with the help of a Combes--Thomas estimate for the resolvent kernel of $K_{\Tc} - V$ it is possible to show that Proposition~\ref{DHS1:KTV_Asymptotics_of_EV_and_EF} also holds for potentials $V$ without a definite sign. This approach requires more effort, and we therefore refrain from giving a general proof here. It can be found in \cite[Chapter 6]{Diss_Marcel}.
\end{bem}

Let us recall the decay properties of the eigenfunction $\alpha_*$ corresponding to the lowest eigenvalue of the operator $K_{\Tc}-V$. Since $\alpha_* = K_{\Tc}^{-1} V\alpha_*$ and $V\in L^\infty(\Rbb^3)$, we immediately have $\alpha_*\in H^2(\Rbb^3)$. Furthermore, for any $\nu \in \Nbb_0^3$, by \cite[Proposition 2]{Hainzl2012},
\begin{align}
\int_{\Rbb^3} \dd x \; \bigl[ |x^\nu \alpha_*(x)|^2 + |x^\nu \nabla \alpha_*(x)|^2 \bigr] < \infty. \label{DHS1:Decay_of_alphastar}
\end{align}
In fact, more regularity of $\alpha_*$ is known, see \cite[Appendix A]{Hainzl2012} but \eqref{DHS1:Decay_of_alphastar} is all we use in this paper. Before we give the proof of Proposition~\ref{DHS1:KTV_Asymptotics_of_EV_and_EF} in Section~\ref{DHS1:sec:new2} below, we prove two preparatory statements.

Let $e\in (-\infty, 2\Tc)$ and denote the kernel of the resolvent $(e - K_{\Tc})^{-1}$ by $\Gcal^e(x-y)$.

\begin{lem}
\label{DHS1:Gcal estimates}
For $e \in (-\infty, 2\Tc)$ and $k\in \Nbb_0$ the functions $|\cdot|^k\Gcal^e$ and $|\cdot|^k \nabla \Gcal^e$ belong to $L^1(\Rbb^3)$.
\end{lem}

\begin{proof}
The function $(e - K_{\Tc}(p))^{-1}$ and its derivatives belong to $L^2(\Rbb^3)$.
Therefore, we have
\begin{align}
\Vert \, |\cdot|^k \Gcal^e \Vert_1 &\leq \Bigl( \int_{\Rbb^3} \dx \; \bigl| \frac{|x|^k}{1 + |x|^{k+2}}\bigr|^2 \Bigr)^{\nicefrac 12} \; \Vert (1 + |\cdot |^{k+2})\Gcal^e\Vert_2 < \infty. \label{DHS1:Gcal_estimates_5}
\end{align}
This proves the first claim and the second follows from a similar argument.
\end{proof}

\subsection[Phase approximation]{Phase approximation for $K_{\Tc,B}$}

\begin{prop}
\label{DHS1:Resolvent_perturbation}
Let $V$ and $|\cdot|^2V$ belong to $L^\infty(\Rbb^3)$. There is $B_0>0$ such that for $0 \leq B\leq B_0$ and $e\in (-\infty, 2\Tc)$, we have
\begin{align}
\Bigl\Vert \Bigl[ \frac{1}{e - K_{\Tc,B}} - \frac{1}{e - K_{\Tc}}\Bigr] V^{\nicefrac 12} \Bigr\Vert_\infty + \Bigl\Vert \pi^2 \Bigl[ \frac{1}{e - K_{\Tc,B}} - \frac{1}{e - K_{\Tc}}\Bigr] V \Bigr\Vert_\infty \leq C_e B. \label{DHS1:Resolvent_perturbation_2}
\end{align}
\end{prop}

\begin{proof}
To prove this result, we apply a phase approximation to the operator $K_{\Tc,B}$. We pursue the strategy that we used in the proof of Lemma \ref{DHS1:gB-g_decay} and define
\begin{align}
\Scal_{B}^{e}(x,y) := \e^{\i \frac{\Bbold}{2}\cdot (x\wedge y)} \; \Gcal^{e}(x-y). \label{DHS1:Scalz_definition}
\end{align}
Let $\Scal_{B}^e$ be the operator defined by the kernel $\Scal_{B}^e(x,y)$. We claim that
\begin{align}
(e - K_{\Tc,B} )\Scal_{B}^e = \Idbb - \Tcal_{B}^{e} \label{DHS1:Interplay SAzTAz}
\end{align}
with the operator $\Tcal_{B}^e$ defined by the kernel
\begin{align}
\Tcal_{B}^e (x,y) := \e^{\i \frac \Bbold 2 \cdot (x\wedge y)} \bigl[ (K_{\Tc}(\pi_{x,y}) - K_{\Tc}) \frac{1}{e - K_{\Tc}}\bigr] (x,y) \label{DHS1:Resolvent_KT_phase_approximation}
\end{align}
and $\pi_{x,y} = -\i\nabla_x + \Abold (x-y)$. To prove \eqref{DHS1:Interplay SAzTAz}, it is sufficient to note that
\begin{align*}
K_{\Tc,B} \; \e^{\i \frac \Bbold 2 \cdot (x\wedge y)} = \e^{\i \frac \Bbold 2 \cdot (x\wedge y)} \; K_{\Tc}(\pi_{x,y}). 
\end{align*}
Using Lemma~\ref{DHS1:KT_integral_rep} and Lemma~\ref{DHS1:Gcal estimates}, a straightforward computation shows that
\begin{align}
\Vert \Tcal_B^e\Vert_\infty \leq C_e\, B 
\label{DHS1:TcalTB_bound}
\end{align}
holds for $B$ small enough.

With the operator $\Scal_B^e$ in \eqref{DHS1:Scalz_definition} we write
\begin{align}
\frac{1}{e - K_{\Tc,B}} - \frac{1}{e - K_{\Tc}} = \frac{1}{e - K_{\Tc,B}} - \Scal_B^e + \Scal_B^e - \frac{1}{e - K_{\Tc}} \label{DHS1:Resolvent_perturbation_1}
\end{align}
For the first term \eqref{DHS1:Interplay SAzTAz} implies
\begin{align*}
\frac{1}{e - K_{\Tc,B}} - \Scal_B^e = \Scal_B^e \;  \sum_{n=1}^\infty (\Tcal_B^e)^n
\end{align*}
and since $\Scal_B^e$ is a bounded operator with norm bounded by $\Vert \Gcal^e\Vert_1$, \eqref{DHS1:TcalTB_bound} yields
\begin{align*}
\Bigl\Vert \frac{1}{e - K_{\Tc,B}} - \Scal_B^e\Bigr\Vert_\infty \leq C_e   B.
\end{align*}
To estimate the second term on the right side of \eqref{DHS1:Resolvent_perturbation_1}, we use $|\e^{\i \frac \Bbold 2 \cdot (x\wedge y)} - 1|\leq B |x - y| |y|$ and bound the kernel of this term by
\begin{align*}
\bigl| \bigl[ \Scal_B^e - \frac{1}{e - K_{\Tc}}\bigr] V^{\nicefrac 12}(x,y)\bigr| &\leq B \;  |x-y| |\Gcal^e(x-y)| \; |y||V^{\nicefrac 12} (y)|.
\end{align*}
We further bound $|y||V^{\nicefrac 12}(y)| \leq \Vert \,|\cdot|^2V\Vert_\infty^{\nicefrac 12}$, which shows that
\begin{align*}
\Bigl\Vert \Bigl[\Scal_B^e - \frac{1}{e - K_{\Tc}}\Bigr] V^{\nicefrac 12}\Bigr\Vert_\infty &\leq B\; \Vert \, |\cdot|^2V\Vert_\infty^{\nicefrac 12} \; \Vert \, |\cdot|\Gcal^e\Vert_1 \leq C_e  B.
\end{align*}
This completes the proof of the first estimate in \eqref{DHS1:Resolvent_perturbation_2}. 

To prove the second estimate, we note that
\begin{align*}
\pi^2 \Bigl[ \frac{1}{e - K_{\Tc,B}} - \frac{1}{e - K_{\Tc}} \Bigr] V = \pi^2 \frac{1}{e - K_{\Tc,B}} [ K_{\Tc,B} - K_{\Tc}] \frac{1}{e - K_{\Tc}} V^{\nicefrac 12}.
\end{align*}
Since $\pi^2(e - K_{\Tc,B})^{-1}$ is a bounded function of $\pi^2$, we know that the operator norm of the operator in the above equation is uniformly bounded in $B$. Thus, it suffices to show that $[K_{\Tc,B} - K_{\Tc}] \frac{1}{e - K_{\Tc}} V$ satisfies the claimed operator norm bound. To this end, we use \eqref{DHS1:Lower-bound-final_2} and obtain two terms. Since $\pi^2 - p^2 = \Bbold \wedge x \cdot p + \frac 14|\Bbold \wedge x|^2$, the estimate for the first term reads
\begin{align*}
\bigl[ (\pi^2- p^2) \frac{1}{e - K_{\Tc}} V^{\nicefrac 12}\bigr](x,y) \leq \bigl[ B \cdot |x| |\nabla \Gcal^e(x-y)| + B^2 \; |x|^2 |\Gcal^e(x-y)| \bigr] |V(y)|.
\end{align*}
The $L^1(\Rbb^3)$-norm in $x-y$ of the right side is bounded by
\begin{align*}
C B\, \bigl[\bigl( \Vert \, |\cdot| \nabla \Gcal^e\Vert_1  + \Vert \, |\cdot|^2\Gcal^e\Vert_1 \bigr) \Vert V\Vert_\infty + \Vert \nabla  \Gcal^e\Vert_1 \, \Vert \, |\cdot| V\Vert_\infty + \Vert \Gcal^e\Vert_1 \, \Vert \, |\cdot|^2 V\Vert_\infty\bigr],
\end{align*}
which is finite by Lemma \ref{DHS1:Gcal estimates}. With the help of \eqref{DHS1:g0_decay_along_speaker} the remaining term can bounded similarly. This proves the claim.
\end{proof}


\subsection[Asymptotics]{Asymptotics for eigenvalues and eigenfunctions}
\label{DHS1:sec:new2}

We are now prepared to give the proof of Proposition~\ref{DHS1:KTV_Asymptotics_of_EV_and_EF}.

\begin{proof}[Proof of Proposition \ref{DHS1:KTV_Asymptotics_of_EV_and_EF}]
We start with the upper bound of part (a). By the variational principle for $e_0^B$ we have
\begin{align}
e_0^B \leq \langle \alpha_*, (K_{\Tc,B} - V)\alpha_*\rangle = \langle \alpha_* , (K_{\Tc} - V)\alpha_*\rangle + \langle \alpha_* , (K_{\Tc,B} - K_{\Tc})\alpha_*\rangle, \label{DHS1:KTV_Asymptotics_1}
\end{align}
where the first term on the right side equals $0$ by the definition of $\alpha_*$. We use \eqref{DHS1:g0_decay_along_speaker}, Lemma \ref{DHS1:KT_integral_rep}, and \eqref{DHS1:Decay_of_alphastar}, and argue as in the proof of \eqref{DHS1:Lower_Bound_B_remainings_1} to see that the second term is bounded by $CB$.

In the next step, we show the lower bound of part (a) and part (b) at the same time. Thus, for $n = 0,1$ we aim to show
\begin{align}
e_n^B \geq e_n^0 - C_nB \label{DHS1:KTV_Asymptotics_10}
\end{align}
for the lowest and next-to-lowest eigenvalue $e_0^B$ and $e_1^B$, respectively. For notational convenience, we give the proof for general $n\in \Nbb_0$ and we order the eigenvalues $e_n^B$ increasingly. Let $\alpha_n^B$ be the eigenfunction to $e_n^B$ for $n\geq 0$ and note that $\alpha_*^B = \alpha_0^B$. 

Now, we switch to the Birman--Schwinger picture: $e_n^B$ being the $(n+1)$-st to smallest eigenvalue of $K_{\Tc,B} - V$ is equivalent to 1 being the $(n+1)$-st to largest eigenvalue of the Birman--Schwinger operator $V^{\nicefrac 12} (K_{\Tc,B} - e_n^B)^{-1} V^{\nicefrac 12}$ corresponding to $e_n^B$. Accordingly, the min-max principle, see, e.g., \cite[Theorem 12.1 (5)]{LiebLoss}, implies
\begin{align}
1 = \max_{\substack{u_0, \ldots, u_n\in L^2(\Rbb^3) \\ u_i \perp u_j, \; i\neq j}} \min\Bigl\{ \Bigl\langle \Phi , V^{\nicefrac 12} \frac{1}{K_{\Tc,B} - e_n^B} V^{\nicefrac 12} \Phi\Bigr\rangle : \Phi \in \mathrm{span} \{u_0, \ldots, u_n\}, \; \Vert \Phi\Vert_2 =1\Bigr\}. \label{DHS1:KTV_Asymptotics_5}
\end{align}
We obtain a lower bound on \eqref{DHS1:KTV_Asymptotics_5} by choosing the functions $u_i$, $i=0, \ldots, n$ as the first $n+1$ orthonormal eigenfunctions $\varphi_i^B$ satisfying
\begin{align}
	V^{\nicefrac 12} \frac{1}{K_{\Tc,B} - e_n^B} V^{\nicefrac 12} \varphi_i^B = \eta_i^B \, \varphi_i^B, \qquad \qquad i = 0, \ldots, n, \label{DHS1:KTV_Asymptotics_8}
\end{align}
where $\eta_i^B \geq 1$ denote the first $n$ eigenvalues of the Birman--Schwinger operator in \eqref{DHS1:KTV_Asymptotics_8} ordered decreasingly. In particular, we have $\eta_n^B = 1$, as well as the relations $\varphi_n^B = V^{\nicefrac 12} \alpha_n^B$ and $\alpha_n^B = (K_{\Tc,B} - e_n^B)^{-1} V^{\nicefrac 12} \varphi_n^B$.


Denote $e_n := e_n^0$ and apply the resolvent equations to write
\begin{align}
V^{\nicefrac 12}\frac{1}{K_{\Tc,B} - e_n^B}V^{\nicefrac 12} &= V^{\nicefrac 12}\frac{1}{K_{\Tc} - e_n}V^{\nicefrac 12} + (e_n^B - e_n) \, \Qcal_n^B + \Rcal_n^B \label{DHS1:KTV_Asymptotics_4}
\end{align}
with
\begin{align*}
\Qcal_n^B &:= V^{\nicefrac 12}\frac{1}{K_{\Tc,B} - e_n} \,  \frac{1}{K_{\Tc,B} - e_n^B}V^{\nicefrac 12}, &
\Rcal_n^B &:= V^{\nicefrac 12}\Bigl[\frac{1}{K_{\Tc,B} - e_n} - \frac{1}{K_{\Tc}-e_n}\Bigr]V^{\nicefrac 12}.
\end{align*}
By Proposition \ref{DHS1:Resolvent_perturbation}, we have $\Vert \Rcal_n^B\Vert_\infty \leq C_n B$. Furthermore, we may assume without loss of generality that $e_n^B \leq e_n$, because otherwise there is nothing to prove. We combine this with \eqref{DHS1:KTV_Asymptotics_5} for $B =0$ and \eqref{DHS1:KTV_Asymptotics_4}, which yields
\begin{align}
1 
&\geq \min\Bigl\{ \Bigl\langle \Phi , V^{\nicefrac 12} \frac{1}{K_{\Tc,B} - e_n^B} V^{\nicefrac 12} \Phi\Bigr\rangle : \Phi \in \mathrm{span} \{\varphi_0^B, \ldots, \varphi_n^B\}, \; \Vert \Phi\Vert_2 =1\Bigr\} \notag\\
&\hspace{30pt}- [e_n^B - e_n] \min \bigl\{ \langle \Phi, \Qcal_n^B\Phi\rangle : \Phi \in \mathrm{span} \{\varphi_0^B, \ldots,  \varphi_n^B\}, \; \Vert \Phi\Vert_2 =1\bigr\} - C_nB.
\label{DHS1:KTV_Asymptotics_2}
\end{align}
We observe that the first term on the right side equals 1. To be able to conclude, we therefore need to show that there is a constant $c>0$, independent of $B$, such that
\begin{align}
\min \bigl\{ \langle \Phi, \Qcal_n^B\Phi\rangle : \Phi \in \mathrm{span} \{\varphi_0^B, \ldots, \varphi_n^B\}\bigr\} \geq c. \label{DHS1:KTV_Asymptotics_3}
\end{align}
Then, \eqref{DHS1:KTV_Asymptotics_2} implies $-[e_n^B-e_n] \leq C_nB$, which proves \eqref{DHS1:KTV_Asymptotics_10}.

We will prove that \eqref{DHS1:KTV_Asymptotics_3} holds with $c = \Vert V\Vert_\infty^{-1}$. To that end, we write
\begin{align*}
\langle \Phi, \Qcal_n^B \Phi\rangle &= \Bigl\langle \Phi, V^{\nicefrac 12} \frac{1}{K_{\Tc,B}- e_n^B} (K_{\Tc,B} - e_n^B) \frac{1}{K_{\Tc,B}- e_n} \frac{1}{K_{\Tc,B} - e_n^B} V^{\nicefrac 12} \Phi \Bigr\rangle,
\end{align*}
apply $-e_n^B \geq -e_n$, and infer
\begin{align*}
\langle \Phi , \Qcal_n^B\Phi\rangle &\geq \Bigl\Vert \frac{1}{K_{\Tc,B} - e_n^B} V^{\nicefrac 12} \Phi\Bigr\Vert_2^2. 
\end{align*}
Since $\Phi\in \mathrm{span}\{\varphi_0^B, \ldots, \varphi_n^B\}$, there are coefficients $c_i^B\in \Cbb$, $i = 0, \ldots, n$ such that we have $\Phi = \sum_{i=0}^n c_i^B \varphi_i^B$. We use the eigenvalue equation in \eqref{DHS1:KTV_Asymptotics_8} for $\varphi_i^B$ as well as $\langle \varphi_i^B, \varphi_j^B \rangle = \delta_{i,j}$ to see that
\begin{align*}
\Vert V\Vert_\infty \Bigl\Vert \frac{1}{K_{\Tc,B} - e_n^B} V^{\nicefrac 12} \Phi\Bigr\Vert_2^2 &\geq \Bigl\langle \frac{1}{K_{\Tc,B} - e_n^B} V^{\nicefrac 12} \Phi, V \frac{1}{K_{\Tc,B} - e_n^B} V^{\nicefrac 12} \Phi \Bigr\rangle \\
&= \sum_{i,j=0}^n \ov{c_i^B} c_j^B \, \eta_i^B\eta_j^B\, \langle \varphi_i^B, \varphi_j^B\rangle = \sum_{i=0}^n |c_i^B|^2 \; (\eta_i^B)^2 \geq 1.
\end{align*}
Here, we used that $\eta_i \geq 1$ and $\Vert \Phi\Vert_2 =1$.  This shows \eqref{DHS1:KTV_Asymptotics_3} and completes the proof of \eqref{DHS1:KTV_Asymptotics_10}.

The case $n =0$ in \eqref{DHS1:KTV_Asymptotics_10} yields the lower bound of part (a), since $e_0 = 0$. For $n = 1$, we have $e_1 = \kappa$, showing part (b) with the help of part (a).

To prove part (c), we write $\varphi_0 := \varphi_0^0$. The chosen phase of $\varphi_0^B$ in the lemma is such that $\langle \varphi_0, \varphi_0^B\rangle$ is real and nonnegative. We write $\varphi_0^B = a_B \varphi_0 + b_B \Phi$ with $\langle \Phi, \varphi_0\rangle =0$, $\Vert \Phi\Vert_2 = 1$ and $|a_B|^2 + |b_B|^2 = 1$. By construction, $a_B = \langle \varphi_0, \varphi_0^B\rangle$. Furthermore, by \eqref{DHS1:KTV_Asymptotics_4},
\begin{align}
1 = \bigl\langle \varphi_0^B , V^{\nicefrac 12} \frac{1}{K_{\Tc,B} - e_0^B} V^{\nicefrac 12} \varphi_0^B\bigr\rangle = \bigl\langle \varphi_0^B , V^{\nicefrac 12} \frac{1}{K_{\Tc}}  V^{\nicefrac 12} \varphi_0^B\bigr\rangle + \langle \varphi_0^B,\Tcal_B\varphi_0^B\rangle \label{DHS1:KTV_Asymptotics_9}
\end{align}
with $\Tcal_B := e_0^B\, Q_0^B + \Rcal_0^B$. Thus, 
\begin{align*}
1 \leq a_B^2 + |b_B|^2 \bigl\langle \Phi, V^{\nicefrac 12} \frac{1}{K_{\Tc}} V^{\nicefrac 12} \Phi\bigr\rangle + 2a_B \Re \bigl[ b_B \bigl\langle \varphi_0, V^{\nicefrac 12} \frac{1}{K_{\Tc}} V^{\nicefrac 12} \Phi\bigr\rangle \bigr] + \Vert \Tcal_B\Vert_\infty.
\end{align*}
By parts (a) and (b) and Proposition \ref{DHS1:Resolvent_perturbation}, we know that $\Vert \Tcal_B\Vert_\infty \leq CB$. Furthermore, the term in square brackets vanishes, since $V^{\nicefrac 12} K_{\Tc}^{-1} V^{\nicefrac 12} \varphi_0 = \varphi_0$ and $\langle \varphi_0, \Phi\rangle =0$. Using the orthogonality of $\Phi$ and $\varphi_0$ once more as well as the fact that 1 is the largest eigenvalue of $V^{\nicefrac 12} K_{\Tc}^{-1} V^{\nicefrac 12}$, we see that there is an $\eta < 1$ such that $\langle \Phi, V^{\nicefrac 12} K_{\Tc}^{-1} V^{\nicefrac 12} \Phi\rangle \leq \eta$. It follows that
\begin{align*}
1 \leq a_B^2 + |b_B|^2 \eta + CB.
\end{align*}
Since $a_B^2 + |b_B|^2 = 1$, this implies $|b_B|^2 \leq CB$ as well as $a_B^2 \geq 1 - CB$. Since $a_B \geq 0$, we infer $1 - a_B \leq CB$. 

The next step is to improve the estimate on $b_B$ to $|b_B|\leq CB$. To this end, we combine the two eigenvalue equations of $\varphi_0^B$ and $\varphi_0$. With $\Tcal_B$ as in \eqref{DHS1:KTV_Asymptotics_9}, we find
\begin{align*}
\varphi_0^B - \varphi_0 &= V^{\nicefrac 12} \frac{1}{K_{\Tc}} V^{\nicefrac 12} (\varphi_0^B - \varphi_0) + \Tcal_B \varphi_0^B.
\end{align*}
Testing this against $\Phi$, we obtain
\begin{align*}
b_B = \langle \Phi, \varphi_0^B - \varphi_0 \rangle = \bigl\langle V^{\nicefrac 12} \frac{1}{K_{\Tc}} V^{\nicefrac 12} \Phi, \varphi_0^B - \varphi_0\bigr\rangle + \langle \Phi, \Tcal_B \varphi_0\rangle.
\end{align*}
We apply Cauchy-Schwarz on the right side and use that $\Vert V^{\nicefrac 12} K_{\Tc}^{-1} V^{\nicefrac 12} \Phi\Vert_2 \leq\eta$, which follows from the orthogonality of $\Phi$ and $\varphi_0$. We also use $\Vert \varphi_0^B - \varphi_0\Vert_2 \leq (1 - a_B) + |b_B|$. This implies
\begin{align*}
|b_B| \leq \eta \, (1 - a_B) + \eta \, |b_B| + CB,
\end{align*}
from which we conclude that $|b_B| \leq CB$.

It remains to use these findings to prove the claimed bounds for $\alpha_*^B - \alpha_*$. According to the Birman--Schwinger correspondence, we have $\alpha_*^B = (K_{\Tc,B} - e_0^B)^{-1} V^{\nicefrac 12} \varphi_0^B$. Thus, since $\varphi_0 = V^{\nicefrac 12} \alpha_*$,
\begin{align}
\alpha_*^B - \alpha_* &= \frac{1}{K_{\Tc,B} - e_0^B} V^{\nicefrac 12} \bigl[ a_B\varphi_0 + b_B \Phi\bigr] - \frac{1}{K_{\Tc}} V^{\nicefrac 12} \varphi_0 \notag\\
&= a_B\Bigl[ \frac{1}{K_{\Tc,B} - e_0^B} - \frac{1}{K_{\Tc}}\Bigr] V \alpha_* + b_B \frac{1}{K_{\Tc,B}-e_0^B} V^{\nicefrac 12} \Phi - (1 - a_B) \frac{1}{K_{\Tc}} V \alpha_*. \label{DHS1:KTV_Asymptotics_6}
\end{align}
The proof of the norm estimates for $\alpha_*^B - \alpha$ and $\pi^2(\alpha_*^B - \alpha_*)$ is obtained from Lemma \ref{DHS1:Gcal estimates}, Proposition \ref{DHS1:Resolvent_perturbation}, and the estimates of part (a) on $e_0^B$.

Part (d) follows from part (c). This ends the proof of Proposition~\ref{DHS1:KTV_Asymptotics_of_EV_and_EF}.
\end{proof}

\begin{center}
\textsc{Acknowledgements}
\end{center}

A. D. gratefully acknowledges funding from the European Union’s Horizon 2020 research and innovation programme under the Marie Sklodowska-Curie grant agreement No 836146 and from the Swiss National Science Foundation through the Ambizione grant PZ00P2 185851. It is a pleasure for A. D. to thank Stefan Teufel and the Institute of Mathematics at the University of T\"ubingen for its warm hospitality.

\printbibliography[heading=bibliography]


\vspace{1cm}

\setlength{\parindent}{0em}

(Andreas Deuchert) \textsc{Institut für Mathematik, Universität Zürich}

\textsc{Winterthurerstrasse 190, CH-8057 Zürich}

E-mail address: \href{mailto:  andreas.deuchert@math.uzh.ch}{\texttt{andreas.deuchert@math.uzh.ch}}

\vspace{0.3cm}

(Christian Hainzl) \textsc{Mathematisches Institut der Universität München}

\textsc{Theresienstr. 39, D-80333 München}

E-mail address: \href{mailto: hainzl@math.lmu.de}{\texttt{hainzl@math.lmu.de}}

\vspace{0.3cm}

(Marcel Maier, born Schaub) \textsc{Mathematisches Institut der Universität München}

\textsc{Theresienstr. 39, D-80333 München}


\setlength{\parindent}{17pt}
\end{document}